\documentclass[10pt,twoside,a4paper]{mythesis}
\usepackage{epsf}
\makeatletter
 \def\Tr{\mathop{\operator@font Tr}\nolimits}
\makeatother

\def\bea{\begin{eqnarray}}
\def\eea{\end{eqnarray}}
\def\nonu{\nonumber}
\def\[{\left[}
\def\]{\right]}
\def\({\left(}
\def\){\right)}

\def\gm{\gamma^{\mu}}
\def\gn{\gamma^{\nu}}

\def\g5{\gamma_{5}}

\def\GS{{\Gamma_{\rm S}}}
\def\GP{{\Gamma_{\rm P}}}

\def\DelP{\Delta_{\rm P}}
\def\DelS{\Delta_{\rm S}}
\def\Fs{F^{({\rm s})}}
\def\Fp{F^{({\rm p})}}
\def\fslash{\hat}
\def\pslash{\fslash p}
\def\kslash{\fslash k}

\def\qslash{\fslash q}

\def\wslash{\fslash w}
\def\dslash{\fslash \partial}

\def\fermionprop{\epsfxsize=3cm \epsffile[-15 -5 130 -5]{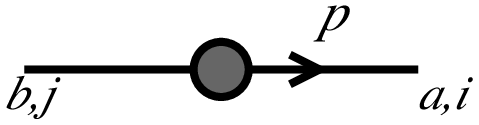}}
\def\barefermionprop{\epsfxsize=3cm \epsffile[-15 -5 130 -5]{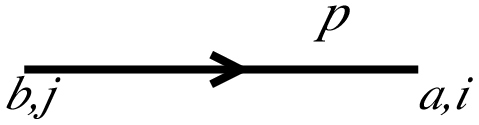}}
\def\photonprop{\epsfxsize=3cm \epsffile[-15 -5 130 -5]{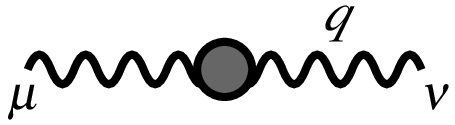}}
\def\barephotonprop{\epsfxsize=3cm \epsffile[-15 -5 130 -5]{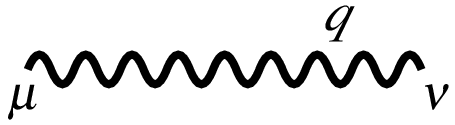}}
\def\scalarprop{\epsfxsize=3cm \epsffile[-15 -5 130 -5]{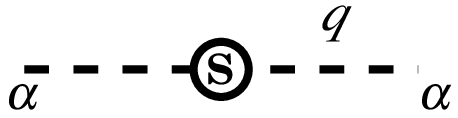}}
\def\bosonprop{\epsfxsize=3cm \epsffile[-15 -5 130 -5]{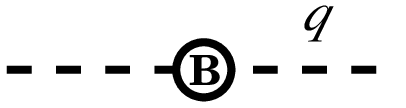}}
\def\barescalarprop{\epsfxsize=3cm \epsffile[-15 -5 130 -5]{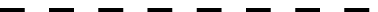}}
\def\pseudoprop{\epsfxsize=3cm \epsffile[-15 -5 130 -5]{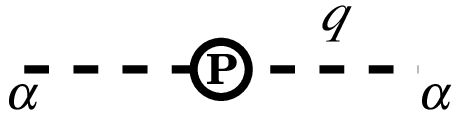}}

\def\vertex{\epsfxsize=3cm \epsffile[-15 -5 130 -5]{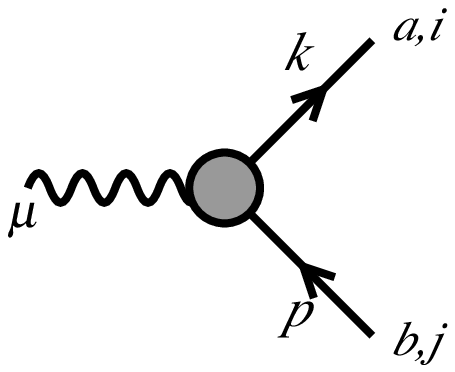}}
\def\barevertex{\epsfxsize=3cm \epsffile[-15 -5 130 -5]{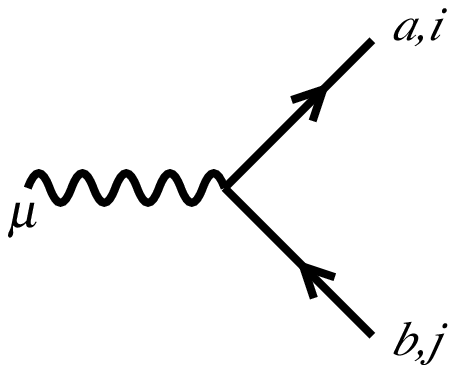}}
\def\scalarvertex{\epsfxsize=3cm \epsffile[-15 -5 130 -5]{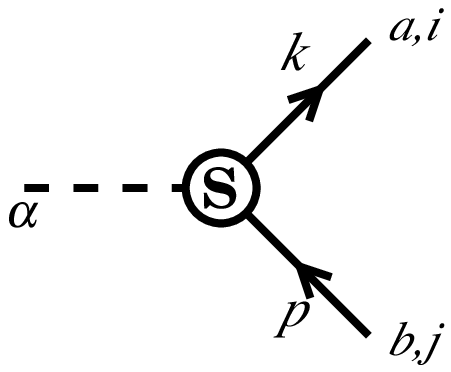}}
\def\bosonvertex{\epsfxsize=3cm \epsffile[-15 -5 130 -5]{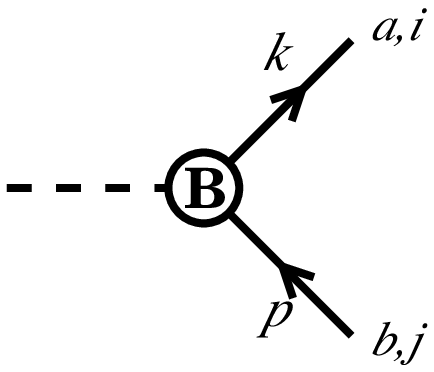}}
\def\barescalarvertex{\epsfxsize=3cm \epsffile[-15 -5 130 -5]{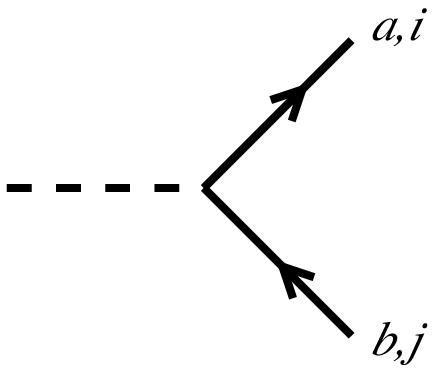}}
\def\pseudovertex{\epsfxsize=3cm \epsffile[-15 -5 130 -5]{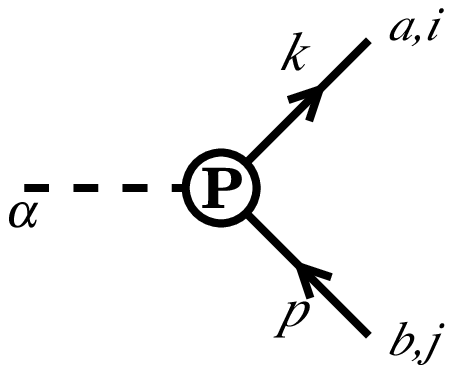}}
\def\axialvectorvertex{\epsfxsize=3cm \epsffile[-15 -5 130 -5]{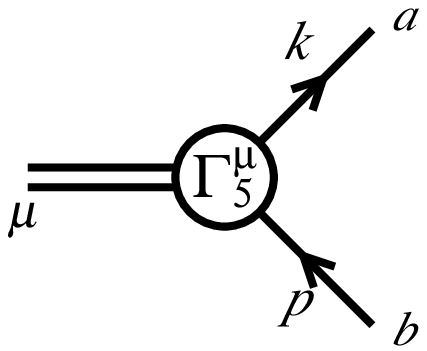}}
\def\bareaxialvectorvertex{\epsfxsize=3cm \epsffile[-15 -5 130 -5]{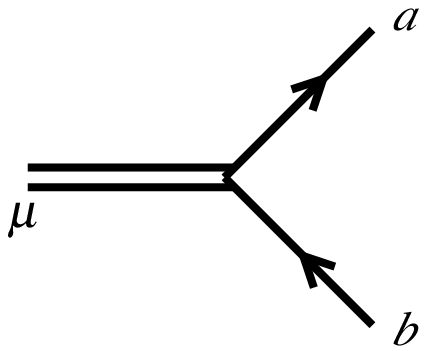}}

\def\bethesalpeter{\epsfxsize=3cm \epsffile[-15 -5 130 -5]{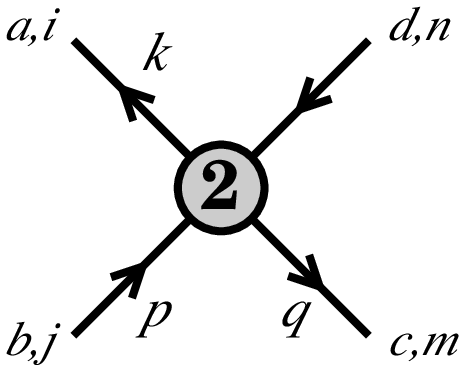}}

\begin{document}
\frontmatter 
\title{\huge Dynamical Symmetry Breaking in the Gauged Nambu--Jona-Lasinio Model}
\author{\Large Manuel Reenders}
\date{
Ph.D. Thesis \\
University of Groningen \\
26 March 1999}

\maketitle
\pagebreak
\thispagestyle{empty}

\noindent
{
\par {\it To the people of Ukraine}
}
\vfill
\par {Supervisor}
\vskip0.2cm
\par Prof.~Dr.~M.~Winnink
\vskip0.4cm
\par{Co-supervisor}
\vskip0.2cm
\par Dr.~V.~P.~Gusynin
\vskip0.4cm
\par {Referees}
\vskip0.2cm
\par Prof.~Dr.~T.~C.~Dorlas
\vskip0.2cm
\par Prof.~Dr.~V.~A.~Miransky
\vskip0.2cm
\par Prof.~Dr.~M.~R.~Pennington

\vskip0.8cm
{
\vskip0.4cm
\par ISBN 90-367-1042-1
}

\thispagestyle{empty}

{
\parskip=3pt
\begin{center}

{\large
\par Rijksuniversiteit Groningen}
\vskip0.5cm plus 1fil
{\Large
\par Dynamical Symmetry Breaking in 
\smallskip
\par the Gauged Nambu--Jona-Lasinio Model}
\smallskip
{\sc}
\vskip1.cm plus 1fil
{\large\par Proefschrift}
\bigskip
\bigskip
\par ter verkrijging van het doctoraat in de
\par Wiskunde en Natuurwetenschappen
\par aan de Rijksuniversiteit Groningen
\par op gezag van de
\par Rector Magnificus, Dr.~D.~F.~J.~Bosscher,
\par in het openbaar te verdedigen op
\par vrijdag 26 maart 1999
\par om 14.15 uur
\vskip1.5cm plus 1fil
door
\bigskip
\bigskip
{\large
\par Klaas Hommo Immanuel Reenders}
\bigskip
\par geboren op 8 april 1970
\par te Groningen
\vfill

\end{center}
\hbox{}
}

\pagebreak
\thispagestyle{empty}

\noindent
{\large
\par Promotor 
\par Prof.~Dr.~M.~Winnink
\vskip0.4cm
\par Referent
\par Dr.~V.~P.~Gusynin
}
\vfill

\tableofcontents
\mainmatter
%
%
%
\nonumchapter{Introduction and summary}
In this thesis we perform an analytic study of the gauged Nambu--Jona-Lasinio 
(GNJL) model. The GNJL model is a ``toy-model'' of dynamical chiral symmetry 
breaking (D$\chi SB$).
The famous Nambu--Jona-Lasinio (NJL) model \cite{najola61} is one of 
the first models exhibiting D$\chi$SB
in relativistic quantum field theory and was inspired by  
models of superconductivity. 

D$\chi$SB is based on the phenomenon that 
strongly attractive interactions between massless fermions give rise to the formation 
of fermion--anti-fermion bound states and generation of a fermion mass.
In this way dynamical symmetry breaking is a special case of spontaneous 
symmetry breaking where a symmetry of a particular model 
is broken by the appearance of a vacuum expectation value (VEV) of a composite
operator ({\em e.g.} $\langle\bar\psi\psi\rangle$) instead of the VEV of a 
fundamental field  ({\em e.g.} the VEV of the Higgs field in the standard model).
If the broken symmetry is continuous, such as the chiral symmetry,
dynamical symmetry breaking gives rise to the Goldstone mechanism, {\em e.g.} 
see Refs.~\cite{itzzub,mirbook}.

Probably the most well-known realization of D$\chi$SB 
and the Goldstone mechanism is provided by the low-energy dynamics of 
hadrons.
Even long before the formulation of QCD (the microscopic theory
of the strong interactions)
the NJL model formed a basis for the phenomenological description of 
the chiral dynamics of hadrons.  

Chiral symmetry is a symmetry of models with massless fermions.
In such models ``left-handed'' spinors and ``right-handed'' spinors\footnote{Left-handedness, respectively right-handedness, refers to the helicity properties
of the fermions.}
do not interact and consequently their corresponding quantum  numbers
are conserved.
To be more precise, 
chiral symmetry is the invariance of a model under 
independent unitary transformations of left-handed and right-handed 
spinors, for example $U_L \times U_R$.

Whenever a mass term is added to the Lagrangian of the model an interaction 
between left-handed and right-handed spinors is introduced, 
giving fermions a mass, and the chiral symmetry is lost.
This is called explicit chiral symmetry breaking.
However, in the case of strong interactions 
between the fermions it is possible that 
a mass is generated dynamically and a chiral
condensate, $\langle \bar\psi\psi\rangle$, is formed.

The chiral condensate is the ``order parameter'' of chiral symmetry breaking
and plays a quite similar role as the total magnetization
in models of magnets in statistical mechanics. 
In this context, the presence of an explicit mass parameter is analogous to 
the presence of an external magnetic field in {\em e.g.} certain Ising models.

Due to the existence of chiral (axial) Ward--Takahashi 
identities the generation of a fermion mass ({\em i.e.} the appearance of a 
gap in the fermion spectrum) 
coincides with the appearance of spinless bound states.
This, then, is the dynamical realization of the Goldstone mechanism.

Suppose that the space of fermion mass operators is spanned by the ``orthogonal'' 
operators $\bar\psi\psi$ and $\bar\psi i\g5\psi$.
Then we choose the chiral symmetry breaking 
(the long range ordering) to be in the direction of $\bar\psi\psi$.
The massless bound states or Nambu--Goldstone (NG) bosons
are given by correlations ``transverse'' to the direction of ordering. 
These states are called the pseudoscalar composites or the $\pi$ bosons.
The massive spinless scalar bound states ({\em i.e.} the ``Higgs'' bosons)
describe correlations ``longitudinal'' to the direction of ordering
and are referred to as the scalar composites or the $\sigma$ bosons.

Moreover, in analogy with statistical mechanics, the generated mass $m_\sigma$ 
of the $\sigma$ should be considered as the inverse of the correlation length,
{\em i.e.} $m_\sigma \sim 1/\xi$.
The correlation length describes the spatial extent of fluctuations of a 
physical quantity ({\em e.g.} see Refs.~\cite{goldenfeld,amit,zinjus}).
In quantum field theories (QFTs) 
the correlation length describes the range of forces (interactions).
Of course a well-known example is given by QED.
The Coulomb interaction has ``infinite'' range, which is reflected 
by the fact that the photon, the carrier of the electromagnetic interaction, is massless.
The range of the electroweak force is given in terms the masses of
the $W$ and $Z$ gauge bosons, $\xi\sim 1/M_W$, respectively $\xi\sim 1/M_Z$.

The equation for the fermion mass is called the gap equation and describes
the self-interaction (self-energy) of the fermion fields.
The gap equation is a non-linear eigenvalue equation.
For values of the coupling constants above some critical coupling 
the gap equation has a nonzero solution for the fermion mass even 
in the absence of an explicit external mass. This solution leads
to a nonvanishing chiral condensate and consequently the chiral 
symmetry is broken.
For values of the coupling below the critical value 
the gap equation only has the trivial solution with zero fermion mass
and the chiral symmetry is unbroken.
Exactly at the critical coupling 
the fermion mass is zero and the condensate vanishes; 
the gap equation describes a continuous phase transition.
This is reflected by the non-analyticity of the gap equation
in the coupling constant which causes the change of phase.
The chiral condensate continuously changes from zero in the symmetric phase
to nonzero values in the broken phase.
As was pointed out by Yang and Lee \cite{yale52} 
this non-analyticity is characteristic for phase transitions
and can only occur in systems describing an infinite number of 
degrees of freedom.\footnote{Relativistic quantum field theories 
are systems with an infinite number of degrees of freedom.}

The realization of D$\chi SB$ is inherently nonperturbative 
and typically requires that coupling constants describing the ``fundamental''
interactions between fermions are of order one.

In the NJL model dynamical chiral symmetry breaking was realized
by strong attractive four-fermion interactions (think of $(\bar\psi\psi)^2$)
which were incorporated in a mean-field approach known
as the Hartree--Fock approximation.
In mean-field approximations the 
composite operators such as $\bar\psi\psi$ 
are replaced by their VEVs ($\bar\psi\psi\rightarrow\langle\bar\psi\psi\rangle$) 
and fluctuations about that value are ignored.
However the NJL model turned out to be ``in the language of the sixties''
a nonrenormalizable model. It has a limited range of applicability
because of the explicit dependence on the ultra-violet cutoff $\Lambda$.\footnote{The cutoff is required to render the theory finite.}

It was realized by the Kiev group in Refs.~\cite{fomi76,fogumisi76}
that QED in the quenched-ladder approximation 
also exhibits dynamical chiral symmetry breaking for values of the bare gauge coupling 
$\alpha_0>\alpha_c\sim 1$, where $\alpha_0=e_0^2/4\pi$.
This was argued to be intimately connected with the  
``fall into the center'' phenomenon.
By considering the Dirac equation describing a light fermion in a static 
Coulomb potential, it was shown that, when the Coulomb attraction dominates 
the centrifugal barrier for $\alpha_0\geq \alpha_c$, 
the system ``collapses'' 
({\em i.e.} a drastic rearrangement of the ground state (vacuum) occurs). 
Later it was pointed out in Refs.~\cite{fogumisi83,mi85a} that the 
critical coupling $\alpha_c$ above which the chiral symmetry is 
broken should be considered as an ultra-violet (UV) fixed point.

An important step was performed by Bardeen, Leung, and Love
in Refs.~\cite{balelo86,leloba86}.
These authors realized that, in quenched-ladder QED, 
four-fermion interactions 
have a so-called scaling dimension 4 instead of 6 at the critical gauge 
coupling $\alpha_0=\alpha_c$. Consequently these operators mix with the gauge 
interaction which also has dimension 4 in four space-time dimensions;
this means that QED is not a closed theory at the chiral phase transition.
The scaling dimension of an operator is important since it determines the 
possible range of the interaction which the operator describes.
Their model is referred to as the gauged Nambu--Jona-Lasinio (GNJL) model.

The origin and physical significance of scaling dimensions is undoubtedly 
best explained by the renormalization-group methods of Wilson \cite{wi71,wiko74}.
Wilson stressed the importance of defining a low-energy effective action 
or microscopic action from which a generating functional can be constructed.
The Wilsonian effective action is an action describing local interactions
defined at some microscopic length scale $a\sim 1/\Lambda$.
The interactions between the degrees of freedom corresponding to distances 
shorter than $1/\Lambda$ are not defined and such dynamics 
are ``effectively'' described by the coupling constants of the 
microscopic action.
In this way the high energy frequencies of the fields with $E\geq \Lambda$  
are ``coarse grained'' or ``integrated out'' into the coupling constants and 
the coupling constants implicitly depend on the cutoff $\Lambda$.

The interesting question is what type of macroscopic 
physics characterized by an ``infra-red (IR)'' energy scale $E$ 
with $E\ll \Lambda$
can be realized from the microscopic action.
The transformation of the microscopic action to the macroscopic action
is governed by the renormalization-group (RG) transformation
of the coupling constants of the model.
If a macroscopic model can be derived from a microscopic model 
in such a way that the macroscopic dynamics can be formulated independently 
of the scale $\Lambda$ and only depend on the coupling constants of 
the microscopic model, the microscopic model is said to be renormalizable.

Wilson pointed out that such a macroscopic theory can only be formulated
if the RG transformation exhibits UV fixed points.
UV fixed points are ``singular'' points\footnote{
UV fixed points are specific roots of the $\beta$ functions.} 
of the RG transformation
at which the model becomes scale ({\em i.e.} conformal) invariant. 
Wilson pointed out that natural candidates for UV fixed points
are critical points governing a continuous phase transition.
Since, at a continuous phase transition, the correlation 
length $\xi$ is infinite, the model is 
scale invariant and the dynamics is extremely sensitive to
external perturbations.

A mathematical formulation of the RG transformation 
and UV fixed points for QED was already given by Gell-Mann and Low in 
Ref.~\cite{gemalo54}.
But the physical significance of such fixed points 
and the interpretation of infinities appearing in Feynman diagrams 
were not fully understood in the fifties and sixties.   

The most crucial observation of Wilson is that in the RG transformation, 
{\em i.e.} the ``coarse graining'' process, new types of local interactions
are generated and that the new interactions can be classified
as either irrelevant or relevant interactions.
In this way the RG transformation is a nontrivial improvement of 
the coarse graining methods of Kadanoff \cite{ka66}.
By definition, the effect of an irrelevant interaction\footnote{ 
Irrelevant interactions are also referred to as nonrenormalizable.}
of a microscopic action is suppressed by positive powers of $E/\Lambda$
in the macroscopic action. 
This is closely related to the Appelquist--Carazzone decoupling theorem 
\cite{apca75}.
Hence only relevant interactions are important in determining
what kind of macroscopic dynamics emerges from microscopic models.
The effect of irrelevant interactions can always be absorbed
by adapting the coupling constants of relevant and marginal interactions.
Marginal interactions are in between irrelevant and relevant and are considered
the most interesting ones; in four dimensions 
they comprise the gauge interactions.

In analogy with statistical mechanics, the continuous chiral phase transition 
can be classified in terms of critical exponents which describe the scaling of
various macroscopic quantities ({\em e.g.} the chiral condensate, correlation 
length, effective potential, chiral susceptibility) close to or at the 
critical point.
It is considered a strong indication of the existence of a nontrivial continuum 
limit ($\Lambda\rightarrow \infty$), 
if so-called hyperscaling relations between these various critical exponents 
are satisfied.
The RG of Wilson provides the framework for explanation and computation
of the critical exponents and makes an intimate connection
between critical phenomena (in particular continuous phase transitions)
in statistical mechanics and the renormalization and existence of the
continuum limit in quantum field theories.

Especially, close to a phase transition, 
the RG methods of Wilson show that it is not 
possible, a priori (without solving the equations of motion), 
to determine which interactions are relevant or irrelevant;
particular interactions can acquire anomalous dimensions and 
interactions which are irrelevant 
in a certain region of coupling constant space
might become relevant in another region.

The coupling constants which govern the chiral phase transition 
are similar to the inverse of the temperature in models
of statistical mechanics.

Perturbatively the four-fermion interaction is irrelevant, which is supposedly 
due to the fact that the four-fermion coupling constant has negative 
dimension (in mass units).
However, as was shown by
Bardeen, Love, and Leung \cite{balelo86, leloba86}
the solutions to the gap equation in the GNJL model suggest
that the four-fermion interactions acquire a large anomalous dimension.
The anomalous dimension $\gamma_m$ of the mass operator $\bar\psi\psi$
equals one ($\gamma_m=1$) at $\alpha_0=\alpha_c$.

The GNJL model literally is the gauged version of the NJL model.
It is based on induced chiral invariant and gauge invariant 
four-fermion interactions.
The gauge group can be either a $U(1)$ gauge symmetry or some
non-Abelian gauge symmetry.
In case of GNJL model there are two attractive interactions:
an attractive four-fermion interaction and a gauge interaction (with 
gauge coupling $\alpha_0$).
In this 
thesis we will consider the GNJL model with $U(1)$ gauge symmetry,
in that case the GNJL model is a generalization of
of quantum electrodynamics (QED).
The microscopic Lagrangian for a single fermion flavor reads
\bea
{\cal L}=-\frac{1}{4}F_{\mu\nu}^2
+\bar\psi (i\gamma^\mu D_{\mu}-m_0)\psi
+\frac{G_0}{2}\[(\bar\psi\psi)^2+(\bar\psi i\gamma_5\psi)^2\],
\eea
where $D_{\mu}=\partial_\mu-i e_0 A_\mu$, and where $G_0$  
is the positive four-fermion coupling.
The Lagrangian is determined by the following three dimensionless 
coupling constants:
\bea
\mu_0=m_0/\Lambda,\qquad \alpha_0=e_0^2/4\pi,\qquad g_0=G_0\Lambda^2/4\pi^2,
\eea
where $\Lambda$ is the ultraviolet cutoff.
The GNJL model can be conveniently analyzed in terms of so called auxiliary fields 
($\sigma=-G_0 \bar\psi\psi$ and $\pi=-G_0\bar\psi i\g5 \psi$)
describing scalar and pseudoscalar degrees of freedom.
In such a formulation, the four-fermion interactions are described by 
the interactions of the auxiliary fields with the fermion fields.
Then the connected two-point Green function of the $\pi$ field
describes the NG-boson 
and the connected two-point Green function of the $\sigma$ field describes the
``Higgs'' boson.

In the quenched-ladder (quenched-planar) approximation
the chiral phase structure was computed in Refs.~\cite{komiya89,apsotawij88}.
There is a critical curve in the coupling constant plane $(\alpha_0,g_0)$
which separates a chiral symmetric phase from a phase in which 
the chiral symmetry is dynamically broken.

In Ref.~\cite{miya89} the anomalous dimension $\gamma_m$
along the critical curve was computed (in the broken phase)
and it was found that $1\leq \gamma_m \leq 2$, where
$\gamma_m=1$ at $\alpha_0=\alpha_c$ and $\gamma_m=2$ at $\alpha_0=0$.

The particle spectrum consists of pseudoscalar $(\pi)$ and scalar $(\sigma)$ 
bound states which become relevant degrees of freedom (their masses are light)
in the vicinity of the critical curve.
In the D$\chi$SB phase there is a massless 
pseudoscalar (NG-boson) and a massive scalar.
Since the phase transition is second order along the critical curve
there are scalar and pseudoscalar resonances on the symmetric side of 
the curve whose masses also vanish as the critical curve 
is approached~\cite{aptewij91}.
The phase transition at NJL point ($\alpha_0=0$) 
is believed to be of the mean field type (up to logarithmic corrections) 
and to correspond to that of a trivial theory, {\em e.g.} 
see Ref.~\cite{mirbook}.

In the intermediate region ($0<\alpha_0< \alpha_c$),
the phase transition is quite similar to the pure NJL case except that 
the critical exponents satisfy nonmean-field hyperscaling relations 
which supports the view that the quenched approximation 
has a nontrivial continuum limit~\cite{kohakoda90,balomi90}.

At, what is referred to as the QED point, $\alpha_0=\alpha_c$, 
the phase transition is rather special and is characterized by a scaling law 
with an essential singularity (instead of a power law) for the fermion 
dynamical mass.
There is an abrupt change in the particle spectrum when the phase boundary 
is crossed.
This peculiar phase transition has 
properties similar to the
Berezinsky-Kosterlitz-Thouless phase transition \cite{be70,koth73}
and is referred to as the conformal phase transition (CPT)~\cite{miya97}.
The crucial point is that the symmetry is broken by marginal operators 
instead of by relevant operators.

Whereas from the ``perturbative'' point of view the correlation length
of a four-fermion interaction is typically the inverse of the cutoff 
$\xi\sim 1/\Lambda$,
close to a critical point in the GNJL model we find that
\bea
\xi \sim 1/m_\sigma,\qquad m_\sigma =(\Delta g_0)^\nu,
\eea
where $\Delta g_0=|g_0-g_c|$ 
measures the distance of $g_0$ to critical point $g_c$ and $\nu$ 
is a (positive) critical exponent.
Thus, when $\Delta g_0 \rightarrow 0$, the correlation length goes to infinity.
This is interpreted as a dimensional transmutation of the four-fermion
operator $(\bar\psi \psi)^2$ which has a canonical dimension of $6$
to a nonperturbatively induced dynamical dimension $d_{\rm dyn} \leq 4$
(hence relevant (marginal)) at the critical point:
\bea
{\rm dim}(\bar\psi \psi)^2 =6-2\gamma_m\leq 4,\qquad 
{\rm dim}(\bar\psi \psi) =3-\gamma_m.
\eea

The factorization property ${\rm dim}(\bar\psi \psi)^2=2{\rm dim}(\bar\psi \psi)$
is implicit in the quenched-ladder approximation and is crucial
for the nonperturbative renormalizability of the GNJL model \cite{kohakoda90};
it leads to the requirement that the critical exponent $\gamma$ of the
so-called chiral susceptibility equals one, $\gamma=1$.

The interest in the GNJL has been stimulated by its importance for
constructing new scenarios of dynamical realization of 
electroweak symmetry breaking. Examples are strong extended technicolor models 
({\em e.g.} see Refs.~\cite{ho85,apeitawij89a,apeitawij89b,miya89})
and the top-quark condensate model 
({\em e.g.} see Refs.~\cite{mitaya89a,mitaya89b,bahili90}).
It was argued in Refs.~\cite{ho81,yabama86,bamaya86,akya86}
that large anomalous dimensions could solve the problem
of flavor changing neutral currents in technicolor 
and extended technicolor models.
For a general introduction into these models and their predecessors
we refer to the book by Miransky~\cite{mirbook}.

Since these models are based on strong interactions, giving rise to
the appearance of a large anomalous dimension for the mass operator 
$\bar\psi\psi$, these models are inherently nonperturbative.
The origin of the four-fermion interactions
is assumed to be given by physics at or above some high energy scale $\Lambda$.
Two natural candidates for the high energy scale or ultra-violet cutoff 
$\Lambda$ are the scale of grand unified theories (GUTs), 
{\em i.e.} $\Lambda\sim 10^{15}-10^{16}$ GeV, 
and the Planck scale $\Lambda\sim 10^{19}$ GeV.
The relevant physics at these scales should somehow be given by GUTs,
sypersymmetric field theories, supergravity, string theory, or perhaps
$M$-theory. 
However, the main point is that as long as the model is close to criticality
the low-energy effective dynamics is 
rather insensitive to the details of the interactions at $\Lambda$.
This is known as universality.
For instance the exchange of a heavy gauge boson 
with a mass of the order of the cutoff $\Lambda$ could induce
such a local four-fermion interacting for energies $E\leq \Lambda$.

The GNJL model in the quenched approximation has been studied extensively on 
the lattice (in so-called noncompact formulation, {\em e.g.} 
see Ref.~\cite{kodako88a}), by means of nonperturbative 
renormalization group (NPRG) methods, and with Schwinger--Dyson (SD) techniques.
In the quenched approximation fermion-loops are omitted, and consequently
the gauge coupling $\alpha_0$ does not run.
In addition to the quenched approximation, 
the SDEs are studied by taking the gauge interaction in the ladder or planar
approximation; this means that crossed photon graphs and vertex corrections 
are omitted.

A very useful and popular method is the computation
of the low-energy effective action for the composite fields
which is obtained after integrating out the fermion and photon fields
in the generating functional.
Such methods are capable of computing macroscopic quantities such as 
the effective Yukawa coupling, $\lambda_Y$, 
the mass of the $\sigma$ boson, $m_\sigma$, 
and the ``pion'' decay constant, $f_\pi$.

The comparison of the lattice simulations and NPRG methods 
show that qualitatively the ladder approximation is reliable,
and rather small quantitative differences arise because of the
neglect of vertex corrections and crossed photon exchange diagrams.
The advantage however of the quenched-ladder SD formalism with respect
to the numerical methods is that results can be obtained analytically.

The first three chapters of this thesis cover most of the important
achievements and results on dynamical chiral symmetry breaking in 
the GNJL model and QED, which we feel are necessary to understand the 
results of chapters~\ref{chap4} and \ref{chap5}. 

In chapter~\ref{chap1} we review 
the path-integral formalism for the generating functional 
and the Schwinger--Dyson equation (SDE).
We briefly discuss the basic features governing phase transitions
and critical phenomena in statistical mechanics and discuss
the RG of Wilson, the Goldstone mechanism, and the auxiliary
field method for the GNJL model.
In chapter~\ref{chap2} we derive both the Schwinger--Dyson equations
and the so-called partial conserved axial currents (PCAC).
Chapter~\ref{chap3} is devoted to the gap equation, scaling laws,
the critical curve etcetera.
Moreover in chapter~\ref{chap3} we compare the SD quenched-ladder approach with 
other nonperturbative techniques.
The results and conclusions
of this thesis are the following.

In chapter~\ref{chap4}:
in quenched-ladder approximation, treating the four-fermion interaction
in the mean-field setting, we have computed analytically the 
scalar Yukawa vertex in specific kinematical regimes.
The Yukawa vertex describes the interaction between fermions and the
composite states, {\em i.e.} the spinless $\sigma$ and $\pi$ bosons.
By making use of Chebyshev expansions we are able
to derive asymptotic expressions 
for the Yukawa vertex as functions of the in-going momentum of the 
fermions and out-going momentum of the $\sigma$ boson.
This allows us to derive an analytic expression for the scalar propagator
(the $\sigma$ boson propagator), which is valid along the entire 
critical curve.

Our results are in agreement with previous work of Appelquist 
{\em et al.}~\cite{aptewij91}, and related work \cite{kotaya93}, 
who used a resummation technique which applies only for relatively small 
values of $\alpha_0$.
Near the critical gauge coupling $\alpha_0=\alpha_c$
the light resonances in the symmetric phase disappear 
from the particle spectrum, and the scaling 
law with an essential singularity, which is characteristic for 
the CPT of Miransky and Yamawaki (\cite{miya97}), is
recovered by analytic continuation of the $\sigma$ boson propagator
across the critical curve.
The results of chapter~\ref{chap4} were published in Ref.~\cite{gure98}.

In chapter~\ref{chap5} we study the ``unquenched'' GNJL model\footnote{That means
we turn back on fermion loops.} 
and reinvestigate the screening of 
the gauge coupling $\alpha_0$ by computing the vacuum polarization.
We consider the GNJL model with global $U_L(N)\times U_R(N)$ chiral symmetry,
where $N$ is the number of fermion flavors.\footnote{The 
global $U_L(N)\times U_R(N)$ chiral symmetry comprises the largest
set of relevant chiral invariant four-fermion operators 
({\em e.g.} see Ref.~\cite{aomosuteto97}).}
The interesting question is whether the GNJL model exhibits
ultra-violet fixed points
or the gauge coupling $\alpha_0$,
\bea
\beta_\alpha(\alpha_0)=0,\qquad \beta^\prime_\alpha(\alpha_0)<0,\qquad
\beta_\alpha(\alpha_0)\equiv
\Lambda \frac{{\rm d}\alpha_0}{{\rm d}\Lambda}.
\label{introbetadef}
\eea
The function $\beta_\alpha$ describes the renormalization group flow of $\alpha_0$.
If such an UV fixed point exists the model has a nontrivial (interacting) 
continuum limit. 

Many works on QED conclude, following Landau
and Pomeranchuk \cite{lapo55}, that QED
is trivial (non-interacting) in four dimensions; 
the continuum limit describes a free-field theory
(there is no root of the $\beta$ function 
satisfying Eq.~(\ref{introbetadef})). 
This is a result of the total screening of the gauge coupling 
by virtual fermion pairs (fermion loops) from the vacuum.
However we think that the main ingredient for settling the charge screening
problem is missing,
namely the contribution of the composite $\sigma$ and $\pi$ bosons
and their effective interactions with fermions.

The observation  that hyperscaling relations are satisfied 
(\cite{kohakoda90,balomi90}) clearly indicates that the correlations corresponding 
to the four-fermion interactions are not of the mean-field type. 
According to Landau's mean-field theory only specific mean-field exponents 
can describe the scaling near the critical point.  
Therefore any deviation from the mean-field exponents signals the 
inconsistency of the mean-field approach and suggests 
that fluctuations of the composite $\sigma$ and $\pi$ fields about their mean 
value cannot be ignored.

We argue that in case of the GNJL model close to the chiral phase
transition the virtual fermion loops causing charge screening are suppressed 
due to the appearance of attractive relevant four-fermion interactions, where
we recall that the four-fermion interactions are represented by the 
exchange of $\sigma$ and $\pi$ bosons.
The standard picture of the QED-vacuum as a medium of virtual electric
dipoles formed by virtual fermion--anti-fermion pairs 
breaks down due to the strong self-interactions of the fermions.
Moreover the problem is flavor dependent;
the larger the number of fermion flavors, the more the charge is screened.
On the other hand with $N$ fermion flavors there are $2N^2$
chiral invariant four-fermion interactions.
Thus increase of $N$ increases the role of four-fermion interactions.
The four-fermion interactions induce a nontrivial 
Yukawa interaction, this is also referred to as the nondecoupling of 
the composites, see Ref.~\cite{mi91}.
The spinless composites states are relevant degrees of freedom for a large 
range of momenta $m_\sigma \leq p \leq \Lambda$.

We incorporate the $\sigma$ and $\pi$ exchanges using the 
skeleton expansion \cite{bjodre} for the irreducible Bethe--Salpeter kernel
and the $1/N$ expansion of 't Hooft \cite{tho74}.
The skeleton expansion is formulated in terms of 
``fully dressed'' vertices and propagators for which
we substitute the asymptotic expressions for the scalar Yukawa vertex 
and $\sigma$ boson propagator derived in chapter~\ref{chap4}.
The basic idea of the $1/N$ expansion is that for $N$ large
only the planar ({\em i.e.} ladder) exchanges of $\sigma$ and $\pi$ bosons
are dominant.
Then by making use of a resummation technique 
of Johnson {\em et al.} \cite{jowiba67}
we can compute the $\beta$ function of the gauge coupling $\alpha_0$ 
defined by Eq.~(\ref{introbetadef}) 
and search for ultra-violet fixed points.
We find that UV fixed points exist for values of $N$ above a critical value 
$N_c\approx 54$.
The largest UV fixed point is $\alpha_0\approx 0.14$.

Assuming that our results are qualitatively correct, we have realized, 
in an inherently nonperturbative manner\footnote{Perturbatively 
the GNJL model is a trivial theory.}, a 
nontrivial ({\em i.e.} interacting) continuum limit for a nonasymptotically 
free gauge field theory in four dimensions.
This means that the low-energy dynamics characterized by an energy scale $E$
($E\ll \Lambda$) can be formulated independent
of the ultra-violet cutoff $\Lambda$ by tuning the bare couplings
to their critical values. 
Thus the trade-off is that bare couplings have to be fine-tuned
to their critical values (the UV fixed points). 

This fine-tuning, which is also known in the literature as the gauge 
hierarchy problem ($m_\sigma\ll \Lambda$), 
is not (yet) explained by the model itself.
Studies of dynamical symmetry breaking in supersymmetric models
provide a promising solution, {\em e.g.}
see Ref.~\cite{clloba90}.
The crucial question is; why is the system close to criticality?
An interesting resolution is that of self-organizing criticality, see
Ref.~\cite{bowe92}. 

The realization of such a nontrivial theory requires a large number of fermion 
flavors $N>N_c\approx 54$ and the fine-tuning of the four-fermion coupling $g_0$ 
and gauge coupling $\alpha_0$ to their UV fixed points.
Clearly our claims should be verified in the future by quantitatively
more reliable techniques such as lattice simulations and 
nonperturbative renormalization group methods.
Especially we believe that the determination of the critical value $N_c$ 
is rather sensitive to the various approximations used, hence
$N_c$ is a so-called nonuniversal quantity.

\section*{List of short-hand notations}
\addcontentsline{toc}{section}{List of short-hand notations}
\begin{tabular}[h!]{ll}
BS & Bethe--Salpeter\\
CJT & Cornwall--Jackiw--Tomboulis\\
CP & Charge Parity \\
CPT & Conformal Phase Transition \\
D$\chi$SB & Dynamical Chiral Symmetry Breaking \\
EOS & Equation of State \\
ETC & Extended technicolour \\
GNJL & Gauged Nambu--Jona-Lasinio\\
GUT & Grand Unified Theory \\
IR & Infra-red\\
JWB & Johnson--Willey--Baker \\
NG & Nambu--Goldstone\\
NJL  &      Nambu--Jona-Lasinio\\
NPRG & Nonperturbative Renormalization Group \\
QCD & Quantum Chromodynamics \\
QED &  Quantum Electrodynamics \\
QFT & Quantum Field Theory \\
PCAC & Partial Conserved Axial Currents \\
PCDC & Partial Conserved Dilaton Currents \\
RG & Renormalization Group\\
SDE & Schwinger--Dyson Equation\\
UV & Ultra-violet\\
VEV & Vacuum Expectation Value \\
WTI & Ward--Takahashi Identity 
\end{tabular}

%
\chapter{Phase transitions and quantum field theory}\label{chap1}
This chapter provides the basic framework of this thesis.
After introducing the path-integral formalism
for relativistic quantum field theory, we
discuss the feature of critical phenomena (such as continuous phase transitions) 
and scaling in statistical mechanics.
We review the renormalization group methods of Wilson 
and the Goldstone mechanism.
This chapter is concluded with the introduction of 
the GNJL model and the auxiliary field method.
\section{Review of the path-integral formalism}\label{sec_rev}
Relativistic quantum field theory (QFT) 
is governed by expectation values of opera-tor-valued fields on Minkowski space-time. 
The field at each space-time point is referred to as a degree 
of freedom, thus any particular finite region of space-time 
deals with an infinite number of degrees of freedom.
The ``vacuum'' expectation values are particular expectation values 
of the fields.
These so-called Green functions (``or correlation functions'') describe 
the behavior of elementary particles.
The interactions are given by an action $S$, and the dynamics is derived from 
a generating functional $Z$. 
From this generating functional the entire (infinite) set of Green functions 
can be constructed.
The generating functional is analogous to the partition function or Gibbs integral
in statistical mechanics.
The generating functional is given as a functional path-integral 
over all field configurations weighted by the ``classical'' 
action times $i/\hbar$, henceforth we will use units in which $\hbar=1$.
This so-called path-integral formalism was introduced by Feynman \cite{fe48,fe49}
for quantum mechanics and quantum field theory.
  
In what follows only one scalar field is considered for simplicity.
The generating functional $Z$ as the function of some arbitrary ``external''
source $J(x)$ can be written as
\bea 
Z[K,J]=N\int{\cal D}\phi\,\exp\{iS[K,\phi,J]\}, \label{Zgen}
\eea 
where $N$ is some normalization constant such that $Z[K,J=0]=1$,
and where the action (or weight) $S[K,\phi,J]$ is
the sum of some particular fundamental action $S_f$, and the
source action $S_J$ describing the
coupling to the arbitrary source $J$:
\bea
S[K,\phi,J]=S_{f}[K,\phi]+S_J[J,\phi]. 
\eea 
The source action, $S_J$, is of the form
\bea
S_J[J,\phi]\sim \int{\rm d}^dx\,  J(x)\phi(x),\label{sourceaction}
\eea
in $d$ dimensions.
We consider actions of local operators ${O}_i$ of $\phi$: 
\bea
S_{f}[K,\phi]=\int{\rm d}^dx\, \sum_i K_i O_i\[\phi(x)\],
\eea
where an example of a local (Euclidean) operator $O$ is
\bea
O\[\phi(x)\]=-\phi(x)\Delta \phi(x),
\eea
with $\Delta$ the Laplacian.
The generating functional $Z$ is a functional of the external sources $J(x)$ 
and couplings $K$; $Z$ is by 
construction analytic in the sources $J$, see 
Refs.~\cite{glimmjaffe,zinjus}, and Eq.~(\ref{Zgenanaly}).
Green functions are defined as time-ordered vacuum expectation values (VEV)
of a specific number of fields, {\em e.g.},
\bea
iG^{(n)}(x_1,x_2,\cdots, x_n)&\equiv& 
\langle 0|T\( \phi(x_1)\phi(x_2)\cdots \phi(x_n)\)|0\rangle\nonu\\
&=&\frac{\delta}{i\delta J(x_n)} \cdots 
\frac{\delta}{i\delta J(x_2)}\frac{\delta}{i\delta J(x_1)} 
Z[K,J]\biggr|_{J=0},
\eea 
{\em i.e.} 
the generating functional generates VEVs of time-ordered products of fields.
Hence the Green functions can be constructed by functionally differentiating
$Z$ with respect to sources.
Since the generating functional is analytic in $J(x)$,
the generating functional can be expressed as a Taylor series in sources,
\bea
Z[K,J]=\sum_{n=0}^\infty \frac{i^{n}}{n!} 
\int {\rm d}^d x_1\cdots {\rm d}^dx_n\, 
iG^{(n)}(x_1,\cdots,x_n) J(x_1)\cdots J(x_n), \label{Zgenanaly} 
\eea 
where the Green function depend implicitly on the couplings $K$.
However, although the above construction of functional integration 
might appear to be rather straightforward and intuitive, this is deceptive.  
It is a highly nontrivial mathematical problem how to define the 
functional integration measure, ${\cal D}\phi$, 
and functional integration properly. 
In Ref.~\cite{olivier} the problem of functional integration is addressed,
and a new perspective on functional integration is presented. 
The idea of de~Mirleau is to define such functional integration just as one
would define integration, {\em i.e.}, as a solution of a differential equation,
with appropriate boundary conditions.

For a fixed Euclidean action $S$, consider the following linear functional
\bea
\langle f\rangle\equiv N\int{\cal D}\phi\, f \exp\{-S\[\phi\]\}, 
\eea
where $f$ and $S$ belong to the same algebra, and $N$ is a number.
The functional integration is defined by the following properties:
\begin{enumerate}
\item{The Schwinger--Dyson equation (SDE): 
$\langle(\delta S) f\rangle-\langle \delta f\rangle=0$,}
\item{Positivity: $f\geq 0\Longrightarrow \langle f\rangle\geq 0$,}
\item{Normalization: $\langle 1\rangle=N\int{\cal D}\phi\,\exp\{-S\[\phi\]\}$,}
\end{enumerate}
where the $\delta$ represent derivatives
(mutually commuting (bosons) or anti-commuting (fermions))
of the particular functional space or algebra, 
({\em e.g.} $\delta\sim \delta/\delta \phi(x)$).
The normalization condition 3 is straightforward for finite 
dimensional integration, but is not always applicable to the 
infinite dimensional case.
We refer to \cite{olivier} for an extensive discussion of this point.
The Schwinger--Dyson equation (SDE) was independently
formulated by Dyson \cite{dy49} and Schwinger \cite{sc51}. 

Let us consider the Euclidean version of generating functional 
$Z$ of Eq.~(\ref{Zgen}) which is defined as
\bea
Z[K,J]&\equiv&\langle \exp\{-\int{\rm d}^dx\,J(x)\phi(x)\} \rangle_{S_f},
\label{Zpartdef}\\
\langle f[\phi] \rangle_{S_f}&\equiv& 
N\int{\cal D}\phi\,f[\phi] \exp\{-S_f[K,\phi,J]\}.
\eea
Then, the SDE for $Z$ can be written as
\bea
\biggr\{
\frac{\delta S_f\[K,\phi\]}{\delta \phi(x)}\biggr|_{\phi(x)=\delta/\delta J(x)} 
+ J(x)\biggr\}Z\[K,J\]=0. \label{defSDE}
\eea
Such an equation was in fact formulated
already by Feynman in Ref.~\cite{fe48}.
The SDE for any Green function corresponding 
to this particular model (\ref{Zpartdef}) can be obtained by first 
differentiating Eq.~(\ref{defSDE}) 
with respect to appropriate combinations of sources ($J(x)$, $J(y),\dots$) 
and then putting all sources equal to zero.

From the generating functional one can construct the following functional 
(in Euclidean formulation):
\bea 
W[K,J]&\equiv&\log Z[K,J]\nonu\\
&=&\sum_{n=1}^\infty \frac{1}{n!} 
\int {\rm d}^d x_1\cdots {\rm d}^dx_n\, 
G^{(n)}_C(x_1,\cdots,x_n) J(x_1)\cdots J(x_n), \label{connGreenfies}
\eea
which is the generating functional for connected Green 
functions, $G^{(n)}_C$, {\em i.e.} Green functions which satisfy 
the cluster property, see Ref.~\cite{zinjus}\footnote{In short, 
the cluster property states that the vacuum expectation value
of a product of local operators factorizes when their
space-like separations become large \cite{itzzub}.}.
The functional $W$ is now convex in the couplings $K$, and
analytic in the sources $J$; $W$ is extensive, {\em i.e.} proportional to the
total volume, and is analogous to the free energy in statistical mechanics.

The Green functions given by Eq.~(\ref{Zgenanaly}) and the connected 
Green functions of Eq.~(\ref{connGreenfies}) are not defined on the entire 
space-time manifold, because of the existence of so called short distance 
singularities, {\em e.g.} the limit
\bea
\lim_{x_1\rightarrow x_2} iG^{(n)}(x_1,x_2,x_3,\cdots, x_n) \label{Gsing}
\eea
is ill defined. 
Regularization ({\em e.g.} Pauli--Villars), which we will discuss in 
section~\ref{WRG}, permits a setting for the treatment of Eq.~(\ref{Gsing}). 
\section{Continuous phase transitions}\label{sec_phasetrans}
Phase transitions can occur in systems containing an infinite number 
of degrees of freedom exhibiting nonanalytic behavior 
({\em e.g.} in certain VEVs) in one (or more) of 
its parameters such as the temperature. 
A model in statistical mechanics, {\em e.g.} the Ising model, 
is given in terms of a partition function
\bea
Z_{\Omega}\[K\]\equiv \Tr \exp\{-\beta H_{\Omega}\},\label{Zpart}
\eea
where $\beta\equiv 1/k_B T$ ($k_B$ is Boltzman's constant).
The partition function is defined over a finite region $\Omega$ 
with a volume $V(\Omega)\sim L^d$, surface $S(\Omega)\sim L^{d-1}$, 
and with $N(\Omega)$ degrees
of freedom, where $L$ is characteristic length scale, 
and $d$ the dimension of the system. 
The Hamiltonian for the system is written as
\bea
\beta H_\Omega=\sum_{i}^n K_i O_i\[S\],
\eea
where $K=(K_1,K_2,\dots,K_n)\sim (T,h,\dots)$ (with $T$ the temperature,  
$h$ the external magnetic field etc.)
are
coupling constants and the (local) operators $O$ are functionals
of the dynamical degrees of freedom, which in case of the Ising model are
spins $S_j$ ($1\leq j\leq N(\Omega)\sim (L/a)^d$) 
on lattice sites with a lattice spacing
$a$. The lattice spacing $a$ is then referred to as the microscopic length scale, since interactions on a scale smaller than $a$ are not defined.
The trace, $\Tr$, in Eq.~(\ref{Zpart}) represents the sum over 
all possible configurations of spins,
and is analogous to the functional integration in Eq.~(\ref{Zgen}).

The free energy is defined by
\bea
F_\Omega\[K\]=-k_B T\log Z_\Omega\[K\].\label{freeenergy}
\eea
Assuming that the thermodynamic limit,
\bea
V(\Omega),\, S(\Omega),\, N(\Omega) \quad \longrightarrow \quad\infty,\qquad 
N(\Omega)/V(\Omega)=\mbox{constant},
\eea
for the free energy exists, we can define
the bulk free energy density $f_b$ (the free energy per 
lattice site):
\bea
f_b[K]=\lim_{N(\Omega)\rightarrow \infty}
\frac{F_\Omega\[K\]}{N(\Omega)}.
\eea
The so-called extensive thermodynamic behavior is now described by $f_b[K]$,
and its derivatives with respect to coupling parameters such 
as the temperature $T$ or external magnetic field $h$.
An example of a macroscopic quantity 
is the magnetization $M$,
\bea
M\equiv  \frac{\partial f_b[K]}{\partial h} \propto \langle S\rangle.
\eea
The bulk free energy $f_b$ is a convex function of the coupling 
parameters.

A phase of a system is defined by regions of analyticity of $f_b[K]$.
Possible non-analyticities of $f_b[K]$ 
occur at points, 
lines or hyperplanes in parameter space. 
We define $D_S$ as the dimension of a particular 
non-analyticity region of $f_b[K]$.
Suppose that $f_b[K]$ is given in terms of $D$ coupling constants.
If the dimension $D$ of the space of coupling constants, and the dimension 
$D_S$ of the non-analyticity is such that the co-dimension 
\bea
C=D-D_S=1,
\eea 
then this non-analyticity is referred to as a phase-boundary.
Thus, a phase boundary is defined by non-analyticities of 
$f_b[K]$ which separate phases.
Whenever, by changing the coupling constants $K$, the phase boundary is crossed, 
the model is said to undergo a phase transition.

The bulk free energy $f_b[K]$, because it is convex,
is continuous everywhere.
Phase transitions are classified into first order,
respectively, continuous phase transitions depending on whether the
first derivatives of $f_b$ with respect to $K$ are continuous or not.
\begin{enumerate}
\item{We speak of a first order phase transition, if one (or more) 
of the first derivatives $\partial f_b/\partial K_i $ of the bulk free energy 
is discontinuous at the phase boundary (critical point). 
It means that an order parameter changes discontinuously across 
the phase boundary.}
\item{We speak of a continuous phase transition, if 
all first derivatives of the bulk free energy are continuous
at the phase boundary, meaning that an order parameter changes 
continuously across the phase boundary.
The non-analyticity of the phase boundary is reflected by 
discontinuities of certain higher order (than one) derivatives 
of the order parameter at the phase boundary.}
\end{enumerate}

In case of the Ising model the order parameter is the magnetization $M$,
which in absence of an external magnetic field is zero
for temperatures $T\geq T_c$, and $M$ is nonzero for $T<T_c$ (long range order),
where $T_c$ is the critical temperature.
Although (for specific Ising models) the magnetization $M$ changes 
continuously from zero ($T\geq T_c$) to nonzero values for $T< T_c$ 
(thus the phase transition is continuous, since $M$ is 
a first derivative of $f_b$), 
the magnetization $M$ as function of $T$ is nonanalytic at $T=T_c$, {\em i.e.}
the left and right derivatives of $M$ with respect to $T$ differ at $T=T_c$.

For quantum field theories the analogue of the free energy
is the generating functional $W$ ({\em e.g.} Eq.~(\ref{connGreenfies}))
of connected Green functions. 
The above classification of phase transitions applies to QFTs too. 
An order parameter can usually be given in terms of a first derivative of 
$W$, and we speak of a continuous phase transition whenever the order parameter 
changes continuously across the phase boundary.
The phase boundary is then characterized by the non-analyticities of $W$
in certain couplings.

An important quantity in the study of continuous phase transitions is the
correlation length.
The correlation length $\xi$ is the characteristic length scale 
of fluctuations of a physical quantity such as the local magnetization 
$S_i$ around its average $\langle S_i\rangle$.
Let us consider the ``connected'' two-point correlation function
defined by
\bea
G_C(\vec r_i-\vec r_j)
=\langle S_iS_j\rangle -\langle S_i\rangle \langle S_j\rangle,
\eea
which can be obtained by differentiating the free energy
with respect to external fields (or sources).
The correlation function $G_C$ describes local fluctuations parallel 
to the long range ordering in the system,
and the properties of $G_C$ are:
\bea
G_C(\vec r)&\sim& \frac{{\,\rm e}^{-r/\xi}}{r^{(d-1)/2} \xi^{(d-3)/2}},
\qquad r\equiv|\vec r|\gg \xi,\nonu\\
G_C(\vec r)&\sim& \frac{1}{r^{d-2+\eta}},
\qquad r\equiv|\vec r|\ll \xi,\label{GCcordef}
\eea
where $\xi$ is called the correlation length, 
and $\eta$ is the anomalous dimension.
\subsection{The scaling hypothesis}
A very useful mathematical tool in the study of continuous phase transitions 
and critical phenomena is provided by the existence of scaling laws. 
For a system close to a critical point separating different phases, 
a set of critical exponents is introduced to describe 
both the scaling of thermodynamic quantities, such as a the magnetization 
$M$, specific heat, etc., 
as well as local ordering, correlations, and fluctuations given in terms of
correlation functions and a correlation length $\xi$.

The scaling laws are motivated by assuming that the only relevant 
length scale near criticality is the correlation length $\xi$.
The renormalization group (RG) methods of Kadanoff and Wilson, which will 
be discussed later on, show that this assumption is incorrect.
Apart from the correlation length another length scale is crucial 
for understanding critical phenomena, namely the microscopic length 
scale, {\em e.g.} 
the lattice spacing $a$.

There is a threefold way of considering the phenomenon of scaling, namely,
the scaling form for the so-called equation of state (EOS), 
the scaling form for the singular part of the free energy,
and the scaling form for the correlation function.
\subsection{The scaling hypothesis for the equation of state}
The ``equation of state'' for an order parameter 
(which characterizes a certain phase of the system) can be obtained by
differentiating the free energy of the system with respect to some external 
local source or external field, to which the order parameter is coupled.
In the context of magnetic systems, the EOS of a magnet is written as
\bea
h=f_e(M,T),\label{eosmagnet}
\eea
where $h$ is the external magnetic field, $M$ the spontaneous magnetization, 
and $T$ the temperature 
($f_e$ follows from the thermodynamics of some microscopic model).
It is implicitly assumed that there is some microscopic model,
{\em e.g.} an Ising model with spins $S_i$ on a lattice (with lattice spacing $a$),
producing the function $f_e$ of Eq.~(\ref{eosmagnet}) relating 
the macroscopic quantities $M$, $T$ and $h$ in the thermodynamic limit.
The magnet undergoes a phase transition at $T=T_c$, where for temperatures 
$T<T_c$ the system exhibits long range order, {\em i.e.} the spins degrees of
freedom become highly correlated and align to form a nonzero magnetization $M$,
even in absence of an external magnetic field $h=0$. 
For temperatures $T>T_c$ thermal fluctuations dominate the free-energy 
and spins are randomly oriented giving a zero magnetization $M=0$ at $h=0$.
This is a typical example of a continuous phase transition in a thermodynamic 
system; the order parameter $M$ changes continuously from zero for $T\geq T_c$
to nonzero values at temperatures $T<T_c$, 
in the limit $h\rightarrow +0$.

Based on experimental results, Widom \cite{wi65} discovered that the EOS can be 
written as a function of a single variable near a critical point $T=T_c$:
\bea
h=M^\delta {\cal F}_{e}\(t M^{-1/\beta}\), \label{widom1}
\eea
where $t=T-T_c$ is the temperature relative to the critical point.
Thus Eq.~(\ref{widom1}) is a relation between the scaled variables
$hM^{-\delta}$ and $t M^{-1/\beta}$.
The function ${\cal F}_{e}$ is regular at $t=0$, thus we can write
\bea
h={\cal F}_{e}\(0\)M^\delta +{\cal F}^\prime_{e\pm}\(0\) 
t M^{\delta-1/\beta}, \qquad (t=T-T_c)\label{eosexp}
\eea
where the minus sign in ${\cal F}^\prime_{e\pm}$ corresponds to
$t=T-T_c<0$, and the plus sign to $t>0$.
This notation is used to reflect the fact that the function 
${\cal F}_{e}$ though continuous is not analytic at $t=0$ ($T=T_c$).
The critical exponent $\delta$ follows from Eq.~(\ref{eosexp}) at $t=0$:
\bea
M\sim h^{1/\delta},\qquad t=0. \label{deltaexpdef}
\eea
The critical exponent $\beta$ follows from Eq.~(\ref{eosexp}) 
at zero external field $h=0$:
\bea
M\sim t^\beta,\qquad h=0. \label{betaexpdef}
\eea
The critical exponent $\gamma$ describes the scaling of the
magnetic susceptibility $\chi$, which is defined as
\bea
\chi\equiv \frac{\partial M}{\partial h}\biggr|_{h=0},\label{magsusdef}
\eea
close to the critical point:
\bea
\chi \sim t^{-\gamma}.\qquad (h=0) \label{gammaexpdef}
\eea
Differentiating Eq.~(\ref{eosexp}) with respect to $h$ and putting $h=0$ yields
the scaling relation
\bea
\gamma=\beta(\delta-1). \label{widscallaw1}
\eea
Thus the critical exponent $\gamma$ is not really a new critical exponent.
\subsection{The scaling hypothesis for the free energy}
The scaling laws for the ``singular part'' $f_s$ of the free energy $f_b$ 
is written as
\bea
f_s(t,h)=t^{2-\alpha}{\cal F}_{f}\(h t^{-\Delta}\), \label{singfree}
\eea
near criticality, where $F_{f}$ is regular at $h=0$.
The critical exponent $\alpha$ is related to the specific heat. 
The magnetization $M$ is obtained by differentiation of $f_s$ 
with respect to the external field $h$:
\bea
M\equiv -\frac{1}{k_BT} \frac{\partial f_s}{\partial h}\sim 
t^{2-\alpha-\Delta}{\cal F}^\prime_{f}\(h t^{-\Delta}\).
\eea
Since the scaling of $M$ near criticality is given by the exponent 
$\beta$ Eq.~(\ref{betaexpdef}), we find the scaling law 
\bea
\beta=2-\alpha-\Delta.\label{freeElaw1}
\eea
Subsequently the magnetic susceptibility $\chi$, Eq.~(\ref{magsusdef}), 
follows from one more differentiation of $f_s$ with respect to $h$:
\bea
\chi\sim t^{2-\alpha-2\Delta}{\cal F}^{\prime\prime}_{f}\(h t^{-\Delta}\).
\eea
Close to criticality $\chi$ should tend towards $t^{-\gamma}$ 
as $h\rightarrow 0$, see Eq.~(\ref{gammaexpdef}).
Hence we find another scaling law
\bea
2-\alpha-2\Delta=-\gamma.\label{freeElaw2}
\eea
After eliminating $\Delta$ in Eqs.~(\ref{freeElaw1}) and (\ref{freeElaw2})
we obtain the Rushbrooke scaling law
\bea
\alpha+2\beta+\gamma=2.\label{widscallaw2}
\eea
\subsection{The scaling hypothesis for the two-point correlation functions}
The connected two-point correlation function $G_C$ 
longitudinal ({\em i.e.} parallel) to the direction of ordering 
is defined in Eq.~(\ref{GCcordef}).
The scaling hypothesis for the two-point correlation function reads
\bea
G_C(r,t,h)=\frac{1}{r^{d-2+\eta}} {\cal F}_G\(r t^\nu, h t^{-\Delta}\),
\label{scalhypprop}
\eea
which, in momentum space, after a Fourier transformation, 
has the following form at zero external field $h=0$
\bea
\hat G_C(p,t)\sim \int {\rm d}^d r\,{\rm e}^{ipr} G_C(r,t,0)=
\frac{1}{p^{2-\eta}} \hat{\cal F}_G\(t^\nu/p\).
\eea
In analogy with Eq.~(\ref{GCcordef}), the quantity
\bea
\xi\sim t^{-\nu},\qquad h=0, \label{nuexpdef}
\eea
is defined as the correlation length.
One can derive that the magnetic susceptibility should be proportional to
the (longitudinal) two-point correlation function in momentum space
\bea
\hat G_C(p=0,t)\sim \chi, 
\eea
thus $p\ll t^\nu$.
Then the limit 
\bea
\lim_{p\rightarrow 0}\frac{1}{p^{2-\eta}}\hat{\cal F}_G\(t^\nu/p\)
\sim t^{-\gamma},
\eea
gives the scaling law
\bea
\gamma=\nu(2-\eta).\label{widscallaw3}
\eea

The Josephson scaling law follows from the requirement that 
the singular part of the free energy $f_s$ is independent
of microscopic length scales in the limit $t\rightarrow 0$ 
and at zero external field $h=0$, thus the only length scale is then 
given by the correlation length $\xi\sim t^{-\nu}$, thus
\bea
f_s(t,0)\sim \xi^{-d}\sim t^{\nu d} \sim t^{2-\alpha},
\eea 
where the last step follows from the definition of the critical 
exponent $\alpha$ in Eq.~(\ref{singfree})
Hence
\bea
d \nu =2-\alpha. \label{widscallaw4}
\eea
Since it involves the dimension $d$ of the system explicitly, 
Eq.~(\ref{widscallaw4}) is referred to as a hyperscaling law. 

The six critical exponents $\alpha$, $\beta$, $\gamma$, $\delta$, 
$\eta$ and $\nu$ are related by the four (hyper)-scaling laws
(\ref{widscallaw1}), (\ref{widscallaw2}), (\ref{widscallaw3}), 
and (\ref{widscallaw4}), thus only two of the critical 
exponents are independent.

The critical exponents are referred to as universal quantities;
they don't depend much on microscopic details of the underlying theory
describing macroscopic quantities.
Universality means that the critical exponents are determined by only a 
few global features of system; 
the symmetries of the Hamiltonian or action, the dimensionality, and whether 
or not the forces are short-ranged.

The above mentioned scaling hypothesis for the EOS, 
the singular part of the free energy and the correlation function, 
were for a long time predominantly experimentally motivated,
and only Landau theory or mean field theory was capable of computing explicitly
critical exponents.
However in Landau's mean field theory, only one set of 
critical exponents could be accounted for.
These are referred to as mean field exponents:
\bea
&&\alpha=0,\qquad \beta=1/2,\qquad \gamma=1,\nonu\\
&&\delta=3,\qquad \eta=0,\qquad \nu=1/2,
\eea 
and the hyperscaling laws are only satisfied at the critical dimension $d=4$.
Any other set of critical exponents satisfying the hyperscaling
laws would certainly be inconsistent with mean field theory.
But it is known experimentally that certain ferromagnetic systems could be 
consistently described by critical exponents satisfying hyperscaling laws, 
which were clearly incompatible with the mean field exponents.

The problem of mean field theory is that it is based on the assumption
that the only relevant length scale near criticality is the correlation 
length $\xi$.
This appears to be wrong.
In order to account for nonmean field critical exponents, 
and anomalous scaling laws ({\em i.e.} the critical exponent $\eta\not=0$), 
the microscopic length scale $a\sim 1/\Lambda$ plays a crucial role  
in the dimensional analysis\footnote{
Dimensional analysis, in this context, means
the analysis of scale transformations.} 
even when the correlation length is 
much larger than the microscopic scale or lattice spacing $a$.

Kadanoff \cite{ka66} presented an intuitive picture by 
``coarse graining'' degrees of freedom, and invented block spin 
transformations, in which a particular set of 
spin degrees of freedom is described by a single spin degrees of freedom.
His crucial assumption is that the Hamiltonian for the new coarse grained degrees
of freedom is of the same form as the Hamiltonian for the original degrees
of freedom, except that the couplings constants of the Hamiltonian 
now run as functions of a ``new'' larger microscopic length scale corresponding to 
the coarse grained degrees of freedom.
Kadanoff's block spin transformations, and the concept of running 
coupling constants provided a clear motivation for the scaling hypothesis 
of Widom {\em et al.}.
However Kadanoff's method could neither give rise to the critical exponents 
nor the so-called non-universal quantities, such as ${\cal F}_e(0)$ and 
${\cal F}_e^\prime(0)$ of Eq.~(\ref{eosexp}).
The flaw in Kadanoff's picture is the assumption that the course 
grained Hamiltonian is of the same form as the original 
Hamiltonian; he assumed that if the fundamental Hamiltonian described
only nearest neighbour spin-interactions, the coarse grained Hamiltonian 
would also have only nearest neighbour
 spin interactions.
This assumption is in general not valid. 
Wilson refined the concept of block spin transformation and coarse graining,
and his theory ``the renormalization group'' (RG) provides 
a constructive method to derive the course grained Hamiltonian, 
which he realized was not necessarily of the same form as the fundamental 
Hamiltonian.
In course graining degrees of freedom new types of interactions can appear, 
and need therefore to be taken into account. 
In this way, Wilson's RG methods are capable of explicitly computing 
the critical exponents, and non-universal quantities.
In the next section, we will discuss the RG of Wilson in the context
of quantum field theory.
\section{Wilson's renormalization group}\label{WRG}
This section is an introduction to the essential concepts of 
the renormalization group methods of K.G.~Wilson, Refs.~\cite{wi71,wiko74}. 
Large parts of this section are derived and compiled from the following 
references. Excellent reviews on the principles of Wilson's RG 
in statistical mechanics are given in 
Refs.~\cite{goldenfeld,wiko74}, and for quantum field theories, 
in Refs.~\cite{mirbook,zinjus,po84,amit}, see also \cite{apca75}.
For literature on the fundamental axioms of Euclidean and Minkowskian quantum 
field theory, and functional integration techniques we refer to  
Ref.~\cite{glimmjaffe}, and for a modern view of functional integration 
see Ref.~\cite{olivier}. 

Wilson elaborated and completed Kadanoff's argument, and his RG method
provides the constructive framework in which to understand critical phenomena,
universality and renormalization. Wilson's RG allows for explicit computation 
of the relationship between coupling constants at different length scales, at 
least approximately.
The basic meaning of a RG transformation is the redefinition or the 
change (running) of the coupling constants under a change of scale, 
and a rescaling of the fields, and to allow for the possibility that 
new local operators are generated during the RG transformation.

We will present the renormalization group (RG) in three steps.
The first step is a concrete realization of the coarse-graining 
transformations, and in the language of QFT it means that we have to 
construct a low-energy effective action or regulated action defining 
a particular generating functional from which 
the RG transformations can be derived (in principle).
The second step is the identification of the origin of critical 
phenomena and singular behavior with the existence of fixed points 
of the RG transformations. Also we discuss the concept of relevant, marginal, 
and irrelevant interactions (operators), and the flow of coupling
constants close to a particular fixed point.
Finally, in step three we discuss renormalization, and the continuum limit
of QFTs.
\subsection{Step 1. Regularization and coarse graining}\label{step1} 
Let us consider the following generating functional
described by some ``fundamental'' action $S_f$ in $d$ dimensional 
Euclidean space time, which is obtained after a standard 
Wick rotation \cite{wi54}
\bea
Z\[K,J\]&=&N\int{\cal D}\phi\,\exp\{-S_f\[K,\phi\]
-S_J\[J,\phi\]\},\\
S_f\[K, \phi\]&=& \int {\rm d}^dx\, 
\sum_{i=1}^{n_f} K_i O_i\[\phi(x)\], \label{funaction}
\eea
where the $K=(K_1,K_2,\dots,K_{n_f})$ are a set of 
$n_f$ coupling constants (some of which have a dimension),
and where the $O_i$ are some local operators of the elementary fields
$\phi=(\phi_1,\phi_2,\dots,\phi_a)$, thus 
\bea
{\cal D}\phi=\prod_{i=1}^a{\cal D}\phi_i.
\eea
In the Euclidean formulation, we suppose that the $O_i$ are 
positive operators, and that the couplings $K_i$ take only nonnegative values,
so that the action is positive and the generating functional ``well-defined'',
see also the discussion in section~\ref{sec_rev} and \cite{olivier}.
The source action $S_J$ is of the form Eq.~(\ref{sourceaction}).

The initial step in Wilson's approach is to introduce a cutoff $\Lambda$ in the 
action.
A convenient way to do that is by using the generalized Pauli--Villars
regularization (see Ref.~\cite{pavi49,zinjus} and {\em e.g.} \cite{fu94} 
and references therein).
We define the ``regularized action'',
\bea
S^{reg}_\Lambda\[K,\phi\]=
N_{pv}(\Lambda)\int {\cal D}\phi_{pv}\, 
\exp\{-S^{pv}_\Lambda\[K,\phi,\phi_{pv}\]\},\label{regaction}
\eea
where the fields $\phi_{pv}$ are a set of Pauli--Villars fields.
An example of such an action $S^{pv}_\Lambda$ for scalar 
field theory is 
\bea
S^{pv}_\Lambda\[K,\phi,\phi_{pv}\]&=&\int {\rm d}^dx\,
\bigg[\frac{1}{2}\phi(-\Delta+m^2)\phi
-\frac{1}{2}\phi_{pv}(-\Delta+\Lambda^2)\phi_{pv}\nonu\\
&+&V(\phi+\phi_{pv})\bigg],
\eea
where the initial action $S_f$ was
\bea
S_f\[K,\phi\]=\int {\rm d}^dx\,
\bigg[\frac{1}{2}\phi(-\Delta+m^2)\phi+V(\phi)\bigg].
\eea
The mass $m$ and the potential $V$ are defined in terms of the coupling
constants $K$, and yet unspecified local operators.

The basic principle of Pauli-Villars regularization methods is
that any internal propagator of $\phi$ is replaced after the regularization
by the sum of the $\phi$ and $\phi_{pv}$ propagators:
\bea
\frac{1}{p^2+m^2}\longrightarrow \frac{1}{p^2+m^2}-\frac{1}{p^2+\Lambda^2}.
\eea
For modes of $\phi$ corresponding to energies $p^2\ll \Lambda^2$
the propagator is of the form
\bea
\frac{1}{p^2+m^2} -\frac{1}{\Lambda^2}
+\frac{1}{\Lambda^2}{\cal O}(p^2/\Lambda^2),
\eea
whereas for modes of $\phi$ corresponding to energies $p^2\gg \Lambda^2$
the propagator vanishes as fast as $\Lambda^2/p^4$ 
making Feynman diagrams finite.
In the continuum limit $\Lambda\rightarrow \infty$ the 
original propagator is recovered.
The Pauli--Villars field $\phi_{pv}$ does not correspond to a ``physical'' 
particle, since the action cannot be based on a Hermitian Hamiltonian, see 
Chapt. 7 of Zinn--Justin \cite{zinjus}.

In the abstract example given above, we introduced only a single 
Pauli--Villars field; in general more Pauli--Villars fields are needed 
to render all Feynman diagrams finite.
The Pauli--Villars regularization seems to work very well for Abelian gauge 
theories describing the interaction between fermions such as QED, since it 
is possible to regularize in a gauge invariant manner keeping the 
Lorentz covariance. Moreover the Pauli--Villars regularization is supposed to work
also nonperturbatively.
A problem turns up when considering symmetries of an initial action $S_f$.
After regularization it might be the case that a continuous symmetry of 
the initial action is lost in the regularization process by the appearance of 
operators, which violate that particular symmetry.
Such a violation of the symmetry is called an anomaly.
An example in fermionic gauge field theories is the well-know Adler--Bell--Jackiw 
$U(1)$-axial anomaly, Refs.~\cite{ad69,beja69}, see also section~\ref{secwti}.

Let us consider the regularized action Eq.~(\ref{regaction}) in more detail.
The functional integral over Pauli--Villars field $\phi_{pv}$ introduces
a smearing of interactions at modes of $\phi$ with energy scale $E\sim \Lambda$.
Due to such smearing effects the action $S^{reg}_\Lambda$ 
is nonlocal in the fields $\phi$.
Suppose we split the fields $\phi$ into high- and low-energy parts:
\bea
\phi=\phi_h+\phi_l,\label{projhighlow}
\eea
where the energy $E=|p|>\Lambda$ for $\phi_h$ and $E\leq \Lambda$ for $\phi_l$.
Then, since the Pauli--Villars construction is such that the high energy 
modes $\phi_h$ are canceled by the 
Pauli--Villars fields, the action is independent of $\phi_h$.
The remaining fields $\phi_l$ are of energy $E<\Lambda$, and the 
action $S^{reg}_\Lambda$ is approximately local in low energy 
modes with $E\ll \Lambda$.
The idea is that the nonlocal action $S^{reg}_\Lambda$ can be expanded 
in terms of local operators $O_i$:
\bea
S^{reg}_\Lambda\[K,\phi\] =S^{eff}_\Lambda\[\kappa(\Lambda),\phi_l\]
=\int{\rm d}^dx\,\sum_{i=1}^\infty \kappa_i(\Lambda) 
\Lambda^{d-d^{c}_i} O_i\[\phi_l(x)\],
\label{effaction}
\eea
where we have introduced the dimensionless running coupling constants  
\bea
\kappa(\Lambda)=(\kappa_1(\Lambda),\kappa_2(\Lambda),\dots, 
\kappa_\infty(\Lambda)),
\eea
and new local interactions $O_i\[\phi_l(x)\]$, $i=n_f,\dots,\infty$.
Since the action is dimensionless, $d^{c}_i$ is the canonical 
dimension (in units of mass) of the operator $O_i$,
\bea
{\rm dim}\[O_i\]=d^{c}_i.\label{candimdef}
\eea
Now that we have an action spanned by a space of an infinite number 
of couplings and 
operators $O_i$, the question arises: how can this be useful?

The crucial observation is that operators and corresponding coupling constants
can be classified as relevant or irrelevant.   
It will turn out that it is sufficient to consider only the (in practice) 
finite set of operators $O_i$ ($1 < i \leq n$) and corresponding 
coupling constants, which are relevant with respect 
to a particular fixed point, (see later discussion in this section), 
so we write
\bea
S^{eff}_\Lambda\[\kappa(\Lambda),\phi_l\]
\approx\int{\rm d}^dx\,\biggr\{
\sum_{i=1}^n \kappa_i(\Lambda) \Lambda^{d-d^{c}_i}
O_i\[\phi_l(x)\]+\mbox{irrelevant}\biggr\},
\label{effactionrel}
\eea
and the property of the``irrelevant'' part describing the irrelevant operators 
and couplings is that it gives rise to corrections to Green functions in 
terms of positive powers 
of $E/\Lambda$, and the ``irrelevant'' part becomes negligible 
for $E$ sufficiently smaller than $\Lambda$ ($E\ll \Lambda$).

The apparent straightforward decomposition of $\phi$'s into $\phi_l$ 
and $\phi_h$, Eq.~(\ref{projhighlow}),  
is in practice nontrivial and rather technical.
The functional space of fields $\phi$ is chosen such that the $\phi(x)$ 
have Fourier transforms:
\bea
\phi(x)\sim \int{\rm d}^dp\, {\,\rm e}^{ipx}\hat \phi(p).
\eea
Then think of $\phi_l$ in terms of Fourier modes $\hat\phi(p)$ 
with $E=|p|\leq \Lambda$, and $\phi_h$ as modes $\hat\phi(p)$ 
with $E=|p|> \Lambda$:
\bea
\phi_l\sim \int_{|p|\leq \Lambda}{\rm d}^dp\, {\,\rm e}^{ipx}\hat \phi(p),
\qquad
\phi_h\sim \int_{|p|> \Lambda}{\rm d}^dp\, {\,\rm e}^{ipx}\hat \phi(p).
\label{philphih}
\eea
In this way, the measure $N(\Lambda)\int{\cal D}\phi_{l}$ introduces
a momentum cutoff $\Lambda$ in the SDEs for the momentum space Green functions.
For technicalities involved in the ``projection'' 
leading to Eq.~(\ref{projhighlow}), 
and the cutoff $\Lambda$, we refer to \cite{glimmjaffe} (Chap. 8).

If we compare this local ``low-energy effective'' action $S^{eff}_\Lambda$, 
Eq.~(\ref{effactionrel}),
with the fundamental or initial action $S_f$ of (\ref{funaction}), we see that 
new coupling constants $\kappa_i(\Lambda)$ 
and new operators $O_i$ with $n\geq i>n_f$ have been introduced, and that the 
couplings are now defined as functions of $\Lambda$.
In principle the couplings $\kappa_i(\Lambda)\Lambda^{d-d^{c}_i}$ 
can and will differ from the initial
coupling constants $K_i$ ($i=1,\dots,n_f$) of $S_f$.
These new couplings or new interactions are a results of the nonlocality of the regularized action
$S^{reg}_\Lambda$.

The fundamental notion introduced by Wilson is that, 
through coarse graining the degrees of freedom, new local
operators appear with corresponding couplings constants, and
the coupling constants ``run'' as a function of the cutoff $\Lambda$ 
(the cutoff is analogous to the microscopic length scale in lattice models 
$a\sim 1/\Lambda$).
The high energy modes $\phi_h$ are integrated out thereby introducing new 
interactions and running coupling constants.

From the local low energy effective action $S^{eff}_\Lambda$, 
Eq.~(\ref{effaction}), we construct a generating functional 
\bea
Z_\Lambda\[\kappa(\Lambda),J_\Lambda\]
\equiv N(\Lambda)\int{\cal D}\phi_{l}\,
\exp\{-S^{eff}_{\Lambda}\[\kappa(\Lambda),\phi_{l}\]-S_J\[J_\Lambda,\phi_{l}\]\},
\label{geneffdef}
\eea
where the external sources $J_{\Lambda}$ are introduced to generate the 
low-energy Green functions ($E\ll \Lambda$). 
The sources $J_\Lambda$ explicitly depend on $\Lambda$, 
and in momentum space we suppose that the sources $J_\Lambda(p)$ vanish for 
momenta $p>\Lambda$. We only probe physics at a scale $E<\Lambda$.
The generating functional $Z_\Lambda$ should be defined via the 
implementation of the SD scheme as discussed in section~\ref{sec_rev}, 
with a normalization
$N(\Lambda)$ so that $Z_\Lambda\[\kappa(\Lambda),J_\Lambda=0\]=1$.

We can repeat the above coarse graining and consider
now an effective action at a cutoff $\Lambda^\prime<\Lambda$.
Split the fields $\phi_l$ into:
\bea
\phi_l=\phi_{l'}+\phi_{h'},\label{philprime}
\eea
where the energy $E$ is $\Lambda>E>\Lambda'$ for $\phi_{h'}$ and 
$E\leq \Lambda'$ for $\phi_{l'}$.
Then we define a coarse grained local effective action at scale $\Lambda'$ as
follows:
\bea
N(\Lambda') \exp\{-S^{eff}_{\Lambda'}\[\kappa(\Lambda'),\phi_{l'}\]\}=
N(\Lambda)\int{\cal D}\phi_{h'}\,
\exp\{-S^{eff}_{\Lambda}\[\kappa(\Lambda),\phi_{l}\]\},\label{coarsegrainact}
\eea
where 
\bea
S^{eff}_{\Lambda'}\[\kappa(\Lambda'),\phi_{l'}\]\approx
\int{\rm d}^dx\,\sum_{i=1}^{n} \kappa_i(\Lambda')
{\Lambda'}^{d-d^{c}_i}
O_i\[\phi_{l'}(x)\].
\label{effactionprime}
\eea
In principle the right-hand side 
of Eq.~(\ref{coarsegrainact}) corresponds to a 
nonlocal action for energy scales scales $E\sim \Lambda'$, however again we 
assume it is well approximated for energies $E<\Lambda'$ by
the local action $S^{eff}_{\Lambda'}$.
Furthermore, by definition, the coarse grained effective action 
$S^{eff}_{\Lambda'}$ should yield the same generating functional, thus we have 
the identity
\bea
Z_{\Lambda'}\[\kappa(\Lambda'),J_{\Lambda'}\]
=Z_{\Lambda}\[\kappa(\Lambda),J_{\Lambda}\], 
\label{genfieinv}
\eea
since
\bea
N(\Lambda') \int{\cal D}\phi_{l'}\,\exp\{
-S^{eff}_{\Lambda'}\[\kappa(\Lambda'),\phi_{l'}\]\}
=N(\Lambda)\int{\cal D}\phi_{l}\,
\exp\{-S^{eff}_{\Lambda}\[\kappa(\Lambda),\phi_{l}\]\}.
\eea
Hence the generating functional $Z_\Lambda$ is independent of $\Lambda$,
{\em i.e.}
\bea
\Lambda\frac{{\rm d} Z_\Lambda}{{\rm d}\Lambda}=0. \label{basicRG}
\eea

The identities (\ref{genfieinv}) and (\ref{basicRG})
define a RG transformation ${\cal R}$ in the space of coupling constants
(thus in the space of actions) mapping couplings $\kappa(\Lambda)$ to 
couplings $\kappa(\Lambda')$:
\bea
\kappa(\Lambda')={\cal R}\[\Lambda'/\Lambda,\kappa(\Lambda)\].
\eea
The RG transformation ${\cal R}$ can be derived from 
the generating functional $Z_\Lambda$ given in terms of
an low energy effective action $S^{eff}_{\Lambda}$.
Since $Z_\Lambda$ is defined via the SDE, positivity, 
and normalization (section~\ref{sec_rev}),
the solutions (or approximate solutions) of SDEs for the Green function 
of $\phi_l$ ultimately provide us with the RG transformation ${\cal R}$.
In what follows we discuss the properties of ${\cal R}$.
\subsection{Step 2. The origin of singular behavior 
and anomalous scaling}\label{subsec_step2}
The RG transformation ${\cal R}$ maps couplings $\kappa(\Lambda)$ to 
$\kappa(\Lambda')$,
\bea
\kappa(\Lambda')={\cal R}\[\Lambda'/\Lambda,\kappa(\Lambda)\],
\qquad \Lambda'\leq \Lambda \leq 1,
\eea
where the cutoff's $\Lambda'$ and $\Lambda$ have been redefined as 
dimensionless quantities related to the unit energy scale.
The transformation ${\cal R}$ for different $\Lambda\leq 1$ form
a semi-group.
Two successive transformations with $\Lambda_1$ and $\Lambda_2$ should be
equivalent to a combined scale change of $\Lambda_1\Lambda_2$:
\bea
\kappa(\Lambda_0)&=&{\cal R}\[1,\kappa(\Lambda_0)\],
\qquad \Lambda_0\leq 1,\label{RG0}\\
\kappa(\Lambda_1)&=&{\cal R}\[\Lambda_1/\Lambda_0,\kappa(\Lambda_0)\],
\qquad \Lambda_1\leq \Lambda_0,\label{RG1}\\
\kappa(\Lambda_1\Lambda_2)&=&{\cal R}\[\Lambda_2,\kappa(\Lambda_1)\]\nonu\\
&=&{\cal R}\[\Lambda_2,{\cal R}\[\Lambda_1/\Lambda_0,\kappa(\Lambda_0)\]\],
\qquad \Lambda_2 \leq 1,\label{RG2}
\eea
and thus (taking $\Lambda_0=1$)
\bea
{\cal R}\[\Lambda_1\Lambda_2,\kappa_0\]=
{\cal R}\[\Lambda_2,{\cal R}\[\Lambda_1,\kappa_0\]\],
\qquad \kappa_0\equiv \kappa(1).\label{RG3}
\eea
By construction the ${\cal R}$ satisfies the above group properties,
where we recall that the transformation ${\cal R}$ follows 
from Eq.~(\ref{genfieinv}).

The differential form of the RGE can be obtained by
taking $\Lambda_2=(1-\epsilon)$ in Eq.~(\ref{RG2}) 
and taking the limit $\epsilon\rightarrow +0$:
\bea
\Lambda_1\frac{{\rm d}\kappa_i(\Lambda_1)}{{\rm d}\Lambda_1}
=\frac{\partial {\cal R}_i\[\Lambda_2,\kappa(\Lambda_1)\]}{\partial \Lambda_2}
\biggr|_{\Lambda_2=1}\equiv \beta_i(\kappa(\Lambda_1)), \label{diffRG}
\eea
where the $\beta_i$ function is defined with respect to the particular coupling 
$\kappa_i$ of the set of couplings $\kappa=(\kappa_1,\kappa_2,\dots)$.
The differential RG equation (\ref{diffRG}) follows from the differential 
identity, Eq.~(\ref{basicRG}), for the generating functional.

A crucial step, in understanding anomalous scaling behavior and critical phenomena,
is the recognition of the importance of fixed points of the RG transformation.
A fixed point of the RG transformation is a point $\kappa_c$ in 
coupling constant space satisfying
\bea
\kappa_c={\cal R}\[\sigma,\kappa_c\]\quad \Longrightarrow \quad
\beta(\kappa_c)=0,
\eea
for any $\sigma< 1$.
A fixed point is a property of the transformation ${\cal R}$, that,  
in general, corresponds to nonanalytic behavior, hence it can only appear 
in a system with infinitely many degrees of freedom \cite{yale52}.
Fixed points are not necessarily isolated points. 
It is possible to have lines or hypersurfaces of fixed points,
in fact in a large part of this thesis we consider a critical line or curve.
Moreover, a RG transformation can have several isolated fixed points.

The fixed points are either classified 
as critical fixed points, or as Gaussian fixed points.
In the context of quantum field theory, we refer to a critical 
fixed point or an ultraviolet (UV) fixed point, 
if the continuum limit ({\em i.e.} taking the cutoff $\Lambda\rightarrow \infty$)
corresponds to
an interacting theory, in which a macroscopic quantity such as a
correlation length $\xi\sim 1/m$ (where $m$ is a mass) can be defined.
An UV fixed point of a $\beta$ function of coupling $\kappa_i$
is a point $\kappa_c$ with the properties:
\bea
\beta(\kappa_c)=0,\qquad
\frac{\partial\beta_i(\kappa)}{\partial\kappa_i}\biggr|_{\kappa=\kappa_c}< 0. 
\label{UVfixpointdef} 
\eea
We speak of a trivial or Gaussian fixed point if in the continuum limit
the interactions between fields (and thus particles) vanish, and
the theory reduces effectively to a free field or Gaussian theory.
In general a Gaussian fixed point can be classified as being an IR fixed point, 
{\em i.e.}, 
\bea
\beta(\kappa_c)=0,\qquad
\frac{\partial\beta_i(\kappa)}{\partial\kappa_i}\biggr|_{\kappa=\kappa_c}\geq 0.
\label{IRfixpointdef} 
\eea
Thus close to a trivial or Gaussian fixed point 
(corresponding to noninteracting theories), a macroscopic quantity such 
as a correlation length is not defined (there are no correlations at all).
An UV fixed point corresponds to interacting theories, and a correlation 
length $\xi$ can be defined; right at the critical fixed point the correlation
length diverges $\xi=\infty$ ($m=0$), see section~\ref{step3}.

The critical fixed points will describe singular critical 
behavior, such as phase transitions, and can correspond to phase boundaries, 
separating two distinct
phases of a model (whether or not a phase boundary exists depends 
on the so-called co-dimensions of the fixed point (see Ref.~\cite{goldenfeld}).)
Thus knowledge of the location of and nature of fixed points of a RG transformation
enables the phase diagram to be determined, and vice versa.
We will show now that the critical exponents follow from the RG transformation
in the neighbourhood of a critical fixed point.

Let us consider the RG transformation close to a critical point $\kappa_c$,
\bea
\kappa_i(\Lambda)=\kappa_{ci}+\delta\kappa_i(\Lambda),
\eea
since
\bea
\kappa_n(\Lambda')&=&{\cal R}_n\[\Lambda'/\Lambda,\kappa(\Lambda)\]\nonu\\
&=&\kappa_{cn}
+\sum_m\delta\kappa_m(\Lambda)
\frac{\partial {\cal R}_n\[\Lambda'/\Lambda,\kappa\]}{\partial \kappa_m}
\biggr|_{\kappa=\kappa_{c}}
+{\cal O}\((\delta \kappa)^2\)\nonu\\
&\approx&\kappa_{cn}+\delta\kappa_n(\Lambda'),\qquad \Lambda'<\Lambda.
\eea
Hence\footnote{Using the Einstein summation convention.}
\bea
\delta \kappa_n(\Lambda')=M_{nm}(\Lambda'/\Lambda) \delta\kappa_m(\Lambda),
\label{linRG}
\eea
where the matrix $M$
is defined as
\bea
M_{nm}(\Lambda'/\Lambda)\equiv
\frac{\partial {\cal R}_n\[\Lambda'/\Lambda,\kappa\]}{\partial \kappa_m}
\biggr|_{\kappa=\kappa_{c}}. \label{Mcoupldef}
\eea
The semi-group property of Eq.~(\ref{RG3}) implies that
\bea
M_{nl}(\Lambda''/\Lambda')M_{lm}(\Lambda'/\Lambda)=M_{nm}(\Lambda''/\Lambda).
\eea
The equation (\ref{linRG}) is called the linearized RG 
transformation in the vicinity of the fixed point.
The matrix $M$ is real, but in general not symmetric.
However, we assume that $M$ is diagonalizable, with real eigenvalues.
Therefore, for simplicity,  we assume that for a particular 
set of couplings $\tilde \kappa$ the matrix $\tilde M$ is diagonal:
\bea
\tilde M_{nm}(\Lambda'/\Lambda)=\lambda^{(n)}(\Lambda'/\Lambda)\delta_{mn},
\label{RGM}
\eea
so that
\bea
\delta \tilde \kappa_n(\sigma\Lambda)= \lambda^{(n)}(\sigma)
\delta\tilde \kappa(\Lambda),\qquad \sigma=\Lambda'/\Lambda.\label{eigenvaldelkap}
\eea
With Eq.~(\ref{RGM}) this implies that
\bea
\lambda^{(n)}(\Lambda''/\Lambda')\lambda^{(n)}(\Lambda'/\Lambda)
=\lambda^{(n)}(\Lambda''/\Lambda).
\eea
Thus, the solution to the eigenvalue $\lambda^{(n)}$ is 
\bea
\lambda^{(n)}(\sigma)=\sigma^{\eta_n}, \label{eigvalsol}
\eea
where the exponent $\eta_n$ is independent of $\sigma$, in general the exponent
is a function of the critical point, {\em i.e.} 
$\eta_n\rightarrow\eta_n(\kappa_c)$.
Now we can distinguish three cases, since $\sigma<1$:
\begin{enumerate}
\item[(1)]{$\lambda^{(n)}(\sigma)>1$, {\em i.e.} $\eta_n<0$, 
which corresponds to an eigenvalue of a ``relevant'' coupling $\tilde \kappa_n$.}
\item[(2)]{$\lambda^{(n)}(\sigma)=1$, {\em i.e.} $\eta_n=0$, 
which corresponds to an eigenvalue of a ``marginal'' coupling $\tilde \kappa_n$.}
\item[(3)]{$\lambda^{(n)}(\sigma)<1$, {\em i.e.} $\eta_n>0$, 
which corresponds to an eigenvalue of an ``irrelevant'' coupling 
$\tilde \kappa_n$.}
\end{enumerate}
The marginal eigenvalues usually correspond to logarithmic corrections to scaling.
Eq.~(\ref{eigenvaldelkap}) implies then that for a relevant coupling
$\delta\tilde \kappa(\sigma\Lambda)$ increases as $\sigma$ decreases;
for a marginal coupling $\delta\tilde \kappa(\sigma\Lambda)$ stays fixed 
(up to logarithmic corrections); and for an irrelevant coupling 
$\delta\tilde \kappa(\sigma\Lambda)$ shrinks. 
Thus the irrelevant couplings flow towards the fixed point for $\sigma\ll 1$, 
whereas the relevant couplings flow from the fixed point.
The terms relevant, marginal, and irrelevant couplings (eigenvalues) are always 
defined with respect to a particular fixed point.

So we have seen that from the RG transformation ${\cal R}$ 
we obtain the fixed points $\kappa_c$, the RG flows near a particular fixed 
point can be decomposed into relevant, marginal, and irrelevant couplings, 
and from the eigenvalues $\lambda^{(n)}$ close to the fixed point we can 
derive the ``critical exponents'' $\eta_n$. The critical exponents 
$\eta_n$, for many practical explicit cases, can be shown to be related 
to the critical exponents 
defined in section~\ref{sec_phasetrans}. 
Thus the RG transformation ${\cal R}$ enables us to calculate
the critical points and critical exponents.
In a mean field approach the exponents have rational values, whereas in general 
they can have irrational values.
In a case where critical exponents differ from the canonical 
({\em i.e.} mean field) values we speak of anomalous scalings, or 
anomalous dimensions.
The dynamics (which determine ${\cal R}$) of the system under consideration
are such that fluctuations around some mean-value of operators cannot be neglected.

Let us discuss some examples.
Close to a fixed point, it follows straightforwardly from 
Eq.~(\ref{eigenvaldelkap}) that the $\beta$ function for a coupling $\kappa_n$ 
of a set of couplings $\kappa(\Lambda)$\footnote{For 
which the matrix $M$ (Eq.~(\ref{Mcoupldef})) is diagonal.}, 
can be written as 
\bea
\beta_n(\kappa)\equiv \Lambda\frac{{\rm d}\kappa_n}{{\rm d}\Lambda}
\approx \eta_n(\kappa_c) \delta\kappa_n+{\cal O}\((\delta \kappa)^2\),
\qquad \delta \kappa_n=\kappa_n-\kappa_{cn},
\eea
where $\eta_n(\kappa_c)<0$ implies that $\kappa_n$ is a relevant coupling,
and $\eta_n(\kappa_c)>0$ implies that $\kappa_n$ is irrelevant.
Now differentiating $\beta_n$ with respect to $\kappa_n$,
\bea
\frac{\partial \beta_n(\kappa)}{\partial \kappa_n}\biggr|_{\kappa=\kappa_{c}}
\approx \eta_n,
\eea
leads, with Eqs.~(\ref{UVfixpointdef}) and (\ref{IRfixpointdef}),
to the conclusion that $\kappa_{c}$ is an UV fixed point for a relevant 
coupling, and $\kappa_{c}$ is an IR fixed point in case of an irrelevant coupling.

More complicated are in general the marginal couplings.
Consider the following RG of a single marginal coupling $\kappa$:
\bea
\kappa(\Lambda')&=&{\cal R}\[\Lambda'/\Lambda,\kappa(\Lambda)\],\\
{\cal R}\[\Lambda'/\Lambda,\kappa(\Lambda)\]&=&
\frac{\kappa(\Lambda)}{1+\kappa(\Lambda)\(\kappa_{c}-\kappa(\Lambda)\)
\log\Lambda/\Lambda'}. \label{RGmargex}
\eea
Then, when $\kappa(\Lambda)$ is ``fine-tuned'' sufficiently close to $\kappa_c$ so that
we can write
\bea
\delta\kappa(\Lambda')=\(1+\kappa(\Lambda)^2\log\Lambda/\Lambda'\) 
\delta \kappa(\Lambda),
\eea
clearly the eigenvalue of this equation is not of the form 
Eq.~(\ref{eigvalsol}) due to the logarithmic correction, which is typical
for marginal couplings.
The $\beta$ function can be obtained from Eq.~(\ref{diffRG}) and (\ref{RGmargex}), and reads
\bea
\beta(\kappa)=\kappa^2(\kappa_c-\kappa).
\eea
Thus the $\beta$ function has two fixed points, $\kappa=0$, respectively, 
$\kappa=\kappa_c$. According to the definitions given by 
Eqs.~(\ref{UVfixpointdef}) and (\ref{IRfixpointdef})
$\kappa=\kappa_c$ is an UV fixed point, and $\kappa=0$ an IR fixed point.

In practical applications of the RG method to quantum field theories, 
it turns out that the actual computation of critical 
fixed points is highly complicated, even more so when more than one coupling 
is relevant or marginal, therefore it
is sometimes useful to address the problem in the following manner.
Consider a model with two coupling constants $\kappa_1$ and $\kappa_2$,
then the fixed point $\kappa_c=(\kappa_{c1},\kappa_{c2})$ is a solution of
\bea
\beta_1(\kappa_{1},\kappa_{2})&=&0,\label{beta1}\\
\beta_2(\kappa_{1},\kappa_{2})&=&0.
\eea
Suppose we first solve $\beta_1$ for $\kappa_{1}$, and we assume 
that the solution can be expressed in terms of a 
function $f$ with the property
\bea
\kappa_{1}=f(\kappa_{2})\,\rightarrow\, \beta_1(f(\kappa_2),\kappa_2)=0.
\eea
Then $\kappa_{c2}$, provided that it exists, follows from 
the solution(s) of
\bea
\beta_2(f(\kappa_{2}),\kappa_{2})&=&0,
\eea
and $\kappa_{c1}=f(\kappa_{c2})$. The above sketched method is particularly useful
when the function $\beta_1$ has a much simpler structure 
than the function $\beta_2$, {\em i.e.} the function $f$ is rather easy 
to determine
from Eq.~(\ref{beta1}).
\subsection{Step 3. Renormalization and fine-tuning}\label{step3}
The renormalization of a QFT can be considered as the RG transformation 
of a set of so-called bare couplings $\kappa(\Lambda)$ defined at an UV 
cutoff $\Lambda$ to a set of couplings $\kappa(\mu)$, where 
$\mu\ll \Lambda$ is some energy scale related to the
resolution of an experimental apparatus (think of particle accelerators).
Thus the couplings $\kappa(\mu)$ are assumed to be related to 
experimentally measurable quantities (of course only interacting particles 
are measurable).
Then if the limit 
\bea
\kappa(\mu)=
\lim_{\Lambda\rightarrow \infty}{\cal R}\[\mu/\Lambda,\kappa(\Lambda)\],
\label{RGcontlim}
\eea
(with $\mu$ fixed) exists and can be arranged in such a way
that the set of couplings $\kappa(\mu)$ describe an interacting 
theory\footnote{At least one $\kappa_i$ of the set of couplings 
$\kappa(\mu)$ should have a non-Gaussian value.},
then this particular QFT is said to have a nontrivial continuum limit.
That is, the set of couplings $\kappa(\mu)$ (``macroscopic quantities'') 
obtained from experiments
are independent of any microscopic length scale $1/\Lambda$.
When a nontrivial continuum limit does not exists, it does not necessarily imply
that the corresponding QFT is useless, such a model can still
be used as a low energy effective field theory for physics corresponding to 
energies 
$E\leq \Lambda$, where the cutoff $\Lambda$ now has physical implications.
For gauge theories, the Ward-Takahashi identities
(or Slavnov--Taylor identities for non-Abelian gauge theories)
reflecting the local gauge symmetry are crucial for a non-trivial 
continuum limit to exist. 

The ``arrangement'' of the continuum limit, as mentioned previously depends
on the so-called fine-tuning of relevant and marginal interactions.
In order that the couplings $\kappa(\mu)$ coincide with {\em e.g.} experimental
values, one has to choose particular values for the relevant and marginal ``bare'' 
couplings $\kappa(\Lambda)$ which lie in the neighbourhood of a critical fixed 
point.
This is referred to as fine-tuning. 

Now that we have a method to identify couplings $\kappa$ and, consequently, 
the corresponding operators $O$ as relevant, marginal and irrelevant, 
the question arises; 
why are irrelevant interactions irrelevant?
The answer is: irrelevant interactions or couplings do not have to be fine-tuned,
and their effect can completely be absorbed by suitable modifications of the
fine-tuning of relevant and marginal couplings.
As a result, the effect of irrelevant interactions in the infrared region is suppressed
by factors $\(\mu/\Lambda\)^\eta$ with $\eta > 0$.
From Eqs.~(\ref{eigenvaldelkap}) and (\ref{eigvalsol}), 
and assuming $\kappa_i(\Lambda)$ is sufficiently close to $\kappa_{ci}$
we obtain that 
\bea
\beta_i(\kappa)\approx\eta_i(\kappa_i-\kappa_{ci})\quad \Longrightarrow \quad
\kappa_i(\mu)\approx \kappa_{ci}
+\(\mu/\Lambda\)^{\eta_i} (\kappa_i(\Lambda)-\kappa_{ci}).\label{kmusol}
\eea
Thus, if ${\eta_i}>0$ ({\em i.e.} $\kappa_i$ is irrelevant) 
the value of $\kappa_i(\mu)$ is insensitive to the bare value 
$\kappa_i(\Lambda)$ for $\mu\ll \Lambda$, and $\kappa_i(\mu)$ approaches the
IR fixed point $\kappa_{ci}$; the effect of $\kappa_i(\Lambda)$ is suppressed
by a factor $\(\mu/\Lambda\)^{\eta_i}$.
This also means that the IR fixed point of an irrelevant coupling is determined
in terms of the relevant and marginal couplings.

However, if ${\eta_i}<0$ ({\em i.e.} $\kappa_i$ is relevant)
the second term on the right-hand side of Eq.~(\ref{kmusol})
dominates, for $\mu\ll \Lambda$, and the only way to get a small value for $\kappa(\mu)$
in the infrared is to fine-tune $\kappa_i(\Lambda)$ precisely to the UV fixed point
$\kappa_{ci}$.
Marginal couplings (${\eta_i}=0$), because they usually are described 
by logarithmic scaling, are less restricted in the fine-tuning process.

The relevant and marginal couplings with UV fixed points correspond to 
interacting theories, and a correlation length $\xi=1/m$ can be defined; 
right at the critical fixed point the correlation
length diverges $\xi=\infty$ (thus zero mass $m=0$).
Usually the correlation length can be identified with a physical observable,
and in order to get a finite nonzero value for $m$, this requires
fine-tuning of bare couplings.
Suppose we have a model with one coupling $\kappa$ describing interactions, 
then
\bea
\Lambda\frac{{\rm d}\kappa}{{\rm d}\Lambda}=\beta(\kappa)\quad\Longrightarrow\quad
\log \Lambda/m=
\int\limits_{\kappa(m)}^{\kappa(\Lambda)}\frac{{\rm d}\kappa}{\beta(\kappa)},
\qquad m=1/\xi.
\eea 
Thus, keeping the correlation length $\xi=1/m$ fixed, in the continuum limit 
$\Lambda \rightarrow \infty$ the integral on the left-hand side diverges 
to $+\infty$.
Hence, in order that the right-hand side diverges too, 
$\kappa(\Lambda)$ has to be fine-tuned to $\kappa_c$ for an UV fixed point 
(provided that $\beta$ has a first order zero (or higher) at $\kappa_c$).

In order to pin-point whether a coupling is irrelevant or not, 
the critical exponent $\eta_i$ is usually written as
\bea
\eta_i=d^{c}_i-d-\gamma_i, \label{anodef}
\eea
where $d^c_i$ is the canonical dimension defined in Eq.~(\ref{candimdef}), 
$d$ dimension of the space-time manifold, and
$\gamma_i$ defined as the anomalous dimension of the corresponding operator $O_i$.
The existence of an anomalous dimension is a result of (non-trivial) dynamics near a critical point, and therefore the quantity 
\bea
d^{dyn}_i\equiv d^{c}_i-\gamma_i
\eea
is called the ``dynamical'' dimension of the operator $O_i$.
By neglecting the effect of dynamics, thus neglecting $\gamma_i$, 
$\eta_i$ is determined by the canonical dimension $d^c_i$ of the 
operator $O_i$ (this is referred to as naive dimensional analysis).
By so comparing the effective actions (\ref{effaction}) 
and (\ref{effactionprime}), we expect the couplings $\kappa$ to have the scaling relations  
\bea
\kappa(\Lambda')  \sim 
\kappa(\Lambda) \(\Lambda/\Lambda'\)^{d-d^c} \quad \Longrightarrow \quad
\beta(\kappa(\Lambda))=(d^c-d)\kappa(\Lambda).
\eea
In this context, operators with $d^c>d$ are labeled naively irrelevant,
and the anomalous dimension $\gamma$, Eq.~(\ref{anodef}), can turn 
naively irrelevant operators into relevant (or marginal) ones, and vice versa.

Since the generating functional is functional integral of the field
degrees of freedom, we can rescale the fundamental fields 
$\phi=(\phi_1,\phi_2,\dots)$.
This is equivalent to a redefinition of couplings and sources, see Zinn-Justin
Chapt.~8 \cite{zinjus}.
The couplings which correspond to a rescaling of the fundamental degrees
of freedom ({\em i.e.} the fields) 
are called $Z$ factors, and the remaining couplings $g$
are ``physical'' coupling constants.
For instance, suppose for a theory with two fundamental fields 
$\phi=(\phi_1,\phi_2)$, the action $S^{eff}_\Lambda$ Eq.~(\ref{effaction}) 
contains a term 
$\propto \kappa_i O_i\[\phi_1,\phi_2\]$, then we define:
\bea
g_i(\Lambda) Z_{\phi_1}^{n/2}(\Lambda,g(\Lambda))  
Z_{\phi_2}^{m/2}(\Lambda,g(\Lambda)) 
\equiv \kappa_i(\Lambda), 
\eea
where the operator $O_i$ contains $n$ powers $\phi_1$, and $m$ powers of 
$\phi_2$, and where $g=(g_1,g_2,\dots)$ is the set of ``physical'' 
coupling constants.
With such a redefinition of coupling constants, 
the connected Green functions now implicitly depend on couplings $g$ only.
Renormalizability is now the statement
that a connected Green function\footnote{We introduce the notation
$\langle\rangle_{connected}$ for connected time-ordered vacuum expectation 
values.} 
\bea
&&G^{(n,m)}_C[x_1,\dots,x_n,y_1,\dots,y_m,\Lambda,g(\Lambda)]\nonu\\
&&\qquad\sim\langle0|T\( \phi_{1l}(x_1)\cdots\phi_{1l}(x_n)
\phi_{2l}(y_1)\cdots\phi_{2l}(y_m) \)|0\rangle_{connected},
\eea
satisfies the identity
\bea
&&Z^{n/2}_{\phi_1}(\Lambda',g') Z^{m/2}_{\phi_2}(\Lambda',g')
G^{(n,m)}_C[x_1,\dots,x_n,y_1,\dots,y_m,\Lambda',g'] \nonu\\
&&\qquad=
Z^{n/2}_{\phi_1}(\Lambda,g) Z^{m/2}_{\phi_2}(\Lambda,g)
G^{(n,m)}_C[x_1,\dots,x_n,y_1,\dots,y_m,\Lambda,g], \label{rendef}
\eea
where $g'=g(\Lambda')$, and $g=g(\Lambda)$.
Thus
\bea
0=
\Lambda\frac{{\rm d}}{{\rm d} \Lambda} 
\left\{ Z^{n/2}_{\phi_1}(\Lambda,g) Z^{m/2}_{\phi_2}(\Lambda,g)
G^{(n,m)}_C[x_1,\dots,x_n,y_1,\dots,y_m,\Lambda,g]\right\}.\label{diffrendef}
\eea
The identities Eq.(\ref{rendef}) and (\ref{diffrendef}) follow in theory
from the RG invariance of the generating functional, 
Eq.~(\ref{genfieinv}), and the SDE (\ref{defSDE}).
They are supposedly exact\footnote{Though a proof of the indentities 
Eq.(\ref{rendef}) and (\ref{diffrendef}) is highly nontrivial, and cannot be given
for most practical applications of QFT beyond perturbation theory.}, 
if the generating functional is spanned by the entire (infinite) set 
of operators $O_i$.
However, the RG method tells us we only need to take into account the 
relevant and marginal operators, and then the identity
is an approximate one, with corrections of the irrelevant type
({\em e.g.} $(E/\Lambda)^n$, $n\geq 0$, and $E\ll \Lambda$ 
in momentum space).
As an example consider 
a connected two-point Green function (see Eq.~(\ref{connGreenfies})):
\bea
G^{(2)}_C(x,y)&=&G^{(2)}_C[x,y,\Lambda,g]\sim
\frac{\delta^2 
\log Z_\Lambda\[Z_\phi,g,J_\Lambda\]}{\delta J_\Lambda(y)\delta J_\Lambda(x)} 
\nonu\\
&\sim& \int_{p\leq \Lambda}{\rm d}^4p\,{\,\rm e}^{-ip(x-y)} 
G^{(2)}_C[p,\Lambda,g],
\eea
where $G^{(2)}_C[p,\Lambda,g]$ is the Fourier transform of
$G^{(2)}_C[x,y,\Lambda,g]$.

Analogous to Eq.~(\ref{rendef}), we have the identity 
\bea
Z_\phi(\Lambda',g')G^{(2)}_C[p,\Lambda',g']
=Z_\phi(\Lambda,g) G^{(2)}_C[p,\Lambda,g], \qquad p\leq \Lambda'<\Lambda.
\eea
Thus
\bea
&&0=\Lambda\frac{{\rm d}}{{\rm d} \Lambda} \left\{ Z_\phi(\Lambda,g) 
G^{(2)}_C[p,\Lambda,g]\right\}\quad\Longrightarrow\\
&&0=\[\eta(g)+
\sum_i\beta_i(g)\frac{\partial }{\partial g_i(\Lambda)}
+\Lambda\frac{\partial}{\partial \Lambda}\] G^{(2)}_C[p,\Lambda,g],\label{calsym}
\eea
where we have defined
\bea
\beta_i(g)\equiv
\Lambda\frac{{\rm d} g_i}{{\rm d}\Lambda},\qquad
\eta(g)Z_\phi\equiv\Lambda\frac{{\rm d} Z_\phi}{{\rm d}\Lambda}.
\eea
Equation~(\ref{calsym}) is an example of a 
Callan--Symanzik equation, see Ref.~\cite{ca70, sy70}.
If we consider the asymptotic behavior, or short distance behavior, {\em i.e.}
\bea
\xi^{-1}\ll p \ll \Lambda, \label{asymshortdef}
\eea
of the connected two-point function $G^{(2)}_C$:
\bea
G^{(2)}_C[p,\Lambda,g]\approx \frac{1}{\Lambda^2}\(\frac{\Lambda}{p}\)^{2-\eta(g)}
\[1+a\(\frac{m}{p}\)^{2-\eta}
+b\(\frac{p}{\Lambda}\)^{\sigma_1}+\cdots\],
\eea
where $a,\,b,$ are nonuniversal constants, the exponent $\sigma_1$ is positive 
($\sigma_1>0$), and $m=1/\xi$.
Then such an asymptotic behavior approximately satisfies the Callan-Symanzik 
Eq.~(\ref{calsym}) ({\em i.e.} up to corrections of the irrelevant type)
provided the physical couplings are close to their 
critical values, {\em i.e.} $\beta_i(g)\approx 0$.

Naive dimensional analysis implies that $Z_\phi$ is of the marginal type, 
$\eta=0$, hence $\eta$ is referred to as twice the anomalous dimension 
of the field $\phi$, or in fact the anomalous dimension of the kinetic term. 
The above derivation of the Callan--Symanzik equation for a 
two-point Green function can straightforwardly be generalized for 
arbitrary $n$-point Green functions.

In the context of renormalizability,
we conclude this section by mentioning some properties
of gauge theories.
As mentioned previously,
gauge symmetries are continuous local symmetries of the classical action,
and give rise to Ward--Takahashi identities.
These so-called WTIs follow in principle from the SDEs, and are crucial in 
determining whether or not a nontrivial continuum limit exists.

A problem connected with the gauge symmetry is the gauge fixing procedure.
The problem occurs in the functional integration, 
where the integration of paths, which are mutually connected 
by gauge transformations, and hence have the same functional weight,
give rise to uncontrollable infinities. 
To control these infinities, one has to 
gauge fix the generating functional.
The BRST formalism, see for an introduction Ref.~\cite{henteit}, 
takes care of this gauge fixing mechanism, and for QED (and thus the GNJL model)
it is similar to the Gupta--Bleuler formalism.
\section{The Goldstone mechanism}\label{sec_goldstone}
The Goldstone theorem was first suggested by Goldstone \cite{go61}
and later proved by Goldstone, Salam, and Weinberg \cite{gosawe62}. 
Many textbooks on QFT discuss the Goldstone mechanism extensively;
we refer to Refs.~\cite{mirbook,itzzub}.
The first example of the Goldstone mechanism in relativistic
particle physics is given by the Nambu--Jona-Lasinio model, \cite{najola61}, 
where the chiral symmetry is spontaneously broken.

Spontaneous symmetry breaking is the phenomenon that an invariance 
of the action of a model ({\em i.e.} a symmetry on the dynamical level) 
is not an invariance of its vacuum.
If the broken symmetry is a continuous one, it gives rise to the Goldstone 
mechanism.

In classical field theory, a continuous symmetry described by a Lie group $G$
of an action implies, according to Noether's theorem, the existence 
of conserved local currents $j_\mu^a(x)$
\bea
\partial^\mu j_{\mu}^a(x)=0,
\eea 
with $a=1,2,\dots,n$, where $n$ is the number of the group generators.  
Then the space integrals of the time component of the currents $j^\mu_a$
define conserved charges $Q^a$ corresponding to continuous symmetry 
(by implementation of the Stokes theorem neglecting surface terms):
\bea
Q^a\equiv \int{\rm d}^3x\, j_0^a(x)\quad \longrightarrow \quad
\frac{{\rm d}}{{\rm d} t}Q^a(t)=0.
\eea
The charges $Q^a$ can be considered as the generators of the Lie group $G$,
{\em i.e.} the action or Lagrangian is invariant under unitary transformations
of the fundamental field $\phi$ 
\bea
U(\theta)\in G,\qquad
\phi \rightarrow \phi'= U^{\dagger}(\theta)\phi U(\theta),
\qquad  U(\theta)=\exp\{i\theta^a Q^a\}.
\eea
The infinitesimal transformations $\theta^a\ll 1$ 
of operators $O_i$ of fields $\phi$ are generated by $Q_a$:
\bea
\delta O_i\[\phi\] = -i\theta^a\{Q^a,O_i\[\phi\]\},\label{Qgendef}
\eea
where $\{\,,\,\}$ is the Poisson bracket. In quantum field theory the Poisson 
bracket is replaced by a commutator if $Q$ is a bosonic operator
or by an anti-commutator if $Q$ is fermionic .

In a quantum field model with a conserved
current operator $j^a_\mu(x)$,
spontaneous symmetry breaking is now characterized by the condition that there 
exist an operator $O_i$ for which
\bea
\langle 0|\delta O_i\[\phi\]|0 \rangle \not=0, \label{spontsymbrdef}
\eea 
thus at least one of the charges $Q^a$ of Eq.~(\ref{Qgendef}) 
does not annihilate the vacuum; the vacuum is not invariant 
under the continuous symmetry transformation.
If the operator $O_i\[\phi\]$ breaking the invariance of the vacuum
is a composite operator, instead of being a single field, we speak of 
a dynamically broken symmetry.

The Goldstone theorem states the following.
Consider a Lorentz and translational invariant local field theory, 
with conserved currents $j_a^\mu$ relating to a Lie group $G$. Assume that 
this symmetry is spontaneously broken so that Eq.~(\ref{spontsymbrdef}) holds. 
Then there are massless particles (The Nambu--Goldstone (NG) bosons) 
with the same quantum numbers as the operators $O_i[\phi]$, which couple both
to the currents $j^a_\mu$ and operators $O_i[\phi]$.
The NG--bosons are indeed bosons if the $O_i[\phi]$ is a bosonic operator, 
however in general this need not be the case.

The degeneracy of the vacuum in a theory with a spontaneously broken symmetry
is connected with existence of a phase transition in the model, 
hence vacuum degeneracy can only occur in systems with infinitely many 
degrees of freedom. 

In the framework of SDEs a particular continuous symmetry gives rise to 
so-called Green-Ward--Takahashi or Ward--Takahashi identities, 
and the existence of a massless particle follows from the low-energy 
limits ($p\rightarrow 0$) of corresponding Ward--Takahashi identities.
These low-energy relations are very useful
when a symmetry is not completely exact, but explicitly broken
by a small external field (think of the external field $h$ in magnetic systems,
or a bare particle mass $m_0$ in case of chiral symmetry).
If the spontaneous symmetry breaking mechanism is much stronger than
the explicit symmetry breaking, then one can derive, in case of chiral symmetries
(see next section),
the partial-conserved-axial-current (PCAC)
relations.

Let us discus now the specific case of dynamical chiral symmetry breaking.
\subsection{Chiral symmetry}
Chiral symmetry is a special continuous symmetry of massless fermions.
In nature, most fermions that we know of do have masses 
(even neutrinos are not (anymore) believed to be completely massless), hence the 
chiral symmetry is a broken symmetry.
Consider the following Lagrangian in $4$ dimensions,
\bea
{\cal L}=\bar\psi(x) \fslash D \psi(x), \label{exampleL}
\eea
where $\fslash D=\gamma^\mu D_\mu$ is a covariant derivative supposedly 
containing gauge interactions.
Then, the Lagrangian is invariant under the chiral transformations
\bea
\psi \rightarrow \psi^\prime={\,\rm e}^{i\theta \g5}\psi,\qquad
\bar\psi \rightarrow \bar\psi^\prime=\bar\psi{\,\rm e}^{i\theta \g5},
\label{chiralsymmetry}
\eea
where $\theta$ is some arbitrary parameter, and the $\g5$ matrix is defined as
\bea
\g5 =i\gamma^0\gamma^1\gamma^2\gamma^3=\g5^{\dagger} ,\qquad \{\gamma^\mu,\g5\}=0, 
\qquad \g5^2=1.
\eea
The chiral symmetry implies that the Lagrangian can be written
in terms of ``left-handed', and ``right-handed'' spinors,
\bea
\psi_L=\(\frac{1+\g5}{2}\)\psi,\qquad
\psi_R=\(\frac{1-\g5}{2}\)\psi,
\eea
so that
\bea
{\cal L}=\bar\psi_L \fslash D \psi_L+\bar\psi_R\fslash D \psi_R.
\eea
A mass term $m_0\bar\psi\psi$ is not invariant under the chiral transformations,
and the Lagrangian with mass term would read
\bea
{\cal L}=\bar\psi_L \fslash D \psi_L+\bar\psi_R \fslash D \psi_R
-m_0\bar\psi_L\psi_R-m_0\bar\psi_R\psi_L.
\eea
Thus a mass term mixes the left-handed and right-handed spinors. 
The chiral symmetry gives rise to a conserved axial current:
\bea
j^{5\mu}=\bar\psi \gamma^5 \gamma^\mu \psi,\qquad
\partial_\mu  j^{5\mu}=0.
\eea
With a bare mass term, the chiral symmetry is explicitly broken and
the current is not conserved
\bea
\partial_\mu  j^{5\mu}= m_0\bar\psi i\gamma^5 \psi. \label{explicitbreak}
\eea
The current $j^{5\mu}$ defines a charge $Q^5$ as the generator
of the chiral symmetry.

In the framework of spontaneous symmetry breaking given by 
Eq.~(\ref{spontsymbrdef}),
let us consider a composite pseudoscalar operator 
$O\[\bar\psi,\psi\]=\bar\psi(x) i\g5 \psi(x)$.
Under the infinitesimal chiral transformations (\ref{chiralsymmetry})
({\em i.e.} $\theta\ll 1$), 
the operator transforms as
\bea
\delta \(\bar\psi i\g5 \psi\)=-2\theta \bar\psi\psi.
\eea
If now, in absence of bare masses in the Lagrangian (\ref{exampleL}), 
\bea
\langle 0|\delta \(\bar\psi i\g5 \psi\)|0\rangle\not=0  
\quad \Longrightarrow \quad \langle 0| \bar\psi\psi|0\rangle\not=0, 
\label{dynchsym}
\eea
we conclude that the chiral symmetry is dynamically broken
({\em i.e.} spontaneously broken by a composite operator).
The Goldstone theorem implies the existence of a massless pseudoscalar
particle (a pion).

By using the chiral symmetry, we can always rotate in such a way 
that the spontaneous symmetry breaking or ``long range order'' is 
in the same direction (in the space spanned by the operators $\bar\psi\psi$ 
and $\bar\psi i\g5\psi$) 
as the explicit symmetry breaking caused by a bare mass 
term $m_0\bar\psi\psi$ term, {\em i.e.}
\bea
\langle 0| \bar\psi i\g5\psi|0\rangle=0,\qquad
\langle 0| \bar\psi\psi|0\rangle\not=0.
\eea
Thus the ``long-range order'' is in the direction
of the bare mass operator $\bar\psi\psi$.
Then correlations 
\bea
\langle 0|T\( \bar\psi\psi(x)\bar\psi\psi(0)\) |0\rangle_{connected}
\eea
of the operator $\bar\psi\psi$ are referred to as longitudinal, 
and correlations
\bea
\langle 0|T\( \bar\psi i\g5\psi(x)\bar\psi i\g5\psi(0)\)|0 \rangle_{connected}
\eea
of the operator $\bar\psi i\g5\psi$ 
are transverse to $\bar\psi\psi$.
The massless NG--boson is described by the correlations transverse to the 
direction of ordering.
It means that it costs a small amount of energy to change the
direction of ordering in the transverse direction.

\section{The gauged Nambu--Jona-Lasinio model}
Let us introduce the gauged Nambu--Jona-Lasinio (GNJL) model.
The GNJL model is the gauged version of the first model 
describing dynamical breaking of chiral
symmetry in particles physics, {\em i.e.}  the Nambu--Jona-Lasinio 
model \cite{najola61}. 
If the gauge symmetry is $U(1)$,
the GNJL model can also be consider as QED with additional four-fermion 
interactions.
Bardeen, Leung, and Love proposed to study the GNJL model
in Refs.~\cite{balelo86,leloba86}.
They argued that, due to the appearance of 
large anomalous dimension for the four-fermion operators,
the formally irrelevant operators become marginal at the critical gauge coupling
$\alpha_0=\alpha_c$ of QED. 

The interest in the GNJL model has been 
stimulated by its importance for constructing extended technicolour 
models (ETC) and top-quark condensate models.
For an extensive introduction to such models see \cite{mirbook}.

The gauged NJL model is described by the Lagrangian
\bea
{\cal L}_1=\bar\psi (i\gm D_{\mu}-m_0)\psi-\frac{1}{4}F_{\mu\nu}F^{\mu\nu}
+\frac{G_0}{2}\[(\bar \psi\psi)^2+(\bar\psi i\g5\psi)^2\],\label{gnjl1}
\eea
where $D_{\mu}=\partial_{\mu}+ie_0 A_{\mu}$ is the covariant derivative, 
and where $G_0$ is the dimensionful coupling 
constant\footnote{$G_0$ is commonly referred to as the Fermi coupling constant.} 
of the chirally 
invariant four-fermion interaction. 
Since the canonical dimension of fermion fields is $3/2$, the dimension of $G_0$
is negative in terms of energy units,
\bea
{\rm dim}(G_0)=-2.
\eea
In other words, the canonical dimension $d^c$, Eq.~(\ref{candimdef}),
of the local operator
$(\bar\psi\psi)^2$ is six, and naively it corresponds to an 
irrelevant operator.
In the absence of a fermion mass term $m_0$ which breaks the chiral symmetry
explicitly, the Lagrangian (\ref{gnjl1}) possesses a $U(1)$ gauge symmetry
and a global $U_L(1)\times U_R(1)$ chiral symmetry.
It is straightforward to extend 
and
consider the GNJL model
describing $N$ fermion flavors with a $U_L(N)\times U_R(N)$ 
(or $U_V(N)\times U_A(N)$) chiral symmetry. 
The Lagrangian of the GNJL model with $N$ fermion flavors 
is described in appendix~\ref{unsymm},
and the SDEs of this model will be given in chapter~\ref{chap2}.
In chapter~\ref{chap5} the large $N$ limit of the GNJL model 
will be analyzed.
\section{The auxiliary field method}
The four-fermion interaction of (\ref{gnjl1}) is quartic in the fermion fields, 
which is rather inconvenient for the description of the quantum field model
in terms of Feynman graphs and Schwinger-Dyson equations.
The quartic terms give rise to two-loop SDEs for even the two-point 
Green functions such as the fermion propagator.
The problem can be circumvented elegantly by introducing auxiliary
chiral fields $\sigma$ and $\pi$, which are (real) spinless scalar fields. 
We can rewrite the Lagrangian Eq.~(\ref{gnjl1}) as follows
\bea
{\cal L}_2=\bar\psi i\gamma^{\mu}D_{\mu}\psi
-\frac{1}{4}F_{\mu\nu}F^{\mu\nu}
-\bar \psi(\sigma+i\gamma_5\pi)\psi
-\frac{1}{2G_0}\[(\sigma-m_0)^2+\pi^2\].\label{gnjl2}
\eea
The Euler-Lagrange equations for the auxiliary fields $\sigma$, $\pi$ are
\bea
\sigma=m_0-G_0 \bar\psi\psi,\qquad \pi=-G_0\bar\psi i\g5\psi,
\label{constraints}
\eea
which ensure the equivalence of the Lagrangians ${\cal L}_1$ and ${\cal L}_2$
on-shell. 
The Lagrangian ${\cal L}_2$ is now only quadratic in the fermion fields.
The constraints (\ref{constraints}) represent the fact
that the auxiliary fields $\sigma$ and $\pi$ describe the scalar 
and pseudoscalar fermion-anti-fermion degrees of freedom given by the
composite operators $\bar\psi \psi$ and $\bar\psi i\g5 \psi$.

In terms of the path integral formalism the equivalence of the Lagrangians 
(\ref{gnjl1}) and (\ref{gnjl2}) emerges from the following equality for 
the generating functional $Z$:
\bea
Z&=&N\int {\cal D}\psi{\cal D}\bar\psi {\cal D}A\,
\exp\[i\int{\rm d}^4x\,{\cal L}_1(\psi,\bar\psi,A) \]\nonu\\
&=&N^\prime\int {\cal D}\psi{\cal D}\bar\psi {\cal D}A{\cal D}\sigma{\cal D}\pi
\exp\[i\int{\rm d}^4x\,{\cal L}_2(\psi,\bar\psi,A,\sigma,\pi) \],
\label{Hub-Strat}
\eea
where $N$ and $N^\prime$ are some normalization constants, and it 
is implicitly understood that the integration measure of the gauge field $A$ includes factors connected with the gauge fixing.
The equality in Eq.~(\ref{Hub-Strat}) is the so-called Hubbard-Stratonovich
trick or the auxiliary field method, reintroduced into quantum field theory
by Gross and Neveu \cite{grne74}.

The model described by ${\cal L}_2$ is invariant under the combined 
chiral transformations Eq.~(\ref{chiralsymmetry}),
and 
\bea
\left(
\begin{array}{c}
\sigma \\ \pi
\end{array}
\right)
&\rightarrow&
\left(\begin{array}{c}
\sigma^\prime\\ \pi^\prime
\end{array}\right)
=
\left(
\begin{array}{cc}
\cos 2\theta & \sin 2\theta\\ 
-\sin 2\theta & \cos 2\theta
\end{array}
\right)
\left(
\begin{array}{c}
\sigma \\ \pi
\end{array}
\right),
\eea
for the auxiliary or composite fields.
If $m_0=0$,
the chiral symmetry gives rise to the conserved axial current $j^{5\mu}$:
\bea
j^{5\mu}=\bar\psi \gamma^5 \gamma^\mu \psi,\qquad
\partial_\mu  j^{5\mu}=0.
\eea
with a bare mass the chiral symmetry is explicitly broken 
as given by Eq.~(\ref{explicitbreak}).

After performing a regularization of the generating functional, 
Eq.~(\ref{Hub-Strat}), and introducing the ultra-violet cutoff $\Lambda$,
we have three ``physical'' bare running coupling constants, 
$m_0$, $g_0$, $\alpha_0$:
\bea
m_0\equiv m(\Lambda),\qquad g_0=G_0\Lambda^2/4\pi^2\equiv g(\Lambda),\qquad 
\alpha_0=e_0^2/4\pi\equiv\alpha(\Lambda),\label{barecoupdef}
\eea
and four $Z$ factors or renormalization constants
$Z_2$, $Z_3$, $Z_\sigma$, and $Z_\pi$:
\bea
&&Z_2^{1/2}(\Lambda'/\Lambda)\psi_{(\Lambda')}(x)
=\psi(x), \quad 
Z_3^{1/2}(\Lambda'/\Lambda)
A^\mu_{(\Lambda')}(x)=A^\mu(x),\label{Z1Z2Z3def}\\
&&Z_\sigma(\Lambda'/\Lambda) 
\sigma_{(\Lambda')}(x)
=\sigma(x),\quad
Z_\pi(\Lambda'/\Lambda) \pi_{(\Lambda')}(x)=\pi(x), \label{Zsigmapidef}
\eea 
where $\Lambda'/\Lambda\leq 1$, and the fields $\psi$, $A^\mu$, $\pi$, 
$\sigma$ are the bare fields defined at the UV cutoff $\Lambda$, 
{\em e.g.} compare with $\phi_l$ of Eq.~(\ref{projhighlow}),
and the fields $\psi_{(\Lambda')}$ are ``renormalized fields''
defined at the scale $\Lambda'$, {\em e.g.} compare with $\phi_{l'}$ of
Eq.~(\ref{philprime}).

%
%
\chapter{SDEs of the GNJL model}\label{chap2}
In this chapter we introduce the Schwinger--Dyson 
(SDE) equations, and the Ward identities for the GNJL model.
In section \ref{sec_genfiegnjl} after introducing the generating functional, 
we present a derivation of SDEs for the Green functions in momentum space.
In section \ref{secwti} we review the vector and axial Ward 
identities, and derive the SDE for the axial-vector vertex.
In addition the partial conserved axial current (PCAC) relation
are discussed briefly.
Finally some basic properties of particles, resonances and tachyons 
are mentioned in section \ref{sec_particles}.
\section{The generating functional}\label{sec_genfiegnjl}
Suppose that we have some fundamental action $S_f$ analogous to 
Eq.~(\ref{funaction}) ({\em e.g} the action of QED) for which the Wilsonian 
effective action (\ref{effactionrel}) given 
at some cutoff $\Lambda$ is the GNJL model. 
We assume that the GNJL model action can be hypothetically obtained 
from $S_f$ after a suitable Pauli--Villars regularization as described in 
subsection~\ref{step1}.
Thus we write in Minkowski space
\bea
S^{eff}_\Lambda\[\kappa(\Lambda),\phi_l\]
=\int{\rm d}^4x\,\[{\cal L}_{\rm GNJL}+{\cal L}_{\rm GF}\],\label{wilactgnjl}
\eea
where we omit the infinite sum of irrelevant interactions.
The GNJL model Lagrangian ${\cal L}_{\rm GNJL}$ 
describes the local interactions of the fields, 
and for the GNJL model with $N$ fermion flavors we have
\bea
{\cal L}_{\rm GNJL}&=&-\frac{1}{4}F_{\mu\nu}F^{\mu\nu}
+\sum_{i=1}^N\bar\psi_i i\gamma^\mu D_\mu\psi_i
-\sum_{i,j=1}^N\sum_{\alpha=0}^{N^2-1}
\bar\psi_i\tau^\alpha_{ij}(\sigma^\alpha+i\gamma_5\pi^\alpha)\psi_j\nonu\\
&-&\frac{1}{2G_0}\sum_{\alpha=0}^{N^2-1}\[ (\sigma^\alpha)^2+(\pi^\alpha)^2)
-2 c^\alpha\sigma^\alpha\],\label{laggnjl}
\eea
where $\tau^\alpha$ are the generators of $U(N)$, 
$c^\alpha=\Tr\[ M \tau^\alpha\]$, and $M_{ij}=m^{(i)}\delta_{ij}$ 
is the diagonal mass matrix, see also appendix~\ref{unsymm}.
The term ${\cal L}_{\rm GF}$ in Eq.~(\ref{wilactgnjl}) represents
the gauge fixing part of the Lagrangian.
An additional condition is needed, the Gupta--Bleuler gauge fixing.
We use a standard covariant gauge fixing with a constant gauge parameter $\xi$, 
{\em i.e.}
\bea
{\cal L}_{\rm GF}=-1/2\xi \(\partial_\mu A^\mu\)^2.\label{lagGF}
\eea
In case of the GNJL model
the coupling constants and fields are
\bea
\kappa(\Lambda)&=&(m_0,g_0,\alpha_0,\xi,Z_2,Z_3,Z_\sigma,Z_\pi),\\
\phi_l&=&\(A,\bar\psi,\psi,\sigma,\pi\),\label{gnjlphil}
\eea
The set of bare coupling constants (except $\xi$)
and renormalization constants
are given by Eqs.~(\ref{barecoupdef})--(\ref{Zsigmapidef}), and 
the set of fields $\phi_l$ are the ``low'' energy modes 
analogous to $\phi_l$ of Eqs.~(\ref{projhighlow}) and (\ref{philphih}).

From Eq.~(\ref{wilactgnjl})
we construct the generating functional analogous to Eq.~(\ref{geneffdef}),
\bea
Z[{\bf J}]=N(\Lambda)\int
{\cal D}\phi_l\, \exp\left\{iS^{eff}_\Lambda\[\kappa(\Lambda),\phi_l\]+
iS_J\[{\bf J},\phi_l\]\right\}, \label{genfie_GNJL}
\eea
where the dependence of $Z$ on the cutoff $\Lambda$ and the couplings
$\kappa(\Lambda)$ is taken implicitly.
The functional ``measure'' is defined as
\bea
{\cal D}\phi_l\,
\equiv
{\cal D}A{\cal D}\bar\psi{\cal D}\psi{\cal D}\sigma{\cal D}\pi.
\label{measgnjl}
\eea
We implicitly take the product of measures over all flavor, 
Dirac, and Lorentz indices of the fields.
The ``source'' action ${\cal S}_J$ is given by
\bea
S_J\[{\bf J},\phi_l\]=\int{\rm d}^4x\, \[J_\mu A^\mu +\sum_{i=1}^N \[\bar\eta_i \psi_i 
+\bar\psi_i \eta_i\]
+\sum_{\alpha=0}^{N^2-1}
\[ J_\sigma^\alpha \sigma^\alpha+ J_\pi^\alpha \pi^\alpha\]\],\label{sactgnjl}
\eea
where
\bea
{\bf J}&=&(J,\eta,\bar\eta,J_\sigma, J_\pi).
\eea
The sources ${\bf J}$ are defined in a similar manner as the sources
$J_\Lambda$ of Eq.~(\ref{geneffdef}).
The source action defines the response of the system to arbitrary perturbations 
for which the sources form a suitable basis. 
$J$, $\eta$ and $\bar \eta$ are the usual sources of QED, and 
$J_\sigma$ and $ J_\pi$ are the sources which couple to the 
auxiliary fields, $\sigma$ and $\pi$.

As was discussed in section~\ref{sec_rev},
the starting point to derive the Schwinger-Dyson equations (SDE) is the 
following formal identity for the generating functional (\ref{genfie_GNJL}):
\bea
&&\int{\cal D}\phi_l\,
\exp\left\{iS^{eff}_\Lambda+iS_J\right\}
\frac{\delta}{\delta \phi_l(x)} \(S^{eff}_\Lambda+S_J\)=0,
\label{basicGNJLSDE}
\eea
where $\phi_l$ is either one of the five fields of the model
$\phi_l=A,\psi,\bar\psi,\sigma,\pi$.
From Eq.~(\ref{basicGNJLSDE}) 
the SDE for the propagators and vertices in momentum space can be obtained.

Taking into account that the fields $\bar\psi$, $\psi$ and 
the sources $\eta$ and $\bar\eta$
are Grassman variables, we obtain five
functional differential equations for the generating functional $Z$.
The functional derivates in Eq.~(\ref{basicGNJLSDE}) 
follow from Eqs.~(\ref{laggnjl}), 
(\ref{lagGF}), and (\ref{sactgnjl}).
If we then replace the functional integrand by differentiations with respect 
to source terms the equation (\ref{basicGNJLSDE}) 
is equivalent to the following set of functional differential equations:
\bea 
0&=&\biggr\{\[
\partial_\lambda\partial^\lambda g^{\mu\nu}
-\frac{(\xi-1)}{\xi}\partial^{\mu}\partial^{\nu}\]
\frac{\delta}{i\delta J^{\nu}(x)}\nonu\\ 
&&- e_0\gamma^\mu_{ba}
\frac{\delta}{i\delta \bar\eta_{ai}(x)}\frac{\delta}{i\delta\eta_{bi}(x)}
+J^\mu(x)\biggr\}Z[{\bf J}],\label{SDE1}\\
0&=&\biggr\{(i\dslash)_{be}
\frac{\delta}{i\delta \eta_{bj}(x)} +e_0\gm_{be}
\frac{\delta}{i\delta J^{\mu}(x)} \frac{\delta}{i\delta \eta_{bj}(x)}
\nonu\\
&&+ \tau^\alpha_{ij}
\[\frac{\delta}{i\delta J_\sigma^\alpha(x)} 
+i\g5\frac{\delta}{i\delta J_\pi^\alpha(x)} \]_{be}
 \frac{\delta}{i\delta \eta_{bi}(x)}
+\bar\eta_{ej}(x)
\biggr\}Z[{\bf J}],\label{SDE2}\\
0&=&\biggr\{(i\dslash)_{ea}\frac{\delta}{i\delta \bar\eta_{ai}(x)} -e_0\gm_{ea}
\frac{\delta}{i\delta J^{\mu}(x)} \frac{\delta}{i\delta \bar\eta_{ai}(x)}
\nonu\\
&&-\tau^\alpha_{ij}
\frac{\delta}{i\delta \bar\eta_{aj}(x)}
\[\frac{\delta}{i\delta J_\sigma^\alpha(x)}
+i\g5\frac{\delta}{i\delta J_\pi^\alpha(x)} \]_{ea}+\eta_{ei}(x)
\biggr\}Z[{\bf J}],\label{SDE3}\\
0&=&\biggr\{ -\frac{1}{G_0}\frac{\delta}{i\delta J_\sigma^\alpha(x)}
-\tau^\alpha_{ij}{\bf 1}_{cd}\frac{\delta}{i\delta \bar\eta_{dj}(x)}
\frac{\delta}{i\delta\eta_{ci}(x)}\nonu\\
&&+\frac{c^\alpha}{G_0}+J_\sigma^\alpha(x)
\biggr\}Z[{\bf J}],\label{SDE4}\\
0&=&\biggr\{-\frac{1}{G_0}\frac{\delta}{i\delta  J_\pi^\alpha(x)}
-\tau^\alpha_{ij}i{\g5}_{cd}\frac{\delta}{i\delta \bar\eta_{dj}(x)}
\frac{\delta}{i\delta\eta_{ci}(x)} + J_\pi^\alpha(x)
\biggr\}Z[{\bf J}].\label{SDE5}
\eea
We use the standard summation convention, {\em i.e.} 
a respective sum over double indices (flavor, Lorentz, Dirac indices).

SDEs for time-order vacuum expectation values can be 
obtained by differentiating the above expression with respect to 
various source terms.
The above set of five functional differential equations determine the entire 
structure of our model and should be considered as the equations of motion.

As was pointed out in subsection~\ref{step1},
the low energy modes $\phi_l$ of Eq.~(\ref{gnjlphil})
give rise to a momentum cutoff $\Lambda$
in the Fourier transforms of these modes, see Eq.~(\ref{philphih}).
Consequently the cutoff $\Lambda$ enters the expressions
for the momentum space Green functions.
Therefore we introduce the notation
\bea
\int_\Lambda {\rm d}^4p\, =\int_{|p|\leq \Lambda} {\rm d}^4p\,.
\eea
This is referred to as the ``hard'' cutoff regularization.
\subsection{Condensates}
Condensates are described as nonzero vacuum expectation values of local 
operators at a single space time
point, {\em e.g.} $\langle0| \phi(x)|0\rangle$. 
Because of the translational invariance of the vacuum,
a condensate
is independent of the space-time point $x$, 
$\langle\phi(x)\rangle=\langle\phi(0)\rangle$.

Most condensates are zero, for instance the vacuum expectation value of a 
single vector field $A_\mu$ will be zero due to charge conjugation.
The SDEs for the chiral condensates can be derived from SDE (\ref{SDE4}) 
by putting the sources to zero.
The chiral symmetry is used in such a way that all pseudoscalar 
condensates are zero,
\bea
\langle 0|\pi^\alpha|0\rangle=0.\label{pseudocond}
\eea
For the scalar condensates we then find
\bea
\langle 0|\sigma^\alpha|0\rangle=c^\alpha+G_0 \sum_{i=1}^N
\tau^\alpha_{ii} \Tr\[iS^{(i)}(0)\],\label{scalcond}
\eea
where $iS^{(i)}(0)$ the fermion propagator in coordinate space and
$\Tr\[iS^{(i)}(0)\]$ is the fermionic chiral condensate 
of a specific flavor $i$
\bea
-\langle 0|\bar\psi_i \psi_i |0\rangle\equiv \Tr\[iS^{(i)}(0)\]=
\int_{\Lambda} \frac{{\rm d}^4k}{(2\pi)^4}\, 
\Tr\[iS^{(i)}(k)\].\label{condtrace}
\eea
One should note that when the bare masses of the fermions are identical 
(the $N$ fermions are degenerate)
the mass matrix $M$ is proportional to the identity matrix  
and all condensates vanish except $\langle 0|\sigma^0|0\rangle$.
\subsection{The SDE for the photon propagator}
\begin{figure}[ht!]
\epsfxsize=9cm
\epsffile[-40 440 360 540]{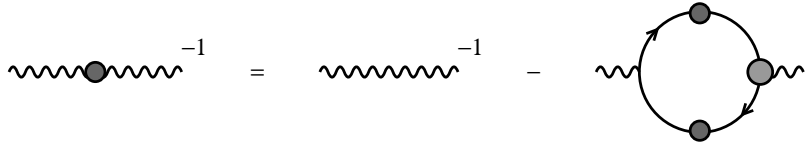}
\caption{The SDE for the photon propagator.}
\label{fig_sde_phot}
\end{figure}

The SDE for the gauge boson, the photon, is derived from Eq.~(\ref{SDE1})
by differentiating with respect to $J_\mu(y)$, and
in momentum space reads
\bea
-iD^{\mu\nu-1}(q)&=&-i\[-q^2 g^{\mu\nu}+\frac{(\xi-1)}{\xi} q^\mu q^\nu\]
-i\Pi^{\mu\nu}(q),\label{sdephot}
\eea
where $\Pi_{\mu\nu}(q)$ is the vacuum polarization defined as
\bea
i\Pi^{\mu\nu}(q)&\equiv&
(-1)\sum_{i=1}^N \int_\Lambda \frac{{\rm d}^4k}{(2\pi)^4}\,\nonu\\
&\times&
\Tr\[(-ie_0)\gamma^\mu iS^{(i)}(k)(-ie_0)\Gamma^{(i)\nu}(k,k-q)iS^{(i)}(k-q)\].
\label{vacpoltensdef}
\eea
The vector Ward-Takahashi, which will be introduced in the next section,
ensures that the vacuum polarization tensor is transverse, 
$q_\mu\Pi^{\mu\nu}(q)=0$, hence we write
\bea
\Pi^{\mu\nu}(q)=(-q^2 g^{\mu\nu}+q^\mu q^\nu)\Pi(q^2),\label{vacpoldef}
\eea
in terms of the vacuum polarization $\Pi(p^2)$.
The SDE for the photon propagator is depicted in Fig.~\ref{fig_sde_phot}.
For the ``Feynman rules'' we refer to appendix~\ref{feynmanrules}.

\subsection{The SDE for the fermion propagator}
The SDE for the fermion propagator (see definition in appendix \ref{gfprpv}, 
Eq.~(\ref{deffpmom})) in momentum space follows from Eq.~(\ref{SDE3})
by differentiating with respect to $\eta_{bj}(y)$ and setting sources 
to zero.\footnote{Or, equivalently, from Eq.~(\ref{SDE2}) by 
differentiation with respect to $\bar\eta$}
After Fourier transforming the SDE in coordinate space to momentum space, we obtain
the SDE for the fermion propagator
\bea
-iS^{(i)-1}(p)&=&-i(\pslash-m^{(i)})
-(-i{\bf 1})
G_0\int_{\Lambda} \frac{{\rm d}^4k}{(2\pi)^4}\, 
\Tr\[iS^{(i)}(k)\]
\nonu\\
&-&\int_{\Lambda} \frac{{\rm d}^4k}{(2\pi)^4}\,
(-ie_0)\gamma^\mu iS^{(i)}(k)(-ie_0)\Gamma^{(i)\nu}(k,p)
iD_{\nu\mu}(k-p)\nonu\\
&-&\sum_{j=1}^N\sum_{\beta=0}^{N^2-1}
\int_{\Lambda} \frac{{\rm d}^4k}{(2\pi)^4}\,
(-i{\bf 1})\tau^\beta_{ij} 
iS^{(j)}(k)(-i)\Gamma^{\beta}_{{\rm S}ji}(k,p)
i\DelS^{(\beta)}(k-p)\nonu\\
&-&\sum_{j=1}^N\sum_{\beta=0}^{N^2-1}
\int_{\Lambda} \frac{{\rm d}^4k}{(2\pi)^4}\,
(\g5)\tau^\beta_{ij} iS^{(j)}(k)(-i)\Gamma^{\beta}_{{\rm P}ji}(k,p)
i\DelP^{(\beta)}(k-p), \label{sde_ferm}
\nonu\\
\eea
where Dirac indices have been omitted.
In the derivation of the above equation we have used 
Eqs.~(\ref{pseudocond}) and (\ref{scalcond}) and the Fierz identity 
(see for instance \cite{mirbook} and appendix~\ref{unsymm}).

\begin{figure}[ht!]
\epsfxsize=9cm
\epsffile[0 460 400 560]{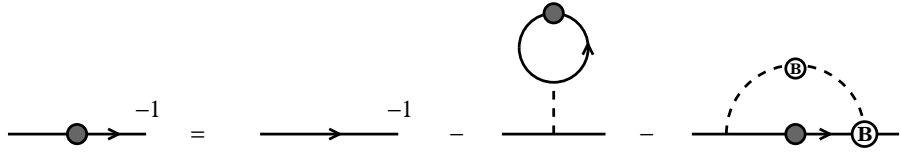}
\caption{The SDE for the fermion propagator.}
\label{fig_sde_ferm}
\end{figure}

The general structure of the fermion propagator of flavor $(f)$ is
\bea
S^{(f)-1}(p)= \frac{\pslash-\Sigma_{(f)}(p^2))}{{\cal Z}_{(f)}(p^2)}, 
\label{struc_ferm}
\eea
since we have rotated all parts proportional to $i\g5$ to zero.
The scalar function ${\cal Z}$ is called the fermion wave function, and $\Sigma$
the fermion mass function.\footnote{Another decomposition is quite common, 
namely $S^{-1}(p)= \pslash A(p^2)-B(p^2) $.}
The SDE for the fermion propagator is depicted in Fig.~\ref{fig_sde_ferm}.
\subsection{SDEs for the scalar and pseudoscalar propagators}
SDEs for the scalar and pseudoscalar follow from Eqs.(\ref{SDE4}) and (\ref{SDE5})
by differentiating with respect to appropriate sources.
Inverting the equations subsequently to momentum space by using the definitions
given in appendix \ref{gfprpv}, we find the following SDEs.
The SDE for the scalar propagator is
\bea
-i\DelS^{(\alpha)-1}(q)&=& (-iG_0)^{-1} -i\Pi_{\rm S}^{(\alpha)}(q^2),
\label{sdescal}
\eea
where the scalar vacuum polarization is defined as
\bea
i\Pi_{\rm S}^{(\alpha)}(q^2)&\equiv&
(-1) \sum_{i=1}^N\sum_{j=1}^N \int_\Lambda\frac{{\rm d}^4k}{(2\pi)^4}\,\nonu\\
&\times&\Tr\[(-i{\bf 1})\tau^\alpha_{ij} iS^{(j)}(k) (-i)
\Gamma^{\alpha}_{{\rm S}ji}(k,k-q)iS^{(i)}(k-q)\],\label{SDscalvacpol}
\eea
where the factor (-1) is a consequence of the fermion loop.

\begin{figure}[ht!]
\epsfxsize=9cm
\epsffile[-40 440 360 540]{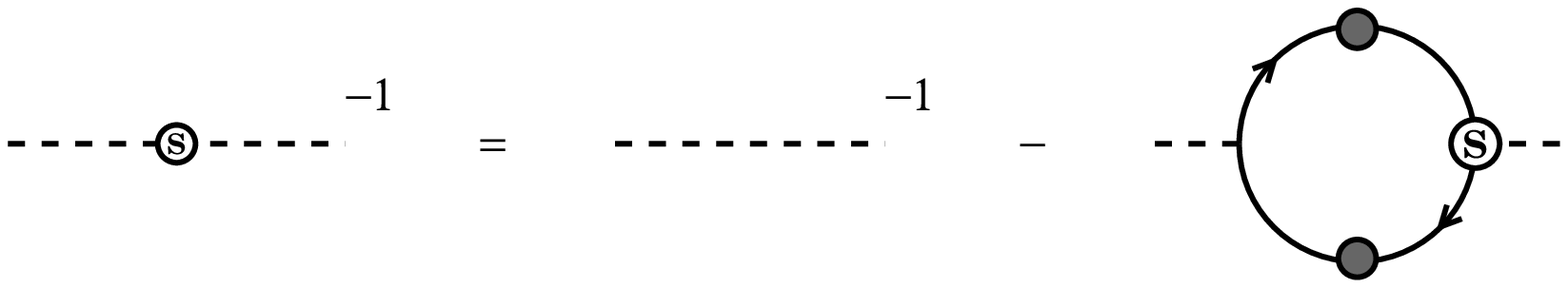}
\caption{The SDE for the scalar propagator.}
\label{fig_sde_scal}
\end{figure}

\begin{figure}[ht!]
\epsfxsize=9cm
\epsffile[-40 440 360 540]{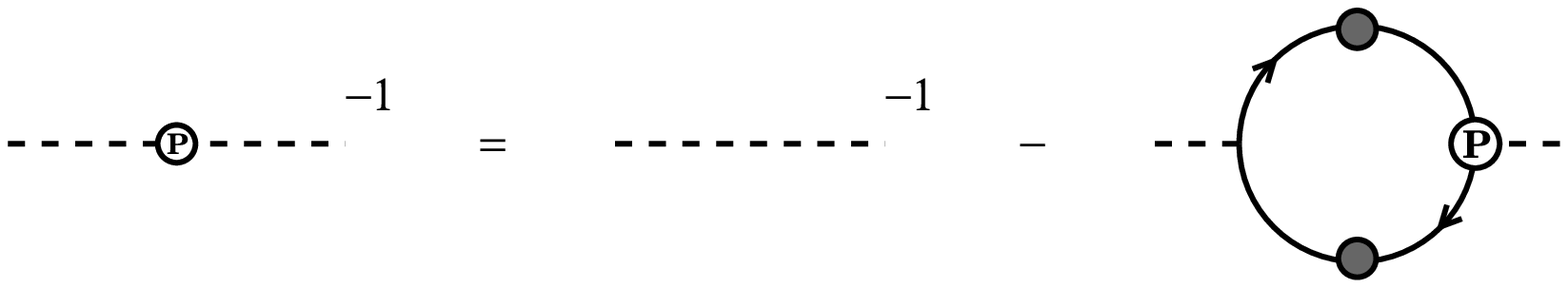}
\caption{The SDE for the pseudoscalar propagator.}
\label{fig_sde_pseudo}
\end{figure}

The SDE for the pseudoscalar propagator is
\bea
-i\DelP^{(\alpha)-1}(q)&=&
(-iG_0)^{-1}
-i\Pi_{\rm P}^{(\alpha)}(q^2).\label{sdepseudo}
\eea
The pseudoscalar vacuum polarization is defined as
\bea
i\Pi_{\rm P}^{(\alpha)}(q^2)&\equiv&
(-1)\sum_{i=1}^N\sum_{j=1}^N \int_\Lambda\frac{{\rm d}^4k}{(2\pi)^4}\,\nonu\\
&\times&
\Tr\[(\g5)\tau^\alpha_{ij} iS^{(j)}(k) (-i)
\Gamma^{\alpha}_{{\rm P}ji}(k,k-q)iS^{(i)}(k-q)\].
\eea
These SDEs for scalar respectively pseudoscalar propagator ar depicted
in Fig.~\ref{fig_sde_scal} and Fig.~\ref{fig_sde_pseudo}.
\subsection{SDEs of the vertices}
\begin{figure}[hb!]
\epsfxsize=9cm
\epsffile[-10 440 390 540]{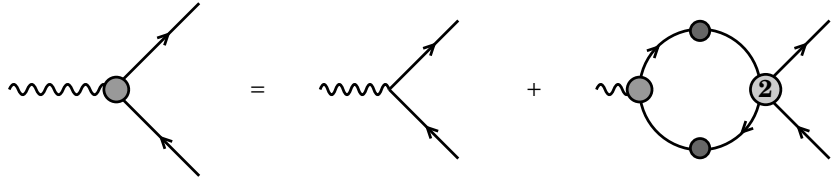}
\caption{The SDE for the photon-fermion vertex.}
\label{fig_sde_vertex}
\end{figure}

In this subsection the SDEs of the proper vertices of the GNJL
model are introduced.
Important in the derivation of the SDE is the four-point function
$K^{(0)}$. In appendix~\ref{gfprpv} we define the 
1-boson irreducible four-point function $K^{(1)}$ in Eq.~(\ref{K1def}) 
and the 2-fermion 1-boson irreducible Bethe--Salpeter kernel $K^{(2)}$ in 
Eq.~(\ref{K2def}).
\paragraph{The photon-fermion vertex:}
The SDE for the photon-fermion vertex is obtained from Eq.~(\ref{SDE1}) 
by differentiation with respect to $\eta$ and $\bar\eta$, and transforming the
equation to momentum space.\footnote{An alternative way is to differentiate 
Eq.~(\ref{SDE2}) with respect to $\bar\eta$ and $J_\mu$. In this way the SDE for the vertex is given in term of the two-photon two-fermion scattering kernel 
or compton kernel, see Ref.~\cite{cabrplgr84}.}
Using the definition of $K^{(2)}$ and the SDE for the photon
propagator, Eq.~(\ref{sdephot}), we obtain
\bea
(-ie_0)\Gamma^{(i)\mu}_{ab}(k+q,k)&=&
(-ie_0)\gamma^\mu_{ab}+\sum_{f=1}^N
\int_{\Lambda}\frac{{\rm d}^4 k_1}{(2\pi)^4}\nonu\\
&\times& \[iS^{(f)}(k_1+q)
(-ie_0)\Gamma^{(f)\mu}(k_1+q,k_1)iS^{(f)}(k_1)\]_{dc}\nonu\\
&\times&(-ie_0^2)K^{(2)}_{^{cd,ab}_{ff,ii}}(k_1,k_1+q,k+q),\label{photfermvertsde}
\eea
where we have used that for $i\neq j$,
\bea
0&=&\sum_{f=1}^N
\int\frac{{\rm d}^4 k_1}{(2\pi)^4}\,
\[(-ie_0)\gamma_{\mu}iS^{(f)}(k_1)
iS^{(f)}(k_1+q)\]_{dc}\nonu\\
&\times&
(-ie_0^2)K^{(1)}_{^{cd,ab}_{ff,ij}}(k_1,k_1+q,k+q).
\eea
The SDE for the photon-fermion vertex is depicted in Fig.~\ref{fig_sde_vertex}.

The photon-fermion vertex has a rich tensor structure; it is both a
spinor matrix and a four-vector, and therefore it can be decomposed into
12 spin-vector amplitudes and 12 coefficients. 
The photon fermion vertex is written as a sum of a longitudinal and 
transverse part
\bea
\Gamma^\mu(k,p)=\Gamma^\mu_L(k,p)+\Gamma^\mu_T(k,p),\label{vertlongtrans}
\eea
where the longitudinal part satisfies a Ward--Takahashi identity
\bea
q_{\mu}\Gamma^\mu_L(k+q,k)=S^{-1}(k+q)-S^{-1}(k).\qquad 
\eea
This vector-Ward--Takahashi (which will
be reviewed in the next section)
fixes four of the coefficients constituting the longitudinal part. 
The eight remaining amplitudes are referred to as transverse, since 
they vanish after projection with in-going photon momentum
\bea
q_\mu\Gamma^\mu_T(k+q,k)=0.\label{wtilong}
\eea 
The transverse part is written as a product of eight Ball--Chiu
tensors \cite{bach80}, $T^\mu_i$, 
and eight scalar coefficient functions, $\tau_i((k+q)^2,k^2,q^2)$,
\bea
\Gamma^\mu_T(k+q,k)=\sum_{i=1}^8 T^\mu_i(k+q,k)\, \tau_i(k+q,k),
\label{transdef}
\eea
where we write for notational convenience the scalar functions 
as $\tau_i(k+q,k)\equiv \tau_i((k+q)^2,k^2,q^2)$.
A modification of the $T^\mu_4$ was proposed in \cite{kirepe95}.
The tensors are listed below 
\bea
T^{\mu}_1(k+q,k)&=&k^{\mu} q^2-q^{\mu} (k\cdot q),\nonu\\
T^{\mu}_2(k+q,k)&=& (2\kslash+\qslash) T^{\mu}_1(k+q,k),\nonu\\
T^{\mu}_3(k+q,k)&=& q^2\gm-q^{\mu}\qslash,\nonu\\
T^{\mu}_4(k+q,k)&=& q^2\[\gm(2\kslash+\qslash)-2k^{\mu}-q^{\mu}\]
+2q^{\mu}\sigma_{\nu\rho}k^{\nu}q^{\rho},\nonu\\
T^{\mu}_5(k+q,k)&=& \sigma^{\mu\rho}q_{\rho},\label{transbasis}\\
T^{\mu}_6(k+q,k)&=& \gm((k+q)^2-k^2)-(2k^{\mu}+q^{\mu})\qslash,\nonu\\
T^{\mu}_7(k+q,k)&=& \frac{(k+q)^2-k^2}{2}\[\gm(2\kslash+\qslash)
-2k^{\mu}-q^{\mu}\]\nonu\\
&+&(2k^{\mu}+q^{\mu})\sigma_{\nu\rho}k^{\nu}q^{\rho},\nonu\\
T^{\mu}_8(k+q,k)&=& -\gm\sigma_{\nu\rho}k^{\nu}q^{\rho}
+k^{\mu}\qslash-q^{\mu}\kslash,\nonu
\eea
where
\bea
\sigma^{\mu\nu}=\frac{1}{2}\[\gm,\gn\].
\eea
In the Feynman gauge, $\xi=1$, the transverse scalar functions, $\tau_i$
have been computed to one-loop order by Ball and Chiu in \cite{bach80}, 
the extension to a general covariant gauge $\xi$ was done 
by Kizilers{\"u} {\em et al.} \cite{kirepe95}.

\begin{figure}[t!]
\epsfxsize=9cm
\epsffile[-10 440 390 540]{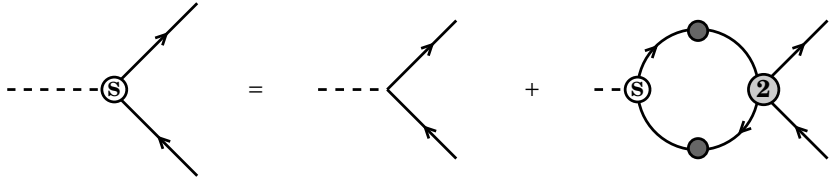}
\caption{The SDE for the scalar vertex.}
\label{fig_sde_scalarvertex}
\end{figure}

\begin{figure}[t!]
\epsfxsize=9cm
\epsffile[-10 440 390 540]{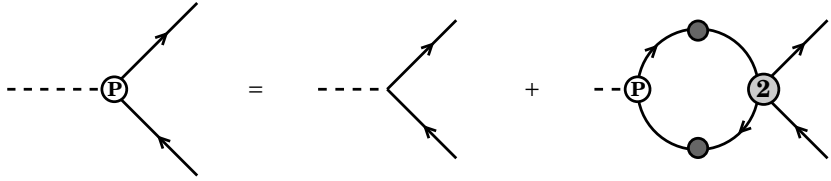}
\caption{The SDE for the pseudoscalar vertex.}
\label{fig_sde_pseudovertex}
\end{figure}

A general constraint on the eight $\tau_i$'s comes from
$C$-parity transformations.
The full vertex must transform under charge conjugation $C$ in the same way as
the bare vertex, so that
\bea
C\Gamma_\mu(k,p)C^{-1}=-\Gamma^T_\mu(-p,-k).\label{CPvert}
\eea 
Thus Eq.~(\ref{CPvert}) together with
\bea
C\gamma_\mu C^{-1}=-\gamma_\mu^T
\eea
gives, using Eqs.~(\ref{transdef}) and (\ref{transbasis}),
\bea
\tau_i(k^2,p^2,q^2)&=&\tau_i(p^2,k^2,q^2),\quad\mbox{for}\quad i=1,2,3,4,5,7,8,
\nonu\\
\tau_6(k^2,p^2,q^2)&=&-\tau_6(p^2,k^2,q^2).
\eea
\paragraph{The scalar and pseudoscalar vertex:}
The SDE for scalar vertex is obtained from Eq.~(\ref{SDE4}) by
differentiation with respect to $\eta$ and $\bar\eta$, and Fourier transforming the
equation to momentum space.
Using the definition of $K^{(2)}$ and the SDE for the scalar
propagator, Eq.~(\ref{sdescal}), we find
\bea
(-i)\Gamma^{\alpha}_{{\rm S}^{ab}_{ij}}(k+q,k)&=&
(-i{\bf 1})_{ab}\tau^\alpha_{ij}+\sum_{k=1}^N\sum_{l=1}^N
\int_{\Lambda}\frac{{\rm d}^4 k_1}{(2\pi)^4}\nonu\\
&\times& 
\[iS^{(k)}(k_1+q)(-i)\Gamma^{\alpha}_{{\rm S}kl}(k_1+q,k_1)
iS^{(l)}(k_1)\]_{dc}\nonu\\
&\times&(-ie_0^2)
K^{(2)}_{^{cd,ab}_{lk,ij}}(k_1,k_1+q,k+q),\label{BSscalvert}
\eea
and in a similar manner the SDE for the pseudoscalar vertex is obtained,
\bea
(-i)\Gamma^{\alpha}_{{\rm P}^{ab}_{ij}}(k+q,k)&=&
(\g5)_{ab}\tau^\alpha_{ij}+\sum_{k=1}^N\sum_{l=1}^N
\int_{\Lambda}\frac{{\rm d}^4 k_1}{(2\pi)^4}\nonu\\
&\times&
\[iS^{(k)}(k_1+q)
(-i)\Gamma^{\alpha}_{{\rm P}kl}(k_1+q,k_1)
iS^{(l)}(k_1)\]_{dc}\nonu\\
&\times&(-ie_0^2)
K^{(2)}_{^{cd,ab}_{lk,ij}}(k_1,k_1+q,k+q).\label{sdepseudovertex}
\eea
These SDEs are depicted in Fig.~\ref{fig_sde_scalarvertex} and
Fig.~\ref{fig_sde_pseudovertex}.

If we consider for simplicity one fermion flavor,
then, in momentum space, the scalar and pseudoscalar vertices 
can be decomposed over four spinor structures\footnote{This follows from 
Lorentz covariance.} with scalar functions 
in the following way
\bea
\GS(k+q,k)&=&\Fs_1(k+q,k)+
(\qslash \kslash-\kslash\qslash)\Fs_2(k+q,k)\nonu\\
&+&(\kslash+\qslash) \Fs_3(k+q,k)+ \kslash \Fs_4(k+q,k),\label{vertfiesdefscal}\\
\GP(k+q,k)&=&(i\g5)\bigg[\Fp_1(k+q,k)+
(\qslash \kslash-\kslash\qslash)\Fp_2(k+q,k)\nonu\\
&+&(\kslash+\qslash) \Fp_3(k+q,k)+\kslash \Fp_4(k+q,k)\bigg],
\label{vertfiesdefpseudo}
\eea
where the scalar functions $\Fs_i$ and $\Fp_i$ depends on the squares of 
the Minkowski momenta, $(k+q)^2$, $k^2$, $q^2$, {\em i.e.,}
$F(k+q,k)\equiv F((k+q)^2,k^2,q^2)$.

Now $C$ invariance  
\bea
C\GS(k,p)C^{-1}=\Gamma_{\rm S}^T(-p,-k),\qquad
C\GP(k,p)C^{-1}=\Gamma_{\rm P}^T(-p,-k).
\label{Cscalvert} 
\eea
implies that
the functions $\Fs_1$, $\Fp_1$, and $\Fs_2$ and $\Fp_2$ 
are symmetric under interchange of fermion momenta
\bea
\Fs_i(k^2,p^2,q^2)=\Fs_i(p^2,k^2,q^2), \qquad
\Fp_i(k^2,p^2,q^2)=\Fp_i(p^2,k^2,q^2),
\eea
with $i=1,\,2$.
\section{Ward--Takahashi identities}\label{secwti}
In this section we review the vector and chiral Ward-Takahashi identities (WTIs),
which reflect the gauge invariance and the chiral symmetry of the GNJL model. 
WTIs represent the symmetry structure of the model in terms of relations 
between Green functions.

The WTIs follow from the SDEs Eq.~(\ref{SDE1})--Eq.~(\ref{SDE5}) 
provided the model is properly regularized.\footnote{This is similar 
to the derivation of Noether currents from Euler--Lagrange 
equations in classical field theory. Contrary to what sometimes is believed, 
WTIs are not additional constraints imposed on the Green functions other 
than already imposed by the SDEs.}
But usually the Ward identities are derived in a more heuristic way
by performing a change of the functional integration variable 
in the generating functional. It is then assumed that 
local unitary transformations leave the path-integral measure invariant.
The invariance of the measure under unitary axial and vector symmetry 
transformations gives rise to 
the vector WTIs and chiral WTIs.
\subsection{Vector Ward-Takahashi identities}
In Abelian gauge field theories the $U(1)$ WTI corresponding to the 
local symmetry plays an essential role in the renormalizability of the model.
The WTIs constrain the independent scalings of the model and are crucial 
for renormalizability.
  
The $U(1)$ part of the vector symmetry is local, and we assume that the 
functional measure ${\cal D}\phi_l$ of Eq.~(\ref{measgnjl})
is invariant under local $U_V(1)$ transformations. This statement
gives a functional differential equation for the generating functional
consistent with the SDEs as mentioned previously.
Hence we get for $U_V(1)$ transformations the following functional differential 
equation:
\bea
0=\biggr\{
\frac{1}{\xi} \partial_\nu\partial^\nu
\partial_\mu \frac{\delta}{i\delta J_\mu}
+\partial_\mu J^\mu+e_0i\bar\eta_{bj} 
\frac{\delta}{i\delta \bar\eta_{bj}}-e_0i\eta_{ai} 
\frac{\delta}{i\delta \eta_{ai}}\biggr\}_x Z\[{\bf J}\],
\label{vect_wti}
\eea
from which the Ward--Takahashi identities for Green functions can be generated. 
The above expression is to be evaluated at space-time point $x$.
Differentiation of Eq.~(\ref{vect_wti}) with respect to $J_\nu$
gives the well-known identity for photon propagator 
\bea
q_\mu D^{\mu\nu}(q)=-\xi\frac{q^\nu}{q^2}.
\eea
This condition for the photon propagator implies that the vacuum 
polarization defined in Eq.~(\ref{vacpoltensdef}) is transverse
\bea
q_\mu \Pi^{\mu\nu}(q)=0.
\eea
Hence we can write the photon propagator as
\bea
D_{\mu\nu}(q)=-\frac{1}{q^2}\(g_{\mu\nu}-\frac{q_\mu q_\nu}{q^2}\)
\frac{1}{1+\Pi(q^2)}-\xi\frac{q_\mu q_\nu}{q^4},
\eea
where $\Pi(q^2)$ is the vacuum polarization defined in Eq.~(\ref{vacpoldef}).

Differentiating Eq.~(\ref{vect_wti}) with respect to $\eta$ and $\bar\eta$
and Fourier transforming the equation to momentum space
yields the WTI for the photon-fermion vertex,
\bea
(k-p)_\mu \Gamma^{(i)\mu}(k,p)=S^{(i)-1}(k)-S^{(i)-1}(p).\label{wtivertex}
\eea
This identity relates the longitudinal part, Eq~(\ref{vertlongtrans}), 
of the vertex to the fermion propagator.

\paragraph{Other vector symmetries:}
For the sake of completeness we also mention here 
the $SU_V(N)$ WTI.
This WTI follows from $SU_V(N)$ symmetry of the GNJL model with $N$ 
fermion flavors, see appendix \ref{unsymm}, 
and the functional differential equation reads
\bea
0&=&\biggr\{-\partial_\mu(i\gamma^\mu)_{ba} \tau^\alpha_{ji}
\frac{\delta^2}{i\delta\bar\eta_{ai} i\delta\eta_{bj} }
+\tau^\alpha_{ji}\[\bar\eta_{bj} \frac{\delta}{i\delta \bar\eta_{bi} }-
\eta_{ai} \frac{\delta}{i\delta \eta_{aj} }\]\nonu\\
&+&\frac{f^{\alpha\beta\gamma} c^\gamma}{G_0} \frac{\delta}{i\delta J_\sigma^\beta}
+f^{\alpha\beta\gamma} J_\sigma^\gamma \frac{\delta}{i\delta J_\sigma^\beta}
+f^{\alpha\beta\gamma}  J_\pi^\gamma \frac{\delta}{i\delta J_\pi^\beta}
\biggr\}_{x} Z\[{\bf J}\]. \label{vectsun_wti}
\eea
\subsection{Chiral Ward--Takahashi identities}
Of great importance for the study of dynamical chiral symmetry
breaking are the chiral WTIs.
These identities are similar to the vector WTIs, and can be derived accordingly.
The functional differential equation which generates chiral WTIs is
\bea
0&=&\biggr\{
\delta_{\alpha 0} A+
\partial_\mu(\gamma^\mu\g5)_{ba} \tau^\alpha_{ji}
\frac{\delta^2}{i\delta\bar\eta_{ai} i\delta\eta_{bj} }
+(i\g5)_{ba}\tau^\alpha_{ji}
\[\bar\eta_{bj} \frac{\delta}{i\delta \bar\eta_{ai} }+
\eta_{ai} \frac{\delta}{i\delta \eta_{bj} }
\]\nonu\\
&+&\frac{g^{\alpha\beta\gamma} c^\gamma}{G_0} \frac{\delta}{i\delta J_\pi^\beta}
+g^{\alpha\beta\gamma} J_\sigma^\gamma \frac{\delta}{i\delta J_\pi^\beta}
-g^{\alpha\beta\gamma}  J_\pi^\gamma \frac{\delta}{i\delta J_\sigma^\beta}
\biggr\}_{x} Z\[{\bf J} \], \label{chir_wti}
\eea
where $\alpha=0,\dots,N^2-1$.
The first term on the right-hand side, $A$, 
is the anomaly operator which is only present in the $U_A(1)$
Ward identity ($\alpha=0$), it is the well known Adler--Bell--Jackiw 
axial-anomaly \cite{ad69,beja69} and can be derived from 
Eqs.~(\ref{SDE2}), (\ref{SDE3}), once 
these equations are properly regularized by Pauli--Villars fields.
The naive assumption is that the functional measure is invariant under local 
unitary transformations, however it was shown by
Fujikawa \cite{fu79, fu80} that this is not the case, proper regularization 
of the functional measure shows that the under $U_A(1)$ transformations 
\bea
\psi(x)&\rightarrow& \psi'(x)={\,\rm e}^{i\theta(x)\g5}\psi(x),\\
\bar\psi(x)&\rightarrow& \bar\psi'(x)=\bar\psi(x){\, \rm e}^{i\theta(x)\g5},
\eea
the measure (\ref{measgnjl}) transforms
as 
\bea
{\cal D}\phi_l \rightarrow {\cal D}\phi_l \exp\[-2i\int{\rm d}^dx\,\theta(x) A(x)\],
\eea
where $A$ is the defined as the anomaly. 
It was shown by Hams \cite{anthony} that
when the generating functional is regularized with Pauli--Villars fields, 
the SDEs do indeed generate the anomaly.

In four dimensions, the anomaly term is of the form:
\bea
A(x)\sim \frac{\alpha_0}{4\pi}\epsilon^{\mu\nu\rho\sigma}
F_{\mu\nu}F_{\rho\sigma},
\eea
where $F$ is the field strength tensor and $\epsilon$ the totally antisymmetric
Levi-Civita tensor.
The anomaly terms represents the explicit breakdown of $U_A(1)$ symmetry 
due to quantum corrections. 

In this thesis we completely neglect the role of the axial-anomaly, 
since at the moment it is not clear how to implement it properly
in nonperturbative studies of chiral symmetry breaking. 
Secondly the dynamical breakdown of the $U_A(1)$ symmetry, 
with neglect of anomaly, 
serves as a very useful model for the breakdown of the more (complicated)
involved $SU_A(N)$ symmetries, which are free of anomalies.
\subsection{The axial vector vertex and PCAC}\label{cwti_pcac}
In this subsection some interesting relations following
from the chiral WTI (\ref{chir_wti}) are reviewed.
From Eq.~(\ref{chir_wti}) we can derive the identity
for the axial-vector vertex.
For simplicity the case with a single fermion flavor case ($N=1$) is considered 
with the neglect of the ABJ-anomaly 
and the axial-vector vertex is defined as
\bea
\[iS(k)(-i)\Gamma^{\mu}_{5}(k,p) iS(p)\]_{ab}
&\equiv& \int{\rm d}^4x{\rm d}^4y\,{\,\rm e}^{ikx-ipy}\nonu\\
&\times& \langle 0|T\(\bar\psi(0)
(-i\gamma^\mu \g5)\psi(0) \psi_a(x)\bar\psi_b(y)\)
|0\rangle\nonu\\
&=&\int{\rm d}^4x{\rm d}^4y\,{\,\rm e}^{ikx-ipy}(-i\gamma^\mu \g5)_{dc}
iD^{(4)}_{cd,ab}(0,0,x,y).\nonu\\
\eea
and the axial vertex as
\bea
\[iS(k)(-i)\Gamma_{5}(k,p) iS(p)\]_{ab}
&\equiv& \int{\rm d}^4x{\rm d}^4y\,{\,\rm e}^{ikx-ipy}\nonu\\
&\times& \langle 0|T\(\bar\psi(0)\g5\psi(0) \psi_a(x)\bar\psi_b(y)\)
|0\rangle\nonu\\
&=&\int{\rm d}^4x{\rm d}^4y\,{\,\rm e}^{ikx-ipy}(\g5)_{dc}
iD^{(4)}_{cd,ab}(0,0,x,y).\nonu\\
\eea
Using the definitions given in appendix \ref{gfprpv}
the following SDE for the axial-vector vertex can be derived:
\bea
(-i)\Gamma^{\mu}_{5ab}(k+q,k)&=&(-i\gamma^\mu \g5)_{ab}+
\int_{\Lambda}\frac{{\rm d}^4 k_1}{(2\pi)^4}\,
\[iS(k_1+q)(-i\gamma^\mu \g5)iS(k_1)\]_{dc}\nonu\\
&\times&
(-ie_0^2)
K^{(0)}_{cd,ab}(k_1,k_1+q,k+q), \label{axvec_vert}
\eea
which is depicted in Fig.~\ref{fig_sde_axialvector}, with the Feynman rules 
given in appendix \ref{feynmanrules}.
\begin{figure}[t!]
\epsfxsize=9cm
\epsffile[-10 440 390 540]{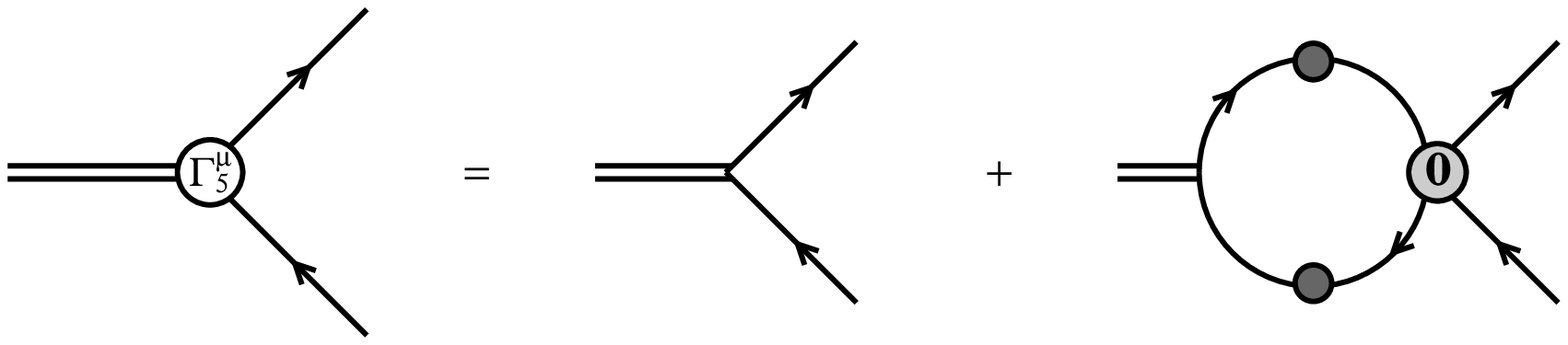}
\caption{The SDE for the axial-vector vertex.}
\label{fig_sde_axialvector}
\end{figure}
The SDE for the axial vertex is
\bea
(-i)\Gamma_{5ab}(k+q,k)&=&(\g5)_{ab}+
\int_{\Lambda}\frac{{\rm d}^4 k_1}{(2\pi)^4}\,
\[iS(k_1+q)(\g5)iS(k_1)\]_{dc}\nonu\\
&\times&
(-ie_0^2)
K^{(0)}_{cd,ab}(k_1,k_1+q,k+q).
\eea
The combination of the previous equation with the SDE for the pseudoscalar vertex
Eq.~(\ref{sdepseudovertex}) 
yields the following relation between the pseudoscalar vertex and 
axial vertex\footnote{Rewrite Eq.~(\ref{sdepseudovertex}) 
in terms of the scattering kernel $K^{(0)}$.}:
\bea
(-i)\Gamma_{5}(k+q,k)= (-iG_0)^{-1} i\DelP(q) (-i)\Gamma_{{\rm P}}(k+q,k).
\eea
From this relation it is clear that 
when the pseudoscalar propagator $i\DelP$ describes a particle
with a physical mass pole at $q^2=m_\pi^2$, the axial vertex $\Gamma_5$ also 
has a pole at $q^2=m_\pi^2$.

\begin{figure}[t!]
\epsfxsize=9cm
\epsffile[-10 440 390 540]{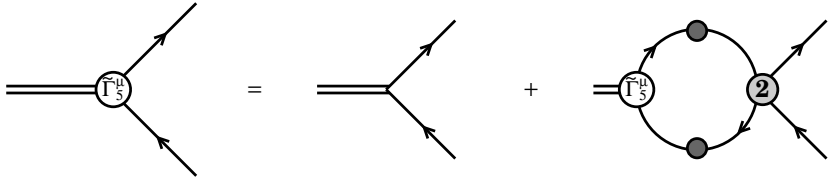}
\caption{The SDE for the regular axial-vector vertex.}
\label{fig_sde_regaxialvector}
\end{figure}

\begin{figure}[t!]
\epsfxsize=8.8cm
\epsffile[15 440 415 540]{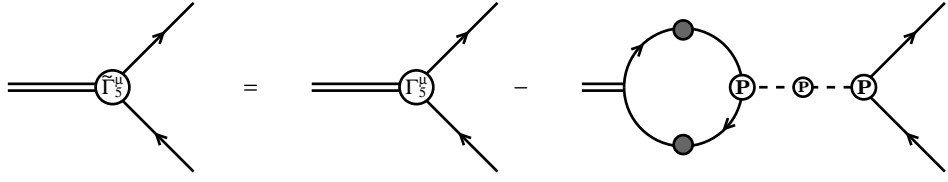}
\caption{Definition of the regular axial-vector vertex.}
\label{fig_sde_regaxialvectordef}
\end{figure}

If we now differentiate Eq.~(\ref{chir_wti}) with respect 
to $\bar\eta(y)$ and $\eta(z)$ and transform the equation to momentum space we 
find the chiral WTI for the axial-vector vertex:
\bea
(k-p)_\mu \Gamma^\mu_{5}(k,p)&=&
S^{-1}(k) (\g5)+(\g5)S^{-1}(p)\nonu\\
&-&2 i m_0 (-iG_0)^{-1}i\DelP(k-p) \Gamma_{{\rm P}}(k,p)\label{axvectwti}\\
&=&
S^{-1}(k) (\g5)+(\g5) S^{-1}(p)-2 i m_0 \Gamma_{5}(k,p). \nonu 
\eea
Hence the axial-vector vertex is related to the fermion propagator and the
pseudoscalar vertex via the chiral WTI.
Also we can show that the axial-vector vertex contains a pole at the physical
mass of the pseudoscalar by rewriting the SDE (\ref{axvec_vert}).
Using the definitions of the kernels $K^{(1)}$ and $K^{(2)}$,
Eqs.~(\ref{K1def}) and (\ref{K2def}), we obtain from Eq.~(\ref{axvec_vert}) 
\bea
(-i)\tilde\Gamma^{\mu}_{5ab}(k+q,k)&=&(-i\gamma^\mu \g5)_{ab}+
\int_{\Lambda}\frac{{\rm d}^4 k_1}{(2\pi)^4}\,\nonu\\
&\times&\[iS(k_1+q)(-i)\tilde\Gamma^{\mu}_{5}(k+q,k)iS(k_1)\]_{dc}\nonu\\
&\times&
(-ie_0^2)K^{(2)}_{cd,ab}(k_1,k_1+q,k+q), \label{axvec_vertreg}
\eea
see Fig.~\ref{fig_sde_regaxialvector}.
Since the Bethe--Salpeter kernel $K^{(2)}$ is defined to contains
no four-fermion poles, {\em i.e.,} all scalar and pseudoscalars
have been subtracted in Eq.~(\ref{K1def}), this ``regular'' axial-vertex 
$\tilde \Gamma^\mu_5$
now contains no poles at scalar or pseudoscalar masses
and is defined as
\bea
(-i)\tilde\Gamma^{\mu}_{5ab}(p+q,p)=
(-i)\Gamma^{\mu}_{5ab}(p+q,p)
-\Pi^\mu_{5}(q)i\DelP(q)(-i)\Gamma_{{\rm P}}(p+q,p),\label{regvertdef}
\eea
where the pseudovector polarization $\Pi^{\mu}_{5}$ is 
\bea
\Pi^\mu_{5}(q)&\equiv&
(-1)\int_{\Lambda} \frac{{\rm d}^4k}{(2\pi)^4}\,
\Tr\[(-i)\Gamma_{{\rm P}}(k,k+q) iS(k+q) (-i\gamma^\mu \g5) iS(k)\]\nonu\\
&=&\int_{\Lambda} \frac{{\rm d}^4k}{(2\pi)^4}\,
\Tr\[(-i\gamma^\mu \g5) iS(k+q) (-i)\Gamma_{{\rm P}}(k+q,k)iS(k)\]\nonu\\
&=&q^\mu \Pi_5(q^2),
\label{pseudovecvac}
\eea
since the expression can only be proportional to $q^\mu$ times a scalar 
function $\Pi_5$ of $q^2$. The change of sign results from reversing the
fermion-loop or in fact from reversing direction of $q$.
The definition of $(-i)\tilde\Gamma^{\mu}_{5}$ is depicted in 
Fig.~\ref{fig_sde_regaxialvectordef}.
The last term on the right-hand side of (\ref{regvertdef})
contains a pole at the pseudoscalar mass and is subtracted.
The residue of the pole is thus directly related to
the pseudoscalar vertex and the pseudovector polarization 
(\ref{pseudovecvac}).

A chiral WTI exists for this pseudovector polarization
which can be obtained directly from (\ref{chir_wti})
by differentiation with respect to source $ J_\pi(y)$ of $\pi$ 
and a transformation to momentum space. The chiral WTI reads
\bea
q_\mu \Pi^\mu_{5}(q)=q^2 \Pi_5(q^2)=
\frac{2m_0}{G_0}+2\langle \sigma \rangle \DelP^{-1}(q).\label{residueaxial}
\eea
Since the pseudovector polarization contains no ``pseudoscalar'' pole
it will be finite at $q^2=0$. Hence at $q^2=0$ we have
\bea
\DelP(q^2=0)=-\frac{G_0 \langle \sigma \rangle}{m_0}. \label{masspiwti}
\eea 
Using now the fact that 
\bea
\lim_{q\rightarrow 0} 
q_\mu \tilde\Gamma^{\mu}_{5}(p+q,p)=0,
\eea
since possible mass pole terms have been subtracted, 
and using the chiral WTI (\ref{axvectwti}), we obtain a relation
between the pseudoscalar vertex at zero boson momentum 
and the fermion mass function,
\bea
\Gamma_{\rm P}(p,p)=(i\g5) \frac{\Sigma(p^2)}{\langle \sigma \rangle},
\label{pseudowti}
\eea
with $\Sigma$ defined in Eq.~(\ref{struc_ferm}).
In the above derivation it was assumed that the fermion wave function 
is one, ${\cal Z}(p^2)=1$ 
\footnote{This assumption is believed to be reliable for the 
quenched-ladder approximation in the Landau gauge, $\xi=0$}.

The identities (\ref{masspiwti}) and (\ref{pseudowti})
are usually mentioned, in the literature, together with the zero boson momentum
expressions for the scalar propagator and scalar vertex
\bea
\DelS(q^2=0)&=&-G_0\frac{\partial \langle \sigma \rangle}{\partial m_0},
\label{masssiganal}\\
\Gamma_{\rm S}(p,p)&=&({\bf 1}) 
\frac{\partial \Sigma(p^2)}{\partial \langle \sigma \rangle},
\label{scalanal}
\eea
were again ${\cal Z}(p^2)=1$ is assumed.
However, these identities are not a straightforward consequence of
the chiral Ward identity (\ref{chir_wti}), 
but follow from the (presummed) analyticity of the generating functional
$Z[{\bf J}]$, Eq.~(\ref{genfie_GNJL}), 
in the sources ${\bf J}$, and the normalization $Z[0]=1$.\footnote{This 
analyticity in the sources ${\bf J}$ should not be confused with
analyticity in bare parameters such as the gauge coupling $\alpha_0$, since
the chiral phase transition is a consequence 
of nonanalyticity of $Z$ in for instance $\alpha_0$.}
The analyticity in sources requires that differentiations with respect 
to bare parameters ($b_0$) of $Z$ are zero at vanishing sources, {\em i.e.},
\bea
\frac{\partial^n Z[{\bf J}]}{\partial b_0^n}\biggr|_{{\bf J}=0}=0,\qquad 
b_0=m_0,\,\alpha_0,\,\xi\,,\dots.
\eea
Eq.~(\ref{masssiganal}) can be derived 
from the identity
\bea
\frac{\partial^2 Z[{\bf J}]}{\partial m_0^2}\biggr|_{{\bf J}=0}=0,
\eea
and, by making use of Eqs.~(\ref{scalcond}), (\ref{condtrace}) 
and (\ref{sdescal}), Eq.~(\ref{scalanal}) is obtained.

The relations given in this subsection are very important for the 
low-energy effective dynamics or the PCAC dynamics.
Suppose that scalar and pseudoscalar propagators have the following
low-energy or infra-red structure:
\bea
\DelS(q)=\frac{\tau_\sigma}{q^2-m_\sigma^2},\qquad
\DelP(q)=\frac{\tau_\pi}{q^2-m_\pi^2}.\label{delplow}
\eea
Then the mass-pole of the pseudoscalar at $m_\pi$ also appears as a pole 
in the axial-vector vertex, $\Gamma^\mu_5$, 
and the residue $\Pi_5(q^2)$ of Eq.~(\ref{pseudovecvac})
is related to the pion decay constant $f_\pi$ in the following 
way, see {\em e.g.} Refs.~\cite{mirbook,report256,ha98b,marota98}:
\bea
\Pi_5(0)=\frac{f^2_\pi}{2\langle\sigma\rangle},
\label{decayconstdef}
\eea
The chiral WTI (\ref{residueaxial}) using Eq.~(\ref{delplow}) 
at $q^2=0$ implies that
\bea
\Pi_5(0)=
\frac{2\langle\sigma\rangle}{\tau_\pi}
\quad \Longrightarrow\quad
f_\pi=\frac{2\langle\sigma\rangle}{\tau_\pi^{1/2}}
,\qquad \tau_\pi\sim Z_\pi^2(0), \label{fpi}
\eea
where $Z_\pi$ is the renormalization constant of
the pseudoscalar field, Eq.~(\ref{Zsigmapidef}). 
An approximation to $f_\pi$ can be obtained by
assuming that, for small $q$, $\GP(k+q,k)\approx \GP(k,k)$ 
in Eq.~(\ref{pseudovecvac}), this
reproduces the well-known Pagels--Stokar's \cite{past79} formula 
for $f_\pi^2$, 
\bea
f_\pi^2=\(-\frac{i}{2\pi^2}\)\int\frac{{\rm d}^4k}{\pi^2}\,
\frac{\Sigma(k^2)}{\[k^2-\Sigma^2(k^2)\]^2}\[\Sigma(k^2)
-\frac{k^2}{2}\Sigma^\prime(k^2)\].
\eea

In the D$\chi$SB phase, in the chiral limit $m_0\rightarrow 0$,
Eq.~(\ref{pseudowti}) together with Eq.~(\ref{residueaxial})
relates the pion decay constant $f_\pi$, the fermion dynamical mass $\Sigma$
and the vertex $\Gamma_{\rm P}$, and
can be considered as the generalization of the Goldenberger--Treiman
relation in the GNJL, see for instance \cite{mirbook,rowi94}.
Eq.~(\ref{pseudowti}) relates the pseudoscalar vertex to the dynamical mass 
function. Moreover Eq.~(\ref{masspiwti}) tells us that
if we take the chiral limit, {\em i.e.,} the bare mass $m_0\rightarrow 0$,
the pseudoscalar propagator will describe a massless particle, $m_\pi=0$, 
{\em i.e.} a Goldstone boson.

From the chiral WTI one can also derive the relation between 
UV behavior of the scalar and pseudoscalar propagators
and the scalar and pseudoscalar vertices. 
Is to be expected that such Green functions are asymptotically degenerate, 
since the UV off-shell behavior of Green functions
is not determined by the low-energy potential describing the dynamically broken
chiral symmetry; it is determined 
by the behavior of the ``potential'' far above the minimum ($q\gg m_\sigma$), 
where it still is chirally symmetric (there, $\DelS$ and $\DelP$ are degenerate).
\section{Particles}\label{sec_particles}
In the GNJL model the particle spectrum contains in addition to 
the usual spin-$1/2$
fermions, and spin-$1$ gauge-bosons also the spinless composite states of 
fermion and anti-fermion, the scalars and pseudoscalars.
Such particles or states are classified either 
as massive respectively massless stable particles, 
unstable particles (resonances), or as unphysical states called tachyons.
Suppose we can write a particle propagator phenomenologically as
\bea
\Delta(p)\sim \frac{1}{p^2-\mu^2},
\eea
where for the time being we ignore spin-structure. The expression
is valid for momenta close to the singularity $p^2\sim |\mu|^2$.
Stable particles are characterized by a real positive mass 
singularity $\mu^2 \geq 0$, resonances by a complex singularity $\mu^2$ is complex, 
and tachyons by an imaginary pole $\mu^2 < 0$.

Masses and decay widths play an important role in the analytic structure
of Green functions in the infra-red region. 
Where, by infra-red region, we refer to momenta close to being on-shell, 
{\em i.e.} $p^2\sim |\mu|^2$.
Intermediate and ultra-violet regions correspond to momenta which are 
highly off-shell or virtual, $-p^2 \gg |\mu|^2$.
In coordinate space, this means correlations are considered 
at distances much shorter than the correlation length $\xi\sim 1/\mu$.

It is believed that analysis of SDEs in Euclidean formulation gives reliable 
results in the ultra-violet regions and intermediate regions,
{\em i.e.,} regions where the only relevant scale is the ultra-violet 
cutoff $\Lambda$. 
However to obtain results describing a correct analytic structure 
(from the physical point of view) of Green functions in the infra-red region
requires in general a more delicate analysis, 
and, generally speaking, results are much more sensitive to the various 
approximations used, see, for a thorough discussion of complex branch-point
singularities 
of fermion propagators in the framework of SDEs in QED and QCD, 
the thesis by Maris \cite{pieter}.

Below, a short overview is given of the various particles states
that we will encounter throughout this thesis.
\paragraph{Stable particles:}
A stable physically observable particle, is believed to have an asymptotic
in and out state and a K{\"a}llen--Lehmann representation\footnote{Contrary to QCD
where there is no such thing as an asymptotic in or out quark propagator, 
and hence there is no K{\"a}llen--Lehmann representation 
due to color confinement (breakdown of the cluster property), 
see \cite{rowi94}.}, see \cite{bjodre}.
A stable particle such as a fermion or a gauge boson in an Abelian theory
is described by a particle propagator with a Lorentz structure 
defined by their spin, and has a real mass singularity $p^2=\mu^2 \geq 0$ 
The mass singularity is a branch point on the positive real axis of the complex 
$p^2$-plane. 
\paragraph{Resonances:}
A resonance is an unstable particle or bound state with
a lifetime $\tau\sim 1/\Gamma$, where $\Gamma$ is
the width (or decay rate) of the resonance. 
A resonance of the Breit--Wigner type \cite{landlifQM, newton, pieter} 
is described by a complex pole in the particle or bound state propagator:
\bea
\Delta(p) \sim \frac{1}{p^2-\[M-(i/2) \Gamma\]^2},
\eea
which is supposed to be valid for $p^2$ 
close to the resonance pole, $p^2\sim M^2$. 
The resonance mass $M$ and width $\Gamma$ are defined to be real and positive.
Unitarity of the S-matrix requires that the complex pole of the propagator 
lies on a second Riemann sheet of the complex plane of $p^2$, 
for an explanation of this point see for instance \cite{newton}.
When we parametrize the resonance pole with a radius $R$ and angle $\theta$
\bea
p^2=\[M-(i/2)\Gamma\]^2\equiv R\exp(-i\theta),
\eea
then the condition 
\bea
0\leq \theta < \pi
\eea
ensures that the pole lies on the second Riemann sheet of $p^2$, where 
we have used that the width over mass ratio is positive,
\bea
\frac{\Gamma}{M}=\frac{2\sin\theta}{1+\cos\theta} \geq 0.
\eea
\paragraph{Tachyons:}
Tachyons are described by an imaginary mass pole, and are called unphysical 
states, they in fact correspond to particles moving faster than 
the speed of light.
Although such tachyonic particle propagators sometimes occur as solutions 
of the SDEs, they usually correspond to unstable vacuum configurations,
{\em i.e.,} they correspond to a local maximum instead of a local minimum of the action, 
and rearrangement of the vacuum 
will remove the tachyonic states from the physical particle spectrum.
For a thorough discussion of tachyonic solutions of Bethe--Salpeter 
equations, see \cite{mirbook} and references therein.
\subsection{The Wick rotation}
The Schwinger--Dyson equation are formulated in Minkowski space, given
by the indefinite metric $g^{\mu\nu}$. A four-momentum 
$p^2_{\rm Mink}=p_0^2-\vec p^2$ can either be positive or negative.
However actual calculations of SDEs are preferably 
performed in Euclidean formulation, which has a definite positive metric, 
$p^2_{\rm Eucl} \geq 0$.
The formulation in Euclidean space considerably 
simplifies the actual computation of Feynman graphs and 
henceforth the analysis of SDEs. 
Once Green functions have been obtained in Euclidean space 
(for space-like Minkowski momenta), they are rotated back to Minkowski space, 
after which they can be analytically continued into the time-like (physical) 
momentum region.

The transformation from Minkowski space to Euclidean space is 
called the Wick rotation, see Ref.~\cite{wi54}, and rotates 
the zero-component or energy component
of the four-momentum to the imaginary axis, $p_0 \rightarrow ip_0$.
We then replace all Minkowski four-momenta with definite positive 
Euclidean momenta
\bea
-p^2_{\rm Mink}=p_0^2+\vec p^2\equiv p^2_{\rm Eucl}.
\eea
The Wick rotation is allowed only if the Green functions do not have singularities 
in the second and fourth quadrant of the appropriate complex momentum plane. 
It is assumed for all practical purposes that this is the case in this thesis.

%
%
\chapter{Dynamical chiral symmetry breaking}\label{chap3}
Dynamical symmetry breaking is a special case of spontaneous symmetry breaking,
where a continuous symmetry is broken through the appearance of a 
non-vanishing vacuum expectation value of a composite operator.
Dynamical symmetry breaking is governed by the formation of bound states.
As was mentioned in the introduction, 
the Bardeen-Cooper-Schrieffer (BCS) theory of superconductivity is one of 
earliest models of dynamical symmetry breaking and 
the first model with dynamical chiral symmetry breaking (D$\chi$SB) 
in high energy particle physics is the Nambu--Jona-Lasinio model \cite{najola61}.  

In this chapter we will discuss the chiral phase transition corresponding to
the dynamical breakdown of the chiral 
symmetry in the GNJL model in the quenched ladder approximation treating
the four-fermion interaction in the Hartree-Fock or mean field approximation.
The quenched ladder approximation will be discussed in the next section.
A combination of strong attractive four-fermion interaction and gauge interaction
will turn out to break the chiral symmetry dynamically by the formation of a 
chiral condensate, where the chiral condensate 
Eq.~(\ref{condtrace}) provides the order parameter of the DCSB.
Since the chiral symmetry is a continuous symmetry, the DCSB gives rise to the
Goldstone mechanism (see section~\ref{sec_goldstone}), 
and PCAC dynamics through the formation of spinless 
bound states, 
see subsection~\ref{cwti_pcac}.
DCSB leads to a nonzero fermion dynamical mass 
$m_{\rm dyn}$ and to the appearance of pseudoscalar massless Nambu--Goldstone 
bosons composed of a massive fermion and a massive antifermion.
Following Ref.~\cite{mirbook},
the massless NG bosons are characterized by a large binding energy,
and it is argued that the dynamics producing such bound states should 
be rather strong.

The setup of this chapter is the following.
First, the quenched ladder approximation is discussed, 
subsequently the gap equation for the fermion dynamical mass $m_{\rm dyn}$
following from the SDE for the fermion propagator
is reviewed. In section~\ref{sec_eos},
we derive the EOS for the GNJL model in analogy with Eq.~(\ref{eosmagnet}).
We discuss the solutions to the gap equation and the EOS,
obtain the critical fixed point,
and derive the scaling laws along with the critical exponents.
In section~\ref{sec_cpt}, the concept of the conformal phase transition
is introduced.
Finally we discuss the RG flow and fine-tuning
of the four-fermion coupling $g_0$, and compare the results
obtained in the quenched ladder approximation with various
lattice results.
\section{The quenched ladder approximation}\label{sec_thequenched}
In the SD approach, the gap equation has been studied extensively 
in the so-called quenched-ladder, {\em i.e.} the quenched-planar approximation.
In a quenched approximation only diagrams without fermion loops
are taken into account, {\em i.e.}
the fermion loops in the vacuum polarization of 
the photon propagator
are neglected\footnote{The quenched
approximation can be considered as the limit of zero fermion flavors, $N=0$,
in the SDE for the photon propagator.}.
The quenched approximation can be considered 
as an approximation to a theory with an UV fixed point;
at such a point the gauge coupling does not run, 
and the $\beta$-function of $\alpha_0$ vanishes:
\bea
\beta_\alpha(\mu_0,g_0,\alpha_0)\equiv 
\Lambda\frac{{\rm d}\alpha_0}{{\rm d}\Lambda}=0. \label{basicquenched}
\eea
This is a rather crucial assumption, and the possible existence of an UV 
fixed point of Eq.~(\ref{basicquenched}) will be investigated 
in chapter \ref{chap5}.

The ladder or planar approximation is a replacement 
of the full photon-fermion vertex $\Gamma^\mu$ of Eq.~(\ref{photfermvertsde})
by the bare vertex $\gamma^\mu$ (photon lines don't cross),
the replacement of the BS kernel $K^{(2)}$, Eq.~(\ref{K2def}), 
by a single photon exchange graph.
Thus the ladder approximation is based on the following assumptions
\bea
\Gamma^{\mu}_{ab}(k,p)&\approx&\gamma^\mu_{ab},\\
K^{(2)}_{cd,ab}(p,p+q,k+q)
&\approx&D_{\mu\nu}(k-p)\gamma^\mu_{ad}\gamma^\nu_{cb}. 
\eea
The ladder or planar approximation is assumed to be valid in
the Landau gauge only ($\xi=0)$ where ${\cal Z}_{(f)}=1$, since then the 
vector WTI (\ref{wtivertex}) and chiral WTI (\ref{axvectwti})
are satisfied asymptotically.

In addition to the quenched ladder approximation, the four-fermion
interactions are treated in the Hartree-Fock approximation.\footnote{The 
Hartree-Fock approximation is a specific mean field approach.} 
This means that quantum corrections
corresponding to four-fermion interactions are neglected beyond tree level
in Feynman diagrams,
and only the coupling to the chiral condensate 
represented by the tadpole graphs of Fig.~\ref{fig_sde_ferm}
is taken into account.
Hence, on the level of SDE for the fermion propagator, 
in the mean field approach contributions corresponding to exchanges of virtual
scalars and pseudoscalars composites ($\sigma$ and $\pi$ bosons) 
are neglected.
\section{The Gap equation}
The SDE (\ref{sde_ferm}) for the fermion mass function $\Sigma$ defined by 
Eq.~(\ref{struc_ferm}) is commonly referred to as the gap equation in 
analogy with the equation for the energy gap near the Fermi surface 
of one-fermion excitations in the  Bardeen--Cooper--Schrieffer (BCS) theory 
of superconductivity.
Pioneering studies on the gap-equation in gauge theories
were performed in Refs.~\cite{mana74,fomi76,fogumisi76}.
These studies find a critical coupling $\alpha_c\sim 1$ above which a 
nontrivial solution for the fermion dynamical mass exists in the chiral limit,
{\em i.e.} zero bare mass $m_0=0$.
As was shown in Refs.~\cite{fuku76,atbl79,fi80}, the solution for 
the fermion propagator has moving branch-point singularities 
in the complex $p^2$-plane, instead of a real singularity 
(on physical grounds a real singularity is required). 
It is generally believed that the complex branch-points are an artifact of 
the ladder approximation. 
Extensive discussion of the gap equation in QED, QCD and the GNJL are given 
in {\em e.g.} \cite{fogumisi83,komiya89,ta89,mirbook,report256,kotaya93,pieter}.
In quenched QED, a nontrivial continuum limit can be reached
\cite{fogumi78,mi80,fogumisi83,mi85a,mi85c}. 

After taking the trace of the SDE (\ref{sde_ferm}), 
performing a Wick rotation to Euclidean momentum $x=p^2 \geq 0$, 
and using Eq.~(\ref{scalcond}), we obtain the well-known gap equation 
for the fermion mass function $\Sigma$
\bea
\Sigma(x)&=&\langle \sigma \rangle 
+\frac{\lambda_0}{x}
\int\limits_0^{x}{\rm d}y\,\frac{y\Sigma(y)}{y+\Sigma^2(y)}
+\lambda_0
\int\limits_x^{\Lambda^2}{\rm d}y\,\frac{\Sigma(y)}{y+\Sigma^2(y)},\label{gap_eq}\\
\langle \sigma \rangle &=&
m_0+\frac{g_0}{\Lambda^2}
\int\limits_0^{\Lambda^2}{\rm d}y\,\frac{y\Sigma(y)}{y+\Sigma^2(y)},
\label{condensate2}
\eea
where we recall that the dimensionless four-fermion coupling $g_0$ is defined
as
\bea
\frac{g_0}{\Lambda^2}\equiv \frac{G_0}{4\pi^2},\qquad
\lambda_0=\frac{3\alpha_0}{4\pi}, \label{g0def}
\eea
with $\Lambda$ the ultraviolet cutoff.
The gap equation (\ref{gap_eq}) can be written as a nonlinear second-order
differential equation
\bea
x\frac{{\rm d}^2}{{\rm d}x^2}\Sigma(x)+2\frac{{\rm d}}{{\rm d}x}\Sigma(x)
+\frac{\lambda_0\Sigma(x)}{x+\Sigma^2(x)}=0,\label{diff_gap_eq}
\eea
with an infrared boundary condition (IRBC), 
\bea
0=\[x^2\frac{{\rm d}}{{\rm d}x}\Sigma(x)\]_{x=0},\label{massdse_irbc}
\eea
and 
an ultraviolet boundary condition (UVBC), 
\bea
m_0=\[\(1+\frac{g_0}{\lambda_0}\)
x\frac{{\rm d}}{{\rm d}x}\Sigma(x)+\Sigma(x)\]_{x=\Lambda^2}. \label{massdse_uvbc}
\eea
At vanishing gauge coupling ($\lambda_0=0$), 
the only possible solution to the gap equation is a constant 
$\Sigma(x)=\Sigma_0$.
Hence, at $\lambda_0=0$, the gap equation reads
\bea
\frac{m_0}{\Lambda}=-(g_0-1)\frac{\Sigma_0}{\Lambda}
+g_0\frac{\Sigma_0^3}{\Lambda^3} \ln \(\frac{\Lambda^2+\Sigma_0^2}{\Sigma_0^2}\).
\label{gapeq_njl}
\eea
This is the famous gap equation of Nambu and Jona-Lasinio \cite{najola61}.

For $\lambda_0>0$, the four-fermion coupling $g_0$ comes into play via
the UVBC, since Eq.~(\ref{diff_gap_eq}) does not depend on $g_0$.
At present no (nontrivial) analytic solution to 
the nonlinear second order differential equation Eq.~(\ref{diff_gap_eq}) 
is known, except for the case of zero gauge coupling, 
$\lambda_0=0$, Eq.~(\ref{gapeq_njl}).
\subsection{The linearized approximation}
After approximating $\Sigma^2(x)$ by $\Sigma^2(0)$
in the denominator of Eq.~(\ref{diff_gap_eq}), the gap equation
becomes linear (though with a nonlinear UVBC) and can be solved straightforwardly 
\cite{fogumi78,fogumisi83,mifo85}. 
In this so-called linearized approximation, the gap equation reads
\bea
x\frac{{\rm d}^2}{{\rm d}x^2}\Sigma(x)+2\frac{{\rm d}}{{\rm d}x}\Sigma(x)
+\frac{\lambda_0\Sigma(x)}{x+\Sigma_0^2}=0,\label{lindiff_gap_eq}
\eea
where 
\bea
\Sigma_0\equiv \Sigma(0).
\eea
Clearly the linearized approximation is a good approximation
in both the infrared ($x\ll \Sigma_0^2$) 
and the ultraviolet ($x\gg \Sigma_0^2$) regions.
Moreover numerical analysis of {\em e.g.} Maris \cite{pieter}
shows that the linearized approximation is valid for 
the entire range of momenta, even for momenta 
close to the branch points ($x\sim \Sigma_0^2$).

In terms of the variable $u=-x/\Sigma_0^2$, the equation 
can be written as a hypergeometric differential equation
\bea
u(1-u)
\frac{{\rm d}^2}{{\rm d}u^2}\Sigma(u)+2(1-u)\frac{{\rm d}}{{\rm d}u}\Sigma(u)
-\lambda_0\Sigma(u)=0.
\eea
The hypergeometric differential equation is usually written as
\bea
z(1-z)w^{\prime\prime}(z)+(c-(1+a+b)z)w^\prime(z)-ab w(z)=0,
\eea
which has the general solution in terms of hypergeometric functions
\bea
w(z)&=&C_1 w_1(z)+C_2 w_2(z),\nonu\\
w_1(z)&=&{_2}F_1(a,b,c;z),\\
w_2(z)&=&z^{1-c}{_2}F_1(a-c+1,b-c+1,2-c;z)\nonu,
\eea
for $c$ not integer $\geq 2$.

For the linearized gap equation, the values of $a$, $b$, and $c$ are
\bea
\begin{array}{llll}
a=(1+\omega)/2,&b=(1-\omega)/2,&c=2, &(\alpha_0\leq \alpha_c)\\
a=(1+i\nu)/2,&b=(1-i\nu)/2,&c=2,& (\alpha_0>\alpha_c)
\end{array}
\eea
where 
\bea
\alpha_c&\equiv& \pi/3,\\
\omega&\equiv&\sqrt{1-\alpha_0/\alpha_c}
=\sqrt{1-4\lambda_0},\label{omegadef}\\
\nu&\equiv&\sqrt{\alpha_0/\alpha_c-1}=\sqrt{4\lambda_0-1}.\label{nudef}
\eea
Thus, the point $\alpha_0=\alpha_c$ ($\lambda_0=1/4$) is rather special, 
since at that point $a$ and $b$ change from
real ($\alpha_0<\alpha_c$) to complex values ($\alpha_0>\alpha_c$).
Therefore, we distinguish between four specific regimes of gauge coupling:
\begin{itemize}
\item[A]{The pure Nambu--Jona-Lasinio (NJL) point: $\alpha_0=0$,}
\item[B]{The intermediate region: $0<\alpha_0<\alpha_c$,}
\item[C]{The QED critical point: $\alpha_0=\alpha_c$,}
\item[D]{The strong QED region: $\alpha_0>\alpha_c$.}
\end{itemize}

For any (real positive) value of $\alpha_0$,
the solution $w_2(u)$ is irregular at $u=0$ ($x=0$), and does not satisfy 
the IRBC (\ref{massdse_irbc}).
Therefore the solution of the linearized gap-equation is
\bea
\frac{\Sigma(x)}{\Sigma_0}=
\left\lbrace
\begin{array}{cc}
\,{_2}F_1\((1+\omega)/2,(1-\omega)/2,2;
-x/\Sigma_0^2\),&(\alpha_0\leq\alpha_c)\\
\,{_2}F_1\((1+i\nu)/2,(1-i\nu)/2,2;
-x/\Sigma_0^2\),&(\alpha_0>\alpha_c)
\end{array}\right.\label{linsolu}
\eea
where $\Sigma_0$ has to be determined from the UVBC.
Thus the IRBC (\ref{massdse_irbc}) determines uniquely the solution
of the Eq.~(\ref{lindiff_gap_eq}), 
leaving the value $\Sigma_0=\Sigma(0)$ as a free parameter.
The UVBC (\ref{massdse_uvbc}) then gives a relation between
$\Sigma_0$, $m_0$, $g_0$, $\alpha_0$, and the ultraviolet cutoff $\Lambda$.

By making use of the identity
\bea
\frac{{_2}F_1(a,b,c;z)}{\Gamma(c)}&=&
\frac{\Gamma(b-a)}{\Gamma(b)\Gamma(c-a)}(-z)^{-a}
{_2}F_1\(a,1-c+a,1-b+a;\frac{1}{z}\)\nonu\\
&+&\frac{\Gamma(a-b)}{\Gamma(a)\Gamma(c-b)}(-z)^{-b}
{_2}F_1\(b,1-c+b,1-a+b;\frac{1}{z}\),
\eea
$a-b$ not integer, and $|\arg(-z)|<\pi$,
we get for $|z|\gg 1$
\bea
\frac{{_2}F_1(a,b,c;-z)}{\Gamma(c)}&\approx&
\frac{\Gamma(b-a)}{\Gamma(b)\Gamma(c-a)}\,z^{-a}
+\frac{\Gamma(a-b)}{\Gamma(a)\Gamma(c-b)}\,z^{-b},\quad |z|\gg 1.
\eea
For large momenta, $x/\Sigma_0^2\gg 1$, and $\alpha_0<\alpha_c$ 
the fermion mass function is expressed as
\bea
\frac{\Sigma(x)}{\Sigma_0}&\approx& 
c(\omega)\(\frac{x}{\Sigma_0^2}\)^{-(1-\omega)/2}
-d(\omega)
\(\frac{x}{\Sigma_0^2}\)^{-(1+\omega)/2}\nonu\\
&+& {\cal O}\(\(x/\Sigma_0^2\)^{(\omega-3)/2}\). 
\label{sigma_asym}
\eea
with
\bea
c(\omega)=\frac{\Gamma(\omega)}{\Gamma\(\frac{1+\omega}{2}\)
\Gamma\(\frac{3+\omega}{2}\)},\qquad
d(\omega)=-c(-\omega) > 0.
\eea
An analogous expression which is valid for $\alpha_0>\alpha_c$ 
is obtained by replacing $\omega$  in Eq.~(\ref{sigma_asym}) by $i\nu$.
Special care should be taken at $\alpha_0=\alpha_c$ ($\omega=0$);
at that point, we should expand Eq.~(\ref{sigma_asym}) around $\omega=0$.

To summarize, in the limit $x/\Sigma_0^2\gg 1$, 
Eq.~(\ref{linsolu}) can be written in the following form:
\bea
\!\!\!\!\!\frac{\Sigma(x)}{\Sigma_0}
&\approx&
 \left\{
\begin{array}{ccc}
&A(\alpha_0)\(\Sigma_0^2/x\)^{1/2}
\sinh \[
\frac{\omega}{2}\ln \frac{x}{\Sigma_0^2}+\omega \delta(\alpha_0) \]/\omega,&
(\alpha_0<\alpha_c)\\
&A(\alpha_c)\(\Sigma_0^2/x\)^{1/2}
\[ \frac{1}{2}\ln \frac{x}{\Sigma_0^2}+\delta(\alpha_c)\],&
(\alpha_0=\alpha_c)\label{Adeltaeq}\\
&A(\alpha_0)
\(\Sigma_0^2/x\)^{1/2}
\sin \[
\frac{\nu}{2}\ln \frac{x}{\Sigma_0^2}+\nu \delta(\alpha_0) \]/\nu,&
(\alpha_0>\alpha_c)
\end{array}\right.
\eea
with $\omega$ and $\nu$ given in Eqs.~(\ref{omegadef}) and (\ref{nudef}),
and where
\bea
A(\alpha_0)&=&2\omega \sqrt{c(\omega)d(\omega)}
=\sqrt{\frac{8\omega \cot(\pi\omega/2)}{\pi(1-\omega^2)}},
\qquad A(\alpha_c)=\frac{4}{\pi}\label{Awofsigma}\\
\delta(\alpha_0)&=&\frac{1}{2\omega}\ln \frac{c(\omega)}{d(\omega)},
\qquad \delta(\alpha_c)=2\ln 2-1.
\eea
Here the real valued functions
$A$ and $\delta$ are the ``amplitude and the phase'' of the
solution, while $\Sigma_0$ is the mass-scale, which depends
on the bare mass $m_0$.
Since the functions $A$ and $\delta$ are symmetric in $\omega$,
they can  be analytically continued to values $\alpha_0>\alpha_c$,
{\em i.e.} $\omega$ is replaced by $i\nu$.
\section{The equation of state}\label{sec_eos}
The relation between $\Sigma_0$ and the chiral condensate
$\langle \sigma \rangle$ can be obtained from the gap equation 
(\ref{gap_eq})
\bea
\[x\frac{{\rm d}}{{\rm d}x} \Sigma(x)\]_{x=\Lambda^2}=\frac{\lambda_0}{g_0}\(m_0-\langle\sigma\rangle\)=
\frac{3\pi\alpha_0}{\Lambda^2}\langle \bar\psi \psi\rangle,\label{cond_gap}
\eea
see Eqs.~(\ref{scalcond}) and (\ref{condtrace}).
From the equation (\ref{cond_gap}) 
for the order parameter $\langle \bar\psi\psi \rangle$
and the UVBC (\ref{massdse_uvbc}), the fermion mass scale $\Sigma_0$ 
can be eliminated.
This yields a single equation between the order parameter 
$\langle \bar\psi\psi \rangle$, and the bare couplings
$g_0$, $\alpha_0$ and the bare mass $m_0$, and the ultraviolet 
cutoff $\Lambda$.
In analogy with statistical mechanics, such an equation 
is referred to as ``the equation of state'' (EOS).

For the intermediate region $\alpha_0<\alpha_c$, 
with the assumption $\Sigma_0\ll \Lambda$, the equation for
the condensate $\langle \bar\psi\psi \rangle$ can be obtained
by substituting Eq.~(\ref{sigma_asym}) in Eq.~(\ref{cond_gap}).
Similarly, the UVBC can be solved by substituting  
Eq.~(\ref{sigma_asym}) in Eq.~(\ref{massdse_uvbc}).
This gives the following EOS for $\alpha_0<\alpha_c$:
\bea
\frac{m_0}{\Lambda}&=&
\frac{\Sigma_0^2}{\Lambda^2}\[
-\Delta g_0\, C(\omega)\(\frac{\Sigma_0}{\Lambda}\)^{-\omega}
+\(\Delta g_0+\omega\) D(\omega)\(\frac{\Sigma_0}{\Lambda}\)^{\omega}\],
\label{mu_0bc1}\\
\frac{\langle \bar\psi \psi\rangle}{\Lambda^3}&=&\frac{\Sigma_0^2}{\Lambda^2}
\frac{1}{4\pi^2}
\[-C(\omega)\(\frac{\Sigma_0}{\Lambda}\)^{-\omega}
+D(\omega)\(\frac{\Sigma_0}{\Lambda}\)^{\omega}\],\quad (\alpha_0< \alpha_c)
\label{condbc1}
\eea
where we have used that $\lambda_0=(1-\omega^2)/4$, and where
\bea
\Delta g_0\equiv g_0-\frac{(1+\omega)^2}{4},
\eea
and
\bea
C(\omega)=\frac{2}{1+\omega} c(\omega),\qquad D(\omega)=-C(-\omega).
\eea
At the critical gauge coupling $\alpha_0=\alpha_c$, 
the EOS is obtained by substituting Eq.~(\ref{Adeltaeq})
in Eqs.~(\ref{massdse_uvbc}) and (\ref{cond_gap}).
This gives 
\bea
\frac{m_0}{\Lambda}&=&\frac{\Sigma_0^2}{\Lambda^2} A(\alpha_c)\[
-\Delta g_0\[\ln \frac{\Lambda^2}{\Sigma_0^2}+2\delta(\alpha_c)\]
+1+2\Delta g_0\],\label{mu_0bc2}\\
\frac{\langle \bar\psi \psi\rangle}{\Lambda^3}
&=&\frac{\Sigma_0^2}{\Lambda^2}\frac{A(\alpha_c)}{4\pi^2}
\[-\ln\frac{\Lambda^2}{\Sigma_0^2}-2\delta(\alpha_c)+2\],
\quad (\alpha_0=\alpha_c)
\eea
where $\Delta g_0=g_0-1/4$ (at $\alpha_0=\alpha_c$).

For values of the gauge coupling larger than the critical value
it is also convenient to use Eq.~(\ref{Adeltaeq}) to express 
the UVBC (\ref{massdse_uvbc}) and Eq.~(\ref{cond_gap}) as goniometric equations,
\bea
\frac{m_0}{\Lambda}&=&
\frac{\Sigma_0^2}{\Lambda^2} \frac{2A(\alpha_0)}{(1+\nu^2)\nu}\[
\(\lambda_0-g_0\)\sin \theta
+\nu \(\lambda_0+g_0\)\cos \theta \],\label{mu_0bc3}\\
\frac{\langle \bar\psi \psi\rangle}{\Lambda^3}&=&
\frac{\Sigma_0^2}{\Lambda^2}\frac{A(\alpha_0)}{2\pi^2(1+\nu^2)\nu}
 \[-\sin\theta +\nu \cos \theta \],\quad (\alpha_0>\alpha_c)
\eea
where $\lambda_0=(1+\nu^2)/4$, and where  
\bea
\theta=\frac{\nu}{2} \ln \frac{\Lambda^2}{\Sigma_0^2}+\nu\delta(\alpha_0),
\quad \nu=\sqrt{\alpha_0/\alpha_c-1}.
\label{thetadef}
\eea
The EOS for the pure NJL model ($\alpha_0=0$) is given by 
Eq.~(\ref{gapeq_njl}), and the equation for the condensate follows
from Eq.~(\ref{condensate2}).
Thus
\bea
\frac{m_0}{\Lambda}&=&\frac{\Sigma_0^2}{\Lambda^2}\[
-\Delta g_0\frac{\Lambda}{\Sigma_0}
+(\Delta g_0+1)\frac{\Sigma_0}{\Lambda} \ln \frac{\Lambda^2}{\Sigma_0^2}\],
\label{gapeq_njl2}\\
\frac{\langle \bar\psi\psi\rangle}{\Lambda^3}&=&\frac{\Sigma_0^2}{\Lambda^2} 
\frac{1}{4\pi^2}
\[-\frac{\Lambda}{\Sigma_0}+\frac{\Sigma_0}{\Lambda}
\ln\frac{\Lambda^2}{\Sigma_0^2}\],
\quad (\alpha_0=0) \label{condnjl}
\eea
where $\Delta g_0=g_0-1$ (at $\alpha_0=0$).
\section{Scaling laws in the chiral limit}
The UVBC (\ref{massdse_uvbc}) is an eigenvalue equation for 
$\Sigma_0$, and depends on the bare parameters 
$\alpha_0$, $g_0$, $m_0$ and the UV cutoff.
In the chiral limit
we replace $\Sigma_0$ by $m_{\rm dyn}$, where $m_{\rm dyn}$ is called
the dynamical mass, {\em i.e.}, $m_{\rm dyn}$ is generated purely 
by the dynamics 
(which is rather strong, since $\alpha_0$, $g_0$ are of order one).

Let us now consider the chiral limit $m_0\rightarrow 0$,
and look for a nontrivial solution to the UVBC.
For the case of the pure NJL model ($\alpha_0=0$),
Eq.~(\ref{gapeq_njl}) has a nontrivial solution
for $g_0 > 1$, for $g_0\leq 1$ $\Sigma_0=0$ is the solution.
In the chiral limit we find the scaling law
\bea
(\Delta g_0\leq 0)\qquad
\frac{m_{\rm dyn}}{\Lambda}&=&0,\nonu\\
(\Delta g_0> 0)\qquad
\frac{m_{\rm dyn}}{\Lambda}&=&\[\frac{\Delta g_0}{g_0 \ln(g_0/\Delta g_0)}\]^{1/2},
\eea
where we assumed that $\Delta g_0=g_0-1 \ll 1$.
For the intermediate region ($0<\alpha_0<\alpha_c$)
we have nontrivial solution for $\Sigma_0$ from Eq.~(\ref{mu_0bc1})
for values of the four-fermion coupling such that $\Delta g_0> 0$.
Since $C(\omega)$ and $D(\omega)$ are both positive in this region 
the solution to $\Sigma_0=m_{\rm dyn}$ is
\bea
(\Delta g_0\leq 0)\qquad
\frac{m_{\rm dyn}}{\Lambda}&=&0,\nonu\\
(\Delta g_0 >0)\qquad\frac{m_{\rm dyn}}{\Lambda}&=&
\[\frac{C(\omega)}{D(\omega)}
\frac{\Delta g_0}{\Delta g_0 +\omega}\]^{1/2\omega},
\label{scallaw1}
\eea
this scaling law has been obtained in 
ref.~\cite{gukumi89a}. 

The sign of $\Delta g_0$ determines whether there is a nontrivial 
solution or not.
At the critical gauge coupling $\alpha_0=\alpha_c$, we get
the scaling law
\bea
(\Delta g_0\leq 0)\qquad
\frac{m_{\rm dyn}}{\Lambda}&=&0,\nonu\\
(\Delta g_0 >0)\qquad\frac{m_{\rm dyn}}{\Lambda}&=&
\exp\(\frac{1/4+g_0}{1/4-g_0}+\delta(\alpha_c)\),
\label{scallaw2}
\eea
where $\Delta g_0=g_0-1/4$.
Thus there is only a nontrivial solution 
at $\alpha_c$ for $g_0 > 1/4$.

For values of $\alpha_0>\alpha_c$ the gauge coupling is strong enough by itself 
to give a nontrivial solution, so even for 
zero four-fermion coupling $g_0$ we obtain a nontrivial solution.
By putting $m_0=0$ in Eq.~(\ref{mu_0bc3}) 
we find the following solutions: 
\bea
\frac{m_{\rm dyn}}{\Lambda}=
\exp\(-\frac{n\pi}{\nu}-\frac{\beta}{\nu}+\delta(\alpha_0)\),\quad 
\nu=\sqrt{4\lambda_0-1},
\eea
with 
\bea
\beta=\tan^{-1}\[\frac{\nu(\lambda_0+g_0)}{(g_0-\lambda_0)}\],
\eea
and where $n$ is a positive integer ($n=1,2,\dots$),
since we have the physical constraint that $\Sigma_0 < \Lambda$.
It is explained in \cite{fogumisi83,mirbook} and references therein 
that the largest value of $m_{\rm dyn}$ ({\em i.e.}, $n=1$)
leads to the stable vacuum.
By analysis of the BS equations for the bound state spectrum 
(in quenched-ladder QED) it is shown that the solutions with $n\geq 2$ 
correspond to tachyonic bound states, whereas the $n=1$ solution describes
a physical particle spectrum containing
massless NG-bosons (pseudoscalars), and massive scalar particles.

\begin{figure}[ht!]
\epsfxsize=9cm
\epsffile[150 330 400 570]{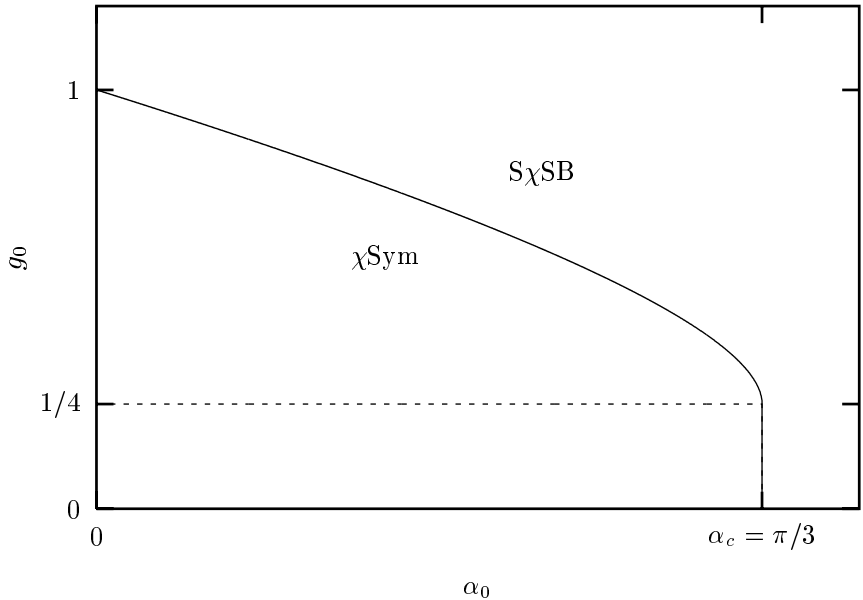}
\caption{The critical curve in the $(\alpha_0, g_0)$ plane seperating
a chiral symmetric phase ($\chi$Sym) from a spontaneous (dynamical) 
chiral symmetry broken phase (S$\chi$SB).}
\label{critcurve}
\end{figure}

Hence, when $\nu\rightarrow 0$ ($\alpha_0\rightarrow \alpha_c$)
we get
\bea
\frac{m_{\rm dyn}}{\Lambda}=
\exp\(\frac{1/4+g_0}{1/4-g_0}
+\delta(\alpha_c)\)
\exp\(-\frac{\pi}{\sqrt{4\lambda_0-1}}\).\label{essent_sing}
\eea 
This is the famous scaling law with the essential singularity,
sometimes referred to as ``Miransky'' scaling.
We will discuss this non-power-like scaling law more thoroughly in 
section~\ref{sec_cpt}.
\paragraph{The critical line in the GNJL model.}
From the considerations above it is clear that 
the GNJL model has a nontrivial chiral phase structure.
The chiral symmetry is dynamically broken for values
$\Delta g_0>0$ or $\alpha_0>\alpha_c$.
We can now draw a critical line in the coupling constant 
plane $(g_0,\alpha_0)$ separating the chiral symmetric phase 
from the chiral broken phase, see Fig.~\ref{critcurve}.
The critical line is
\bea
g_c(\alpha_0)\equiv
\frac{1}{4}\left(1+\sqrt{1-\frac{\alpha_0}{\alpha_c}}\right)^2,\qquad
0\leq\alpha_0< \alpha_c=\frac{\pi}{3}
\label{critline1}
\eea
at $g_0>1/4$, and
\bea
\alpha_0=\alpha_c
\label{critline2}
\eea
at $g_0\leq 1/4$, above which the gap equation for the fermion 
self-energy $\Sigma(p)$ has a nontrivial solution. 
In the continuum formulation using the quenched-ladder approximation,
the critical line was first obtained by Kondo {\em et al.} \cite{komiya89}
and Appelquist {\em et al.} \cite{apsotawij88}.
More recently, a phase plot has been obtained in \cite{azcagagrlapi95b} using 
lattice simulations of the so-called noncompact GNJL model in a mean field approximation. 
The critical line obtained in the continuum formulation  
and the corresponding phase diagram obtained by the lattice 
simulations are in good qualitative 
agreement. 
\section{The critical exponents and scaling}\label{sec_critexp_scal}
An interesting one-to-one correspondence between the 
EOS of a magnet Eq.~(\ref{widom1}), and the
EOS of the order parameter of chiral symmetry breaking in the
GNJL model, {\em i.e.} Eqs.~(\ref{mu_0bc1}) and (\ref{condbc1}), 
was made independently in Refs.~\cite{balomi90,kohakoda90}, and later
in Ref.~\cite{ko91}.

Following Refs.~\cite{balomi90,kohakoda90},
we introduce the following correspondence:
\begin{itemize}
\item{The analog of the magnetization $M$ of Eq.~(\ref{widom1}) 
is the chiral condensate, 
\bea
{\cal M}&\equiv &-4\pi^2\langle \bar\psi\psi\rangle /\Lambda^3,\label{Mdef}
\eea
which is the order parameter.}
\item{The analog of the external magnetic field $h$ is the bare mass $m_0$.}
\item{The analog of the temperature $t=T-T_c$ is $\Delta g_0=g_0-g_c$.}
\item{The analog of the 
magnetic susceptibility $\chi$, Eq.~(\ref{magsusdef}), 
is the chiral susceptibility 
\bea
\chi\equiv\frac{\partial {\cal M}}{\partial m_0}\biggr|_{m_0=0}.\label{chidef}
\eea
}
\end{itemize}
The critical exponents 
$\beta$, $\gamma$, $\delta$, $\nu$ 
near the phase transition in a magnetic system were defined 
in Eqs.~(\ref{betaexpdef}), (\ref{gammaexpdef}), (\ref{deltaexpdef}), 
and (\ref{nuexpdef}). 
Correspondingly, in the GNJL model, the critical exponents $\beta$, $\gamma$, $\delta$, and $\nu$ 
are the defined by the scalings:
\bea
{\cal M}&\sim& (\Delta g_0)^\beta  \quad\mbox{at}\quad m_0=0,
\label{critexp_beta}\\
\chi&\sim&
(\Delta g_0)^{-\gamma}, \label{critexp_gamma}\\
{\cal M}&\sim& m_0^{1/\delta}\quad\mbox{at}\quad \Delta g_0=0,\label{critexp_delta}\\
m_{\rm dyn}
&\sim& (\Delta g_0)^{\nu}\quad\mbox{at}\quad m_0=0,\label{critexp_nu}
\eea
where the dynamical mass $m_{\rm dyn}$ 
can be considered as the inverse correlation length,
$m_{\rm dyn} \sim \xi^{-1}$, and $\chi$ is defined by Eq.~(\ref{chidef}).
Again we mention that the mass $m_\sigma$ of the $\sigma$ boson is the natural
candidate for the inverse correlation length, since the 
$\sigma$ boson propagator is the connected correlation function of 
the field $\sigma$ describing correlations parallel to direction 
of symmetry breaking ({\em i.e.} parallel to the direction
of long-range ordering given by the order parameter $\langle \sigma \rangle$).

However it can be shown that in the broken phase $m_\sigma$ scales 
in the same way as $m_{\rm dyn}$. The symmetric phase
will be analyzed more thoroughly in chapter~\ref{chap4}.

Substituting Eq.~(\ref{condbc1}) in (\ref{mu_0bc1}), 
we can eliminate $\Sigma_0$ and rewrite the EOS, in terms of ${\cal M}$
(Eq.~(\ref{Mdef}), as
\bea
\frac{m_0}{\Lambda}
&=&{\cal F}(0) {\cal M}^{ (2+\omega)/(2-\omega)} 
+{\cal F}^\prime(0) \Delta g_0 {\cal M}+\cdots\nonu\\
&=&{\cal M}^\delta\[ {\cal F}(0) +{\cal F}^\prime(0) 
\Delta g_0 {\cal M}^{-1/\beta}+\cdots\]\sim 
{\cal M}^\delta {\cal F}(\Delta g_0 {\cal M}^{-1/\beta}),\label{EOS2}
\eea
from which the critical exponents $\beta$ and $\delta$ 
can be obtained (compare with Eq.~(\ref{eosexp}))
\bea
\delta&=&\frac{2+\omega}{2-\omega},\qquad 
\beta=\frac{2-\omega}{2\omega}\label{deltabetaexpr},\\
{\cal F}(0)&=&\omega  D(\omega)\[C(\omega)\]^{(2-\omega)/(2+\omega)},
\qquad {\cal F}^\prime (0)=-1.
\eea
The dots in Eq.~(\ref{EOS2}) represent subleading corrections under the
assumption that
\bea
\Delta g_0 \ll \omega.
\eea

The critical exponent $\gamma$ can also be obtained 
from Eq.~(\ref{EOS2}) by differentiation with respect to $m_0$,
\bea
\chi^{-1} \propto \delta {\cal M}^{\delta -1}{\cal F}(0)
+{\cal F}^\prime(0)\Delta g_0. 
\eea
Using that
\bea
{\cal M}^{\delta -1}\sim (\Delta g_0)^{\beta(\delta-1)}=\Delta g_0.
\eea
We obtain 
\bea
\gamma=1.\label{critgamma}
\eea
Hence the hyperscaling equation~(\ref{widscallaw1})
\bea
\gamma=\beta(\delta-1),\label{hyperscal1}
\eea
is satisfied.
 
The critical exponent $\nu$ can be read off directly 
from the scaling law (\ref{scallaw1}), 
\bea
\nu=1/2\omega.\label{critnu}
\eea
Thus the hyperscaling equation
\bea
d\nu =2\beta+\gamma,\label{hyperscal2}
\eea
is also satisfied in $d=4$ dimensions. This scaling law
follows from scaling laws (\ref{widscallaw2}) and (\ref{widscallaw4})
after elimination of the critical exponent $\alpha$.

The four critical exponents $\gamma$, $\beta$, $\delta$ and $\nu$ satisfy
the hyperscaling relations (\ref{hyperscal1}) and (\ref{hyperscal2})
in the intermediate region $0<\alpha_0<\alpha_c$ ($0<\omega<1$), 
hence only two of them are independent.
In the intermediate region the critical exponents have nonmean field values,
and the observation that the exponents satisfy the hyperscaling relations,
suggest that the theory has a nontrivial continuum limit.

The situation changes at the NJL point ($\alpha_0=0$). 
In the limit $\omega\rightarrow 1$, the expressions for the
critical exponents Eqs.~(\ref{deltabetaexpr}), (\ref{critgamma}), and
(\ref{critnu}) yield mean field (or gaussian) exponents  
($\gamma=1$, $\beta=1/2$, $\delta=3$, $\nu=1/2$)\footnote{
Mean field exponents follow from the mean 
field approach (or Landau's theory); an operator 
such as $\bar\psi\psi$ is replaced by its average
value, and fluctuations about that value are ignored, see \cite{goldenfeld}.}.

However this result is not completely correct, since 
we should derive these exponents from Eqs.~(\ref{gapeq_njl2}) and (\ref{condnjl}). 
Doing so, we obtain
\bea
&&\gamma=1,\qquad \beta=\frac{1}{2}+\frac{1}{4}\frac{1}{\ln(1/{\cal M})},\nonu\\
&&\delta=3-\frac{1}{\ln(1/{\cal M})},\qquad 
\nu=\frac{1}{2}+\frac{1}{4}\frac{1}{\ln(1/{\cal M})},
\eea
where ${\cal M}$ is the order parameter defined in Eq.~(\ref{Mdef}).
Hence the mean field critical exponents get logarithmic corrections.
This result is well-known.

Due to the logarithmic corrections, the hyperscaling relations are violated, 
and we have the following inequalities:
\bea
4\nu> 2\beta +\gamma,\qquad \gamma > \beta(\delta-1).\label{scallawineq}
\eea
This violation of hyperscaling is believed to be a sign of triviality
meaning that the effective Yukawa-coupling (which couples
Goldstone bosons to the fermions) vanishes in the continuum limit.
The continuum limit is non-interacting, hence trivial.
This can be seen in the following way. 
Assuming that in the low energy region the correlation length $\xi$  
is the only relevant length scale, we define an effective Yukawa-coupling $g_Y$ 
by the zero-momentum limit of the scattering amplitude of two fermions 
exchanging a scalar bound state in the D$\chi$SB phase
\bea
\frac{g_Y^2}{m_\sigma^2} \sim \xi^2 g_Y^2\sim  \GS(0,0)\DelS(0)\GS(0,0),
\eea
where $\GS(0,0)$ and $\DelS(0)$ are given by the PCAC relations
Eqs.~(\ref{masssiganal}) and (\ref{scalanal}).
Using the definition of the critical exponents 
(\ref{critexp_beta})--(\ref{critexp_nu}),  
$\xi\sim 1/m_{\rm dyn}\sim 1/m_\sigma$, $\langle \sigma \rangle \sim {\cal M}$, 
we find that
\bea
g_Y^2 \sim \frac{1}{\xi^2}\(\frac{\partial \Sigma_0}{\partial \langle \sigma \rangle}\)^2
\frac{\partial \langle \sigma \rangle}{\partial m_0}\biggr|_{m_0=0}
\sim \xi^{(2\beta+\gamma-4\nu)/\nu}. \label{scallawyukawa}
\eea
This expression is related to the definition of 
$g_R\sim g_Y^2$ given in \cite{koko94}, and it is clear that 
the scaling inequalities (\ref{scallawineq}) imply that $g_Y^2\rightarrow 0$ 
when $\xi\rightarrow \infty$.
Thus, only if the hyperscaling relations are satisfied, a nonzero
$g_Y$ might be realized in the continuum limit ($\xi\rightarrow \infty$),
thereby giving rise to a nontrivial interacting theory.
The Goldenberger--Treiman relation reads
\bea
g_Y\sim \frac{m_{\rm dyn}}{f_\pi}, \label{GT}
\eea
with $f_\pi$ given by Eq.~(\ref{fpi}).
The scaling form for $\tau_\sigma\sim Z_\sigma^2(0)$ of Eq.~(\ref{delplow}) can be derived 
from Eq.~(\ref{masssiganal}), 
\bea
\tau_\sigma \sim \xi^{-\eta},\label{compositness}
\eea
thus $\tau_\sigma\rightarrow 0$ for $\xi\rightarrow \infty$, which is
referred to as the ``compositeness'' condition.
Due to the chiral WTI, $\tau_\pi$ of Eq.~(\ref{delplow}) scales, 
close to criticality,
in the same manner as $\tau_\sigma$, {\em i.e.}
$\tau_\pi\sim \xi^{-\eta}$.
Then, with Eq.~(\ref{fpi}) and Eq.~(\ref{widscallaw3}), we obtain that
\bea
f_\pi\sim \xi^{\eta/2-\beta/\nu}=\xi^{-1} \xi^{(4\nu-\gamma-2\beta)/2\nu},
\label{fpiscallaw}
\eea 
and thus Eq.~(\ref{GT}) agrees with Eq.~(\ref{scallawyukawa}).
Moreover Eq.~(\ref{fpiscallaw}) implies the existence of a scaling form
for $\Pi_5(q^2)$ of Eq.~(\ref{pseudovecvac}),
\bea
\Pi_5(q^2)\sim 
\Lambda(m_\sigma^2/\Lambda^2)^\epsilon {\cal F}(q^2/m_\sigma^2),
\eea
where the critical exponent $\epsilon$ should satisfy 
$\epsilon=\beta/2\nu-\eta/2$, and ${\cal F}$ is assumed to be finite at $q^2=0$.

Although the violations of the scaling laws for 
both fermions and magnets (fundamental scalars) is given by the 
inequalities (\ref{scallawineq}), it is said \cite{koko94} that the triviality 
of fermions and magnets are realized in a different manner. 
For instance, the logarithmic corrections to the critical exponent
$\delta$ have opposite sign; for fermions $\delta<3$ and for magnets $\delta>3$.

The important difference is that the NG-bosons (pions etc)
are composites of fermions, whereas the NG-bosons are fundamental in 
scalar fields theory.
This issue is at the basis of determining whether the Higgs boson 
is a ``fundamental'' scalar field or some fermion composite.

If one takes the limit of $\alpha_0\rightarrow \alpha_c$,
$\omega$ vanishes, and the critical exponents $\beta$ and $\nu$ 
blow up.
This can also be seen from the scaling law (\ref{scallaw2})
which cannot be expressed as a power-law dependence Eq.~(\ref{critexp_nu}).
The phase transition at $\alpha_0=\alpha_c$ is rather special
and we will discuss it more extensively in the next section.
\section{The conformal phase transition}\label{sec_cpt}
In the previous section we found that the critical exponents satisfy 
hyperscaling relations in the intermediate region $0<\alpha_0<\alpha_c$.
Moreover, the Cornwall-Jackiw-Tomboulis (CJT)\footnote{See Ref.~\cite{cojato74}.}
effective potential $V$ of
the GNJL model in the intermediate region has been obtained by 
\cite{balomi90,gumi91,gumi92,balo92,kotaya93}, and it can be shown that 
the effective potential for this range of gauge coupling constant
is analogous to the``mexican hat'' potential of the $\sigma$-model
(the prototype potential for second-order phase transitions).
However the phase transition at $\alpha_0=\alpha_c$ cannot be described by a 
$\sigma$-model type of effective potential. 
Although the chiral phase transition is continuous (not first order)
it cannot be classified as a second order (or higher) phase transition 
(\cite{aptewij95,miya97}).
The phase transition at $\alpha_0=\alpha_c$ is an example of a conformal 
phase transition (CPT).

The concept of the CPT was introduced and discussed in 
Ref.~\cite{miya97}. It embodies the classification of specific types of 
phase transitions.
The main feature of the CPT is an abrupt change of the spectrum of light 
excitations (composites)\footnote{By light excitations we mean
excitations which have a small mass $m$ compared with the cutoff $\Lambda$,
$|m|/\Lambda\ll 1$. In other words, the correlation lengths are large.}
as the critical point is crossed, though the
phase transition itself is continuous.
This is connected with the nonperturbative breakdown of the conformal
symmetry (scale invariance) by marginal operators ({\em e.g.}, 
$(\bar\psi\psi)^2+(\bar\psi i\gamma_5\psi)^2$
in the GNJL model), which was illustrated in Ref.~\cite{miya97} 
by a study of the effective potentials in Gross-Neveu and GNJL models. 

The concept of the CPT can be considered as an extension of
the Berezinsky--Kosterlitz--Thouless (BKT) \cite{be70,koth73} phase transition 
(taking place in two dimensions) to higher dimensions.
Witten \cite{wi78} has shown that the phase transition in 
the Gross--Neveu model\footnote{The Gross-Neveu model or the Thirring model
can be consider as the NJL model in two dimensions with $N$ fermion flavors, see
\cite{grne74}.}
in $d=2$ is analogous to the BKT phase transition.
Although due to the Mermin--Wagner theorem \cite{mewa66,co73} 
spontaneous symmetry breaking or long range order is not possible in $d=2$, 
there is a generation of a fermion mass and ``almost long range order''.
The correlations below the critical temperature in the BKT type of models 
decrease with a (non-universal) power-law behavior instead of 
exponentially, hence there is a phase transition though 
the ground state is unique.
Instead of a genuine NG-boson a BKT gapless mode appears.
The correlation length $\xi\sim 1/m$ in the Gross---Neveu model
and in the models with BKT phase transition
have a scaling law with essential singularity, similar to
that of Eq.~(\ref{essent_sing}).
This is the crucial property of the CPT; the presence
of an essential singularity in the mass (energy) gap at the critical
point.

In theories with chiral symmetry breaking in $d>2$, 
the particle-spectrum consists of 
massless NG-bosons ($\pi$), their chiral partners
($\sigma$ bosons), and light\footnote{By light particles we mean particles 
with masses much 
less than the cutoff $\Lambda$.} fermions (and massless gauge bosons)
in the chiral symmetry broken phase.
In case of a $\sigma$-model like phase transition, the particle
spectrum would contain amongst massless fermions (and massless gauge bosons), also
light $\pi$ and $\sigma$ resonances\footnote{The $\pi$ and $\sigma$ bosons can decay to massless fermions and anti-fermions.} in the symmetric phase.
In case of the CPT, the spectrum in the symmetric phase only consists of
massless fermions; light fermion-anti-fermion states are absent.

The absolute value of the mass $m_\sigma$ 
of the $\sigma$ boson is analogous to the inverse correlation length $1/\xi$, 
and the $\pi$ boson plays the role of the Nambu--Goldstone-boson. 
In the symmetric phase, the $\pi$ and $\sigma$ bosons are degenerate, 
({\em i.e.}, $m_\pi=m_\sigma$). 
From the PCAC relations (see section~\ref{cwti_pcac}) it follows that 
the $\pi$ boson mass $m_\pi$ vanishes in the non-symmetric phase 
(with zero bare mass $m_0=0$).
Thus we write, around  
a critical point $z=z_c$,
\bea
m_\pi=0,&& m_\sigma/\Lambda=C_{+}(z)(z-z_c)^{\nu},\qquad z\geq z_c\quad 
\mbox{broken phase},\label{spec_nonsym}\\
m_\pi=m_\sigma,&& m_\sigma/\Lambda= C_{-}(z)(z_c-z)^{\nu},\qquad z< z_c\quad
\mbox{symmetric phase},\label{spec_sym}
\eea
where $z$ is a generic notation for the parameters of the theory, {\em e.g.}
the coupling constant $\alpha_0$ or $g_0$, the number of fermion flavors $N$ or temperature $T$. Moreover the factors $C_{+}(z)$ and $C_{-}(z)$ are 
functions of such parameters.
Furthermore $C_{+}$ is real (stable bound states)
and $C_{-}$ is complex valued (resonances). 
Since in general
\bea
\lim_{z\downarrow z_c} |C_{+}(z)| \not = \lim_{z\uparrow z_c} |C_{-}(z)|,
\eea
the mass $m_\sigma$, although continuous, is nonanalytic in $z=z_c$.
The explanation (see \cite{miya97}) is that
while in the symmetric phase, the $\pi$ and $\sigma$ bosons are described
by BS equations with a zero fermion mass, in the non-symmetric phase they
are described by BS equations with $m_{\rm dyn}\not =0$.
Because of that, BS equations in the non-symmetric phase are not obtained
by an analytic continuation of the equations in the symmetric phase.
Again, this is reflection of the fact that a phase transition
is described by non-analytic behavior in one (or more) of the parameters
of the generating functional. Consequently the Green functions 
(or at least some of them) exhibit non-analyticity around the critical point.

It is clear from Eqs.~(\ref{spec_nonsym}) and (\ref{spec_sym})
that close to criticality, {\em i.e.} $z\sim z_c$, the spectrum comprises 
of light bound states, $|m_\pi|, |m_\sigma| \ll \Lambda$.
Moreover, the fermion dynamical mass $m_{\rm dyn}$, which is given via
the order parameter ${\cal M}\sim \langle \bar\psi \psi\rangle$,
vanishes in the symmetric phase. 
In the non-symmetric phase, the PCAC relations show
that $m_{\rm dyn}$ scales in the same way as $m_\sigma$. 
We recall that, in the chiral limit ($m_0\rightarrow 0$), 
we have the scaling laws
\bea
m_{\rm dyn}\sim m_\sigma,
\qquad {\cal M}&\sim&(z-z_c)^\beta ,\qquad z\geq z_c\quad
\mbox{broken phase},\\
m_{\rm dyn}=0,\qquad {\cal M}&=&0 ,\qquad z< z_c\quad
\mbox{symmetric phase}. 
\eea
for models with $d>2$.
Thus close to criticality we can represent the mass spectrum
of excitations by a universal scaling function $f$
\bea
m_{\rm dyn}, m_\sigma, m_\pi \sim \Lambda f(z),
\eea
around $z=z_c$.  If the scaling function $f$ is of the form 
$f(z)\sim |z-z_c|^\nu$, the spectrum consists of ``light excitations'' 
on both sides of the critical point.

The CPT is characterized by a scaling function $f(z)$ which has 
an essential singularity at $z=z_c$ and satisfies
\bea
\lim_{z\downarrow z_c} f(z)&=&0,\qquad z\geq z_c\quad
\mbox{broken phase},\\
\lim_{z\uparrow z_c} f(z)&\not =&0 ,\qquad z< z_c\quad
\mbox{symmetric phase}. \label{cpt_f_symm}
\eea
The scaling law at $\alpha_0=\alpha_c$ of $m_{\rm dyn}$ in the GNJL model, 
see Eq.~(\ref{essent_sing}), is an example of a scaling law 
with such an essential singularity.
Let us follow \cite{miya97}, and argue that the scaling law with essential
singularity implies an abrupt change in the spectrum consisting 
of light excitations.
The story is as follows.
In the nonsymmetric phase, besides the solutions
with $m_{\rm dyn}\not = 0$, there are solutions with
$m_{\rm dyn}=0$, 
which can be obtained by analytically continuing 
the solutions in the symmetric phase to the nonsymmetric phase.
The solutions with $m_{\rm dyn}=0$ in the broken phase
are unstable or unphysical, in that case the $\pi$ and $\sigma$ 
bosons are tachyons: $m^2_{\rm tach}\equiv m_\pi^2=m_\sigma^2 <0$.
It is shown in \cite{mirbook} that scaling law for the tachyonic masses 
has the same form as the scaling law for the dynamical fermion mass 
$m_{\rm dyn}$ in the broken phase: $|m_{\rm dyn}|\sim |m_{\rm tach}|$.
The crucial point is that the replacement 
of $m_{\rm dyn}\not =0$ by $m_{\rm dyn}=0$ does not change the 
ultraviolet properties of the theory.
The analytic continuation of the scaling law for
$m_\sigma$ Eq.~(\ref{spec_sym}) 
from the symmetric phase ($z<z_c$) to the broken phase ($z>z_c$)
gives the scaling law for the tachyonic mass
\bea
m_\sigma/\Lambda&=& C_{-}(z)(z_c-z)^{\nu}\sim f(z),\qquad z< z_c,\\
m_{\rm tach}/\Lambda&=& C_{-}(z)(z_c-z)^{\nu}\sim f(z),\qquad z> z_c,
\eea
where $m^2_{\rm tach} < 0$, $m_{\rm dyn}\sim |m_{\rm tach}|=|m_\sigma|$.
Since the scaling law $f(z)$ of the $\sigma$-model type
approaches zero from both sides of the critical point $z_c$,
it follows straightforwardly that the spectrum comprises of light excitations
($|m_\sigma|=|m_\pi|\ll \Lambda$) in the symmetric phase.
However in case of the CPT, when $f(z)$ has an essential singularity 
(Eq.~(\ref{cpt_f_symm})) $f(z)$ does not approach zero from the symmetric side
and the mass of the $\sigma$ and $\pi$ resonances will not be small 
($m_\sigma/\Lambda\sim{\cal O}(1)$).
Thus the CPT is characterized by an abrupt change 
in the spectrum of light excitations across the critical point $z=z_c$.

What is the origin of the CPT?
In case of the GNJL model with nonzero $g_0$ and $\alpha_0<\alpha_c$ the 
model contains formally irrelevant four-fermion operators $d_{(\bar\psi\psi)^2}>4$ 
which become relevant operators $d_{(\bar\psi\psi)^2}<4$ 
near the critical curve due to the appearance 
of a large anomalous dimension for these operators. 
The relevant operators break the conformal symmetry (scale invariance) 
explicitly in the symmetric phase (close to criticality), and give rise to 
light $\pi$ and $\sigma$ resonances (which introduce a conformal 
symmetry breaking mass-scale into the theory).
However at $\alpha_0=\alpha_c$ and $g_0<1/4$ the four-fermion operators
are marginal operators, and the theory is supposedly conformal invariant.
Therefore in the symmetric phase it is consistent that we would expect 
only massless fermions and anti-fermions, and absence of 
light unstable $\pi$ and $\sigma$ bound states.
In the symmetric phase the divergence of the dilatation current 
${\cal D}_\mu$ (see \cite{zinjus,mirbook} for exact definitions)
vanishes, {\em i.e.}, $\partial^\mu {\cal D}_\mu=0$.
However dynamical breaking of the chiral symmetry introduces a new mass-scale, 
and introduces a conformal anomaly ($\partial^\mu {\cal D}_\mu\not=0$)\footnote{It is said (\cite{miya97}) that the specific form of the conformal anomaly 
implies a realization of the partial conservation of dilation 
current (PCDC) hypothesis, see for instance \cite{migu89}.}.
Hence, the origin of the CPT lies in the nonperturbative breakdown of conformal 
invariance.

Up till now, the CPT has been found in 2-dimensional Gross-Neveu 
model \cite{wi78,miya97}, the GNJL model and quenched QED4 
at $\alpha_0=\alpha_c$, see for instance \cite{mi85a,mi93}.
More recently, the CPT was considered in $SU(N_c)$ gauge theories 
(with $N_c$ the number of colors) 
with $N$ number of flavors \cite{aptewij96,apratewij98} by concluding
the absence of light excitations in the symmetric phase.
In QED$_3$ without Chern-Simons term 
a CPT like transition was found by \cite{aptewij95} and \cite{gumish98}, 
where Appelquist {\em et al.} considered the spectrum of excitations, and 
Gusynin {\em et al.} constructed the (CJT) effective potential.
We will return to the concept of the CPT, when we discuss 
the propagator of the $\sigma$ boson in section~\ref{again_sec_cpt}.
\section{RG equations and fine-tuning}\label{sec_RGfine}
As was discussed in section~\ref{WRG}, the concept of fine-tuning 
of relevant and marginal interactions
is required for determining whether a nontrivial continuum limit exists.
According to the renormalization group methods the bare couplings of a quantum 
field model run with the UV cutoff $\Lambda$.
Referring to subsection~\ref{step3}, we recall that the bare relevant and marginal 
couplings 
({\em e.g.} $m_0$, $\alpha_0$, $g_0$) have to be fine-tuned
sufficiently close to the critical point in order for scaling behavior 
to set in. 

Near the critical point a new scale is generated, namely the correlation 
length which exist in case of for instance the 
$\sigma$-model type of phase transition on both sides (both phases)
of the system. 
In the broken phase the inverse correlation length is real and can be 
considered as a physical mass of particles, for instance the mass of 
the scalar bound state (the $\sigma$ boson) $m_\sigma$, 
or the mass of fermion $m_{\rm dyn}$, whereas in the symmetric phase the 
correlation length is complex giving rise to resonances
(scalar and pseudoscalar resonances), given by a mass and a width.
It turns out that if hyperscaling relations are satisfied
the absolute value of the mass of the resonance can be considered
as the inverse correlation length, and the mass over width ratio
({\em i.e.}, the complex phase) will only depend on the critical exponents.
In case of a CPT there are no resonances in the symmetric phase, and no new
mass scale is introduced, hence there is absence of fine-tuning.

The fine-tuning depends on the eigenvalues of the couplings
close to the critical point (see Eq.~(\ref{eigenvaldelkap})
and hence on the critical exponents (see Eq.~(\ref{eigvalsol})).
These critical exponents follow in principle from 
the $\beta$ functions for the coupling parameters $\mu_0$, $g_0$ 
and $\alpha_0$, which are defined by
\bea 
\Lambda\frac{{\rm d} \mu_0}{{\rm d} \Lambda}&=&
\beta_\mu(\mu_0,g_0,\alpha_0),\label{betamu}\\
\Lambda\frac{{\rm d} g_0}{{\rm d} \Lambda}&=&
\beta_g(\mu_0,g_0,\alpha_0) ,\label{betag0}\\
\Lambda\frac{{\rm d} \alpha_0}{{\rm d} \Lambda}&=&
\beta_\alpha(\mu_0,g_0,\alpha_0) ,\label{betaa0}
\eea
where $\mu_0$ is the dimensionless bare mass
\bea
\mu_0=m_0/\Lambda.
\eea
The first step is determine the fixed points $(\mu^\star,g^\star,\alpha^\star)$
of the RG equations (\ref{betamu}),
(\ref{betag0}), and (\ref{betaa0}), {\em i.e.}
\bea
\beta_\mu(\mu^\star,g^\star,\alpha^\star)=0,\quad
\beta_g(\mu^\star,g^\star,\alpha^\star)=0,\quad
\beta_\alpha(\mu^\star,g^\star,\alpha^\star)=0.
\eea
In principle these RG equations follow from the regularized SDEs derived
in chapter~\ref{chap2}.
As is well-known, the fixed point for $\beta_\mu$ is $\mu^\star=0$, 
hence we can write
\bea
\Lambda\frac{{\rm d} \mu_0}{{\rm d} \Lambda}\approx
-\(1+\gamma_m\) \mu_0,\label{gammamdef}
\eea
where $\gamma_m$ is the anomalous dimension 
of the mass operator 
$\bar\psi\psi$ evaluated at the fixed point $(\mu^\star,g^\star,\alpha^\star)$, 
see again subsection~\ref{step3}.
Equation~(\ref{gammamdef}) is in correspondence 
with Eqs.~(\ref{kmusol}) and (\ref{anodef}), since the canonical dimension 
of $\bar\psi\psi$ is $d^c=3$ at $d=4$.
The critical exponent $\eta$ (see Eq.~(\ref{anodef}))
of $\mu_0$ is negative (for $\gamma_m>-1$), 
thus the dimensionless bare mass $\mu$ is a relevant coupling, and requires
fine-tuning.

After setting $\mu_0=\mu^\star=0$,
the problem reduces to the determination of the UV fixed points for the subset 
\bea
\beta_g(g_0,\alpha_0)&=&0 ,\label{subbetag0}\\
\beta_\alpha(g_0,\alpha_0)&=&0 ,\label{subbetaa0}
\eea
of the $\beta$ functions.
The quenched ladder approximation
simplifies the solutions of Eqs.~(\ref{subbetag0}), and 
(\ref{subbetaa0}) considerably 
since the quenched hypothesis Eq.~(\ref{basicquenched})
explicitly sets $\beta_\alpha=0$.
In this way, as we will show in a moment, 
the critical fixed point of Eq.~(\ref{subbetag0})
is given by Eq.~(\ref{critline1}):
\bea
g^\star=g_c(\alpha_0),\qquad \beta_g(g_c(\alpha_0),\alpha_0)=0.
\eea
Therefore we can address the problem of finding critical fixed points beyond
the quenched approximation as was proposed 
at the end of subsection~\ref{subsec_step2}.
Thus the analysis of solutions $(g^\star,\alpha^\star)$ of
Eqs.~(\ref{subbetag0}) and (\ref{subbetaa0})
is performed by first assuming that a solution of
\bea
\beta_g(g_0,\alpha_0)=0
\eea
exists and can be expressed as $g^\star=g_c(\alpha_0)$.
Then the equation (\ref{subbetaa0}) should be reconsidered.
In chapter~\ref{chap5} the equation (\ref{subbetaa0})
will be analyzed beyond the quenched approximation,
and we will indeed try to solve 
\bea
\beta_\alpha(g_c(\alpha_0),\alpha_0)=0.
\eea

In what follows, we derive the expression for the anomalous dimension 
$\gamma_m$ of $\bar\psi\psi$ and find an expression for the $\beta$ function, 
$\beta_g$, of $g_0$. 
\subsection{RG flow in the broken phase}
In the broken phase the mass of the scalar $m_\sigma$ scales in the same way
as the dynamical fermion mass.
Hence we assume that in the chiral limit ($m_0\rightarrow 0$) 
the dynamical mass $m_{\rm dyn}\sim \Sigma_0$ 
is related to some physical mass scale
and thus independent of the UV cutoff $\Lambda$.
The dynamical mass is given by the scaling law (\ref{scallaw1}), 
so we assume that
\bea
\Lambda\frac{{\rm d} m_{\rm dyn}}{{\rm d} \Lambda}=0.
\eea
Since in the broken phase $m_{\rm dyn}\not =0$, we get 
\bea
\Lambda\frac{{\rm d} m_{\rm dyn}}{{\rm d} \Lambda}=
m_{\rm dyn}\[1+\frac{\beta_g(g_0,\alpha_0)}{2\Delta g_0(\Delta g_0+\omega)}\],
\eea
where the $\beta$-function of $g_0$ is defined in Eq.~(\ref{betag0}).
The left-hand side is zero, so this gives the following condition 
for the $\beta$-function in the broken phase ($\Delta g_0 > 0$):
\bea
\beta_g(g_0,\alpha_0)=-2\Delta g_0(\Delta g_0+\omega).\label{betabroken}
\eea
An additional fine-tuning condition is that the bare mass $m_0$ 
is supposed to be much smaller than
the dynamically generated mass $m_{\rm dyn}$,
\bea
m_0\ll m_{\rm dyn}\sim \Sigma_0 
\ll \Lambda.\label{finetune1}
\eea
 
An expression for the anomalous dimension $\gamma_m$ of the mass operator
$\bar\psi\psi$ which is defined in Eq.~(\ref{gammamdef}), 
can be obtained by differentiating Eq.~(\ref{mu_0bc1}) 
with respect to $\Lambda$ and assuming 
the fine-tuning condition (\ref{finetune1}).
We obtain
\bea
\frac{{\rm d} m_0}{{\rm d} \Lambda}
&=&-\Delta g_0 C(\omega)\(\frac{\Sigma_0}{\Lambda}\)^{2-\omega}
\[\frac{\beta_g(g_0,\alpha_0)}{\Delta g_0}-(1-\omega)\]\nonu\\
&+&(\Delta g_0+\omega)D(\omega)\(\frac{\Sigma_0}{\Lambda}\)^{2+\omega}
\[\frac{\beta_g(g_0,\alpha_0)}{\Delta g_0+\omega}-(1+\omega)\].
\eea
Using now the equation for the $\beta$-function Eq.~(\ref{betabroken})
we get, for the anomalous dimension in the broken phase,
\bea
\gamma_m=1+\omega+2\Delta g_0,
\eea
near the critical curve.
\subsection{RG flow in the symmetric phase}
In the symmetric phase ($\Delta g_0<0$), the only scale besides the UV cutoff
and the bare mass $m_0$ is the scale set by the mass and width of the
scalar and pseudoscalar resonances, {\em i.e.} $|m_\sigma|=|m_\pi|$ the absolute
value of the complex mass pole of $\DelS$. 
The mass and the width of the scalar and pseudoscalar resonances
(analogous of the correlation length in the symmetric phase)
should be considered as the physical mass scales which in principle
can be obtained from experiment. We set $\mu\sim |m_\sigma|$.

The fine-tuning condition or RG flow of the bare four-fermion coupling $g_0$
in the symmetric phase is obtained from the 
condition
\bea
0=\Lambda\frac{{\rm d}}{{\rm d} \Lambda}
\[\(\frac{\Lambda}{\mu}\)^{2\omega}\(\frac{1}{g_0}-\frac{1}{g_c}\)\]
=-\Lambda\frac{{\rm d}}{{\rm d} \Lambda}
\[\frac{\Delta g_0}{g_0 g_c}\(\frac{\Lambda}{\mu}\)^{2\omega}\],\label{finetune2}
\eea
where $\mu$ is some infrared renormalization scale related to $m_\sigma$ 
(by definition independent of $\Lambda$). 
We assume also that the scale of the bare mass is much smaller
than the physical infrared scale $\mu$, thus in the chiral limit, $\mu\gg m_0$.
At the moment the fine-tuning condition seems a bit arbitrary, however in the next
chapter when we explicitly renormalize the scalar propagator 
in the symmetric phase, the fine-tuning condition (\ref{finetune2})
turns out to be quite natural.

Thus Eq.~(\ref{finetune2}) gives the RG flow or the $\beta$-function 
in the symmetric phase
\bea
\beta_g(g_0,\alpha_0)=-2\omega \frac{g_0}{g_c}\Delta g_0. \label{betasym}
\eea
The anomalous dimension follows from Eq.~(\ref{mu_0bc1}) under the assumption that
\bea
-\Delta g_0\sim \(\mu /\Lambda\)^{2\omega} \gg  
\(\Sigma_0 /\Lambda\)^{2\omega}, \label{addfinetune2}
\eea
since $\Sigma_0$ is the fermion mass in the symmetric phase which 
vanishes in the chiral limit, thus $\Sigma_0\rightarrow 0$ when
$m_0\rightarrow 0$.
We get
\bea
\frac{m_0}{\Lambda}\approx -\Delta g_0 C(\omega) \(\Sigma_0/\Lambda\)^{2-\omega}.
\eea
Differentiating with respect to $\Lambda$ and using Eq.~(\ref{betasym}) gives
\bea
\gamma_m=1-\omega+2\omega \frac{g_0}{g_c}.
\eea
At the critical line $g_0=g_c=(1+\omega)^2/4$ 
the anomalous dimension is
\bea
\gamma_m=1+\omega,\qquad \Delta g_0=0.
\eea
The expression for $\gamma_m$ in the broken phase were firstly obtained
by Miransky and Yamawaki \cite{miya89}.
Later it was shown by Kikukawa and Yamawaki \cite{kiya90}
in $d<4$ dimensional NJL models and by Kondo {\em et al.} \cite{kotaya93}
that the anomalous dimension is continuous (but in fact non-analytic) 
across the critical curve, by identifying the fine tuning of $g_0$. 
\section{Lattice simulations of strong coupling QED}\label{sec_lattice}
Lattice simulations of strong coupling QED and the GNJL models are
usually performed in the noncompact formulation, 
({\em i.e.}, the gauge fields are noncompact).
In such a formulation the pure gauge sector is free of topological excitations. 
The main reason for using the noncompact formulation of the gauge group is the
observation that it appears to have a second order chiral phase transition 
\cite{kodako88a}, whereas the compact formulation exhibits a first order 
chiral phase transition \cite{koda87}.
The latter, therefore, is not a realistic candidate for a relativistic 
continuum quantum field theory.

The scaling law with essential singularity (Miransky scaling)
at the critical coupling $\alpha_0=\alpha_c$
has been studied by Kogut {\em et al.} \cite{kodako88c, kodako89c}.
However later simulations questioned the existence of 
Miransky scaling on the lattice, see for instance the discussion in
\cite{kohakoda90} and \cite{azcagagrlapi95b}.

Early studies of strong QED with four-fermion coupling 
on the lattice were done by \cite{kohakoda90,kokolowa93}.
The conclusions of \cite{kohakoda90} are in agreement with \cite{balomi90};
the critical exponents have nonmean field values and satisfy 
hyperscaling in the intermediate region ($0<\alpha_0<\alpha_c$) 
(in quenched approximation).
Furthermore it was argued in \cite{kohakoda90} that 
from the RG point of view it is essential 
that the critical exponent $\gamma=1$. 
It means one has the factorization
$\eta_{(\bar\psi\psi)^2}=2\eta_{(\bar\psi\psi)}$.
Hence the anomalous dimension of the operator $(\bar\psi\psi)^2$ 
is twice the anomalous dimension of the mass operator $\bar\psi \psi$.
A renormalization of the chiral condensate simultaneously renormalizes
the propagators $\DelS$ and $\DelP$.
Indeed the lattice computations of the critical exponent $\gamma$ 
reported in \cite{kohakoda90,kokolowa93} showed strong evidence for $\gamma=1$.

Moreover is was argued that it is natural for the lattice-regulated form 
of QED to incorporate four-fermion interactions.
The Illinois group (\cite{kohakoda90,kokolowa93}) obtained  
a critical point $(0.44 \alpha_c,0.76)$
in the $(\alpha_0,g_0)$ plane, which fits nicely on the critical line
Eq.~(\ref{critline1}).
However it was not possible at that time to investigate with lattice simulations
the phase transition along the critical line.

Later, the GNJL model was studied on the lattice in noncompact formulation 
using some mean field approach for the fermions, 
by Azcoiti {\em et al.} \cite{azcagagrlapi95b}.
They obtained a critical line qualitatively similar
to the SDE approach Eq.~(\ref{critline1}), 
and nonmean field critical exponents.

Studies of the compact GNJL model \cite{azcagagrla98} (with torus topology)
showed to have both a first order transition as well
a second oder phase transition similar to the continuous transition 
in the noncompact formulation \cite{azcagagrlapi95b,azcagagrla98}.
Also in these studies (\cite{azcagagrlapi95b}) it was shown 
that the critical exponents have nonmean field values 
and satisfy hyperscaling.

Simulations of noncompact full (unquenched) QED on the lattice 
(with flavors, $N=2$ and $N=4$) are controversial \cite{goholirascst97}.
The Illinois group \cite{kokowa93,hakokoresiwa94} (see 
also \cite{kodako89a,kodako89b})
and the Zaragosa 
group \cite{azcagr93,azcagagrlapi95a}, 
find power-law scaling and nonmean field critical exponents, signaling
a possible nontrivial continuum limit for the strong coupling broken phase, 
whereas \cite{goholarascsowi90,gohorascso92,goholirascst97} 
obtain mean field behavior (mean field critical exponents 
with logarithmic corrections).
Thus G{\"o}ckeler {\em et al.} find a vanishing renormalized
gauge coupling and a vanishing effective Yukawa coupling (defined
by the Goldberger--Treiman relation), and they conclude 
that lattice QED is trivial, see for their most recent result
Ref.~\cite{goholirascst98}.

The authors of \cite{kokoha92,kokowa93,hakokoresiwa94} found indications 
that monopole condensation and chiral symmetry transitions were
coincident. These results suggested that theories of fundamental charges
and monopoles provide a natural scenario for nontrivial ultraviolet behavior.
However later analysis of \cite{gohorasc96} 
and the discussion in \cite{hako96} put in doubt the initial ideas on the role of
monopole condensation driving the chiral phase transition. 

In \cite{koko94} it was argued that the triviality of $\lambda\phi^4$ theory
cannot be a good guide to understanding 
the possible triviality of the pure NJL or spinor QED. 
The main reason is that the Nambu-Goldstone bosons in models 
with fundamental fermions which undergo (dynamical) chiral symmetry breaking
are composite, and their propagators have a large anomalous dimension, 
whereas they are fundamental in theories with scalars 
that exhibit a magnetic type of phase transition, 
and the anomalous dimension is small.
The consequence is the following.
Although both the NJL model and $\lambda\phi^4$ theory 
violate hyperscaling due to the appearance of logarithmic corrections, 
and hence are trivial, the bounds on the critical exponents are different.
In NJL type of models the logarithmic violations 
reduce the critical exponent $\delta$ below its mean field 
value $\delta<3$, whereas in scalar models the critical exponent 
exceeds its mean field value $\delta>3$.
Such logarithmic violations\footnote{Logarithmic scaling is rather complicated to investigate on the lattice 
due to the finiteness of the system.} 
of scaling and the direction of violation 
were tested in \cite{kikoko94}, and \cite{azcagagrlapi96}.

Studies of unquenched QED or GNJL model in the continuum 
(both analytic and numerical studies), see Refs.~\cite{ra91,gu90,ko90},
commonly use a one-loop vacuum polarization in the SDE
for the fermion propagator and gap equation.
The effect of the inclusion of screening effects on the phase transition
is that instead of the scaling law with essential singularity, the scaling
appears to be of the mean field type, as one would expected
for a trivial theory (With a one-loop vacuum polarization incorporated
QED is shown to be explicitly trivial).
\section{Beyond the ladder approximation}\label{sec_beyond}
In the previous section we have pointed out the qualitative agreement 
of the SD quenched-ladder approximation 
with the quenched lattice simulations.
But we would like to mention other attempts to study the gap equation beyond 
the ladder approximation in framework of the SD approach.
In Refs.\cite{aplama88,ho89} the validity of the ladder approximation
is analyzed by including the effects of {\em e.g.} crossed photon exchange 
graphs.
Moreover, Holdom \cite{ho89} gives strong arguments
that the scaling law for the fermion dynamical mass $m_{\rm dyn}$ in
the quenched theory (including also crossed photon graphs, and vertex corrections)
is of the same form as the scaling law 
obtained in the quenched-ladder approximation given by Eq.~(\ref{essent_sing}).  
This is analogous to the statement that the anomalous dimension is one 
($\gamma_m=1$)
in quenched QED.

In addition, the nonperturbative renormalization group methods (NPRG) of 
Refs.~\cite{aomosuteto97,aplama88} provide a way to
check the quenched-ladder approximation in the GNJL model
by including the effect of crossed photon exchange graphs, and four-fermion
interactions in the RG flow of coupling constants.
The basic principle of the NPRG methods is to write down the equation for the
so-called shell-mode effective action, which follows from varying 
the cutoff of the Wilsonian
effective action, see {\em e.g.} Eq.~(\ref{effaction}).  
In principle the NPRG method is exact, but unsolvable in practice, since
it involves the computation of the RG flows of an infinite set of
operators
and couplings.
The systematic approximation is to write down the NPRG equations for
a suitable finite set of local operators, which is called
the local potential approximation (LPA). 
The next step is that the solutions obtained in the LPA have to be checked 
for stability with respect to enlargement of the set of operators.

In Ref.~\cite{ao97} the critical line in the full quenched GNJL model
is obtained in a particular LPA,
which incorporates besides crossed photon exchange graphs 
also four-fermion exchanges beyond the mean field approach.
The equation for the critical line obtained by Aoki {\em et al.} reads
\bea
g_c(\alpha_0)=\frac{1}{3} \(1+\sqrt{1-\frac{\alpha_0}{\alpha_c}}\)^2,
\eea
and in comparison with Eq.~(\ref{critline1}) the difference is just
an overall factor $3/4$.
Thus these studies support the reliability of the ladder approximation.
Moreover in Ref.~\cite{ao97} it is shown that the critical exponents
agree rather well with the quenched-ladder results for small values of 
$\alpha_0$.
However, for values of the gauge coupling $\alpha_0$ close to the CPT point 
$\alpha_0=\alpha_c$, a small quantitative difference in the values of the
critical exponents emerges.

Let us mention some other attempts to study D$\chi$SB in quenched QED
beyond the ladder approximation.
In Refs.~\cite{cupe90,cupe91,cupe92,buro93,bape94,bakipe98}
various Ans\"atze for the photon-fermion vertex have been constructed 
and analyzed.
The main motivation for the construction of such Ans\"atze
is the formulation of a gauge independent approach 
to D$\chi$SB.
In the ladder approximation
the full photon-fermion vertex
is approximated by the bare vertex 
and the approximation is believed to be reliable only in Landau gauge $\xi=0$.
Since, there, 
the gauge-dependent anomalous dimension of the fermion 
wave function ${\cal Z}$ (Eq.~(\ref{struc_ferm})) vanishes, and ${\cal Z}=1$.
Consequently
the WTI for the photon-fermion vertex, Eq.~(\ref{wtivertex}), 
is satisfied asymptotically, 
{\em i.e.} for momenta $|p|\gg |m_{\rm dyn}|$. 
Two crucial constraints on the vertex Ans\"atze
are the WTI and power-law behavior for the fermion wave function.
The power-law behavior for ${\cal Z}$ is motivated and derived 
from the multiplicative renormalizability of perturbation theory
for arbitrary gauge-parameters $\xi$.

%
\chapter{Scalar composites in the symmetric phase}\label{chap4}
\section{Introduction}
In the ladder approximation, and using the Hartree-Fock (mean field)
approximation for the four-fermion coupling it has been shown 
\cite{gumi91,gumi92,mi93,kotaya93}  that the GNJL model in four dimensions 
is indeed renormalizable, and that the anomalous dimension
of $\bar\psi\psi$ is large, turning the formally irrelevant four-fermion
operators into relevant operators.
Fine-tuning the coupling $g_0$ to $g_c$ in $S\chi SB$ phase 
in such a way that $m_d/\Lambda\ll 1$,
where $m_d\equiv\Sigma(0)$ is the dynamical mass of a fermion, a nontrivial
continuum limit ($m_d/\Lambda\rightarrow0$) can be reached just as in pure 
quenched QED \cite{fogumi78,fogumisi83,mi85a,mi85c}. The spectrum of such a 
theory contains pseudoscalar ($\pi$) and scalar 
($\sigma$) bound states which become light and dynamically active in the 
vicinity of the critical line. 
Since the phase transition is second order along the part 
Eq.~(\ref{critline1}) of the critical curve, scalar and pseudoscalar 
resonances have been shown to be produced on the symmetric 
side of the curve, whose masses approach
zero as the critical curve is approached \cite{aptewij91}. 
The resonances are of the Breit-Wigner type (see section~\ref{sec_particles})
and described by a complex pole (on a second Riemann sheet) in their 
respective propagators.

The part of the critical curve, Eq.~(\ref{critline2}), 
with $\alpha_0=\alpha_c$ is rather special. For example, an abrupt 
change of the spectrum of light excitations occurs when the line 
$\alpha_0=\alpha_c$, $g_0<1/4$ is crossed: while light scalar and 
pseudoscalar excitations still persist in the broken
phase, there are no such light excitations in the symmetric phase
\cite{mi93,aptewij96}. 
A similar behavior has been revealed also in QED$_3$ \cite{aptewij95}. 
This peculiar phase transition was referred to as a conformal phase 
transition (CPT) \cite{miya97}. 

\begin{figure}[t!]
\epsfxsize=8cm
\epsffile[40 430 370 540]{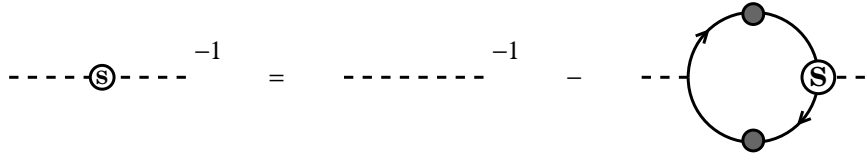}
\caption{The SDE for the scalar propagator $\Delta_{\rm S}(p)$.}
\label{fig_sde_scalar}
\end{figure}

\begin{figure}[t!]
\epsfxsize=8cm
\epsffile[40 430 370 540]{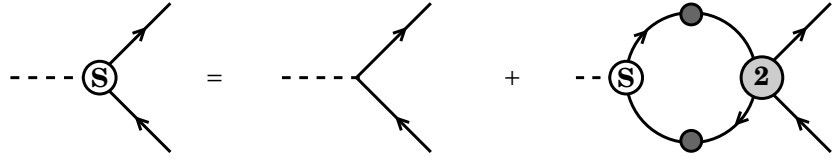}
\caption{The SDE for the scalar vertex ${\Gamma_{\rm S}}(p+q,p)$.
The shaded circle with the $2$ represents the two-fermion one-boson irreducible
fermion-fermion scattering kernel.}
\label{fig_sde_scalvert}
\end{figure}

\begin{figure}[b!]
\epsfxsize=8cm
\epsffile[40 430 370 540]{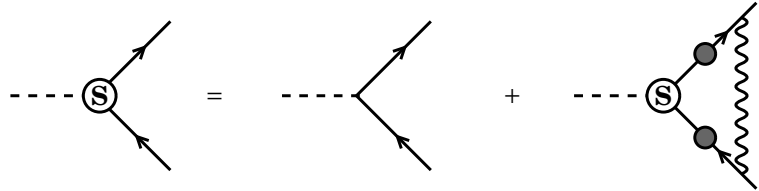}
\caption{The SDE for the scalar vertex in the quenched-ladder 
approximation.}
\label{fig_sde_scalvertladder}
\end{figure}

In this chapter we study scalar composites ($\sigma$ and $\pi$ bosons)
in the symmetric phase of the GNJL model. Computing the scalar propagator, see
Eqs.~(\ref{sdescal})--(\ref{SDscalvacpol}) and Fig.~\ref{fig_sde_scalar},
requires knowledge of the full scalar-fermion-antifermion 
vertex ${\Gamma_{\rm S}}(p+q,p)$ (the Yukawa vertex)
which in turn satisfies the 
Bethe-Salpeter (BS) equation \ref{BSscalvert}
and is displayed in Fig.~\ref{fig_sde_scalvert}. 
We solve this BS equation in the ladder 
approximation (Fig.~\ref{fig_sde_scalvertladder}) but differ 
from the corresponding studies in Refs.~\cite{kotaya93,aptewij91} 
who used an approximation for ${\Gamma_{\rm S}}(p+q,p)$ 
with zero boson momentum ($q=0$). 
A technique of expansion in Chebyshev polynomials is introduced for 
solving the Yukawa vertex with nonzero boson momentum and consequently an 
explicit analytical expression is derived for the propagator of the 
$\sigma$ boson valid along the entire critical curve. 
Our main physical conclusions are the same as in 
Refs.~\cite{mi93,aptewij91,aptewij96}: 
in the region $\alpha_0<\alpha_c$, $g_0>1/4$, 
in the symmetric phase, a spectrum of light resonances exists while 
at $\alpha_0<\alpha_c, g_0<1/4$ there are no light resonances. 
Having obtained an analytical expression for the scalar propagator, we can 
analytically continue it into the region $\alpha_0>\alpha_c$ and find 
light tachyons there, signaling the instability of the symmetric solution.

The plan of the present chapter is as follows. 
First we solve the equation for the Yukawa 
vertex with nonzero boson momentum in section~\ref{sec_scal_vertex} 
keeping only the zero order Chebyshev harmonics. 
In section~\ref{scal_prop} we obtain an analytical expression for the 
$\sigma$ boson propagator valid along 
the entire critical line and analyze its behavior in different 
asymptotical regimes. Section~\ref{sec_review} is devoted to comparing 
our results for the 
Yukawa vertex and boson propagator with the corresponding ones 
in Refs.~\cite{kotaya93,aptewij91}. 
In section~\ref{res_near_crit} we discuss the behavior of the scalar 
propagator near the critical line (\ref{critline1}) in the symmetric phase 
and, in particular, the mass and the width of resonances. The analysis of 
the scalar composites near the critical line (\ref{critline2}) is given in 
section~\ref{again_sec_cpt}, where we show 
the absence of light excitations at $\alpha_0\leq\alpha_c$ while 
analytically continuing the symmetric phase propagator into the region 
$\alpha_0\geq\alpha_c$ leads to the appearance of tachyonic states. 
We discuss this behavior from the viewpoint of the CPT conception 
proposed in Ref.~\cite{miya97} and section~\ref{again_sec_cpt}. 
A summary is given in 
section~\ref{summary} and in appendix~\ref{Chebyshev} we give an 
analysis of the contribution of higher order Chebyshev harmonics into 
the Yukawa vertex equation and scalar vacuum polarization. 
The results of this chapter were published in Ref.~\cite{gure98}.
\section{Scalar vertex in quenched-ladder approximation}
\label{sec_scal_vertex}
In this section we discuss the SDEs for the scalar propagator
and the scalar vertex in the well-known quenched-ladder approximation and 
introduce an approximation scheme for solving the SDE for the scalar 
vertex.
For the time being we consider one flavor of fermions, $N=1$.
Then the SDE for $\sigma$ boson reads
\bea
\Delta_{\rm S}^{-1}(p)=-\frac{1}{G_0}+\Pi_{\rm S}(p^2),\label{delsinv}
\eea
where the scalar vacuum polarization $\Pi_{\rm S}(q^2)$ 
is given by
\bea
\Pi_{\rm S}(p^2)=
i\int_{\Lambda}\frac{{\rm d}^4k}{(2\pi)^4}\,{\rm Tr}
\left[S(k+p){\Gamma_{\rm S}}(k+p,k)S(k) \right]\label{sde_scalvac},
\eea
and for the Yukawa vertex $\Gamma_{\rm S}$ we have
\bea
-i{\Gamma_{\rm S}}_{ab}(p+q,p)&=&
({-i\bf 1})_{ab}+\int_{\Lambda}\frac{{\rm d}^4r}{(2\pi)^4}
\left[iS(r+q) (-i){\Gamma_{\rm S}}(r+q,r) iS(r) \right]_{dc} \nonu\\
&\times&(-ie_0^2)K^{(2)}_{cd,ab}(r,r+q,p+q)
\eea
(see Fig.~\ref{fig_sde_scalvert}), and where $\Lambda$ is the UV cutoff.
We recall that in the symmetric phase of the GNJL the pseudoscalar and scalar 
propagators are degenerate, so are the pseudoscalar vertex and 
scalar vertex.

The ladder approximation is obtained by replacing the Bethe-Salpeter
kernel $K^{(2)}$ of Eq.~(\ref{K2def}) by the one photon exchange graph,
\bea
(-ie_0^2)K^{(2)}_{ab,cd}(k,p,p+q)
=(-ie_0)\gamma^\lambda_{cb}iD_{\lambda\sigma}(q)(-ie_0)\gamma^\sigma_{ad}.
\eea
Furthermore the photon propagator is considered as quenched, {\em i.e.},
vacuum polarization effects are turned off, and thus the gauge
coupling does not run (as explained earlier), we again assume
\bea
\beta_\alpha(\alpha_0)\approx 0.
\eea

In principle the Bethe-Salpeter kernel also contains scalar and 
pseudoscalar exchanges. One question is whether such exchanges can be 
neglected, this question will be addressed in the next chapter.
The answer not only depends on the short-distance behavior 
of the full scalar propagators and Yukawa vertex
which we will try to solve in this chapter, 
but also on the representation of the chiral symmetry.  
In the next chapter we will argue that, 
provided scalars and pseudoscalars are 
considered  both in the adjoint representation of the chiral symmetry, 
the neglect of scalar and pseudoscalar 
exchanges in the kernel $K^{(2)}$ seems reasonable for the SDE 
for the Yukawa vertices ${\Gamma_{\rm S}}$ and ${\Gamma_{\rm P}}$.
Hence we will treat the four-fermion interactions again in the 
Hartree--Fock approximation, a sort of mean field approach, 
and postpone discussions
of validity and self-consistence to the chapter~\ref{chap5}.

The SDE equation for the scalar vertex in the ladder approximation 
can be written as
\bea
{\Gamma_{\rm S}}(p+q,p)&=&{\bf 1}+ie_0^2\int_{\Lambda}\frac{{\rm d}^4r}{(2\pi)^4}
\nonu\\&\times&
\gamma^\lambda S(r+q) 
{\Gamma_{\rm S}}(r+q,r) S(r) 
\gamma^\sigma D_{\lambda\sigma}(r-p)
\label{sde_ladder}
\eea
(see Fig.~\ref{fig_sde_scalvertladder}).
The SDE for the scalar propagator, Eq.~(\ref{sde_scalvac}), is left 
unchanged.
In the symmetric phase, the equation for the scalar vertex,
Eq.~(\ref{sde_ladder}), is a self-contained equation, if we note that 
in the Landau gauge the fermion propagator is $S(p)=1/\hat p$.
We recall that in the Landau gauge the ladder approximation
respects the chiral and vector Ward-Takahashi identities.
The SDEs in ladder approximation for scalar propagator
and vertex, Eqs.~(\ref{sde_scalvac}) and (\ref{sde_ladder}), 
have been studied extensively in the literature 
\cite{kotaya93,aptewij91,gukumi89b,balelo89,balo92}, 
but mainly for the case of zero-transfer boson momentum ($q=0$).

In what follows, we present a method for solving the scalar vertex with 
nonzero boson momentum. The starting point is a general structure of 
the scalar vertex and pseudoscalar vertex. The scalar and pseudoscalar 
vertices in momentum space can be decomposed over four spinor structures 
with dimensionless scalar functions 
as given in Eqs.~(\ref{vertfiesdefscal}) and (\ref{vertfiesdefpseudo}).
In the symmetric phase, the scalar 
and pseudoscalar vertex functions coincide,
{\em i.e.}, we set $F_i\equiv F^{(\rm s)}_i=F^{(\rm p)}_i$, $i=1,2$.
The straightforward consequence is that the scalar and pseudoscalar 
propagators are also identical (degenerate) in the symmetric phase, 
just a reflection of the chiral symmetry.
Thus we have the structure
\bea
\GS(p+q,p)= {\bf 1}\left[ F_1+\(\qslash\pslash-\pslash\qslash\)F_2
+(\pslash+\qslash) F_3+\pslash F_4\right]. \label{vertfiesdef}
\eea
Furthermore, because of the absence of a dynamical mass in the symmetric phase, 
the equations for the scalar functions $F_3$
and $F_4$ decouple from the equations for $F_1$ and $F_2$. Moreover, 
$F_3$ and $F_4$ do not contribute in scalar and pseudoscalar vacuum 
polarizations. In fact the integral equations for these functions 
are homogeneous ones and in the symmetric phase we can always take the 
solution $F_3=F_4=0$ which is a consistent one. 

So the problem is reduced to solving a coupled set of integral equations
for two scalar functions $F_1$, $F_2$, which has the form 
(after making a standard Wick rotation \cite{tony})
\bea
F_i(p+q,p)=\delta_{i1}
+\lambda_0 \sum_{j=1}^2 \int_{\Lambda} {\rm d}r^2\,
\int\frac{{\rm d}\Omega_r}{2\pi^2}\,
K_{ij}(p,q,r) F_j(r+q,r),\quad i=1,2, \label{sde_vertfies}
\eea
where $\lambda_0=3\alpha_0/4\pi$, and
\bea
K_{11}(p,q,r)&=&
\frac{(r^2+q\cdot r)}{(r+q)^2 (r-p)^2},\label{k11}\\
K_{12}(p,q,r)&=&
\frac{2[(q\cdot r)^2-r^2 q^2]}{(r+q)^2 (r-p)^2},\label{k12}\\
K_{21}(p,q,r)&=&\frac{1}{6}
\frac{\kappa(p,q,r)}{(r+q)^2(r-p)^4[(p\cdot q)^2-p^2 q^2]},
\label{k21}\\
K_{22}(p,q,r)&=&\frac{1}{3}
\frac{\kappa(p,q,r)(r^2+q\cdot r)}{(r+q)^2(r-p)^4[(p\cdot q)^2-p^2 q^2]},
\label{k22}
\eea
with
\bea
\kappa(p,q,r)&=&
p^2 p\cdot r q^2-p^2 p\cdot q q\cdot r-2 p\cdot q p\cdot r q\cdot r
+2 p^2 (q\cdot r)^2\nonu\\
&+&2(p\cdot q)^2 r^2-2 p^2 q^2 r^2+p\cdot r q^2 r^2-p\cdot q q\cdot r r^2,
\eea
and $\int {\rm d}\Omega_r$ denotes the usual angular part of
the four-dimensional integration.

The SDE for Yukawa vertex is quite similar to a Bethe-Salpeter equation (BSE) 
for bound state wave function $\chi$ ($\chi\sim S \GS S$).
The essential difference of course is that 
the SDE for $\GS$ is an inhomogeneous equation, whereas
the homogeneous BSE describing bound states is an eigenvalue equation.

The equations (\ref{sde_vertfies}) are still very complicated due to
the fact that the angular part of the integration cannot be performed
in explicit form, since the angular dependence of the Yukawa vertex 
is unknown. Without any further approximations it seems impossible to 
solve the equations analytically. Our primary interest is the scalar 
propagator defined by the vacuum polarization Eq.~(\ref{sde_scalvac}).
The equation for the scalar vacuum polarization is
\bea
\Pi_{\rm S}(q^2)=\frac{1}{4\pi^2}
\int\limits_0^{\Lambda^2}
{\rm d} k^2\,\int\frac{{\rm d}\Omega_k}{2\pi^2}
\biggr[ A_1(k,q)  F_1(k+q,k)
+ A_2(k,q)F_2(k+q,k)\biggr], 
\label{sde_vacpol2}
\eea
where
\bea
A_1(k,q)\equiv \frac{k^2+k\cdot q}{(k+q)^2},\qquad
A_2(k,q)\equiv\frac{2[(k\cdot q)^2-k^2 q^2]}{(k+q)^2}. \label{A1A2def}
\eea
The method to tackle the angular dependence is to expand in terms of 
Chebyshev polynomials of the second kind $U_n(x)$, a method which was 
used before, for instance in Ref~\cite{fogumisi83}.

We define the following Chebyshev expansions.
For the vertex function $F_1$ and $F_2$
we define
\bea
F_1(p+q,p)&=&\sum_{n=0}^\infty f_n(p^2,q^2) U_n(\cos\alpha),\label{chebF1}\\
F_2(p+q,p)&=&\sum_{n=0}^\infty g_n(p^2,q^2) U_n(\cos\alpha),
\eea
for the kernels of the $\Pi_{\rm S}$, Eq.~(\ref{A1A2def})
\bea
A_1(p,q)&=&\sum_{n=0}^\infty a_n(p^2,q^2) U_n(\cos\alpha),\\
A_2(p,q)&=&\sum_{n=0}^\infty b_n(p^2,q^2) U_n(\cos\alpha),
\eea
and for the kernels, Eqs.~(\ref{k11})--(\ref{k22})
\bea
K(p,q,r)=\sum_{n,m,l=0}^\infty K_{nml}(p^2,q^2,r^2) U_n(\cos\alpha)
U_m(\cos\beta)U_l(\cos\gamma),
\eea
where
\bea
\cos \alpha=\frac{p\cdot q}{p q},\quad
\cos \beta=\frac{p\cdot r}{p r},\quad
\cos \gamma=\frac{q\cdot r}{q r}
\eea
(for the coefficients $K_{nml}(p^2,q^2,r^2)$ see appendix~\ref{Chebyshev}).
After that the angular integration can be done explicitly leading to an 
infinite chain of equations for harmonics $f_n(p^2,q^2)$ and $g_n(p^2,q^2)$. 
The important thing is that only the harmonics $f_0$ contains an 
inhomogeneous term in the equation for it (the constant $1$ in 
Eq.~(\ref{sde_vertfies})), while other harmonics can be found iteratively 
once $f_0$ is computed. In other words, for the vertex function $F_1$ the 
scale is set by the bare vertex, {\em i.e.}, such a function has nonhomogeneous 
ultraviolet boundary conditions. For the vertex function $F_2$ there is 
no such inhomogeneous term other than given indirectly by the coupling to 
vertex function $F_1$.

We assume that the scalar vertex function $F_1(p+q,p)$ depends only weakly 
on the angle between fermion and scalar-boson momentum $p\cdot q$,
so that an infinite set of equations for $f_n$ and $g_n$ 
is replaced by the equation for the zeroth-order Chebyshev 
coefficient function $f_0$ which we shall solve exactly. 
The main approximation is to replace
the Yukawa vertex by the angular average of the vertex function $F_1$,
\bea
{\Gamma_{\rm S}}(p+q,p)\approx {\bf 1} 
\int\frac{{\rm d}\Omega_p}{2\pi^2} F_1(p+q,p)
={\bf 1} f_0(p^2,q^2),
\label{canonic}
\eea
since Chebyshev polynomials of the second kind are precisely 
orthogonal with respect to such integration.
Then we write
\bea
f_0(p^2,q^2)\equiv F_{\rm IR}(p^2,q^2)\theta(q^2-p^2)
+ F_{\rm UV}(p^2,q^2)\theta(p^2-q^2).\label{channelapprox}
\eea
The functions $F_{\rm IR}$ 
and $F_{\rm UV}$ are respectively referred to as the IR channel
(infrared), and the UV channel (ultraviolet).

\begin{figure}[ht!]
\epsfxsize=8cm
\epsffile[-120 140 380 430]{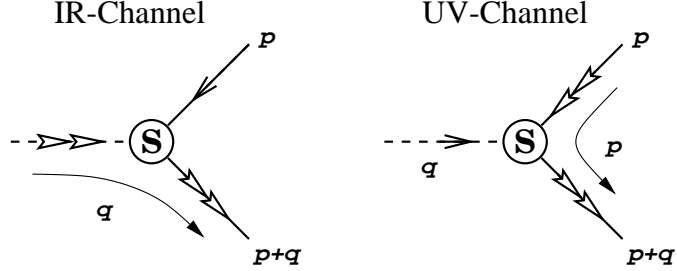}
\caption{The two-channel approximation.}
\label{two_channel_fig}
\end{figure}

If the scalar vertex indeed weakly depends on angle between
scalar-boson and fermion momentum $p$ flowing through 
the Yukawa vertex, these channel functions should have the limits
\bea
\lim_{p^2\gg q^2} {\Gamma_{\rm S}}(p+q,p)&=&
{\bf 1}\lim_{p^2\gg q^2} F_{\rm UV}(p^2,q^2),\\
\lim_{q^2\gg p^2} {\Gamma_{\rm S}}(p+q,p)&=&
{\bf 1}\lim_{q^2\gg p^2} F_{\rm IR}(p^2,q^2),
\eea
{\em i.e.}, the asymptotics of the scalar vertex are independent of the angle 
between $p$ and $q$. Hence the UV channel contains a limit of the Yukawa 
vertex with the boson momentum $q$ that is much smaller than both fermion momenta 
$(q\ll p)$, and the IR channel contains a limit of the vertex 
with the fermion momentum $p$ 
that is much smaller than the boson momentum $(q\gg p)$. The connection between 
the Yukawa vertex ${\Gamma_{\rm S}}$ and these two channel 
functions is illustrated in Fig.~\ref{two_channel_fig}.

The expansion in Chebyshev polynomials is discussed in detail 
in appendix~\ref{Chebyshev}. Moreover, the error, due to our approximation, 
Eq.~(\ref{canonic}), in the computation of the scalar vacuum polarization, 
is estimated in that appendix. 

The zeroth-order Chebyshev or the two channel approximation 
of Eq.~(\ref{channelapprox}) gives the following
equation for the vertex function $f_0(p^2,q^2)$:
\bea
f_0(p^2,q^2)=1+\lambda_0
\int\limits_0^{\Lambda^2} {\rm d}r^2\,N_0(r^2,p^2)
a_0(r^2,q^2) f_{0}(r^2,q^2),
\eea
where
\bea
N_0(r^2,p^2)=\frac{\theta(r^2-p^2)}{r^2}+\frac{\theta(p^2-r^2)}{p^2}
\label{N0secIII}
\eea
and
\bea
a_0(r^2,q^2)=\frac{1}{2}\left[\left(2-\frac{q^2}{r^2}\right)\theta(r^2-q^2)
+\frac{r^2}{q^2} \theta(q^2-r^2)\right]
\label{a0secIII}
\eea
(see also appendix~\ref{Chebyshev}).
The equation for the scalar vacuum polarization 
in this approximation Eq.~(\ref{sde_vacpol2}) takes the form
\bea
\Pi_{\rm S}(q^2)=
\frac{1}{4\pi^2}\int\limits_0^{\Lambda^2}
{\rm d} k^2\, a_0(k^2,q^2) f_0(k^2,q^2 ).\label{vacpol3}
\eea

With the Eqs.~(\ref{N0secIII}) and (\ref{a0secIII}) for $N_0$ and $a_0$,  
respectively, and the definition of the channel functions 
Eq.~(\ref{channelapprox}), we get two coupled integral equations for 
the IR-UV channels:
\bea
(s<t)\qquad F_{\rm IR}(s,t)&=&
1+\lambda_0\int\limits_0^s{\rm d}u\,\frac{u}{2st} F_{\rm IR}(u,t)
+\lambda_0\int\limits_s^t{\rm d}u\,\frac{1}{2t} F_{\rm IR}(u,t)\nonu\\
&+&\lambda_0\int\limits_t^{\Lambda^2}{\rm d}u\, 
\frac{2u-t}{2u^2} F_{\rm UV}(u,t),\label{fireq}\\
(s>t)\qquad F_{\rm UV}(s,t)&=&
1+\lambda_0\int\limits_0^t{\rm d}u\,\frac{u}{2st} F_{\rm IR}(u,t)
+\lambda_0\int\limits_t^s{\rm d}u\,
\frac{2u-t}{2su}F_{\rm UV}(u,t)\nonu\\
&+&\lambda_0\int\limits_s^{\Lambda^2}{\rm d}u\,
\frac{2u-t}{2u^2}F_{\rm UV}(u,t),
\label{fuveq}
\eea 
where $s=p^2$, $t=q^2$, and $u=r^2$. For the vacuum polarization, 
Eq.~(\ref{vacpol3}), we obtain the equation
\bea
\Pi_{\rm S}(t)=
\frac{1}{4\pi^2}
\left[\int\limits_0^t {\rm d}u\,\frac{u}{2t} F_{\rm IR}(u,t)
+\int\limits_t^{\Lambda^2} {\rm d}u\,\frac{2u-t}{2u}F_{\rm UV}(u,t)\right],
\qquad t=q^2.\label{scalvacpol_eq}
\eea
Using Eq.~(\ref{fuveq}) with $s=\Lambda^2$ provides a simple relation
between the vacuum polarization and the UV-channel function
\bea
\Pi_{\rm S}(q^2)=
\frac{\Lambda^2}{4\pi^2}
\frac{1}{\lambda_0}\left[F_{\rm UV}(\Lambda^2,q^2)-1\right],
\label{scalvacpol_eq2}
\eea
which is different from the functional form proposed in Ref.~\cite{kotaya93}.

The integrals equations for $F_{\rm IR}$ and $F_{\rm UV}$
are equivalent to two second order differential equations
with four appropriate boundary conditions.
We get for the IR channel:
\bea
s^2 \frac{{\rm d}^2}{{\rm d}s^2}F_{\rm IR}(s,t)
+2s\frac{{\rm d}}{{\rm d}s}F_{\rm IR}(s,t)+
\frac{\lambda_0}{2t} s F_{\rm IR}(s,t)=0,
\label{eq:IR}
\eea
and for the UV channel
\bea
s^2 \frac{{\rm d}^2}{{\rm d}s^2}F_{\rm UV}(s,t)+
2s\frac{{\rm d}}{{\rm d}s}F_{\rm UV}(s,t)+
\lambda_0\frac{2s-t}{2s} F_{\rm UV}(s,t)=0.
\label{eq:UV}
\eea
The infrared and ultraviolet boundary conditions (IRBC), 
respectively, (UVBC)
are
\bea
\left[s^2\frac{{\rm d}}{{\rm d}s}F_{\rm IR}(s,t)\right]\Biggr|_{s=0}=0,\qquad
\left[F_{\rm UV}+s\frac{{\rm d}}{{\rm d}s}F_{\rm UV}\right]
\Biggr|_{s=\Lambda^2}=1.
\label{BC}
\eea
Moreover we get a continuity and differentiability equation
at $s=t$
\bea
F_{\rm IR}(t,t)=F_{\rm UV}(t,t),\qquad 
\frac{{\rm d}}{{\rm d}s}F_{\rm IR}(s,t)\Biggr|_{s=t}=
\frac{{\rm d}}{{\rm d}s}F_{\rm UV}(s,t)\Biggr|_{s=t}.\label{contdiff}
\eea
The differential equations can be solved straightforwardly.
The equation for $F_{\rm IR}$ can be written as a Bessel equation, and
the equation for $F_{\rm UV}$ as a modified Bessel equation
(see Ref.~\cite{ermaobtr53} for mathematical details).
The general solutions of the differential equations are
\bea
F_{\rm IR}(s,t)&=&c_3(t/\Lambda^2,\omega) 
\left(\frac{t}{s}\right)^{1/2} J_1\left(\sqrt{2\lambda_0 s/t }\right)\nonu\\
&+&c_4(t/\Lambda^2,\omega) 
\left(\frac{t}{s}\right)^{1/2}  Y_1\left(\sqrt{2\lambda_0 s/t}\right),
\label{fir}\\
F_{\rm UV}(s,t)&=&c_1(t/\Lambda^2,\omega) \left(\frac{t}{s}\right)^{1/2} 
I_{-\omega}\left(\sqrt{2\lambda_0 t/s}\right)\nonu\\
&+&c_2(t/\Lambda^2,\omega) \left(\frac{t}{s}\right)^{1/2} 
I_{\omega}\left(\sqrt{2\lambda_0 t/s}\right),\label{fuv}
\eea
with $J_1$ and $Y_1$ the Bessel functions
of first and second kind,  respectively, and where
$I_{\pm\omega}$ are modified Bessel functions and  
$\omega=\sqrt{1-4\lambda_0}$, see also Eq.~(\ref{omegadef}). We note that since 
Eqs.~(\ref{eq:IR}), (\ref{eq:UV}) are scale invariant their solutions
are functions of the ratio $s/t$ and the scale invariance is violated by 
the UV boundary condition (\ref{BC}) only.

The IRBC for the IR channel requires $c_4(t/\Lambda^2,\omega)=0$, since the
Bessel function $Y_n(z)$ is irregular at $z=0$,
the other coefficients are fixed by the remaining three boundary
conditions and the solutions are
\bea
c_1(t/\Lambda^2,\omega)&=&\frac{\pi\gamma(\omega)}{2\sin\omega\pi}
Z^{-1}(t/\Lambda^2,\omega),\\
c_2(t/\Lambda^2,\omega)&=&c_1(t/\Lambda^2,-\omega),\\
c_3(t/\Lambda^2,\omega)&=&Z^{-1}(t/\Lambda^2,\omega),
\eea
where
\bea
Z(q^2/\Lambda^2,\omega)\equiv \frac{\pi}{2\sin\omega\pi}
\left[\gamma(\omega)G(q^2/\Lambda^2,-\omega)
-\gamma(-\omega)G(q^2/\Lambda^2,\omega) \right],\label{Zdef}
\eea
and
\bea
\gamma(\omega)&\equiv&\sqrt{2\lambda_0}\left[
J_1(\sqrt{2\lambda_0})I_\omega^\prime(\sqrt{2\lambda_0})
+J_1^\prime(\sqrt{2\lambda_0})I_\omega(\sqrt{2\lambda_0})\right],\label{gammaeq}\\
G(q^2/\Lambda^2,\omega)&\equiv&\frac{1}{2}\sqrt{\frac{q^2}{\Lambda^2}}
\left[I_\omega\left(\sqrt{\frac{2\lambda_0 q^2}{\Lambda^2}}\right)-
\sqrt{\frac{2\lambda_0 q^2}{\Lambda^2}}I_\omega^\prime
\left(\sqrt{\frac{2\lambda_0 q^2}{\Lambda^2}}\right)\right].
\label{Geq}
\eea
Summarizing, the solution of scalar vertex in terms of channel functions is
\bea
F_{\rm IR}(p^2,q^2)&=&Z^{-1}\left(\frac{q^2}{\Lambda^2},\omega\right) 
\left(\frac{q^2}{p^2}\right)^{1/2}
J_1\left(\sqrt{\frac{2\lambda_0 p^2}{q^2} }\right),
\label{fir2}\\
F_{\rm UV}(p^2,q^2)&=&
\frac{\pi}{2\sin \omega\pi} Z^{-1}\left(\frac{q^2}{\Lambda^2},\omega\right)
\left(\frac{q^2}{p^2}\right)^{1/2}\nonu\\
&\times&\left[ 
\gamma(\omega)I_{-\omega}\left(\sqrt{\frac{2\lambda_0 q^2}{p^2}}\right)
-\gamma(-\omega)I_{\omega}\left(\sqrt{\frac{2\lambda_0 q^2}{p^2}}\right)\right].
\label{fuv2}
\eea

It is easy to verify that at zero boson momentum one gets
\begin{equation}
{\Gamma_{\rm S}}(p,p)\equiv F_{\rm UV}(p^2,q^2=0)
=\frac{2}{1+\omega}\left(\frac{p^2}
{\Lambda^2}\right)^{-(1-\omega)/2},\label{GSqzero}
\end{equation}
which coincides with the zero transfer vertex of Refs.~\cite{kotaya93,aptewij91}.
In the limit of pure NJL model, $\alpha_0\to0\quad (\omega\to1)$, the 
vertex is equal, of course, to the bare vertex, 
$\Gamma_S=1=F_{\rm UV}=F_{\rm IR}$. To study the vertex at critical 
gauge coupling $\alpha_0=\alpha_c\quad (\omega=0)$ we expand the Bessel 
functions in small $\omega$ using the following property of the 
modified Bessel functions:
\bea
I_\omega(x(\omega))\approx I_0(x(0))-\omega K_0(x(0))+{\cal O}(\omega^2),
\eea
where $x(\omega)\propto\sqrt{1-\omega^2}$.
Then the expressions for $F_{\rm IR}$, $F_{\rm UV}$, 
Eqs.~(\ref{fir2}), (\ref{fuv2}), take the following form at 
$\alpha_0\rightarrow\alpha_c$:
\bea
F_{\rm IR}(p^2,q^2)&=&Z^{-1}\left(\frac{q^2}{\Lambda^2},0\right) 
\left(\frac{q^2}{p^2}\right)^{1/2}J_1\left(\sqrt{\frac{p^2}{2q^2} }\right),\\
\label{vertirwzero}
F_{\rm UV}(p^2,q^2)&=&
Z^{-1}\left(\frac{q^2}{\Lambda^2},0\right)\left(\frac{q^2}{p^2}\right)^{1/2}\nonu\\
&\times&
\left[\epsilon_1K_0\left(\sqrt{\frac{q^2}{2p^2}}\right)
-\epsilon_2I_0\left(\sqrt{\frac{q^2}{2p^2}}\right)\right],
\label{vertuvwzero}
\eea
where 
\bea
Z^{-1}\left(\frac{q^2}{\Lambda^2},0\right)&=&
\frac{1}{2}\left(\frac{q^2}{\Lambda^2}\right)^{1/2}
\big[\epsilon_1 K_0(x)+\epsilon_1 x K_1(x)
\nonu\\
&&-\epsilon_2 I_0(x)-\epsilon_2 x 
I_0^\prime(x)\big],\\
x&=&\sqrt{q^2/2\Lambda^2},\nonu
\eea
and where
\bea
\epsilon_1&=&\sqrt{1/2}(
J_1(\sqrt{1/2})I_0^\prime(\sqrt{1/2})+J_1^\prime(\sqrt{1/2})I_0(\sqrt{1/2})),\\
\epsilon_2&=&\sqrt{1/2}(
J_1(\sqrt{1/2})K_0^\prime(\sqrt{1/2})+J_1^\prime(\sqrt{1/2})K_0(\sqrt{1/2})),
\eea
and $K_i$ is the modified  Bessel function of the third kind.
Note that when we expand the UV channel function, ($p^2\gg q^2$),
Eq.~(\ref{vertuvwzero}), we get
\bea
F_{\rm UV}(p^2,q^2)=2\left(\frac{p^2}{\Lambda^2}\right)^{-1/2}
\left[\frac{\epsilon_3-2+\ln(p^2/q^2)}{\epsilon_3
-\ln(q^2/\Lambda^2)} +{\cal O}\left(q^2/p^2\ln(q^2/p^2)\right)\right],
\label{FUVkondo}
\eea
where
\bea
\epsilon_3=2(1-\gamma)+3\ln 2-2\frac{\epsilon_2}{\epsilon_1}\approx 3.2
\eea
and $\gamma$ is the Euler gamma.
The expression (\ref{FUVkondo}) is of the same form as 
obtained in Ref.~\cite{kotaya93} (their formula (2.88)).
\section{Analytic structure of the scalar propagator}
\label{scal_prop}
In the previous section we obtained an analytical expression for the
scalar vertex by assuming that the vertex only weakly depends on the angle
between
boson momentum and fermion momentum.
The expression for the scalar vacuum polarization in such an approximation
takes the form of Eq.~(\ref{scalvacpol_eq}).
The main object of investigation is scalar compositeness near the 
critical line.
In the neighborhood of the critical line,
the tendency of fermion-antifermion pairs to form bound states under
the influence of strong attractive four-fermion forces becomes apparent.

Since we are in the symmetric phase of the GNJL model
(there is no dynamical mass), the only important variable is the scalar boson 
momentum over cutoff $t=q^2/\Lambda^2$.
Equation~(\ref{scalvacpol_eq}) shows that the vacuum polarization depends only 
on the UV channel with fermion momentum at $\Lambda^2$, 
Eq.~(\ref{scalvacpol_eq2}).
Substituting the expressions obtained for the vertex function
Eq.~(\ref{fuv}) in the equation for the vacuum polarization
Eq.~(\ref{scalvacpol_eq2}), we obtain 
\bea
\Pi_{\rm S}(q^2)=
\frac{\Lambda^2}{4\pi^2}\frac{1}{\lambda_0}
\left[\frac{\gamma(\omega)H(q^2/\Lambda^2,-\omega)
-\gamma(-\omega)H(q^2/\Lambda^2,\omega)}{\gamma(\omega)
G(q^2/\Lambda^2,-\omega)
-\gamma(-\omega)G(q^2/\Lambda^2,\omega)} \right],
\label{vacas1}
\eea
where $\gamma$ and $G$ are given by Eqs.~(\ref{gammaeq}) and (\ref{Geq}), 
and where
\bea
H(q^2/\Lambda^2,\omega)\equiv\frac{1}{2}\sqrt{\frac{q^2}{\Lambda^2}}
\left[I_\omega\left(\sqrt{\frac{2\lambda_0 q^2}{\Lambda^2}}\right)+
\sqrt{\frac{2\lambda_0 q^2}{\Lambda^2}}
I_\omega^\prime\left(\sqrt{\frac{2\lambda_0 q^2}{\Lambda^2}}\right)\right].
\label{Heq}
\eea
This expression is valid along the entire critical curve in the symmetric
phase. 
Note that the vacuum polarization is symmetric in $\omega$ which means
that we can analytically continue it to values $\alpha_0>\alpha_c$.

Equation~(\ref{vacas1}) is a rather complicated expression, and we 
will first investigate some specific limits:
\begin{enumerate}
\item[(A)]{
The pure NJL limit, where the gauge interaction is turned off,
{\em i.e.}, the case where $\alpha_0=0$, thus $\omega=1$.}
\item[(B)]{Asymptotic behavior of $\Pi_{\rm S}$, the infrared behavior 
$q^2/\Lambda^2\ll 1$ in such a manner, so that
$(q^2/\Lambda^2)^\omega \gg q^2/\Lambda^2$.}
\item[(C)]{The behavior at the critical gauge coupling
$\alpha_0=\alpha_c$, thus $\omega=0$.}
\item[(D)]{The behavior of $\Pi_{\rm S}$ for $\alpha_0>\alpha_c$, $\omega=i\nu$,
$\nu=\sqrt{\alpha_0/\alpha_c-1}$, {\em i.e.}, analytic
continuation across the critical curve at $\alpha_0=\alpha_c$.}
\end{enumerate}
\subsection{The pure NJL limit}
The pure Nambu--Jona-Lasinio limit is the case where the gauge interaction
is completely turned off, possible bound states in such a model are 
purely due to the four-fermion interaction.
So $\alpha_0\rightarrow 0$, and therefore $\omega\rightarrow 1$.
The scalar vertex at $\alpha_0=0$ is just equal to the bare vertex, 
${\Gamma_{\rm S}}=1=F_{\rm UV}=F_{\rm IR}$.
The pure NJL limit can be correctly obtained from Eq.~(\ref{vacas1}) by making
expansions of the Bessel functions keeping sufficient number of terms, 
and then performing an expansion in $(1-\omega)$ to zeroth order.
Hence with $\alpha_0=0$, $\omega=1$ in Eq.~(\ref{vacas1}), we get
\bea
\Pi_{\rm S}(q^2)= \frac{\Lambda^2}{4\pi^2}
\left[1+\frac{q^2}{2\Lambda^2}\ln \left(\frac{q^2}{\Lambda^2}\right)
-\frac{3q^2}{4\Lambda^2}\right].\label{vacas_njl}
\eea
The log term is a consequence of the ``hard'' fermion loop
which also ruins the renormalizability of the pure NJL model.
\subsection{Asymptotic behavior of the $\sigma$ boson vacuum polarization}
The asymptotic behavior $0<\omega< 1$ so that
$(q^2/\Lambda^2)^\omega \gg q^2/\Lambda^2$ can be obtained 
by considering first the $q^2\ll \Lambda^2$ limit of $Z$, Eq.~(\ref{Zdef}): 
\bea
Z\approx \frac{\pi}{2\sin \omega \pi}
\sqrt{h(\omega)h(-\omega)(1-\omega^2)} 
\left(\frac{q^2}{\Lambda^2}\right)^{1/2}
\sinh \left[\frac{\omega}{2}\ln\left(\frac{\Lambda^2}{ q^2}\right)
+\delta(\omega)\right],
\eea
where
\bea
\delta(\omega)\equiv
\frac{1}{2}\ln\frac{ h(\omega)(1+\omega)}{h(-\omega)(1-\omega)}
-\omega\ln \sqrt{2\lambda_0},\qquad
h(\omega)\equiv \frac{\gamma(\omega) 2^\omega}{\Gamma(1-\omega)}.
\eea
Then the UV-channel function with fermion momentum $p^2=\Lambda^2$
can be expressed, in this limit, as
\bea
F_{\rm UV}(\Lambda^2,q^2)\approx\frac{2}{1+\omega}
+\frac{2\omega}{1-\omega^2}(1-\coth y),
\qquad y={\omega\over2}\ln\left(\frac{\Lambda^2}{q^2}\right)+\delta(\omega).
\eea
Hence
\bea
\Pi_{\rm S}(q^2)&\approx&
\frac{\Lambda^2}{4\pi^2}
\left[\frac{1}{g_c(\omega)}+\frac{8\omega}{(1-\omega^2)^2}(1-\coth y)
\right],\qquad q^2\ll \Lambda^2.
\label{vacasym}
\eea
However, the above expression does not reproduce the correct leading term of 
the NJL limit ($\omega\rightarrow 1$), for that we should use the expression
(\ref{vacas1}). In order to obtain such an expression which contains properly
the pure NJL limit, we have to keep more terms in the expansion of $Z$.
We then get for the scalar vacuum polarization
\bea
\Pi_{\rm S}(q^2)&\approx&
\frac{\Lambda^2}{4\pi^2}\biggr[\frac{1}{g_c(\omega)}
-B(\omega)\left(\frac{q^2}{\Lambda^2}\right)^\omega\hspace{-2mm}
+A(\omega) \frac{q^2}{\Lambda^2} \nonu\\
&&\qquad+{\cal O}\left( (q^2/\Lambda^2)^{2\omega}\right)
+{\cal O}\left( (q^2/\Lambda^2)^{1+\omega}\right)
\biggr],
\label{vacasym2}
\eea
where
\bea
A(\omega)\equiv \frac{1}{2 g_c(\omega)(1-\omega)},
\qquad B(\omega)=\frac{16 \omega}{(1-\omega^2)^2}
\frac{\gamma(-\omega)}{\gamma(\omega)} 
\frac{\Gamma(2-\omega)}{\Gamma(2+\omega)}
\left(\frac{\lambda_0}{2}\right)^\omega.
\label{Bw}
\eea
For Eq.~(\ref{vacasym2}) to be valid, it is assumed that $\omega> 1/2$.
It is straightforward to check that this expression satisfies
the NJL limit ($\omega\rightarrow 1$).

This is a suitable point to refer to appendix~\ref{Chebyshev}
for an analysis of the reliability of Eq.~(\ref{vacasym2}), 
where it is shown that the leading and next-to-leading terms 
in $q^2/\Lambda^2$ of $\Pi_{\rm S}$ ({\em i.e.}, the first two terms on 
the right-hand side of 
Eq.~(\ref{vacasym2})) are indeed correctly obtained by our approximation, 
see Eq.~(\ref{error_expr}).
\subsection{The $\sigma$ boson at critical gauge coupling}
At critical gauge coupling $\alpha_0=\alpha_c$ ($\omega=0$) is the onset 
of the scalar compositeness originating purely from electromagnetic forces. 
The expression for $F_{\rm UV}$, Eq.~(\ref{vertuvwzero}), has the form 
at $\alpha_0=\alpha_c$
\bea
F_{\rm UV}(p^2,q^2)&=&2\left(\frac{p^2}{\Lambda^2}\right)^{-1/2}\nonu\\
&\times&\left[\frac{\epsilon_2 I_0(y)-\epsilon_1  K_0(y)}{
\epsilon_2 I_0(x)-\epsilon_2 x I_1(x)
-\epsilon_1 K_0(x)-\epsilon_1 x K_1(x)}\right], 
\label{fuvwzero}
\eea
where $x=\sqrt{q^2/2\Lambda^2}$ and $y=\sqrt{q^2/2 p^2}$.
This gives 
\bea
\Pi_{\rm S}(q^2)=\frac{\Lambda^2}{4\pi^2}\left[
4+\frac{8\left[\epsilon_2 x I_1(x)+\epsilon_1 x K_1(x)\right]}{
\epsilon_2 I_0(x)-\epsilon_2 x I_1(x)-\epsilon_1 K_0(x)-\epsilon_1 x K_1(x)}
\right]. 
\label{alpcrit}
\eea
\subsection{Analytic continuation across the critical curve}
Since the expression for the scalar vacuum polarization is symmetric in 
$\omega$, Eq.~(\ref{vacas1}), it can be analytically continued to the values
 $\alpha_0> \alpha_c$.
This holds in replacing $\omega$ by $i\nu$ in Eq.~(\ref{vacas1}), where
\bea
\nu=\sqrt{\alpha_0/\alpha_c-1}.
\eea

The four specific limits of the scalar vacuum polarization described above
are very useful for studying the resonance structure of the
bound states which will be done in section~\ref{res_near_crit}, and
for the study of the CPT which will be done in section~\ref{again_sec_cpt}.
But first we shall compare our result for $\Delta_{\rm S}$ with that 
one obtained by other authors in the next section.

\section{Comparison with earlier work}
\label{sec_review}

In this section we discuss the earlier work of Appelquist,
Terning, and Wijewardhana \cite{aptewij91} and related work based on that
by Kondo, Tanabashi, and Yamawaki \cite{kotaya93} on the
scalar composites in the GNJL model.
The method used by these authors to solve the coupled set of
scalar vertex and scalar vacuum polarization is the following.
They consider a Taylor-series expansion about $q=0$ of the
scalar vacuum polarization $\Pi_{\rm S}(q^2)$.
In the ladder approximation, such a series has the property that
the $n$th derivative can be written as
\bea
\left(\frac{{\rm \partial}}{\partial q}\right)^n\Pi_{\rm S}(q^2)
&\sim& \int_\Lambda \frac{{\rm d}^4k}{(2\pi)^4}\,
\sum_{m=0}^{n-1} C_m
{\rm Tr}\biggr[\left(\frac{\partial^m}{\partial q^m}{\Gamma_{\rm S}}(k+q,k)\right)
\nonu\\&\times&
\frac{1}{\hat k}{\Gamma_{\rm S}}(k,k+q)
\frac{\partial^{n-m}}{\partial q^{n-m}}\frac{1}{\hat k-\hat q}\biggr].
\eea
Their basic assumption is then
that derivatives of the scalar vertex ${\Gamma_{\rm S}}$
can be neglected with respect to ${\Gamma_{\rm S}}$ for small $\alpha_0$,
\bea
\frac{\partial^n{\Gamma_{\rm S}}(k+q,k)}{\partial q^n}
\Biggr|_{q=0}\sim\frac{1}{k^n}
\frac{\alpha_0}{4\alpha_c}{\Gamma_{\rm S}}(k,k).\label{dervscalvert}
\eea
Subsequently the Taylor series is resummed using the assumption stated above
to obtain
\bea
\Pi_{\rm S}(q^2)\sim
\int_\Lambda \frac{{\rm d}^4k}{(2\pi)^4}\,
{\rm Tr}\left[{\Gamma_{\rm S}}(k,k) \frac{1}{\hat k}{\Gamma_{\rm S}}(k,k)
\frac{1}{\hat k-\hat q}\right],
\eea
which yields in terms of Euclidean momentum,
\bea
\Pi_{\rm S}(q^2)=
\frac{\Lambda^2}{4\pi^2}
\left[\frac{1}{g_c(\omega)}-b(\omega)\left(\frac{q^2}{\Lambda^2}\right)^\omega
+a(\omega)\frac{q^2}{\Lambda^2}\right],
\label{appelres}
\eea
where
\bea
a(\omega)=\frac{1}{2g_c(\omega)(1-\omega)},
\qquad b(\omega)=\frac{1}{g_c(\omega)\omega(1-\omega^2)},\qquad
g_c=\frac{(1+\omega)^2}{4}.\label{aandb}
\eea
How does their result compare to ours?
From our expression for the asymptotic behavior of $\Pi_{\rm S}$,
Eq.~(\ref{vacasym2}), and the result obtained in
Refs.~\cite{aptewij91} and
\cite{kotaya93}, Eq.~(\ref{appelres}),
the leading power of momentum is the same, namely, $(q^2/\Lambda^2)^\omega$.
However, the $\omega$-dependent factors in front of the leading 
and next-to-leading powers are
different. At the same time the pure NJL limit is obtained correctly 
in Refs.~\cite{aptewij91} and \cite{kotaya93}.
The differences are rather small for values of $\omega$ close to 1, {\em i.e.}
the coefficients are
\bea
A(\omega)=a(\omega),\qquad 
B(\omega)=b(\omega)\left[1+{\cal O}((1-\omega)^2)\right].
\eea
Hence the approach to the NJL point $\omega=1$ of both approximations 
is equal. However, for smaller values of $\omega$ the coefficient $B$ 
obtained in the present chapter starts to deviate from the one obtained 
in Ref.~\cite{aptewij91}. Then for such values of $\omega$, 
$b(\omega)>B(\omega)$.

What is the origin of this difference?
The first point is that the expression derived in Refs.~\cite{aptewij91} and
\cite{kotaya93} is valid for $\alpha_0$ not too large.
Secondly, from their answer Eq.~(\ref{appelres}) it is clear that a Taylor
series of the scalar propagator about $q=0$ is not well defined due to the
noninteger power behavior for values $0<\omega<1$.
This is reflected also in their assumption regarding the derivatives
of the scalar vertex at $q=0$, Eq.~(\ref{dervscalvert}).
The expression for the scalar vertex obtained in
section~\ref{sec_scal_vertex} shows that in general, for $0<\omega<1$,
the assumption (\ref{dervscalvert}) is not true.
Such derivatives of ${\Gamma_{\rm S}}$ are singular at $q=0$ due to the fact
that they depend on noninteger powers of $q$, which can be seen
from Eq.~(\ref{fuv}).

The scalar vertex for small $q\ll p$ is of the form
$\Gamma_{\rm S}(p+q,p)\sim F_{\rm UV}(p^2,q^2)$, and
\bea
F_{\rm UV}(p^2,q^2)
&\approx&
\left(\frac{p^2}{\Lambda^2}\right)^{-1/2+\omega/2}\Biggr\{\frac{2}{(1+\omega)}
+\frac{q^2}{p^2}\left[\frac{1}{4}+\frac{(1-\omega)}{4(1+\omega)}
\frac{p^2}{\Lambda^2}\right]
\nonu\\
&-&2\left(\frac{q^2}{p^2}\right)^\omega 
\frac{\gamma(-\omega)}{\gamma(\omega)}
\frac{\Gamma(1-\omega)}{\Gamma(2+\omega)}\left[1
-\frac{(1-\omega)}{(1+\omega)}
\left(\frac{p^2}{\Lambda^2}\right)^\omega\right]
\left(\frac{\lambda_0}{2}\right)^\omega\nonu\\
&+&{\cal O}\left((q^2/p^2)^\omega (q^2/\Lambda^2)^\omega\right)
+{\cal O}\left((q^2/p^2)^{1+\omega}\right)+\cdots\Biggr\}\label{fuv_asym},
\eea
which is consistent with
Eq.~(\ref{vacasym2}) for $p^2=\Lambda^2$, because of 
Eq.~(\ref{scalvacpol_eq2}).
Of course at $q=0$ our result coincides with that of the other authors.
But for nonzero $q$ this expression clearly shows that for $0<\omega<1$ the 
vertex contains noninteger powers of $q$. Hence the assumption made in 
Refs.~\cite{kotaya93,aptewij91} is not true in general, since higher
derivatives of the scalar vertex with respect to $q$ are singular at $q=0$.

\section{Light scalar resonances near criticality}\label{res_near_crit}

In this section we discuss the behavior of the scalar propagator near the
critical line in the symmetric phase $g_0\leq g_c$.
In the symmetric phase the scalar and pseudoscalar composites, the $\sigma$
and $\pi$ bosons are degenerate. Near the critical curve, a combination of 
strong four-fermion coupling and gauge coupling will tend to bind fermions 
and antifermions into these scalar composites.
Since the chiral symmetry is unbroken the $\sigma$ and $\pi$ bosons decay
to massless fermions and antifermions.
Hence the scalar composites are resonances which are described by a complex
pole in their propagators. The complex pole determines the mass and the
width of the resonances.

In what follows, we redo the computation of the 
complex poles of the $\sigma$ boson which was performed by Appelquist {\em et al.} 
in Ref.~\cite{aptewij91} using the expression for 
$\Delta_{\rm S}$, Eq.~(\ref{vacas1}),
obtained with the two-channel approximation of the Yukawa vertex.
The expressions obtained in section~\ref{scal_prop} for $\Pi_{\rm S}(p^2)$
in various regimes are rotated back to Minkowski momentum
$p^2\rightarrow p^2_M \exp(-i\pi)$.
Then the complex poles are given by
\bea
p_M^2=p_0^2\exp{(-i\theta)},\qquad
\Delta_{\rm S}^{-1}(p_M)=
-\frac{\Lambda^2}{4\pi^2g_0}+\Pi_{\rm S}(p_0^2\exp{(-i\theta)})=0.
\eea
We can also parametrize the location of a pole by a mass and a width, {\em i.e.},
$p_0^2\exp{(-i\theta)}=\left[M_\sigma-(i/2)\Gamma_\sigma\right]^2$,
which yields
\bea
M_\sigma=p_0\left[\frac{1+\cos\theta}{2}\right]^{1/2},\qquad
\frac{\Gamma_\sigma}{M_\sigma}=\frac{2\sin\theta}{1+\cos \theta}.
\eea
If $\theta$ is small, then $\Gamma_\sigma/M_\sigma\approx \theta$.

Near the Nambu--Jona-Lasinio point ($\alpha_0=0$) our expression for 
the vacuum polarization coincides with that obtained by Appelquist {\em et al.}, 
Eq.~(\ref{vacas_njl}),
and we get the following equations for the resonances:
\bea
\frac{c\Lambda^2}{p_0^2}=\cos\theta\[\ln(\Lambda^2/p_0^2)+3/2\],
\qquad c=\frac{2(1-g_0)}{g_0},
\eea
and
\bea
\pi+\theta =\left[\ln(\Lambda^2/p_0^2)+\frac{3}{2}\right]\tan\theta.
\eea
If now $g_0$ is tuned close enough to the critical value $g_c=1$, 
so that $\ln 1/c \gg 1$, the solution is approximately
\bea
p_0^2\approx\frac{2(1-g_0)}{g_0 \ln\left[g_0/2(1-g_0)\right]}\Lambda^2
\eea
and we find a narrow width
\bea
\theta \approx \frac{\pi}{\ln\left[g_0/2(1-g_0)\right]}.
\eea
These results are nothing else than the familiar NJL results.
For intermediate values of the gauge coupling,
$0<\alpha_0<\alpha_c$ ($0<\omega<1$),
we assume the poles of $\Delta_{\rm S}$ are small, $p_0/\Lambda\ll 1$, so that
\bea
\left(p_0/\Lambda\right)^\omega \ll 1. \label{restriction}
\eea
Then, from Eq.~(\ref{vacasym2}) we get the following equation for 
the real part of the pole:
\bea
0\approx-\frac{1}{g_0}+\frac{1}{g_c}-B(\omega)
\left(\frac{p_0^2}{\Lambda^2}\right)^\omega
\cos\omega(\theta+\pi).
\eea
The equation for the imaginary part reads
\bea
0\approx \sin\omega(\theta+\pi),
\eea
where $B(\omega)$ is given by Eq.~(\ref{Bw}).

\begin{figure}[t!]
\epsfxsize=9cm
\epsffile[150 330 400 570]{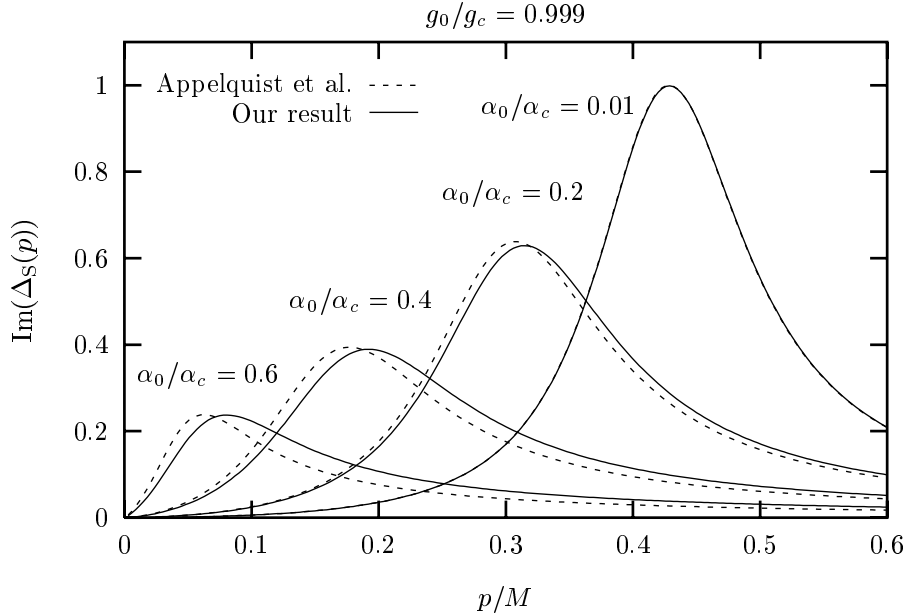}
\caption{A comparison of our result with Appelquist {\em et al.}
Graphs of ${\rm Im}(\Delta_{\rm S}(p))$ for different values of
$\alpha_0/\alpha_c$.
The curves are normalized so that the peak of the
$\alpha_0/\alpha_c=0.01$ curve equals 1.}
\label{appelfig}
\end{figure}

The solution is
\bea
\theta=\frac{\pi(n-\omega)}{\omega},
\eea
and $n$ is odd integer, so that $\cos \omega(\theta+\pi)=-1$,
thus
\bea
p_0\approx 
\Lambda\left[\frac{(1-g_0/g_c)}{g_0 B(\omega)}\right]^{1/2\omega}.
\label{massinter}
\eea
Hence $\theta$ is only small if $\omega\sim 1$ for $n=1$.
The result obtained in Ref.~\cite{aptewij91}, 
see Eq.~(\ref{aandb}), gives
a mass
\bea
p_0\approx 
\Lambda\left[\frac{(1-g_0/g_c)}{g_0 b(\omega)}\right]^{1/2\omega},
\qquad b(\omega)=\frac{1}{g_c\omega(1-\omega^2)}.
\eea
The pole obtained in Ref.~\cite{aptewij91} is of the same order as
Eq.~(\ref{massinter}) for values of $\omega$ close to 1
(see also discussion in the previous section).
For more intermediate values of $\omega$ the poles obtained 
in our approximation are somewhat bigger, since 
$b(\omega) > B(\omega)$ for $0<\omega<1$.
The quantitative difference between the
result of Ref.~\cite{aptewij91} and that
obtained in this chapter for resonance structures
are visualized in Fig.~\ref{appelfig}.
In Fig.~\ref{appelfig} the imaginary part of $\Delta_{\rm S}$ given by 
Eq.~(\ref{vacas1}) and Eq.~(\ref{delsinv}) 
is plotted versus $p/M$ where the tuning of the four-fermion to the critical
line is $g_0/g_c=0.999$, and $M/\Lambda=(1-g_0/g_c)^{1/2}$.
From Fig.~\ref{appelfig} it is clear that
the position of the peak of the resonant curve is slightly shifted 
to the right in our case at a fixed ratio 
$\alpha_0/\alpha_c \sim {\cal O}(1)$ (intermediate or small values 
of $\omega$), while the width over mass ratio remains comparable.
Near the pure NJL point $\omega\sim 1$ both results coincide.

As was pointed out in the previous section, and following from the restriction
Eq.~(\ref{restriction}), these results are only valid
for $\alpha_0/\alpha_c$ small. For larger values of the gauge coupling, 
$\omega\rightarrow +0$, the widths become larger, and Eq.~(\ref{restriction}) 
is no longer satisfied. This can also be seen in Fig.~\ref{appelfig}.
As the ratio $\alpha_0/\alpha_c$ increases the width increases too,
and the position of the peak becomes more difficult to define.
\section{Again: the conformal phase transition}\label{again_sec_cpt}
In this section we analyze the scalar composites near the critical
gauge coupling $\alpha_0=\alpha_c$, with the purpose of investigating the
conformal phase transition, which was introduced in section~\ref{sec_cpt}.
We recall that the main feature of the CPT is an abrupt change of the spectrum 
of light excitations (composites) as the critical point is crossed, though the
phase transition itself is continuous.

In the previous section we encountered a no-CPT, $\sigma$-model-like
phase transition for values of $\alpha_0<\alpha_c$.
The masses of light excitations\footnote{
The scalar composites, {\em i.e.}, the $\pi$ and $\sigma$ bosons are
light resonances, $|m_\pi|=|m_\sigma|\ll \Lambda$.} 
are continuous functions across the critical
curve (though non-analytic); 
there is no abrupt change in the spectrum of light excitations.
In the broken phase the $\pi$ boson becomes a massless
Nambu-Goldstone boson, while the fermion and $\sigma$ boson acquire
a dynamical mass which is small with respect to the cutoff $\Lambda$ near 
criticality.

The pole at the critical gauge coupling $\alpha_0=\alpha_c$ is determined by 
Eq.~(\ref{alpcrit}), which is rather complicated but 
we only need to study the IR limit, so we assume that the pole is small 
$p_0^2\ll \Lambda^2$.
The infrared limit obtained from Eq.~(\ref{alpcrit}) is
\bea
\Pi_{\rm S}(q^2)\approx \frac{\Lambda^2}{4\pi^2}\left[
4+\frac{16}{\ln (q^2/\Lambda^2)-\epsilon_3}+{\cal O}\left(
q^2/\Lambda^2\ln (q^2/\Lambda^2)\right)\right].
\eea
\noindent
We then find zeros of $\Delta_S^{-1}(p)$ at
\bea
0\approx\left(\frac{1}{g_0}-4\right)
+\frac{16(\ln (\Lambda^2/p_0^2)+\epsilon_3)}{
\left(\ln (\Lambda^2/p_0^2)+\epsilon_3\right)^2+(\theta+\pi)^2},
\label{real_ac}
\eea
\bea
0\approx \theta+\pi.\label{imag_ac}
\eea
Thus $\theta\approx -\pi$, and from Eq.~(\ref{real_ac}) it is clear that 
if $g_0\leq g_c=1/4$ both terms on the
right-hand side are positive, and there is no solution for the pole with
$p_0/\Lambda \ll 1$. 
Hence if there is a pole it will be heavy, {\em i.e.}, $p_0/\Lambda={\cal O}(1)$.
Therefore at $\alpha_0=\alpha_c$ no light resonances are present in 
the spectrum for $g_0\leq g_c=1/4$.
The imaginary phase $\theta$ approaches $-\pi$ which means the heavy
pole occurs at ``Euclidean'' momentum, a sign of tachyonic states.

The statement above can be made more explicit.
If we analytically continue the scalar propagator to the values
of $\alpha_0>\alpha_c$, then we end up in the ``wrong vacuum''
and we should get tachynonic states. In the broken phase
($\alpha_0>\alpha_c$), a chiral symmetric solution still exists, but
it is unstable. The $\pi$ and $\sigma$ bosons are tachyons for 
such a solution.
The unstable symmetric solution is obtained by analytic continuation
of the solution in the symmetric phase across the critical curve
(at $\alpha_c$).
The scaling law is determined by the UV properties of the theory and therefore
the scaling law of the tachyonic masses is the same as that
of the fermion and $\sigma$-boson mass in the broken phase.

Tachyons are described by an imaginary mass $m^2<0$.
This means the scalar propagator must have a real pole for Euclidean
momentum. If the pole $p_0$ is small, $p_0\ll \Lambda$,
we analytically continue Eq.~(\ref{vacasym}) to $\alpha_0>\alpha_c$
by replacing $\omega$ by $i\nu$, 
\bea
\omega\rightarrow i\nu\equiv i\sqrt{4\lambda_0-1}.
\eea
We then obtain
\bea
\Pi_{\rm S}(q^2)=
\frac{\Lambda^2}{\pi^2}\frac{1-\nu^2-2\nu\cot y}{(1+\nu^2)^2},
\qquad y=\frac{\nu}{2}\ln\left(\frac{\Lambda^2}{q^2}\right)+\nu\phi(\nu^2),
\eea
where
\bea
\phi(\nu^2)&\equiv&\frac{1}{2i\nu}\ln\frac{h(i\nu)(1+i\nu)}{h(-i\nu)(1-i\nu)}
-\ln \sqrt{2\lambda_0}.
\eea
The tachyonic pole is then given by the root of the equation
\bea
-\frac{\Lambda^2}{4\pi^2g_0}+\Pi_{\rm S}(p_0^2)=0,
\eea
which gives
\bea
\frac{p_0}{\Lambda}=
\exp\left(-\frac{n\pi}{\nu}-\frac{\beta}{\nu}+\phi(\nu^2)\right),
\eea
where $n$ is a positive integer and
\bea
 \beta=\tan^{-1}\frac{\nu g_0}{g_0-2\lambda_0(g_0+\lambda_0)}.
\eea
The tachyon with largest $p_0$ in the physical region 
$p_0< \Lambda$ corresponds to $n=1$.
If we now consider the limit  $\nu\to0\quad (\lambda_0\to 1/4)$, we get
\bea
\beta &\approx&
\frac{2\nu g_0}{g_0- 1/4},\quad
\phi(\nu^2)\approx 1-\gamma-\frac{1}{2}\ln 2-\frac{\epsilon_2}{\epsilon_1}
+{\cal O}(\nu^2).
\eea
In this case,
\bea
\frac{p_0}{\Lambda}\approx 
\exp\left(\frac{2g_0}{1/4-g_0}+\phi(0)\right)
\exp\left(-\frac{\pi}{\sqrt{4\lambda_0-1}}\right),
\label{tachpole}
\eea
which is proportional to the well-know scaling law of quenched QED.
Thus the scalar propagator giving the
tachyon pole equation, Eq.~(\ref{tachpole}), reproduces the scaling law
with essential singularity (compare with Eq.~(\ref{essent_sing})), 
which is another confirmation of the CPT.
\section{Renormalization and the scaling hypothesis}
In what follows, we discuss the results from the viewpoint of 
the renormalizability of the GNJL model.
The renormalization is performed by a suitable redefinition of the 
composite or auxiliary fields $\sigma$ and $\pi$ (see Eq.~(\ref{Zsigmapidef})):
\bea
\sigma_R\equiv \left[Z_\sigma^{(\mu)}\right]^{-1}\sigma,\qquad
\pi_R\equiv\left[Z_\pi^{(\mu)}\right]^{-1}\pi,\label{ren_aux}
\eea
where $\mu$ is related to some physical scale, which should be independent 
of $\Lambda$.
The renormalization factors of the scalar and pseudoscalar fields
$Z_\sigma^{(\mu)}$, respectively, $Z_\pi^{(\mu)}$ can be chosen to coincide
both in the symmetric and broken phase,
and the renormalized auxiliary fields $\sigma_R$ and $\pi_R $ define
the renormalized scalar propagator
\bea
\Delta_{\rm S}^{R}(q)=\left[Z_\sigma^{(\mu)}\right]^{-2} \Delta_{\rm S}(q)
\eea
and the renormalized scalar vertex
\bea
\Gamma_{\rm S}^R(p+q,p)= Z_\sigma^{(\mu)} {\Gamma_{\rm S}}(p+q,p).
\eea
In order to renormalize the scalar propagator and Yukawa vertex 
simultaneously (see also Eqs.~(\ref{vacasym2}) and (\ref{fuv_asym})), 
the wave function renormalization factor at some renormalization 
scale $\mu$ should be of the form
\bea
Z_\sigma^{(\mu)}\propto 
\left(\frac{\mu^2}{\Lambda^2}\right)^{(1-\omega)/2}.
\label{Zscal}
\eea
Freedom in the choice of renormalization scheme allows us to
take the factor $Z$ defined in Eq.~(\ref{Zdef}) as the wave function 
renormalization factor, since
\bea 
Z_\sigma^{(\mu)}=Z(\mu^2/\Lambda^2,\omega)\approx 
\frac{\pi}{2\sin \omega\pi}\frac{\gamma(\omega)}{2}
\frac{(1+\omega)}{\Gamma(1-\omega)}
\left(\frac{\lambda_0}{2}\right)^{-\omega/2}
\left(\frac{\mu^2}{\Lambda^2}\right)^{(1-\omega)/2}. \label{Zasym}
\eea
Hence it follows that four-fermion scattering amplitudes, for
instance one-scalar exchange amplitudes are
renormalization group (RG) invariant, {\em i.e.}
\bea
\Gamma_{\rm S}^R(p_1+q,p_1) \Delta_{\rm S}^{R}(q) 
\Gamma_{\rm S}^R(p_2,p_2+q)\sim
{\Gamma_{\rm S}}(p_1+q,p_1)\Delta_{\rm S}(q){\Gamma_{\rm S}}(p_2,p_2+q),
\label{RGscat}
\eea
({\em e.g.} see again \cite{aptewij91}).
Consider the case where $p_1^2,\,p_2^2 \gg q^2$,
so that the scalar vertices are described by the UV channels,
and suppose that we are sufficiently close to the critical line
(the fine-tuning condition Eq.~(\ref{finetune2}))
\bea
-\frac{\Delta g_0}{g_0 g_c}
\equiv\frac{1}{g_0}-\frac{1}{g_c} \ll \left(\frac{q^2}{\Lambda^2}\right)^\omega
\Longrightarrow -\frac{\Delta g_0}{g_0 g_c}
\sim \left(\frac{\mu^2}{\Lambda^2}\right)^\omega,
\qquad \mu^2\ll q^2,
\eea
where $\Delta g_0\equiv g_0-g_c$,
then from Eqs.~(\ref{delsinv}) and (\ref{vacasym2}) the scalar propagator has the 
asymptotic behavior
\bea
\Delta_{\rm S}(q)\approx -\frac{4\pi^2}{\Lambda^2}\frac{1}{B(\omega)}
\left(\frac{\Lambda^2}{q^2}\right)^\omega. \label{asym_scalprop}
\eea
Such a specific power-law behavior for the scalar propagator 
is essential for the renormalizability of the GNJL model as is shown
in Refs.~\cite{gumi91,aptewij91,gumi92,mi93,kotaya93}.

Thus, from Eq.~(\ref{Zscal}) and Eq.~(\ref{fuv_asym}), we get
\bea
\Delta_{\rm S}^{R}(q)&\propto&
-\frac{1}{\mu^2}\left(\frac{\mu^2}{q^2}\right)^\omega,\\
F_{\rm UV}^R\left(p^2,q^2\right)&\propto&
\frac{2}{(1+\omega)}\left(\frac{p^2}{\mu^2}\right)^{-(1-\omega)/2 }\nonu\\
&\times&
\left[1-2\left(\frac{q^2}{p^2}\right)^\omega
\frac{\gamma(-\omega)}{\gamma(\omega)}
\frac{\Gamma(1-\omega)}{\Gamma(2+\omega)}
\left(\frac{\lambda_0}{2}\right)^\omega \right].
\eea
With these expressions, it is straightforward to check that
Eq.~(\ref{RGscat}) is indeed independent of $\Lambda$ and $\mu$.
Hence the renormalization of the auxiliary fields 
$\sigma$ and $\pi$, Eq.~(\ref{ren_aux}), simultaneously renormalizes 
the Yukawa vertex and the scalar propagator.

The above considerations can be made more explicit by writing
the scalar propagator according to the scaling hypothesis Eq.~(\ref{scalhypprop})
(in Euclidean formulation).
The scaling hypothesis states that the propagator should have the following form:
\bea
\Delta_{\rm S}(q)=\frac{1}{\Lambda^2}\(\frac{\Lambda^2}{q^2}\)^{1-\eta/2} 
{\cal F}_\Delta (|m_\sigma|^2/q^2),\label{scalhypscalprop}
\eea
where $\eta$ is by definition the anomalous dimension.
We write $|m_\sigma|=p_0$, with $p_0$ given by Eq.~(\ref{massinter}), 
where $|m_\sigma|$ is the analog of the inverse correlation 
length $|m_\sigma|=1/\xi$, and $\xi\sim (\Delta g_0)^{-\nu}$.
Then
\bea
\Delta_{\rm S}(q)=-\frac{4\pi^2}{\Lambda^2}\frac{1}{B(\omega)}
\(\frac{\Lambda^2}{q^2}\)^{\omega}
\[1+\(\frac{|m_\sigma|^2}{q^2}\)^\omega-\frac{A(\omega)}{B(\omega)}\(\frac{q^2}{\Lambda^2}\)^{\sigma_1}+\dots \]^{-1}
\eea
where the exponent $\sigma_1 > 0$ for $0< \omega<1$.
From this we read of that the anomalous dimension $\eta$ is given by
\bea
2\omega=2-\eta \Longrightarrow  \eta=2(1-\omega),\label{etaexpr}
\eea
and that the scaling function is (provided $|m_\sigma|^2, q^2 \ll \Lambda^2$)
\bea
{\cal F}_\Delta (x)\approx -\frac{4\pi^2}{B(\omega)}\frac{1}{1+x^\omega}.
\eea
Thus indeed if $q^2\gg |m_\sigma|^2$
we obtain Eq.~(\ref{asym_scalprop}), and when $q^2\ll |m_\sigma|^2$
we get 
\bea
\Delta_{\rm S}(0)=\frac{4\pi^2}{\Lambda^2} \frac{g_0 g_c}{\Delta g_0}
\sim (\Delta g_0)^{-\gamma}\sim
\chi=\frac{\partial {\cal M}}{\partial m_0}\biggr|_{m_0=0},\label{pcacscal}
\eea
supporting the fact that the critical exponent $\gamma=1$.
Within the quenched-ladder approximation,
Eq.~(\ref{pcacscal}) agrees with the PCAC relation (\ref{masspiwti}) 
and the analyticity condition (\ref{masssiganal}). 
The PCAC relation for the pseudoscalar propagator is
\bea
-\frac{\langle \sigma \rangle}{m_0}=\frac{\Delta_{\rm P}(0)}{G_0}.
\eea
Using the EOS Eqs.~(\ref{mu_0bc1}) and (\ref{condbc1}) and the fact that
in the symmetric phase (in the chiral limit)
\bea
-\Delta g_0 \gg \(\Sigma_0/\Lambda\)^{2\omega},
\eea
see Eq.~(\ref{addfinetune2}),
we get that
\bea
\frac{\Delta_{\rm P}(0)}{G_0}=\frac{g_c}{\Delta g_0}.
\eea
Same can be done for $\Delta_{\rm S}(0)$ (of course 
$\DelS=\DelP$ in the symmetric phase), thus
\bea
\frac{\Delta_{\rm S}(0)}{G_0}=\frac{\Delta_{\rm P}(0)}{G_0}
=\frac{g_c}{\Delta g_0}.
\eea
Hence, 
with $G_0=4\pi^2 g_0/\Lambda^2$, Eq.~(\ref{pcacscal}) is obtained.

Since the absolute value of the resonance {\em i.e.}, $|m_\sigma|$
is now the new scale (besides the UV cutoff) 
generated close to the critical curve, it is natural to take it to be 
a physical scale, which by definition should be independent of $\Lambda$.
Using Eq.~(\ref{massinter}), we get the fine-tuning condition
\bea
0=\Lambda \frac{{\rm d} |m_\sigma|}{{\rm d}\Lambda}\sim
\Lambda \frac{{\rm d}}{{\rm d}\Lambda}
\[ \Lambda \(\frac{-\Delta g_0}{g_0}\)^{1/2\omega}\] 
\,\Longrightarrow \, \beta_g(g_0)=-2\omega\frac{g_0}{g_c}(g_0-g_c),
\eea
which is equivalent to the fine-tuning condition given in \cite{kotaya93}, 
and Eq.~(\ref{finetune2}).\footnote{Again it is assumed 
that $\beta_\alpha(\alpha_0)\approx 0$.}
Hence the critical curve $g_0=g_c$ is an UV fixed point $\beta_g(g_c)=0$
($\beta^\prime_g(g_c)< 0$) 
of the renormalization group flow.
\section{Summary of results}\label{summary}
In this chapter we studied the scalar composites near criticality
in the GNJL model.
We obtained an analytic expression for the scalar propagator
describing the composite states which is valid along the entire
critical curve of the GNJL model.
We presented a method for solving the Yukawa vertex in the GNJL in the
quenched-ladder approximation. The crucial assumption was that such a vertex
depends only weakly on the angle between $\sigma$-boson momentum and
fermion momentum.
The method presented here incorporated the infrared boundary condition
in a more natural way than previous attempts in this direction.
Also the observation that derivatives of the Yukawa vertex
are singular at zero $\sigma$-boson momentum transfer is a warning
that derivative expansions and Taylor series could fail.
Moreover this is reflected by the property of the scalar composites
having noninteger power-law behavior, which means that, although these states 
are tightly bound, they are not pointlike.

The conclusion of the comparison of the method presented here
and work done previously on
the $\sigma$-boson propagator is the following.
Qualitatively the results obtained by Appelquist {\em et al.}, Kondo {\em et al.}
and by our approximation are in agreement. Both methods yield
a renormalizable $\sigma$-boson propagator and Yukawa vertex
near criticality and find light resonances for gauge coupling
$0<\alpha_0<\alpha_c$.
Quantitatively, at a fixed value of the gauge coupling,
the scalar composites computed in our case
are slightly heavier with comparable width, as is illustrated in 
Fig.~\ref{appelfig}.

In addition, the scalar composites propagator were examined
for values of the coupling near the critical gauge coupling $\alpha_c$.
Near the critical line $\alpha_0=\alpha_c,g_0<{1\over 4}$ the conformal
phase transition is encountered and the spectrum of light excitations 
(resonances) in the symmetric phase disappears.
Moreover the well-known scaling law with essential singularity,
which is characteristic for the CPT, was recovered by analytic continuation
of the $\sigma$-boson propagator across the critical curve at 
$\alpha_0=\alpha_c$.

%
\chapter{
On the existence of ultra-violet fixed points}\label{chap5}
\section{Beyond the quenched approximation}
The general consensus on $U(1)$ gauge theories such as the GNJL model 
is that they are trivial.
Once fermion-loops are present the electromagnetic charge 
will be screened completely in the continuum limit ($\Lambda\rightarrow \infty$).
But we argue that (pseudo)scalar degrees of freedom play an
important role, and in case of the GNJL model, provided the number of
fermion flavors $N$ exceeds some critical value, ultra-violet (UV) 
fixed points can be realized.
The existence of UV fixed points gives rise to a 
non-trivial theory. 

The crucial mechanism responsible for
the realization of such a nontrivial root
of the $\beta$ function of the gauge coupling is given by the formation 
of tight scalar and pseudoscalar 
resonances in the symmetric phase or bound states in the broken phase
near the critical line.  
In other words, 
the formation, due to the strong coupling chiral symmetry breaking dynamics,
of neutral scalars and pseudoscalar composites 
reduces the screening of charge.

As a perturbation theory, QED has been proven to be a successful gauge
field theory, tested at extremely high precision.
However, in the language of the renormalization group, 
perturbative QED appears to 
be a trivial theory in four dimensions (see Ref.~\cite{lapo55}).
In the continuum limit, taking the cutoff $\Lambda$ to infinity, the effective or 
running gauge coupling vanishes, and the theory becomes a free field 
({\em i.e.} trivial) theory. 

In the renormalized theory, triviality represents itself in the
form of the so-called Landau ghost, which is a singularity
in the RG transformation for the gauge coupling $\alpha_0$
at $\Lambda=\Lambda_{LG}$, see Refs.~\cite{lapo55,fr55,laabha56}.
The renormalized theory should then be considered as an effective theory, 
which is not defined for distances shorter than $1/\Lambda_{LG}$,
see Ref.~\cite{itzzub}.
In QED the Landau ghost $\Lambda_{LG}$ lies far beyond the Planck scale, 
and is not of practical importance.

The triviality of QED originates from the screening of charged particles
by their interactions with virtual fermion anti-fermion pairs from
the vacuum. 
Such charge screening is described by the vacuum polarization $\Pi$, 
Eq.~(\ref{vacpoldef}), and the
implications of the one-loop leading contribution were discussed in
Refs.~\cite{laabha54a,laabha54b,laabha54c}: 
\bea
\Pi(q^2)=\frac{N\alpha_0}{3\pi} \log\frac{\Lambda^2}{q^2},
\eea
where $N$ is the number of fermion flavors.
The QED vacuum is not a perfect insulator and can be considered as a medium 
of dipoles, where the dipoles represent the virtual fermion anti-fermion pairs 
({\em i.e.} the fermion loops in the vacuum polarization).
This logarithmic screening effect is sufficient to 
cause the complete screening of charge in the continuum limit.

The charge screening and charge renormalization give rise to a 
running gauge coupling, {\em i.e.} $\alpha_0$ depends on the cut-off $\Lambda$,
and the roots $\alpha_0$ of the $\beta$ function, 
\bea
\beta_\alpha(\alpha_0)\equiv \Lambda \frac{{\rm d}\alpha_0}{{\rm d} \Lambda}=0
\label{betaadef},
\eea
are essential in the renormalization group analysis. 
Such a root is either a Gaussian fixed point corresponding to
a trivial theory or an UV fixed point ({\em i.e} critical fixed point)
corresponding 
to a nontrivial continuum limit.
In case of perturbative QED, the only zero of $\beta_\alpha$ 
is the origin $\alpha_0=0$, 
the Gaussian fixed point.
In case of asymptotic free theories such as QCD the origin is an UV fixed point.
A nontrivial (interacting) theory arises whenever $\beta_\alpha$ has a root which 
is an UV fixed point.
The first formulation of the renormalization group for the gauge coupling 
$\alpha_0$ of QED and speculations about the existence of a fixed point was 
given in Ref.~\cite{gemalo54}.

Although the conclusions concerning QED are usually drawn from
the one-loop results, the two-loop computations of Ref.~\cite{jolu50}
and the three-loop (quenched) computations of Refs.~\cite{ro66,ro67}
are consistent with the view that QED is trivial.
In this context we should mention that the effort of Johnson, Baker, 
Willey, and Adler
\cite{jowiba67,bajo69,ad72,joba73} to find a finite nontrivial continuum limit 
of QED was inspired by the simplicity and the negative sign of the 
three-loop result of Rosner.
No realization of the scenario presented by Johnson {\em et al.} has been found,
mainly because they were looking for a perturbative resolution.
The problem is that for a possible nontrivial continuum limit of QED,
critical and hence nonperturbative dynamics is required.

Therefore, the discovery of dynamical chiral symmetry breaking in 
the strong coupling phase of QED and the existence of an UV
fixed point in the quenched-ladder approximation sheds new light on 
the nonperturbative nature of QED.       
The fundamental notion is that composite degrees of freedom play an important role
in resolving the scaling behavior of the theory near criticality.
The GNJL model
is the generalization of strong coupling QED analysis taking into account
the composite degrees of freedom.

An interesting attempt to incorporate the dynamics of chiral symmetry breaking 
({\em e.g.} the formation of bound states) was made in Ref.~\cite{kodako88b}. 
Koci\'c {\em et al.} argued that magnetic effects described by parts of 
the transverse vertex (the vertex function $\tau_5$) could prevent the total 
screening of charge. 
However, we think that the analysis of Koci\'c {\em et al.} lacks an 
inhomogeneous term in the equation for the transverse vertex function.
It is precisely the inhomogeneous term that introduces boundary conditions 
which eliminate the nontrivial solution for the transverse vertex function 
obtained by Koci\'c {\em et al.}, henceforth, their scenario is not suitable 
to resolve the triviality problem.

Magnetic effects described by the transverse vertex can only be proportional 
to the fermion dynamical mass; singular solutions of the type proposed by
Koci\'c {\em et al.} are eliminated by ultraviolet boundary conditions.
Since the fermion dynamical mass vanishes close to criticality 
(large correlation length), it is unlikely that magnetic effects
are important for the study of charge screening.

We speculate therefore that some other mechanism is responsible for the possible
realization of an UV fixed point, and the absence of charge screening,
in the unquenched GNJL model.
Formally irrelevant four-fermion interactions become relevant
near the critical line of the chiral phase transition,
\bea
g_c(\omega)=(1+\omega)^2/4,\qquad \omega=\sqrt{1-\alpha_0/\alpha_c},\label{critcurv2}
\eea
and the dynamics of the 
composite states described by the four-fermion interactions might play a crucial 
role in the phenomenon of charge screening.

Dynamical symmetry breaking is an inherently nonperturbative phenomenon,
and, due to the lack of other solvable approximations,
has been predominantly studied in the framework of SDEs in the
quenched-ladder approximation.
As was mentioned in chapter~\ref{chap3}, 
the physical implications and the consistency of the quenched-ladder results 
with many quenched lattice simulations, and with the nonperturbative RG 
techniques, support the view that the qualitative features
of the approach might be realistic and describe properties of the full theory.
Moreover the ladder approach respects the vector and axial Ward--Takahashi 
identities.

The quenched approximation is analogous to the assumption that the full 
gauge boson propagator 
$D_{\mu\nu}(k)$ can be approximated by the bare or canonical propagator (for large momenta),
\bea
D_{\mu\nu}(k)=\(-g_{\mu\nu}+\frac{k_\mu k_\nu}{k^2}\)\frac{1}{k^2},\label{bargb}
\eea
in the Landau gauge. 
The quenched approximation is only consistent when the vacuum polarization 
is finite in the continuum limit, {\em i.e.} 
the logarithmic running of the coupling is absent.
This is the case at an UV fixed point of the $\beta$ function Eq.~(\ref{betaadef}) 
of $\alpha_0$.
The assumption that such a critical fixed point exists, and that it lies somewhere on the critical
curve (\ref{critcurv2}) is the starting point for many studies of dynamical chiral symmetry 
breaking in context of the GNJL model. 
In fact the quenched hypothesis is only consistent when the bare gauge coupling 
$\alpha_0$ is the UV fixed point of the theory, $\beta_\alpha(\alpha_0)=0$.

Attempts to include a logarithmic running of the coupling drastically 
changes the chiral phase transition and the critical line, 
see Refs.~\cite{ra91,gu90,ko90,go90}. 
It was shown in Refs.~\cite{ko90,ko91,gu91,gosa91} that $\gamma_m=2$, 
which is analogous to the vanishing of the anomalous dimension
of the scalar propagator, $\eta=2(2-\gamma_m)=0$ ({\em e.g.} 
see Eq.~(\ref{scalhypscalprop})).
Consequently, the critical exponents turn out to be of the mean field type. 
The quenched theory and the theory with the logarithmically 
running coupling seem to be two different theories.

It is essential to incorporate the composite degrees of freedom and 
their effective interactions 
to properly understand the scaling structure of the full unquenched GNJL model.
In case of the GNJL model, the four-fermion interactions are treated
in a mean field approach known as the Hartree--Fock approximation, see also
section~\ref{sec_thequenched}.
We recall that by mean field approach, we refer to approximations which 
neglect quantum corrections corresponding to four-fermion interactions beyond 
tree level.

The merit of the mean field approximation is the decoupling of SDEs.
Clearly it is necessary to check the consistency between the initial
assumptions and the final result.
As long as four-fermion interactions are irrelevant the mean-field
approach for these operators is justified.
However the conclusions of Refs.\cite{balomi90,kohakoda90}
are that the quenched-ladder approximations with four-fermion interactions 
in the mean field 
approach satisfy the hyperscaling equations for the critical exponents, 
suggesting that the four-fermion operators become relevant due to 
the appearance of large anomalous dimensions.
In other words, the mean field approach
yields nonmean-field exponents, thereby being inconsistent
({\em e.g.} see Refs.\cite{goldenfeld,azcagagrlapi95b}).

The fluctuations of the gauge field for the composite degrees of freedom
are essential and cannot be neglected; they give rise to a large 
anomalous dimension turning irrelevant four-fermion interaction into relevant ones 
implying that such interactions give nontrivial quantum corrections 
(they cannot be neglected beyond tree level.)
This is supported by the computations
of the anomalous scaling laws for the Yukawa vertices and 
the propagators of the scalar and pseudoscalar composite states, {\em i.e.} 
the $\sigma$ and $\pi$ bosons of chapter~\ref{chap4}.
In fact, as was discussed in section~\ref{sec_critexp_scal},
the hyperscaling relations imply the existence of a nontrivial
Yukawa interaction describing the interaction
between fermions and $\sigma$ and $\pi$ composites
in the GNJL model. 

The question is; how to modify the mean field approach? 
An appropriate answer depends on the following three concepts:
skeleton expansions, the $1/N$ expansion (with $N$ the number of fermion flavors),
and the specific form of the chiral symmetry.
Clearly going beyond the quenched approximation seems only interesting
once four-fermion interactions are treated beyond the mean field approach.
\section{Beyond the mean field approach}
Let us first review the concept of the so-called skeleton expansions.
The anomalous dimensions follow from the scaling structure of the 
fully dressed Yukawa vertices and $\sigma$ and $\pi$ propagators, 
hence it is necessary to
incorporate the full Green functions corresponding to the four-fermionic 
composite degrees of freedom
(or at least the leading or asymptotic parts of these functions). 
Therefore it is crucial to study such issues in a renormalization group 
(RG) invariant manner.
This can be done on the level of the Bethe-Salpeter fermion--anti-fermion 
scattering kernels in the skeleton expansion 
(see {\it e.g.} Bjorken and Drell \cite{bjodre}).

Analogous to the pure QED kernels we define the 
one-boson irreducible kernel $K^{(1)}$, and the two-fermion one-boson 
irreducible kernel
$K^{(2)}$, where these kernels now also include the $\sigma$ and 
$\pi$ composites.
For both type of kernels a skeleton expansion 
exists, and the integral equation between $K^{(1)}$ 
and $K^{(2)}$ is known as the Bethe-Salpeter equation.
The skeleton expansion is a series in topologically  distinct Feynman diagrams
with all vertices and propagators fully dressed.
Each term in the skeleton expansion of the BS kernel is RG invariant, 
up to fermion wave function factors, {\em i.e.} the expansion is independent of
the renormalization factors $Z_3$
and $Z=Z_\sigma=Z_\pi$ (see Eqs.~(\ref{Z1Z2Z3def}) and (\ref{Zsigmapidef}))
of, respectively, the gauge field and the composite fields $\sigma$ and $\pi$.
The two $Z^{-1}$ factors with anomalous dimensions of each Yukawa 
vertex cancel with the $Z^2$ factors of the $\sigma$ and $\pi$ propagators.
Since in the quenched approximation the renormalization constant of the gauge 
field is one ($Z_3=1$),
the quenched-ladder approximation 
is consistent from the RG point of view.

The GNJL model, with $N$ number of fermion flavors, is taken 
to be invariant under global $U_L(N)\times U_R(N)$ chiral transformations,
so that both the scalar and pseudoscalar four-fermion interactions are in 
the adjoint representation, and, consequently, 
the number of scalar composites ($N^2$) equals the number of pseudoscalar 
composites ($N^2$). 
In this way, when $N$ is large we can use the $1/N$ expansion introduced 
by 't Hooft \cite{tho74}.
The $1/N$ expansion states that planar, {\em i.e.} ladder,
graphs describe the dominant contribution to Green functions.
The $1/N$ expansion will be discussed in the next section.

With the gauge interaction treated in the ladder approximation
each full Yukawa vertex and $\sigma$, $\pi$ boson propagator can be written as
\bea
\Gamma^\alpha_{{\rm S}^{ab}_{ij}}(k,p)=\tau^\alpha_{ij} \Gamma_{{\rm S}ab}(k,p),
\qquad \Delta_{\rm S}^{(\alpha)}(q)=\DelS(q).
\eea
So that
\bea
&&
\sum_{\alpha=0}^{N^2-1}
\Gamma^\alpha_{{\rm S}^{ab}_{ij}}(k+q,k) \Delta_{\rm S}^{(\alpha)}(q)
\Gamma^\alpha_{{\rm S}^{cd}_{kl}}(p,p+q)\nonu\\
&&= \delta_{il} \delta_{kj}
\Gamma_{{\rm S}ab}(k+q,k) \DelS(q)\Gamma_{{\rm S}cd}(p,p+q),\label{fierzap1}
\eea
where we have used the Fierz identity
\bea
\sum_{\alpha=0}^{N^2-1}\tau^\alpha_{ij} \tau^\alpha_{kl}=\delta_{il}\delta_{kj},
\label{fierz2}
\eea
with $\tau$ the generators of the $U(N)$ symmetry.
Then the first term of the skeleton expansion for $K^{(2)}$ is the following 
single boson exchange term:
\bea
&&(-ie_0^2)K^{(2)}_{^{ab,cd}_{i_1j_1,i_2j_2}}(k,p,p+q)
=\nonu\\
&&\delta_{i_2j_1}\delta_{i_1j_2}
(-ie_0)\Gamma^\lambda_{cb}(p+q,p)iD_{\lambda\sigma}(q)
(-ie_0)\Gamma^\sigma_{ad}(k,k+q)\nonu\\
&&+
\delta_{i_1j_1}\delta_{i_2j_2}
(-i)\Gamma_{{\rm S}cb}(p+q,p)i\DelS(q)
(-i)\Gamma_{{\rm S}ad}(k,k+q)\nonu\\
&&+
\delta_{i_1j_1}\delta_{i_2j_2}
(-i)\Gamma_{{\rm P}cb}(p+q,p)
i\DelP(q)
(-i)\Gamma_{{\rm P}ad}(k,k+q),\label{BSapprox}
\eea 
where the labels $i$s and $j$s are flavor labels.

As a result of the chiral symmetry the contributions of
four-fermion interactions, which are represented by $\sigma$ and $\pi$ exchanges, 
exhibit two distinct features depending on whether they are incorporated 
in SDEs describing quantities connected with so-called zero-spin 
structures\footnote{Such structures are characterized by spinor matrices which commute with the $\g5$ matrix.} 
({\em e.g.} the dynamical mass $\Sigma$, the Yukawa vertices 
$\Gamma_{\rm S}$, $\Gamma_{\rm P}$, and the $\sigma$ and $\pi$ propagators 
$\DelS$, $\DelP$),
or whether the exchanges are included in
SDEs describing nonzero-spin structures (anti-commuting with $\g5$)
({\em e.g.} the vacuum polarization $\Pi$, 
the photon-fermion vertex $\Gamma^\mu$, and the fermion wave function ${\cal Z}$).
Henceforth, we refer to (non)zero-spin functions, and their equations as 
(non)zero-spin channels. 

The chiral symmetry gives rise to the following properties:
\begin{enumerate}
\item{
In spin-zero-channels, the contribution of 
planar diagrams ({\em i.e.} planar in $\sigma$ and $\pi$ exchanges)  
vanishes due to the fact
that the exchange of a $\sigma$ has an opposite sign with respect 
to a $\pi$ exchange.
Why?
Let us consider a planar contribution  
to the scalar vacuum polarization which contains (amongst others)
a $\pi$ exchange.
Both $\g5$ matrices corresponding to this particular planar $\pi$
exchange can be eliminated from the fermion trace of the scalar vacuum 
polarization by moving them to right-hand side of the trace. 
For planar diagrams such a process involves the interchange of
the $\g5$ matrix with an even number of fermion propagators, and
an arbitrary number of Yukawa vertices.
Since the Yukawa vertices commute with the $\g5$ matrix,
and $\g5$ anti-commutes with the 
fermion propagator\footnote{In the symmetric phase $\g5 S=-S\g5$.} $S$,
the process of moving the $\g5$ to the right does not introduce an overall 
minus sign.
Now using that 
$(i\g5)(i\g5)=-(\bf 1)(\bf 1)$, we see that such a specific $\pi$ exchange
is identical to minus the same diagram with the $\pi$ exchange replaced by 
a $\sigma$ exchange.
Since each diagram containing a $\pi$ exchange has a scalar counter part
({\em i.e.} an analogous diagram with a $\sigma$ 
instead of a $\pi$ exchange), the sum of all planar diagrams, 
with a particular number of exchanges, vanishes.
Thus in channels with spin-zero matrix structures ({\em i.e.}
channels which correspond to vertices which commute with $\g5$)
scalar forces appear to be opposite to pseudoscalar forces.}
\item{
In nonzero-spin channels (think of $\Pi$, $\Gamma^\mu$, etc.)
containing vertices which anti-commute with the $\g5$ matrix, 
the situation is different: planar $\sigma$ and $\pi$ exchanges contribute
with identical sign.
Let us now consider a planar contribution  
to the (photon) vacuum polarization containing 
a $\pi$ exchange.
If we again move the $\g5$ matrices to the right-hand side of the
trace, we get an overall minus sign due to the anti-commutation
of $\g5$ with $\gamma^\mu$.
This means that any planar diagram in the vacuum polarization
containing a $\pi$ exchange is identical to the same diagram with 
the $\pi$ exchange replaced by a $\sigma$ exchange.
Hence, in channels connected with spin structures 
anti-commuting with $\g5$, 
$\sigma$ and $\pi$ exchanges behave identical.
In other words, in such channels, pseudoscalar and scalar forces 
act in the same direction (instead of canceling). 
}
\end{enumerate}
The properties described above are, strictly speaking, only valid in 
the symmetric (massless) phase, where 
the $\sigma$ and $\pi$ bosons are degenerate.
However, in the broken phase, the properties are valid whenever
momenta larger than the dynamical mass $\Sigma$ or inverse correlation
length $\xi^{-1}\sim m_\sigma$ are considered, because then the
degeneracy emerges too.

These properties also provide us with a general argument
why the mean field approach for four-fermion operators
for Green functions corresponding to spin-zero channels 
({\em e.g.} $\GS$ and $\DelS$ of chapter~\ref{chap4})
is reliable. For such channels planar contributions vanish
and the next non-vanishing contributions (such as contributions
containing crossed $\sigma$ and $\pi$ exchanges)
are proportional to $1/N$, thus small for large $N$.
This suggests that quantities such as the critical curve, dynamical mass, 
anomalous dimensions etc., are nearly independent of $N$, and are described 
rather well by the mean field approach.

However, such a cancellation of scalars against pseudoscalars degrees of freedom
does not occur in nonzero-spin channels, such as the vacuum polarization $\Pi$.
In the vacuum polarization it turns out that $\sigma$ and $\pi$ exchanges 
are attractive and virtual fermion anti-fermion pairs from vacuum fluctuations are 
constrained by these forces, reducing their capability to screen.

\section{The $1/N$ expansion}\label{sec1/N}
The $1/N$ expansion of 't Hooft (see Refs.~\cite{tho74,coleman}) 
provides us with a useful nonperturbative tool to incorporate four-fermion 
interactions beyond the mean field approach.
As mentioned previously, the $1/N$ expansion states that the planar 
({\em i.e.} ladder) diagrams, with fermions at the edges, 
describe the leading or dominant contributions to Green functions.
The interesting feature of the $1/N$ expansion is that
Feynman diagrams can be classified in terms of two-dimensional surfaces
with specific topology.
Diagrams with other (than planar) topological structures are multiplied
by factors of $1/N$, and in the limit of large $N$, their contribution
can be neglected with respect to planar graphs.

In the formulation of the $1/N$ expansion of 't Hooft, 
the planar graphs coincide with ladder graphs in case of the GNJL model.
One important rule is to draw Feynman graphs with fermion loops 
forming the boundary of the graph (if possible).
In this way, vertex corrections are not necessarily classified as being 
planar, {\em e.g.} see Fig.~9 of section 3.1 of the book by Coleman \cite{coleman}.

In the context of the paper by 't Hooft which deals mainly with QCD, 
we should consider internal or virtual $\sigma$ and $\pi$ exchanges 
analogous to gluon exchanges,
with the important difference that due to the chiral symmetry we have two
types of particles both being in the adjoint representation 
($N^2$ scalars and $N^2$ pseudoscalars).

Suppose we can define an ``effective'' Yukawa coupling $g_Y$ describing
the interaction of scalars and pseudoscalar with fermions.
For the time being we leave unspecified such a coupling.  
Let us consider $N$ large with $g_Y^2 N$ is fixed and of order one
\bea
g_Y^2 N\sim{\cal O}(1).\label{gyasum1}
\eea
If the theory under consideration
is rewritten in terms of the coupling constant 
$g_N^2\equiv g_Y^2 N$, a $1/N$ expansion can be formulated straightforwardly.

Then, by keeping track of the flavor indices within a particular Feynman diagram,
we can count factors of $1/N$.
Each fermion carries a flavor index $(i)$, which runs from $1$ to $N$.
A virtual $\sigma$, $\pi$ exchange carries two flavor indices.
This follows from Eq.~(\ref{fierzap1}) and the Fierz identity (\ref{fierz2}). 
Each virtual $\sigma$, $\pi$ exchange is associated with two Yukawa vertices, 
therefore giving rise to a pair of Kronecker delta functions
connecting the flavor indices of the scattered fermions.
In the context of flavor indices,
either a $\sigma$ or $\pi$ boson can be considered as a propagating
fermion--anti-fermion pair carrying double flavor indices. 

Whenever a trace over a flavor Kronecker delta function enters
into the expression for a particular Feynman diagram,
we speak of an index loop.
An index loop is easily identified by using
the double-line representation of 't Hooft.
A fermion propagator is represented by a single index-line
({\em i.e.} fermion line), whereas each internal scalar, respectively, 
pseudoscalar propagator is represented by a double index-line. 
Consequently, whenever, after drawing a particular Feynman diagram,  
an index-line closes, it forms an index loop giving rise to a factor 
\bea
N=\Tr \delta=\sum_{i=1}^N\delta_{ii}. 
\eea
A particular Feynman diagram containing a number of virtual 
$\sigma$ and $\pi$ exchanges is associated with a factor
\bea
r=g_Y^V N^I,
\eea
where $V$ is the number of Yukawa vertices, and $I$ the number of index loops.
The factor $r$ can be written as
\bea
r=(g^2_Y N)^{V/2} N^\chi = g_N^{V} N^\chi,
\eea
where the Euler characteristic $\chi$ is given by
\bea
\chi=2-2H-B,
\eea
with $H$ the number of ``handles'', and $B$ the number of ``holes'' 
of the two dimensional surface to which the Feynman diagram can be associated.
The Euler characteristic $\chi$ is a topological invariant.
For instance,
the vacuum polarization has the topology of a sphere with a single hole 
({\em i.e.} a disk), 
where the fermion-loop forms the boundary ({\em i.e.} hole) of the graph.
Thus planar diagrams in the vacuum polarization
with $n$ exchanges of $\sigma$'s and $\pi$'s
are associated with a factor
\bea
r=N (g_Y^2 N)^n.
\eea
Other types of diagrams, {\em e.g.} diagrams with one crossed
$\sigma$, respectively, $\pi$ exchange have the topology of
a disk with a handle, hence giving rise to a factor
\bea
r=N^{-1}(g_Y^2 N)^2.
\eea
However, the sum of such type of vacuum polarization
diagrams is zero due to the chiral symmetry.

Since planar diagrams vanish in spin-zero channels such as the SDE for 
the dynamical mass, the first nonzero contribution 
is given by diagrams with crossed 
$\sigma$ and $\pi$ exchanges, and vertex corrections.
Those type of diagrams have one extra handle as compared with planar graphs.
Therefore in zero-spin channels, we should compare {\em e.g.} graphs with a single
handle with the gauge interaction in the planar (ladder) approximation.
Furthermore, each handle correspond to a factor $g_Y^4$, and since $g_Y$  
is small because of the assumption (\ref{gyasum1}),
$g_Y^4$ should be compared with the gauge coupling $\alpha_0$.
Then, if the following inequality holds
\bea
\alpha_0 \gg \lambda_Y^2/\pi,\qquad \lambda_Y=g^2_Y/4\pi,\label{ineq2}
\eea
the gauge interaction dominates over four-fermion interactions, 
and we have an argument supporting the validity of the mean field approach 
in such channels.
\section{Regularization and the fermion wave function}
The inclusion of relevant four-fermion interactions beyond
the mean field approach requires a reinvestigation of the SDE for the 
fermion wave function ${\cal Z}$. 
In QED in the quenched approximation, the fermion wave function has a gauge 
dependent anomalous dimension. In the Landau gauge, this anomalous dimension
vanishes and the fermion wave function equals one (${\cal Z}=1$). 
We conjecture that the inclusion of relevant four-fermion interactions
does not introduce an anomalous dimension for the fermion wave function other
than already introduced by the gauge interactions.
Thus, in the Landau gauge, the wave function ${\cal Z}$ is finite
though it might deviate from unity.
The argument in support of the conjecture stated above is that
only one full Yukawa vertex appears in the self-energy part, which means 
that anomalous dimensions of four-fermion interactions are not canceled.
Only two fully dressed Yukawa vertices and a fully dressed scalar composite 
are renormalization group invariant (anomalous dimensions cancel!).
 
The consequence is that a remnant power of the cutoff (related to anomalous 
dimension of a Yukawa vertex) lowers the degree of divergence of the self-energy 
part from a logarithmic divergence to a finite integral. 
The question is how to compute the finite correction to ${\cal Z}$.

The problem of the self-energy SDEs lies in regulating it.
In case of the fermion-loops appearing in vacuum polarizations
it is possible to introduce Pauli-Villars fields which will 
regulate the fermion loops.
However an additional regularization and fields are needed in order 
to regulate self-energy parts of the fermion, especially
the wave function, since the SDE is naively linearly divergent, 
and shifts in momentum integration variables are tricky. 
A momentum shift invariant regularization is by no means trivial to implement.
At the moment we do not know how to properly regularize the SDE for ${\cal Z}$, 
and how to compute a possible deviation of ${\cal Z}$ from unity in the 
Landau gauge.

We believe that the conjecture holds irrespectively of the regularization method.
Therefore throughout this chapter we assume ${\cal Z}=1$.
Eventually, a deviation from unity might change the results
presented here quantitatively, although this is not expected, since
we will use a nonperturbative method introduced by Johnson, Willey,
and Baker \cite{jowiba67}, which is independent of ${\cal Z}$.
The nice feature of the assumption that ${\cal Z}=1$ is that
with the gauge interaction treated in the quenched-ladder approach
the chiral and vector Ward--Takahashi identities are preserved, since in 
channels with spin-zero the planar $\sigma$ and $\pi$ exchanges cancel 
each other.
\section{Scalars, pseudoscalars, and charge screening}
Since, in the GNJL model, the scalars and pseudoscalars are neutral states which 
therefore do not couple to the photon field, 
their contribution to the vacuum polarization is described indirectly 
in terms of photon-fermion vertex corrections, and fermion self-energy corrections.
Hence, in order to gain some intuition for the role of scalar degrees of 
freedom on the mechanism of charge screening, 
we analyze the two-loop contribution arising from $\sigma$ and $\pi$ exchanges
to the vacuum polarization. 

Let us consider a gauge--Higgs--Yukawa type of interaction described by the 
Lagrangian
\bea
{\cal L}_{GHY}&=&-\frac{1}{4}F_{\mu\nu}F^{\mu\nu}
+\bar\psi i\gamma^\mu \partial_\mu\psi+\frac{1}{2}(\partial_\mu\sigma)^2
+\frac{1}{2}(\partial_\mu\pi)^2\nonu\\
&-&e_0\bar\psi \gamma^\mu A_\mu \psi-g_Y\bar\psi(\sigma+i\gamma_5\pi)\psi
-V(\sigma,\pi), \label{GHY}
\eea
where the potential $V$ contains {\em e.g.} mass terms, and a $\sigma^4$ type 
of interaction ({\em i.e.} a quartic scalar interaction).
For simplicity, we take $N=1$, and  
we ignore the effect of the potential $V$. 

The Lagrangian ${\cal L}_{GHY}$ gives rise to the following bare or free  
propagators for the scalar and pseudoscalar bosons:
\bea
\DelP(p)=\DelS(p)=\frac{1}{p^2}.
\eea

In appendix~\ref{appvacpol}, the two-loop contribution
has been computed for the special case of $N=1$ and is given by 
Eq.~(\ref{2loopwscalar}).
If the scalar and pseudoscalar fields in Eq.~(\ref{GHY}) 
are both in the adjoint representation of $U(N)$, 
the result, for arbitrary $N$, reads
\bea
\Pi(q^2)&\approx& \frac{N\alpha_0}{2\pi}
\(\frac{2}{3}+\frac{\alpha_0}{2\pi}-\frac{N\lambda_Y}{2\pi}\)
\log\(\frac{\Lambda^2}{q^2}\)
+(\alpha_0/\pi){\cal O}(1),\label{2loopwscalarN}
\eea
with $\lambda_Y=g_Y^2/4\pi$.
The $\beta$ function corresponding to such a vacuum polarization is
\bea
\beta_\alpha(\alpha_0,\lambda_Y)=\frac{N\alpha_0^2}{\pi}
\[\frac{2}{3}+\frac{\alpha_0}{2\pi}-\frac{N\lambda_Y}{2\pi}\].\label{betscal}
\eea
The interesting result of this computation is difference in sign
between terms corresponding to photon exchanges, and terms corresponding 
to (pseudo)scalar exchanges. 
Furthermore, we might be tempted to conclude that a nontrivial
root of Eq.~(\ref{betscal}) could be realized whenever $N\lambda_Y/2\pi\sim 2/3$.
However, the complete situation is more involved. 
The RG equation for {\em e.g.} $\lambda_Y$ should be
considered too, {\em i.e.} we should compute the $\beta$ functions
of $\lambda_Y$, and of any quartic scalar coupling.
If and only if a nontrivial (nonzero) UV fixed point for $\lambda_Y$ exists, 
the realization of a zero of Eq.~(\ref{betscal}) becomes a realistic option. 
In other words, such a scenario is only possible if the Yukawa interaction 
$\lambda_Y$ is nontrivial.
The discussion in section~\ref{sec_critexp_scal} 
suggests that in order to obtain a nontrivial Yukawa coupling, the hyperscaling 
laws should be obeyed.

In this context, we mention that 
the renormalization of Gauge-Higgs-Yukawa models, with
a non-Abelian gauge interaction, has been considered
extensively in Ref.~\cite{hakikuna94}.
In that paper, 
it was shown that special nontrivial cases of gauge-Higgs-Yukawa models 
are equivalent to GNJL models with non-Abelian 
(asymptotically free) gauge interaction (see also Ref.~\cite{koshya91}),
hence proving the renormalizability of non-Abelian GNJL models.

Moreover, in appendix~\ref{appvacpol} it is shown explicitly
that when gauge-invariant four-fermion interactions are 
treated perturbatively (thus their anomalous dimensions are small), 
they do not contribute to the vacuum polarization, and consequently $\beta_\alpha$.
In fact, the conclusion can be drawn beforehand, since 
the RG of Wilson states that irrelevant interactions cannot affect
low-energy effective quantities.

Summarizing, we have illustrated that fundamental scalars and 
pseudoscalars in a gauge-Higgs-Yukawa system
tend to decrease charge screening.
On the other-hand, perturbatively irrelevant four-fermion
interaction do not contribute to the vacuum polarization.
The idea is that, nonperturbatively, close to the critical curve in 
the GNJL model, 
the scalar and pseudoscalar Yukawa interactions are nontrivial, 
and kinetic terms for 
the scalar and pseudoscalar composites are effectively induced
via the appearance of a large anomalous dimension.
\section{The vacuum polarization in the $1/N$ expansion}
In studying the vacuum polarization,
the usual assumption is to neglect vacuum polarization corrections to 
internal photon propagators. As explained earlier, such a quenching of 
internal photon propagators is supposed 
to be valid close to a fixed point.
Thus, following Ref.~\cite{jowiba67}, 
with the replacement of full photon propagators by the bare ones 
({\em e.g.} Eq.~(\ref{bargb})), the vacuum polarization should be of the form:
\bea
\Pi(\mu^2)=\Pi[\mu/\Lambda,\alpha_0]
\approx f(\alpha_0)\log \frac{\Lambda^2}{\mu^2}+{\cal O}(1).
\eea
The vacuum polarization contains only a single power of $\log\Lambda$. 
Higher powers of $\log\Lambda$, which correspond to a higher number of 
fermion-loops, are absent. 
Furthermore, the constant term and other small corrections to $\Pi$ are 
irrelevant for the question of screening.
The vacuum polarization defines a running coupling $\alpha(\mu)$, 
\bea
\alpha(\mu)=\frac{\alpha_0}{1+\Pi[\mu/\Lambda,\alpha_0]},
\eea
where we define $\alpha_0=\alpha(\Lambda)$, and the $\beta$ function Eq.~(\ref{betaadef}) 
at $\alpha(\mu)$ is
\bea
\beta_\alpha(\alpha(\mu))\equiv \mu\frac{{\rm d}\alpha(\mu)}{{\rm d}\mu}.
\eea
Since the bare coupling is independent of $\mu$, we have the identity
\bea
\beta_\alpha(\alpha_0)=\lim_{\mu\rightarrow \Lambda}\mu 
\frac{{\rm d}\alpha(\mu)}{{\rm d}\mu}
&=& 2 \alpha_0 f(\alpha_0). \label{betaa0def}
\eea
Thus the function $f(\alpha_0)$ is proportional to the $\beta$ function, and since it is a 
non-singular function in $\alpha_0$ clearly the $\beta$ function has a Gaussian or trivial 
fixed point at $\alpha_0=0$, {\em i.e.}, $\beta_\alpha(0)=0$.
To obtain an UV fixed point it is essential
that, for $\alpha<\alpha_z$ (where $\alpha_z$ is a root of $\beta_\alpha$), 
the $\beta$ function is positive.
Hence, 
the equation (see Eq.~(\ref{UVfixpointdef}))
\bea
\beta_\alpha(\alpha_z)=0,\quad \beta_\alpha^\prime(\alpha_z)<0, 
\label{againUVfixpdef}
\eea
determines an UV fixed point $\alpha_z$.

The function $f(\alpha_0)$ has been studied thoroughly by Johnson {\em et al.} in
Refs.~\cite{jowiba67,bajo69,joba73} and by Adler \cite{ad72} in the context of massless QED, 
these authors write
(according to Johnson {\em et al.} \cite{jowiba67})
\bea
f(\alpha_0)=\frac{N\alpha_0}{2\pi}\[\frac{2}{3}+\Phi(\alpha_0)\],\label{fdef}
\eea
where the term with the $2/3$ is the one-loop vacuum polarization.
Johnson {\em et al.} derived an expression for $\Phi(\alpha_0)$ in terms of the Bethe-Salpeter 
(BS) kernel $K^{(2)}$ as the single unknown Green function.
In defining the BS kernels we follow the definitions of Bj{\"o}rken and Drell 
\cite{bjodre}.
We mention that, although the strong belief of 
Johnson {\em et al.} in the possible existence of an UV fixed point for $\alpha_0$,
might seem poorly motivated from the point of view of Wilson's RG methods
(see Ref~\cite{wiko74})\footnote{Johnson {\em et al.} 
don't explain the dynamical origin of the singular critical behavior, 
which would be required for the realization of an UV fixed point in QED.},  
their methods and techniques are still valid and 
directly applicable to the GNJL model.

The BS kernel $K^{(1)}$ is defined as the one-boson irreducible
fermion-fermion scattering kernel, and the BS kernel $K^{(2)}$
as the two-fermion one-boson irreducible fermion-fermion scattering kernel.
The integral equation between $K^{(1)}$ and $K^{(2)}$ is the Bethe--Salpeter
equation, which is depicted in Fig.~\ref{fig_K2def}.
In appendix~\ref{dervJWB} we give a derivation of the Johnson--Willey--Baker (JWB)
equation for $\Phi$, their result is 
\bea
\Phi(\alpha_0)&=&\frac{\phi_1+\phi_2(2+\phi_2)}{1-\phi_1}+\phi_3,
\label{PHIK2}
\eea
where the functions $\phi_j$ are identical to the functions $f_j$ 
defined by JWB, with
\bea
\!\!\!\!\!\phi_1(\alpha_0)&\equiv& \sum_{{\rm flavors}}-\frac{ie_0^2}{48}
\int\frac{{\rm d}^4p}{(2\pi)^4}\,\nonu\\
&\times&
\Tr\[
\frac{(\gamma^\mu\pslash \gamma^\alpha-\gamma^\alpha\pslash\gamma^\mu)}{2p^4}
K^{(2)}(p,k)(\gamma_\mu\kslash \gamma_\alpha-\gamma_\alpha\kslash\gamma_\mu)
\],\label{phi1}
\\
\!\!\!\!\!\phi_2(\alpha_0)&\equiv& \sum_{{\rm flavors}}-\frac{ie_0^2}{48}
\int\frac{{\rm d}^4p}{(2\pi)^4}\,
\Tr\[
\frac{\pslash\gamma^\mu \pslash}{p^4}
K^{(2)\alpha}(p,k)(\gamma_\mu\kslash \gamma_\alpha-\gamma_\alpha\kslash\gamma_\mu)\],
\label{phi2}
\\
\!\!\!\!\!\phi_3(\alpha_0)&\equiv& \sum_{{\rm flavors}}
\frac{ie_0^2}{48}\int\frac{{\rm d}^4p}{(2\pi)^4}\,
\Tr\[
\frac{\pslash\gamma^\mu \pslash}{p^4}K_\alpha^{(2)\alpha}(p,k) 
\kslash\gamma_\mu\kslash\].\label{phi3}
\eea
The derivatives of the BS kernels $K^{(2)}$ are defined 
in appendix~\ref{dervJWB}.
Furthermore,
Eqs.~(\ref{betaa0def}), (\ref{fdef}), and (\ref{PHIK2}) 
give rise to the following $\beta$ function: 
\bea
\beta_\alpha(\alpha_0)=\frac{N\alpha_0^2}{\pi}
\[\frac{2}{3}+
\frac{\phi_1+\phi_2(2+\phi_2)}{1-\phi_1}+\phi_3
\].\label{betasc2a}
\eea

\begin{figure}[t!]
\epsfxsize=8.5cm
\epsffile[5 300 405 560]{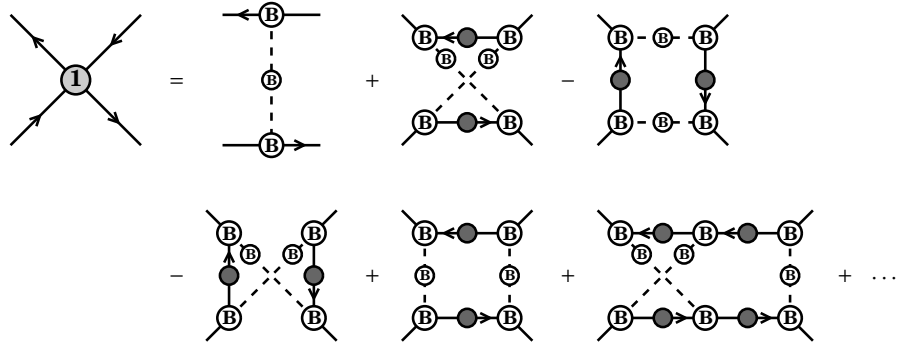}
\caption{Skeleton expansion for $K^{(1)}$.}
\label{fig_skel1}
\end{figure}

\begin{figure}[b!]
\epsfxsize=8.5cm
\epsffile[-20 300 380 540]{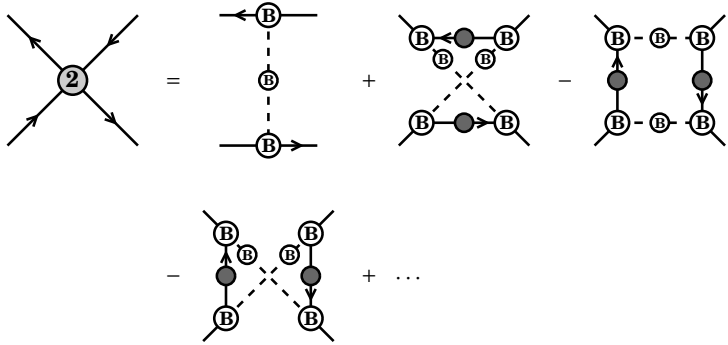}
\caption{Skeleton expansion for $K^{(2)}$.}
\label{fig_skel2}
\end{figure}

For the BS kernels there exists a so-called skeleton expansion 
(see also \cite{bjodre}), which is an expansion in topologically distinct
Feynman diagrams with all vertices and propagators fully dressed. 
The skeleton expansion is a special way of resumming the entire set of Feynman diagrams 
in a consistent manner, {\em i.e.,} without double counting.
The lowest order terms (``lowest'' in terms of loops) 
of the skeleton expansions for $K^{(1)}$, respectively, $K^{(2)}$ are illustrated 
in Fig.~\ref{fig_skel1}, respectively, Fig.~\ref{fig_skel2}.

As was pointed out in section~\ref{sec1/N},
the $1/N$ expansion states that the planar diagrams for
the $\sigma$ and $\pi$ exchanges are dominant.
The approximation for the BS kernel $K^{(2)}$, which
generates the entire set of planar scalar and pseudoscalar skeleton 
diagrams for the vacuum polarization is the following:
the BS kernel $K^{(2)}$ is approximated by its ``lowest'' order 
skeleton graph, {\em i.e.,} 
\bea
K^{(2)}_{^{cd,ab}_{kl,ij}}(p,p+q,k+q)
&=&\frac{\delta_{il}\delta_{kj}}{e_0^2}
\biggr[\Gamma_{{\rm S}cb}(k+q,p+q) \DelS(k-p) \Gamma_{{\rm S}ad}(p,k)\nonu\\
&+&\Gamma_{{\rm P}cb}(k+q,p+q) \DelP(k-p) \Gamma_{{\rm P}ad}(p,k)\biggr]\nonu\\
&+& \delta_{ij}\delta_{kl}  \gamma^\mu_{ad}\gamma^\nu_{cb} D_{\mu\nu}(k-p).
\eea

The decomposition of the Yukawa vertices in terms of scalar functions
is given by Eqs.~(\ref{vertfiesdefscal}) and (\ref{vertfiesdefpseudo}).
First of all,
as was explained in chapter~\ref{chap4}, in the symmetric phase the
Yukawa vertex structure function $F_3$ and $F_4$ are zero, and
the scalar and pseudoscalar vertex functions are identical
\bea
F^{(s)}_1=F^{(p)}_1=F_1,\qquad F^{(s)}_2=F^{(p)}_2=F_2,
\eea
and so are the $\sigma$ and $\pi$ propagators, $\DelS=\DelP$.
Secondly, it was shown in appendix~\ref{Chebyshev} that the structure function 
$F_2$ is rather small compared to the leading structure function $F_1$.
(it is assumed that $F_1$  describes the leading asymptotic behavior of 
the Yukawa vertices).
Therefore, we neglect contributions related to the scalar structure function $F_2$.
Although it might be possible 
that the contribution coming from gauge interactions is smaller, 
than corrections resulting from this structure function $F_2$, we 
keep the gauge interaction in order to compare with
results mentioned in the literature.
Thus, we take for $K^{(2)}$
\bea
K^{(2)}_{^{cd,ab}_{kl,ij}}(p,p+q,k+q)
&\approx&\frac{\delta_{il}\delta_{kj}}{e_0^2}
F_1(k+q,p+q)F_1(p,k)\DelS(k-p) \nonu\\
&\times&\[{\bf 1}_{ad}{\bf 1}_{cb}+{i\g5}_{ad}{i\g5}_{cb}\]\nonu\\
&+&\delta_{ij}\delta_{kl} D_{\mu\nu}(k-p)\gamma^\mu_{ad}\gamma^\nu_{cb}, 
\label{K2approx}
\eea
where $F_1$ is given by the ladder SDE (\ref{sde_vertfies}), and
$\DelS$ by Eq.~(\ref{delsinv}) and (\ref{sde_vacpol2}).

With this truncation for the BS kernel $K^{(2)}$, we can actually 
compute the $\phi_j$ functions (\ref{phi1})--(\ref{phi3}), and subsequently
analyze the $\beta$ function (\ref{betasc2a}).
We recall that, in the vacuum polarization, 
the scalars and pseudoscalars give the same contribution in the functions 
$\phi_j$.
Moreover the sum over flavor indices yields a factor of $N$ in the 
expressions for $\phi_j$ 
for contributions corresponding to $\sigma$ and $\pi$ exchanges.
\section{Analysis of the Johnson--Willey--Baker functions}
In computing the functions $\phi_j$,
we initially neglect the ladder photon exchange in Eq.~(\ref{K2approx}).
Since such contributions were already computed by Johnson {\em et al.}
\cite{jowiba67}, it will be rather easy to include them later on in the
analysis.

It is straightforward to show that, within the proposed approximations,
$\phi_2$ vanishes.
Using Eq.~(\ref{K2approx}), we obtain that
\bea
\Tr\[\pslash\gamma^\mu \pslash
K^{\alpha}(p,k)(\gamma_\mu\kslash \gamma_\alpha-\gamma_\alpha\kslash\gamma_\mu)\]
\propto
\Tr\[\pslash\gamma^\mu \pslash
(\gamma_\mu\kslash \gamma_\alpha-\gamma_\alpha\kslash\gamma_\mu)\]=0.
\eea
Thus,
\bea
\phi_2(\alpha_0)=0.
\eea
What remains is the evaluation of $\phi_1$ and $\phi_3$.
With
\bea
\Tr\[(\gamma_\mu\pslash \gamma_\alpha-\gamma_\alpha\pslash\gamma_\mu)
(\gamma^\mu\kslash \gamma^\alpha-\gamma^\alpha\kslash\gamma^\mu)\]=-96 p\cdot k,
\eea
the equation for $\phi_1$ Eq.~(\ref{phi1}) reads
\bea
\phi_1(\alpha_0)&=&2Ni\int_\Lambda\frac{{\rm d}^4p}{(2\pi)^4}
\frac{p\cdot k}{p^4} [F_1(p,k)]^2\DelS(p-k)\nonu\\
&=&2Ni\int_\Lambda\frac{{\rm d}^4p}{(2\pi)^4}
\frac{(p+k)\cdot k}{(p+k)^4} [F_1(k+p,k)]^2\DelS(p),
\eea
where in last step,
we have performed a supposedly ``harmless''\footnote{The integral is naively logarithmically divergent, therefore translationally invariant.} shift of integration,
and used the fact that $F_1$ is symmetric in the fermion momenta, 
$F_1(p,k)=F_1(k,p)$, because of charge conjugation properties, 
see Eq.~(\ref{Cscalvert}).
The factor $N$ results from the trace of the flavor Kronecker 
$\delta$ function, {\em i.e.} an index line which closes.
After a usual Wick rotation 
\bea
\phi_1(\alpha_0)&=&
\frac{N}{8\pi^2}
\int\limits_0^{\Lambda^2}
{\rm d}p^2\,\int \frac{{\rm d}\Omega_p}{2\pi^2}\,
\frac{p^2(p\cdot k+k^2)}{(p+k)^4}[F_1(k+p,k)]^2\DelS(p).
\eea
Since the integrals for the functions $\phi_j$ are finite,
the cutoff can be taken to infinity (the continuum limit).
This can be written as
\bea
\lim_{\Lambda\rightarrow \infty}\phi_1(\alpha_0)
\approx -N\int\limits_0^{\infty}{\rm d}u\, G_1(u)+{\cal O}\((k/\Lambda)^\sigma_1\),
\label{contlimphi1}
\eea
where $\sigma_1$ is (again) some positive power, $u=p^2/k^2$, and
\bea
G_1(p^2/k^2)\equiv
-\lim_{\Lambda\rightarrow \infty}
\frac{1}{8\pi^2} 
\int \frac{{\rm d}\Omega_p}{2\pi^2}\,
\frac{k^2 p^2(k\cdot p+k^2)}{(k+p)^4}[F_1(k+p,k)]^2\DelS(p).\label{G1def}
\eea
The function $G_1$ is defined with a minus sign to 
make it a positive function, as will be shown to be the case later.
The angular integral can be performed if we define the following 
Chebyshev expansion
\bea
\frac{k^2(k\cdot p+k^2)}{(k+p)^4}=\sum_{n=0}^\infty c_n(k^2,p^2) 
U_n(\cos\alpha),\quad\cos\alpha=\frac{k\cdot p}{kp},
\eea
where
\bea
c_n(k^2,p^2)&=&\frac{2}{\pi}\int\limits_0^\pi{\rm d}\alpha\,
\sin^2\alpha\,U_n(\cos\alpha)\frac{k^2(k\cdot p+k^2)}{(k+p)^4},\\
c_n(k^2,p^2)&=& \frac{(-1)^n}{2}
\[(2+n)\theta(k^2-p^2)\(\frac{p}{k}\)^n
-n\theta(p^2-k^2)\(\frac{k}{p}\)^{n+2}\].\label{cnpr}
\eea
The Chebyshev expansion for the function $F_1$ 
was already introduced in Eq.~(\ref{chebF1}) (and \cite{gure98}), 
and reads 
\bea
F_1(k+p,k)=\sum_{n=0}^\infty f_n(k^2,p^2)U_n(\cos\alpha).
\eea
Thus, following analogous derivations in appendix~\ref{Chebyshev},
the function $G_1$ can be expressed as
\bea
G_1(p^2/k^2)=-\lim_{\Lambda\rightarrow \infty}
\frac{p^2\DelS(p)}{8\pi^2} 
\sum_{l,m,n=0}^\infty C_{lmn} \,c_l(k^2,p^2) f_m(k^2,p^2) f_n(k^2,p^2), \label{G1exp}
\eea
where the constants $C_{lmn}$ are given in Eq.~(\ref{Clmn}).
We approximate $G_1$ by keeping only the
lowest order term in the Chebyshev expansion,
\bea
G_1(p^2/k^2)\approx-\lim_{\Lambda\rightarrow \infty}
\frac{p^2\DelS(p)}{8\pi^2} 
c_0(k^2,p^2) \[f_0(k^2,p^2)\]^2, \label{G1approx}
\eea
where $f_0$ is decomposed into 
the two channel functions $F_{\rm UV}$ and $F_{\rm IR}$, 
see Eq.~(\ref{channelapprox}).
Then, the asymptotics, $k^2\gg p^2$,  respectively, $p^2\gg k^2$,
of $G_1$ are well approximated by the lowest order Chebyshev term (\ref{G1approx}).
Again, this is the two channel approximation for the Yukawa vertices of 
chapter~\ref{chap4}.
However for momenta $k^2\sim p^2$ the channel approximation 
is not necessarily valid. So, how about $G_1(1)$?
Since, from the appendix~\ref{Chebyshev}, Eq.~(\ref{fnfrac}),
it follows that
\bea
f_{2n}(k^2,p^2)\geq 0, \qquad f_{2n+1}(k^2,p^2)\leq 0,
\eea
from Eq.~(\ref{Clmn}) that 
\bea
C_{(2l+1)(2m+1)(2n+1)}=C_{(2l+1)(2m)(2n)}=0, \qquad \forall\quad l,\,m,\,n,
\eea
and from Eq.~(\ref{an}) that
\bea
c_n(k^2,k^2)=\frac{(-1)^n}{2}\quad\longrightarrow\quad
c_{2n}(k^2,k^2)\geq 0, \quad c_{2n+1}(k^2,k^2)\leq 0.
\eea
Hence,
we conclude that all terms of the series
\bea
\sum_{l,m,n=0}^\infty C_{lmn} \,c_l(k^2,k^2) f_m(k^2,k^2) f_n(k^2,k^2),
\eea
of $G_1$ are positive,
and the lowest order term gives a lower bound on the series,
\bea
c_0(k^2,k^2)\[f_0(k^2,k^2)\]^2 \leq
\sum_{l,m,n=0}^\infty C_{lmn} c_l(k^2,k^2) f_m(k^2,k^2) f_n(k^2,k^2).
\eea
Therefore, the approximation Eq.~(\ref{G1approx}) is reliable
for the asymptotics $k^2\gg p^2$, and $p^2\gg k^2$. Moreover, 
Eq.~(\ref{G1approx}) is a lower bound on Eq.~(\ref{G1exp}) at $k^2=p^2$,  
so that at least we won't overestimate the contribution
of scalar and pseudoscalar composites to the vacuum polarization.

The function $G_1$ can now be computed, since $f_0$ is expressed in 
terms of the channel functions $F_{\rm UV}$ and $F_{\rm IR}$ of 
chapter~\ref{chap4}.
Furthermore, from Eq.~(\ref{cnpr}) we see that
\bea
c_0(k^2,p^2)&=& \theta(k^2-p^2).\label{c0}
\eea
and the only nonzero contribution to $G_1$ of Eq.~(\ref{G1approx}) 
comes from the momenta $k^2\geq p^2$. Thus, using Eqs.~(\ref{G1approx}), 
(\ref{c0}), and (\ref{channelapprox}), we find
\bea
G_1(p^2/k^2)&\approx& 
-\lim_{\Lambda\rightarrow \infty}
\frac{p^2\DelS(p)}{8\pi^2} [F_{\rm UV}(k^2,p^2)]^2\theta(k^2-p^2).\label{G1approx2}
\eea
The ultraviolet channel function $F_{\rm UV}$ is proportional to the
factor $Z^{-1}(p^2/\Lambda^2,\omega)$, Eq.~(\ref{Zdef}), and
the scalar propagator is proportional to $Z^2$. Thus 
the $Z$ factors in Eq.~(\ref{G1approx2})
cancel as was expected and the angular integral Eq.~(\ref{G1def})
can indeed be written in terms of a function which depends only on the ratio 
of $p^2/k^2$.

Recall that the scaling form for the scalar propagator is 
\bea
\frac{p^2\DelS(p)}{8\pi^2}\approx-\frac{1}{2B(\omega)} 
\(\frac{p^2}{\Lambda^2}\)^{1-\omega},\label{scalformscalprop}
\eea
where $B(\omega)$ is given by Eq.~(\ref{Bw}), 
and $p^2/\Lambda^2 \ll 1$.
The channel function $F_{\rm UV}$ is given by 
Eq.~(\ref{fuv2}), and using the asymptotic form 
for $Z(p^2/\Lambda^2,\omega)$ given in Eq.~(\ref{Zasym}),
the scaling form for $F_{\rm UV}$ is
\bea
F_{\rm UV}(k^2,p^2)&\approx&
\frac{2}{\gamma(\omega)}\frac{\Gamma(1-\omega)}{(1+\omega)}
\(\frac{\lambda_0}{2}\)^{\omega/2}\(\frac{p^2}{\Lambda^2}\)^{-(1-\omega)/2}\nonu\\
&\times&
\(\frac{p^2}{k^2}\)^{1/2}
\[\gamma(\omega)I_{-\omega}\(\sqrt{\frac{2\lambda_0 p^2}{k^2}}\)-
\gamma(-\omega)I_{\omega}\(\sqrt{\frac{2\lambda_0 p^2}{k^2}}\)
\],\nonu\\\label{scalfuv}
\eea
with $p^2\leq k^2\ll \Lambda^2$.
Inserting Eqs.~(\ref{scalformscalprop}) and (\ref{scalfuv}) 
in (\ref{G1approx2}), 
we obtain for $G_1$
\bea
G_1(u)&=&\frac{\Gamma(2-\omega)
\Gamma(2+\omega)}{8\omega \gamma(\omega)\gamma(-\omega)}
\nonu\\&\times&u\[\gamma(\omega)I_{-\omega}\(\sqrt{2\lambda_0 u}\)
-\gamma(-\omega)I_{\omega}\(\sqrt{2\lambda_0 u}\)\]^2\theta(1-u),
\label{G1esexp}
\eea
where $u=p^2/k^2$.
Thus Eq.~(\ref{contlimphi1}) is
\bea
\lim_{\Lambda\rightarrow \infty}\phi_1(\alpha_0)\approx
-N\zeta_1(\alpha_0),\label{phi1zeta1}
\eea
where
\bea
\zeta_1(\alpha_0)=\int\limits_0^1{\rm d}u\, G_1(u) \geq 0.\label{zeta1}
\eea
The function $G_1$ is positive, hence $\phi_1$
is negative.
The integral over the function $G_1$ can be done explicitly by making 
use of the integral identity 2.15.19.1 in Volume 2 of 
Prudnikov {\em et al.} \cite{prbrma86}.
The result is
\bea
\zeta_1(\alpha_0)&=&\frac{1}{\omega}\frac{\lambda_0}{2}
\biggr\{\frac{1}{(2+\omega)}
\frac{\Gamma(1-\omega) \gamma(-\omega)}{\Gamma(1+\omega)\gamma(\omega)}
\(\frac{\lambda_0}{2}\)^\omega\nonu\\
&\times&{_2}F_3\(2+\omega,1/2+\omega;
3+\omega,1+2\omega,1+\omega;2\lambda_0\)\nonu\\
&-&{_2}F_3\(2,1/2;
3,1+\omega,1-\omega;2\lambda_0\)\nonu\\
&+&\frac{1}{(2-\omega)}
\frac{\Gamma(1+\omega) \gamma(\omega)}{\Gamma(1-\omega)\gamma(-\omega)}
\(\frac{\lambda_0}{2}\)^{-\omega}\nonu\\
&\times&{_2}F_3\(2-\omega,1/2-\omega;
3-\omega,1-2\omega,1-\omega;2\lambda_0\)\biggr\}.\label{zeta1expl}
\eea

The above analysis of the function $\phi_1$ is repeated for the function $\phi_3$. 
The second derivative of the BS kernel, $K_\alpha^{\alpha}(p,k)$ of 
Eq.~(\ref{K2approx}), is
\bea
K_\alpha^{\alpha}(p,k)
\propto \lim_{q\rightarrow 0}
\frac{\partial^2}{\partial q^\alpha \partial q_\alpha}F_1(k+q,p+q). 
\eea
The SDE for $F_1$ (in quenched-ladder approximation) is 
given by Eq.~(\ref{sde_vertfies})
\bea
F_1(k+q,p+q)&=&1-i
\lambda_0\int_\Lambda\frac{{\rm d}^4r}{\pi^2}\,
\frac{(r^2+(k-p)\cdot r)}{r^2(r+k-p)^2(r-p-q)^2} \nonu\\
&\times& F_1(r+k-p,r),
\eea
where we neglect the vertex function $F_2$.
Thus
\bea
\lim_{q\rightarrow 0}
\frac{\partial^2}{\partial q^\alpha \partial q_\alpha}F_1(k+q,p+q)
&=&
-i\lambda_0\int_\Lambda\frac{{\rm d}^4r}{\pi^2}\,
\frac{(r^2+(k-p)\cdot r)}{r^2(r+k-p)^2}F_1(r+k-p,r)\nonu\\
&\times&\lim_{q\rightarrow 0}
\frac{\partial^2}{\partial q^\alpha \partial q_\alpha}
\frac{1}{(r-p-q)^2}. \label{laplsde}
\eea
By making use of the identity
\bea
\frac{\partial}{\partial q^\alpha}
\frac{\partial}{\partial q_\alpha}\frac{1}{q^2}
=-4\pi^2i \delta^4(q),
\eea
we obtain
\bea
K_\alpha^{\alpha}(p,k)&=&
\frac{\delta\delta}{e_0^2}
 \[-4\lambda_0\frac{p\cdot k}{p^2 k^2}\]\[F_1(p,k)\]^2 \DelS(p-k)\nonu\\
&\times&\[{\bf 1}_{ad}{\bf 1}_{cb}+{i\g5}_{ad}{i\g5}_{cb}\],
\eea
in Minkowsky formulation.
Inserting the above expression in Eq.~(\ref{phi3})
the equation $\phi_3(\alpha_0)$
takes the form
\bea
\phi_3(\alpha_0)=-\frac{\alpha_0}{\pi}
\frac{N}{8\pi^2}
\int\limits_0^{\Lambda^2}
{\rm d}p^2\,\int \frac{{\rm d}\Omega_p}{2\pi^2}\,
\frac{p^2}{k^2}\frac{(p\cdot k+k^2)^3}{(p+k)^6}
[F_1(k+p,k)]^2\DelS(p).
\eea
Then 
\bea
\lim_{\Lambda\rightarrow \infty}\phi_3(\alpha_0)
\approx N\int\limits_0^{\infty}{\rm d}u\, G_3(u)+{\cal O}\((k/\Lambda)^\sigma\),
\label{contlimphi3}
\eea
where $u=p^2/k^2$, and
\bea
G_3(p^2/k^2)\equiv
\lim_{\Lambda\rightarrow \infty}
-\frac{\alpha_0}{\pi}\frac{1}{8\pi^2} 
\int \frac{{\rm d}\Omega_p}{2\pi^2}\,
\frac{p^2(k\cdot p+k^2)^3}{(k+p)^6}[F_1(k+p,k)]^2\DelS(p).\label{G3def}
\eea
We define the following Chebyshev expansion
\bea
\frac{(k\cdot p+k^2)^3}{(k+p)^6}
=\sum_{n=0}^\infty d_n(k^2,p^2) U_n(\cos\alpha),\quad 
\cos\alpha=\frac{k\cdot p}{kp},
\eea
where
\bea
d_n(k^2,p^2)&=&\frac{2}{\pi}\int\limits_0^\pi{\rm d}\alpha\,
\sin^2\alpha\,U_n(\cos\alpha)\frac{(p\cdot k+k^2)^3}{(p+k)^6},\\
d_0(k^2,p^2)&=&\(1-\frac{3p^2}{4k^2}\)\theta(k^2-p^2),\label{d0pr}\\
d_n(k^2,p^2)&=& \frac{(-1)^n}{8}
\left\{n+1+\[6+\sum_{l=0}^{n-1}(4+l)\] \(1-\frac{p^2}{k^2}\)\right\}
\theta(k^2-p^2)\(\frac{p}{k}\)^n \nonu\\
&-&\frac{(-1)^n}{8}
\left\{n+1-\[\sum_{l=0}^{n-1}(2-l)\]\(1-\frac{k^2}{p^2}\)\right\}
\theta(p^2-k^2)\(\frac{k}{p}\)^n,\nonu\\
&&n\geq 1.
\eea
The function $G_3$ can be expressed as
\bea
G_3(p^2/k^2)=\lim_{\Lambda\rightarrow \infty}
-\frac{\alpha_0}{\pi}\frac{p^2\DelS(p)}{8\pi^2} 
\sum_{l,m,n=0}^\infty C_{lmn} \,d_l(k^2,p^2) f_m(k^2,p^2) f_n(k^2,p^2).
\label{G3exp}
\eea
We also approximate $G_3$ by keeping only the
lowest order term in the Chebyshev expansion,
\bea
G_3(p^2/k^2)\approx\lim_{\Lambda\rightarrow \infty}
-\frac{\alpha_0}{\pi}\frac{p^2\DelS(p)}{8\pi^2} 
d_0(k^2,p^2) \[f_0(k^2,p^2)\]^2. \label{G3approx}
\eea
Then, again the asymptotics, $k^2\gg p^2$,  respectively, $p^2\gg k^2$,
of $G_3$ are well approximated by the lowest order Chebyshev 
term (\ref{G1approx}).
Moreover, for momenta $k^2=p^2$ 
the approximation Eq.~(\ref{G3approx}) is exact, since
\bea
d_n(k^2,k^2)=0,\quad \forall\quad n\geq 1.
\eea
Therefore, the approximation Eq.~(\ref{G3approx}) is even better
than the analogous approximation, Eq.~(\ref{G1approx}), to $G_1$. 
Furthermore, from Eq.~(\ref{d0pr}) we see that the only nonzero contributions 
to $G_3$ of Eq.~(\ref{G3approx}) 
are given by momenta $k^2\geq p^2$. Thus, using Eqs.~(\ref{G3approx}), 
(\ref{d0pr}), and (\ref{channelapprox}), we find
\bea
\!\!\!\!
G_3(p^2/k^2)&\approx& 
\lim_{\Lambda\rightarrow \infty}
-\frac{\alpha_0}{\pi}\(1-\frac{3p^2}{4k^2}\)
\frac{p^2\DelS(p)}{8\pi^2} [F_{\rm UV}(k^2,p^2)]^2
\theta(k^2-p^2).\label{G3approx2}
\eea
Substituting Eqs.~(\ref{scalformscalprop}) and (\ref{scalfuv}) 
in Eq.~(\ref{G3approx2}),
we obtain for $G_3$
\bea
G_3(u)&=&\frac{\alpha_0}{\pi}\(1-\frac{3u}{4}\)\frac{\Gamma(2-\omega)
\Gamma(2+\omega)}{8\omega \gamma(\omega)\gamma(-\omega)}
\nonu\\&\times&u\[\gamma(\omega)I_{-\omega}\(\sqrt{2\lambda_0 u}\)
-\gamma(-\omega)I_{\omega}\(\sqrt{2\lambda_0 u}\)\]^2\theta(1-u),\label{G3esexp}
\eea
where $u=p^2/k^2$.
Thus Eq.~(\ref{contlimphi3}) is
\bea
\lim_{\Lambda\rightarrow \infty}\phi_3(\alpha_0)\approx
N\zeta_3(\alpha_0),\label{phi3zeta3}
\eea
where
\bea
\zeta_3(\alpha_0)=\int\limits_0^1{\rm d}u\, G_3(u) \geq 0.\label{zeta3}
\eea
The function $\phi_3$
is positive, and can be computed in the same way as $\phi_1$.
The result is
\bea
\zeta_3(\alpha_0)=\frac{\alpha_0}{\pi}\[\zeta_1(\alpha_0)-\tau(\alpha_0)\],
\label{zeta3expl}
\eea
where
\bea
\tau(\alpha_0)&=&\frac{3}{4\omega}\frac{\lambda_0}{2}
\biggr\{\frac{1}{(3+\omega)}
\frac{\Gamma(1-\omega) \gamma(-\omega)}{\Gamma(1+\omega)\gamma(\omega)}
\(\frac{\lambda_0}{2}\)^\omega\nonu\\
&\times&{_2}F_3\(3+\omega,1/2+\omega;
4+\omega,1+2\omega,1+\omega;2\lambda_0\)\nonu\\
&-&\frac{2}{3}{_2}F_3\(3,1/2;
4,1+\omega,1-\omega;2\lambda_0\)\nonu\\
&+&\frac{1}{(3-\omega)}
\frac{\Gamma(1+\omega) \gamma(\omega)}{\Gamma(1-\omega)\gamma(-\omega)}
\(\frac{\lambda_0}{2}\)^{-\omega}\nonu\\
&\times&{_2}F_3\(3-\omega,1/2-\omega;
4-\omega,1-2\omega,1-\omega;2\lambda_0\)\biggr\}.
\eea
\section{UV fixed points}\label{critfixpoints}
In the computation of the functions $\phi_1$, $\phi_2$, and 
$\phi_3$ the ladder (planar) photon exchanges have been neglected.
After reinstating the ladder photon exchange term 
in Eq.~(\ref{K2approx}), 
we obtain, together with Eqs.~(\ref{phi1zeta1}) and (\ref{phi3zeta3}), 
that
\bea
\phi_1(\alpha_0)=\frac{\alpha_0}{2\pi}-N\zeta_1(\alpha_0),
\quad \phi_2(\alpha_0)=0,
\quad\phi_3(\alpha_0)=N\zeta_3(\alpha_0).\label{phi1phi3phot}
\eea
The ladder photon exchange only contributes to $\phi_1$, see again
\cite{jowiba67}.
After substitution of Eq.~(\ref{phi1phi3phot}) in Eq.~(\ref{betasc2a}),
the $\beta$ function reads
\bea
\beta_\alpha(\alpha_0)=\frac{N\alpha_0^2}{\pi}
\[\frac{2}{3}+\frac{\alpha_0/2\pi-N\zeta_1(\alpha_0)}{1-\alpha_0/2\pi
+N\zeta_1(\alpha_0)}+N\zeta_3(\alpha_0)\],\label{betasc2}
\eea
where explicit expressions for $\zeta_1$ and $\zeta_3$ 
are given by Eq.~(\ref{zeta1expl}) and Eq.~(\ref{zeta3expl}).

\begin{figure}[t!]
\epsfxsize=9cm
\epsffile[150 330 400 570]{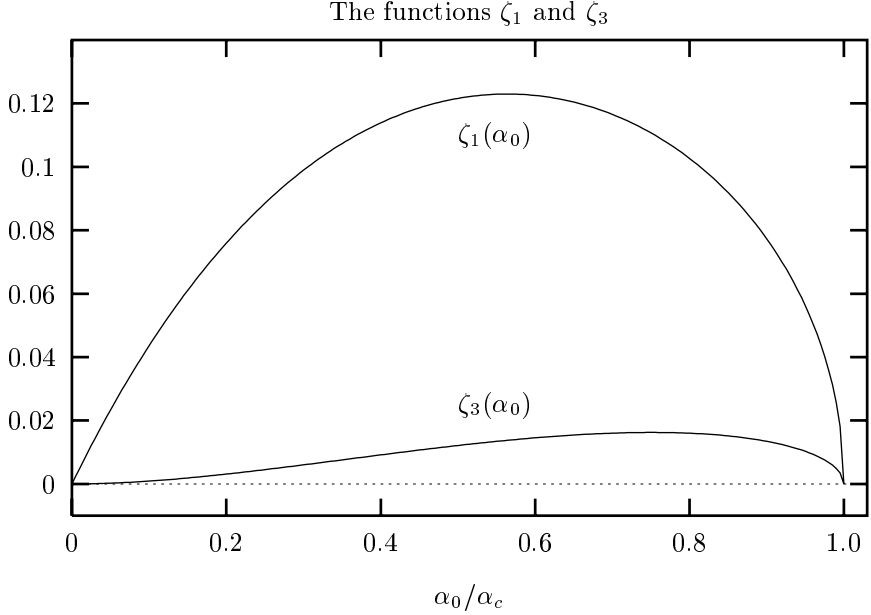}
\caption{The functions $\zeta_1$ and $\zeta_3$ plotted versus
$\alpha_0/\alpha_c$.}
\label{fig_zeta13}
\end{figure}

Let us start analyzing Eq.~(\ref{betasc2}) by first considering the
properties of the functions $\zeta_1(\alpha_0)$ and $\zeta_3(\alpha_0)$. 
These functions have been plotted versus $\alpha_0/\alpha_c$ in 
Fig.~\ref{fig_zeta13}. 
Firstly, it is clear that the $\zeta_1$ and $\zeta_3$, are positive, and 
have a maximum at some intermediate value of $0<\alpha_0<\pi/3$. 
For instance, $\zeta_1$ has a maximum $\zeta_1\approx 0.123$ 
at $\alpha_0/\alpha_c\approx 0.58$ ($\omega\approx 0.65$).
Secondly, the functions $\zeta_1$ and $\zeta_3$ vanish at the pure NJL 
point $\alpha_0=0$, and at the CPT point $\alpha_0=\alpha_c=\pi/3$.
At $\alpha_0=0$, we can consider this is as a reflection of the fact
that hyperscaling breaks down due to logarithmic corrections; 
the ``effective'' Yukawa coupling is trivial, therefore vanishes, 
see again the discussion in section~\ref{sec_critexp_scal}.
At $\alpha_0=\alpha_c$, where the critical exponents 
become singular, the vanishing of $\zeta_1$ and $\zeta_3$
is related to the dynamics of the CPT, see 
sections~\ref{sec_cpt} and \ref{again_sec_cpt}.
There are no light $\sigma$ and $\pi$ 
exchanges in the symmetric phase which consequently implies the absence of
effective Yukawa interactions.\footnote{Moreover at the CPT point four-fermion
interaction are marginal instead of relevant, and start to mix
with the gauge interaction, hence the analysis becomes considerably
more complicated.}

Let us illustrate this by comparing the $\beta$ function (\ref{betasc2})
with the $\beta$ function of the gauge-Higgs-Yukawa model (\ref{GHY})
in the $1/N$ expansion. 
Then, the entire set of planar $\sigma$ and $\pi$ exchanges
is generated by the kernel
\bea
K^{(2)}_{^{cd,ab}_{kl,ij}}(p,p+q,k+q)
&\approx&\delta_{il}\delta_{kj}\frac{g_Y^2}{e_0^2}\DelS(k-p) 
\[{\bf 1}_{ad}{\bf 1}_{cb}+{i\g5}_{ad}{i\g5}_{cb}\],\label{K2approxGHY}\\
\DelS(k-p)&=&\frac{1}{(k-p)^2}. \nonu
\eea
With such a kernel, $\phi_2$ and $\phi_3$ are zero, because the
right-hand side of Eq.~(\ref{K2approxGHY}) does not depend on the momentum $q$.
The equation for $\phi_1$ reads
\bea
\phi_1&=&
2 N g_Y^2 i\int_\Lambda\frac{{\rm d}^4p}{(2\pi)^4}
\frac{p\cdot k}{p^4} \frac{1}{(p-k)^2}.
\eea
After a standard Wick rotation, the integral
can be performed in the standard way:
\bea
\phi_1&=&
-\frac{N \lambda_Y}{2\pi}
\int\limits_0^{\Lambda^2}
{\rm d}p^2\,\int \frac{{\rm d}\Omega_p}{2\pi^2}\,
\frac{p\cdot k}{p^2}\frac{1}{(p-k)^2}\nonu\\
&=&-\frac{N \lambda_Y}{4\pi}\[
\int\limits_0^{k^2}
{\rm d}p^2\,\frac{1}{k^2}\,+
\int\limits_{k^2}^{\Lambda^2}
{\rm d}p^2\,\frac{k^2}{p^4}\]
=-\frac{N \lambda_Y}{4\pi}\[2-\frac{k^2}{\Lambda^2}\].
\eea
Then, taking the cutoff to infinity, we obtain
\bea
\phi_1(\lambda_Y)=-\frac{N \lambda_Y}{2\pi}.
\eea
Again we introduce the ladder photon exchanges by the
replacement
\bea
\phi_1(\lambda_Y)\quad \longrightarrow \quad
\phi_1(\alpha_0,\lambda_Y)=\frac{\alpha_0}{2\pi}-\frac{N \lambda_Y}{2\pi}.
\eea
Hence, in this case, 
the $\beta$ function is
\bea
\beta_\alpha(\alpha_0,\lambda_Y)=\frac{N\alpha_0^2}{\pi}
\[\frac{2}{3}
+\frac{\alpha_0/2\pi-N \lambda_Y/2\pi}{1-\alpha_0/2\pi+N \lambda_Y/2\pi}\].
\label{betajwbGHY}
\eea
Comparing the $\beta$ functions (\ref{betasc2}) and (\ref{betajwbGHY})
leads to the suggestion that $\zeta_1(\alpha_0)$ is analogous to the 
Yukawa coupling $\lambda_Y$ in a gauge-Higgs-Yukawa model,
\bea
\zeta_1(\alpha_0)\sim \frac{\lambda_Y}{2\pi}.\label{identzetalam}
\eea

This is a crucial point.
The general consensus is that for a gauge-Higgs-Yukawa model 
the Yukawa interaction $\lambda_Y$ is trivial, thus
$\lambda_Y\rightarrow 0$ in Eq.~(\ref{betajwbGHY}). 
However, the situation is essentially different for $\zeta_1$ in the GNJL model.
There the ``effective'' coupling $\zeta_1$ 
is formed by the exchange of $\sigma$ and $\pi$ bosons, with
the Yukawa vertices, and (pseudo)scalar propagators fully
dressed ({\em i.e.} the skeleton expansion). 
The cancellation of the $Z$ factors, see Eq.~(\ref{RGscat}),
which is related to the fact that the hyperscaling equations are satisfied,
gives rise to a finite nonzero $\zeta_1(\alpha_0)$ at the
critical curve for $0<\alpha_0<\alpha_c$.
The other nonzero function $\zeta_3$ results from taking into account fully 
dressed Yukawa vertices.

Let us now the discuss the possible existence of UV fixed points.
A necessary but not a sufficient condition for the realization of
an UV fixed point is that $N\zeta_1$ has to be larger than both $N\zeta_3$ and 
$\alpha_0/2\pi$, and $N\zeta_1\sim {\cal O}(1)$.
For large $N$, the contribution of the planar photon exchanges 
(represented by the $\alpha_0/2\pi$ terms) is negligible
with respect to $N\zeta_1$ and $N\zeta_3$.
Moreover Fig.~\ref{fig_zeta13} shows, for $\alpha_0$ small, that 
$\zeta_1$ is considerably larger than $\zeta_3$.
This suggests that only
for flavors $N$ larger than some critical value $N_c$ UV fixed points 
can be obtained.

\begin{figure}[b!]
\epsfxsize=9cm
\epsffile[150 330 400 570]{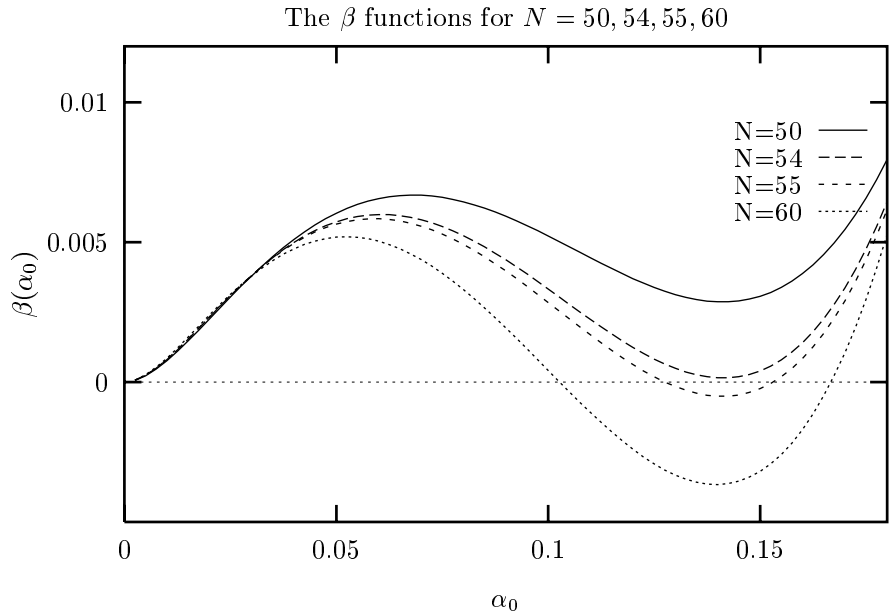}
\caption{Plot of the function $\beta_\alpha$ versus the gauge coupling $\alpha_0$ 
for various values of the fermion flavor number $N$.}
\label{fig_beta5060}
\end{figure}

By substituting the expressions (\ref{zeta1expl}) and (\ref{zeta3expl})
for $\zeta_1$ and $\zeta_3$ in Eq.~(\ref{betasc2}), 
we can straightforwardly analyze the $\beta$ function graphically.
In Fig.~\ref{fig_beta5060} the $\beta$ function 
is plotted for various values of $N$.
Fig.~\ref{fig_beta5060} shows that for values of 
$N>N_c$, with $55>N_c>54$, UV fixed points exists, 
the largest being $\alpha_0\approx 0.14$,
\bea
N=55:&\quad&\beta_\alpha(0.14)\approx 0,\\
N=60:&\quad&\beta_\alpha(0.1)\approx 0.
\eea
The general pattern is clear;  
the larger $N$, with $N>N_c$, the smaller will be the UV fixed point.
Moreover, Fig.~\ref{fig_beta5060} suggests that the UV fixed points 
are first order zeros of $\beta$, and clearly satisfy
Eq.~(\ref{againUVfixpdef}). 

Since we have made use of results obtained in the quenched approximation, 
we mention that the plots of the $\beta$ function are (at the most) 
reliable at or in the vicinity of the UV fixed points 
at which the quenched approach is self consistent.

\begin{figure}[b!]
\epsfxsize=9cm
\epsffile[150 330 400 570]{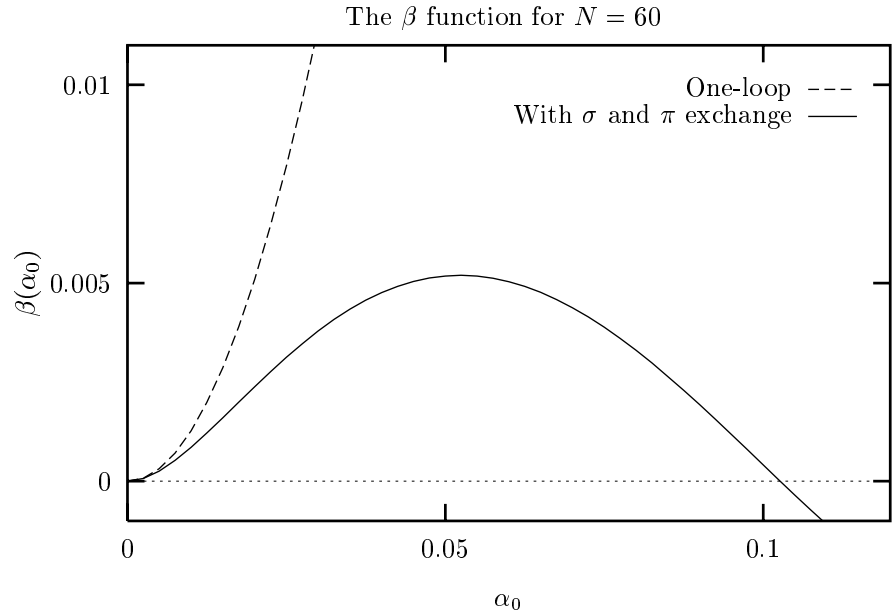}
\caption{The one-loop $\beta$ function for $N=60$
compared with the $\beta$ function including four-fermion interactions.}
\label{fig_beta60}
\end{figure}

In Fig.~\ref{fig_beta60}, the case of $N=60$ fermion flavors is 
compared with the one-loop $\beta$ function of QED.
For very small values of $\alpha_0< 1/100$ indeed the one-loop QED result
coincides with that of the GNJL model, however for larger values of $\alpha_0$
the $\beta$ function (\ref{betasc2}) deviates from the one-loop 
expression, and eventually an UV fixed point is realized 
at $\alpha_0\approx 0.1$.

The analysis shows that a rather large number of flavors, 
\bea
N>N_c\approx 54, 
\eea
is
required to obtain UV fixed points.
From the point of view of the $1/N$ expansion this seems a consistent
result, since other than planar contributions are suppressed 
by at least factors of (say) $1/N_c$. 
However, from the phenomenological point of view,
the result is unsatisfactory, since it implies that the unquenched GNJL model 
(exhibiting UV fixed points) is only practically applicable for 
models which have at least $N_c$ fermion flavors (fractions rounded up). 
Therefore, it is appropriate to discuss how $N_c$ depends on the approximation.

Firstly, we stress that
the second term on the right-hand side
in Eq.~(\ref{betasc2}) containing the $\zeta_1$ function
causes the suppression of charge screening and
is responsible for the possible realization of an UV fixed point.
The denominator in the second term 
is a direct consequence of the resummation of the infinite ladder $\sigma$
and $\pi$ exchanges, and it is mainly due to this
denominator $1+N\zeta_1$ that the critical number of fermion flavors
$N_c$ is large.

Secondly, the existence of an UV fixed point for a specific 
number of fermion flavors $N$ depends on the interplay between 
the functions $\zeta_1$ and $\zeta_3$, which are given in terms of integrals
of the functions $G_1$ and $G_3$. 
Let us recall that the lowest order Chebyshev expansion
for $G_1$, Eq.~(\ref{G1approx}) is a lower bound on $G_1$ 
of Eq.~(\ref{G1exp}), since all terms of the Chebyshev 
expansion are positive at $k^2=p^2$, the same cannot be said about 
$G_3$ of Eq.~(\ref{G3approx}).
Thus keeping more terms in the Chebyshev expansion leads to an 
increase of $\zeta_1$, whereas the effect on $\zeta_3$ is less clear,
because of the alternating Chebyshev series for $\zeta_3$. 
Therefore, an improvement of the computation of $\zeta_1$ 
will most likely lead to a decrease of the critical flavor number $N_c$.

Moreover in computation of the functions $\zeta_1$ and $\zeta_3$ 
we have used Yukawa vertices ($\GS$) and $\sigma$ and $\pi$ propagators ($\DelS$)
which were obtained in the quenched-ladder approximation.
It is an interesting question, whether the improvement
of the ladder approximation for the gauge interaction
({\em e.g.} by including crossed photon exchanges)
leads to a increase of $\zeta_1$, and thus a decrease of $N_c$.

Finally, we recall that we have neglected 
the effect of the Yukawa vertex function $F_2$ (Eq.~(\ref{vertfiesdef})),
but clearly the inclusion of $F_2$ in the analysis 
could change the results quantitatively. 
Whether such an improvement will tend to increase or decrease $N_c$ remains 
unclear at this stage.

\subsection{Close to the NJL point}
Let us analyze the interplay between $\zeta_1$ and $\zeta_3$ in more detail
close to the pure NJL point ({\em i.e.} for small values of $\alpha_0$)
by expanding these functions in $\alpha_0$.
The computation of the lowest order terms in $\alpha_0$
of Eqs.~(\ref{zeta1expl}) and (\ref{zeta3expl}) 
is laborious but straightforward, 
therefore we mention just the final results:
\bea
\zeta_1(\alpha_0)&\approx&\frac{3\alpha_0}{2\pi}+{\cal O}(\alpha_0^2),\\
\zeta_3(\alpha_0)&\approx&\frac{15}{16}
\frac{\alpha_0^2}{\pi^2}+{\cal O}(\alpha_0^3).
\eea
With these expansions in $\alpha_0$, 
the $\beta$ function Eq.~(\ref{betasc2}) reads
\bea
\beta_\alpha(\alpha_0)=\frac{N\alpha_0^2}{\pi}
\[\frac{2}{3}-\frac{(3N-1)\alpha_0/2\pi}{1+(3N-1)\alpha_0/2\pi}
+N\frac{15}{16}\frac{\alpha_0^2}{\pi^2}\].\label{betanjl}
\eea
Let us write the $\beta$ function in terms of the variable
\bea
x\equiv N\alpha_0/\alpha_c,\quad \alpha_c=\pi/3,
\eea
then
\bea
\beta_\alpha(\alpha_0)=\frac{N\alpha_0^2
}{\pi}
\[\frac{2}{3}-\frac{(1-1/3N)x}{2+(1-1/3N)x}
+\frac{5}{48}\frac{x^2}{N}\].\label{betanjlx}
\eea
The equation for the zero of $\beta_\alpha$ is a third order equation,
and a UV fixed point is given by the lowest positive root
satisfying Eq.~(\ref{UVfixpointdef}).
The equation is
\bea
4-(1-\epsilon)x+\frac{15}{8}\epsilon x^2+\frac{15}{16}\epsilon (1-\epsilon)x^3=0,
\qquad \epsilon\equiv \frac{1}{3N}.
\eea
The expression for the relevant root is rather ugly.
However, in the limit $N\rightarrow \infty$ keeping $x$ fixed, the solution is
\bea
x=\frac{N\alpha_0}{\alpha_c}\approx\[4+\frac{94}{3}\frac{1}{N}\], 
\qquad N\rightarrow \infty,\qquad x=\mbox{constant}.
\eea
The critical number of flavors $N_c$, below which no UV fixed point can be 
realized, is given by the following expression
\bea
\frac{1}{N_c}&=&\frac{3}{64}\[2209+285\sqrt{57}
-\sqrt{9505410+1259130\sqrt{57}}\]\quad\Longrightarrow
\nonu\\
N_c&\approx& 45.42.
\eea
This $N_c$ is slightly smaller than the critical number $N_c\approx 54$, 
which was obtained graphically from Fig.~\ref{fig_beta5060}.
The value of the root $x$ at this critical number $N_c$ is
\bea
x\approx 6.32131\quad \Longrightarrow\quad \alpha_0\approx 0.146.
\eea
Summarizing, we find, that for $N>N_c\approx 45$ 
UV fixed points $\alpha_0$ ($\alpha_0< 0.146$) are obtained.
Moreover, with the identification Eq.~(\ref{identzetalam}), 
$\lambda_Y=g_Y^2/4\pi$ and the fact that $\zeta_1$ is of order $\alpha_0$
the inequality (\ref{ineq2}) is satisfied 
for $\alpha_0$ sufficiently small.
\section{Discussion}
Relevant four-fermion interactions in four dimensions are possible at the 
chiral phase transition in the GNJL model.
The main objective of this chapter was to study the effect of such ``relevant''
four-fermion dynamics on the vacuum polarization and to reinvestigate the 
problem of triviality of $U(1)$ gauge theories.

To obtain new results, the four-fermion interactions had to
be incorporated beyond the commonly used Hartree--Fock or 
mean-field approach.
At the chiral phase transition ({\em e.g.} the critical 
curve Eq.~(\ref{critline1})) 
in the quenched 
GNJL model, the appearance of 
a large anomalous dimension is believed to turn naively irrelevant
interactions into relevant ones. 

The crucial feature of the GNJL model is that 
a nontrivial Yukawa interaction ({\em i.e.} an interaction between 
(pseudo)scalars and fermions) exists for $0<\alpha_0<\alpha_c$.
The existence of such a nontrivial Yukawa interaction
requires the cancellation of $Z$ factors\footnote{The $Z$ factor is the 
wave function renormalization constant of the $\sigma$ and $\pi$ fields.} 
in fermion--anti-fermion scattering amplitudes such as the BS 
kernel $K^{(2)}$, and
is analogous to the requirement of hyperscaling 
(see section~\ref{sec_critexp_scal}).
If the hyperscaling equations are satisfied,
then only two of the critical exponents are independent, namely
$\eta$ and $\gamma$, 
with
\bea
\gamma=1,\qquad \eta=2(1-\omega),\qquad \omega=\sqrt{1-\alpha_0/\alpha_c},
\eea
see Eqs.~(\ref{critgamma}) and (\ref{etaexpr}).
We recall that the anomalous dimension
$\eta$ is defined at the critical point, whereas $\gamma$
describes scaling in the neighborhood of the critical point.\footnote{These values of the critical exponents 
are in rather good agreement  
with lattice simulations, and results obtained by nonperturbative
RG techniques, see the discussions 
in sections~\ref{sec_lattice} and \ref{sec_beyond}.}

The skeleton expansion for the BS kernel $K^{(2)}$ provides a natural framework 
to take into account the anomalous dimension of Yukawa vertices and 
$\sigma$ and $\pi$ propagators.
Within the skeleton expansion, $\sigma$ and $\pi$ exchanges are described
in terms of fully dressed Yukawa vertices and $\sigma$ 
and $\pi$ propagators.
The actual computation of the anomalous dimension, and the resolution of 
the scaling form requires a solution of the SDEs for 
Yukawa vertices, and $\sigma$ and $\pi$ bosons. 

In previous work such fully dressed Yukawa vertices and $\sigma$ 
and $\pi$ propagators have been analized in the quenched-ladder
approximation, see chapter~\ref{chap4}, Ref.~\cite{gure98}, and references therein.
To make use of these results consistently, we used the following
approximations.
Firstly, we assumed that all dynamics takes place in a region
close to an UV fixed point, $\beta_\alpha(\alpha_0)\approx 0$.
In that case, the quenched approximation is self-consistent.
Secondly, the gauge-interaction is considered 
in the ladder form, with bare vertices.
Thirdly, we used the $1/N$ expansion (with
$N$ the number of fermion flavors)
which states that planar $\sigma$ and $\pi$ exchanges describe the
leading contribution to Green functions for large $N$.
Then, due to the specific form of the chiral symmetry
with both scalars and pseudoscalar in the adjoint representation,
we argued that in so-called zero-spin channels 
(such as Yukawa vertices and $\sigma$ and $\pi$ propagators) 
the planar $\sigma$ and $\pi$ exchanges cancel each other 
for momenta larger than the inverse correlation length
(in fact in the symmetric phase this cancellation is exact).
The inverse correlation length in the GNJL model is the mass
of the $\sigma$ boson, $\xi^{-1}=m_\sigma$.
Moreover, an important property of the planar (ladder) approximations
is that such truncations respect the vector and chiral Ward--Takahashi 
identities.

The so-called JWB method \cite{jowiba67}, 
which is a nonperturbative framework independent of the fermion wave function 
${\cal Z}$, 
allowed us to compute the contributions of the infinite set of 
planar $\sigma$ and $\pi$ exchanges to the vacuum polarization. 
The result of the computations is that the GNJL model exhibits an UV fixed point, 
$\beta_\alpha(\alpha_0)=0$, for any value of $N$ that exceeds some critical 
value $N_c$ ($N>N_c$).
This critical number of flavors turned out to be $N_c\approx 54$.
The larger the number of fermion flavors, 
the smaller the UV fixed point $\alpha_0$ will be, provided $N>N_c$.

The large value for $N_c$ puts questions to the applicability of the GNJL.
However, we have given a few arguments in the previous section suggesting 
that $N_c$ could be rather sensitive to approximations,
and that an improvement of the approximations and calculations 
will probably lead to a smaller value for $N_c$.

The mechanism responsible for the realization of an UV fixed point 
is illustrated by the observation that contributions of
planar $\sigma$ and $\pi$ exchanges to the vacuum polarization 
have identical sign, and tend to reduce screening. 
The four-fermion interactions describe attractive forces
between virtual fermion--anti-fermion pairs in the vacuum polarization.

The conventional leading term in the vacuum polarization is 
the one-loop correction describing the creation of  
fermion--anti-fermion pairs. These virtual pairs can be considered as 
dipoles causing the screening; the vacuum is a medium of the insulator type.
Such a screening is proportional to the coupling 
$\alpha_0$ and proportional to the number of fermion flavors $N$.
However, if a particular fraction of the total 
amount of fermion--anti-fermion pairs created
are correlated by attractive four-fermion interactions, represented by
$\sigma$ and $\pi$ exchanges, then clearly these
composite neutral states are not capable of screening.
Thus the negative term $N\zeta_1$ in the $\beta$ function (\ref{betasc2}) 
represents the contributions and the attractive nature of 
four-fermion interactions in the vacuum polarization.

It is known that in the quenched approximation, the critical curve and critical 
exponents are independent of the number of fermion flavors. In the quenched-ladder 
approximation using the mean-field approach for the four-fermion interactions 
it is straightforward to derive that this is indeed the case.
Clearly, the mechanism of charge screening is flavor dependent, since the
total number of virtual fermion--anti-fermion pairs is proportional to $N$ and 
the total number of composite scalars and pseudoscalars grows as $2N^2$.
The larger the number of flavors, the stronger the effect of four-fermion
interactions. 
The fixed point appears when the virtual pairs completely loose
the ability to screen. 

The existence of an UV fixed point implies a nontrivial continuum limit 
of the GNJL model with a $U(1)$ gauge interaction.
The analysis presented here suggest that in the full unquenched GNJL model 
the critical line is replaced by an UV fixed point somewhere on the critical 
line depending on the number of fermion flavors.
If the number of fermion flavors is below some specific value, the critical
four-fermion dynamics are not sufficient to yield an UV fixed point. 
In that case the unquenched GNJL model only has a trivial (IR) fixed point.

Finally, let us briefly compare our results with that obtained by 
lattice
simulations and nonperturbative RG techniques.
Although there still is some controversy between different groups
performing lattice computations,
the standard conclusion of the Desy group (G\"ockeler {\em et al.}) 
is that QED is trivial and that the chiral phase transition is of the
mean field type.
Our results agree with these conclusions for $N<N_c$.

Using nonperturbative RG techniques, 
it was found in Refs.~\cite{aomosuteto97,ao97}, that within a particular local 
potential approach the gauge coupling $\alpha_0$ in GNJL model
is trivial. Again for $N<N_c$ the results agree.
However, for large $N$, 
we think that, in order to compare such type of approaches with 
our results, the RG flows of operators corresponding to the dynamical nature
of $\sigma$ and $\pi$ exchanges should be taken into account.
In the context of NPRG techniques, we think this probably involves taking
into account higher (than $6$) dimensional chiral invariant operators containing 
derivatives, such as $(\partial_\mu (\bar\psi\gamma^\mu\psi))^2$.
%
%

\appendix
%
%
\chapter{GNJL model with {\protect\boldmath$U(N)$} symmetry}\label{unsymm}
The gauged NJL model with a global $U_V(N)\times U_A(N)$ symmetry is described 
by the Lagrangian (see \cite{mirbook})
\bea
{\cal L}_1&=&
\bar\psi_i (i\gm D_{\mu}-M)_{ij}\psi_j-\frac{1}{4}F_{\mu\nu}F^{\mu\nu}
\nonu\\
&+&\frac{G_0}{2}\sum_{\alpha=0}^{N^2-1}\[\(\bar \psi_i \tau^\alpha_{ij} \psi_j\)^2
+\(\bar\psi_i \tau^\alpha_{ij}(i\g5)\psi_j\)^2\],\label{gnjlun}
\eea
where $D_{\mu}=\partial_{\mu}+ie_0A_{\mu}$
and where the flavor labels $i,j$ run from $1$ to $N$. 
Of course the symmetry only applies when the (bare) mass matrix 
$M$ vanishes.
The generators, $\tau$, of the $U(N)$ Lie algebra have the following properties:
\bea
&&\tau^{\alpha\dagger}=\tau^\alpha,\qquad
\Tr \tau^\alpha \tau^\beta=\delta^{\alpha\beta},\label{trace_id}\\
&&\sum_{\alpha=0}^{N^2-1}\tau^\alpha_{ij} \tau^\alpha_{kl}
=\delta_{il}\delta_{kj}.\label{Fierz}
\eea 
The last identity is called the Fierz identity.
The Lagrangian is invariant under global $U_V(N)$ transformations \footnote{We use the standard summation convention, {\em i.e.} sum over all double indices.}:
\bea
\psi &\longrightarrow&\psi^\prime_i=\[\exp
\(i\theta^\alpha \tau^\alpha\)\]_{ij}\psi_j,\\
\bar\psi &\longrightarrow&
\bar\psi^\prime_i=\bar\psi_j\[
\exp\(-i\theta^\alpha \tau^\alpha\)\]_{ji}.
\eea
and under global $U_A(N)$ transformations:
\bea
\psi(x) &\longrightarrow&\psi^\prime_i(x)=
\[\exp\(i\g5\theta^\alpha \tau^\alpha\)\]_{ij}
\psi_j(x),\\
\bar\psi(x) &\longrightarrow&
\bar\psi^\prime_i(x)=\bar\psi_j(x)\[\exp\(i\g5\theta^\alpha \tau^\alpha\)\]_{ji}.
\eea

In terms of auxiliary spinless fields, $\sigma^\alpha$ and $\pi^\alpha$, 
the Lagrangian Eq.~(\ref{gnjlun}) is rewritten as
\bea
{\cal L}_2&=&\sum_{i=1}^N\bar\psi_i i\gamma^\mu D_\mu\psi_i
-\frac{1}{4}F_{\mu\nu}F^{\mu\nu}-\sum_{i,j=1}^N\sum_{\alpha=0}^{N^2-1}
\bar\psi_i\tau^\alpha_{ij}(\sigma^\alpha+i\gamma_5\pi^\alpha)\psi_j\nonu\\
&-&\frac{1}{2G_0}\sum_{\alpha=0}^{N^2-1}\[ (\sigma^\alpha)^2+(\pi^\alpha)^2)
-2 c^\alpha\sigma^\alpha\], \label{unaux}
\eea
where the Euler--Lagrange equations for the auxiliary fields are constraints
\bea
\sigma^\alpha&=&-G_0\bar\psi_i\tau^\alpha_{ij} \psi_j,\\
\pi^\alpha&=&-G_0\bar\psi_i\tau^\alpha_{ij}(i\g5)\psi_j,
\eea
and $c^\alpha=\Tr\[ M \tau^\alpha\]$.

The infinitesimal $U_V(N)$ transformations for the Lagrangian 
Eq.~(\ref{unaux}) are
\bea
&&\delta \psi_i(x)= i\theta^\beta \tau^\beta_{ij}\psi_j(x),\qquad
\delta \bar\psi_i(x)= -\bar\psi_j(x)i\theta^\beta \tau^\beta_{ji},\\
&&\delta\sigma^\alpha(x)= f^{\alpha\beta\gamma}i\theta^\beta\sigma^\gamma(x),\qquad
\delta\pi^\alpha(x)=f^{\alpha\beta\gamma}i\theta^\beta\pi^\gamma(x),
\eea
where $\theta$ is small,
and the infinitesimal $U_A(N)$ transformations are
\bea
&&\delta \psi_i(x)= i\g5\theta^\beta \tau^\beta_{ij}\psi_j(x),\qquad
\delta \bar\psi_i(x)= \bar\psi_j(x)i\g5\theta^\beta \tau^\beta_{ji},\\
&&\delta\sigma^\alpha(x)= g^{\alpha\beta\gamma}\theta^\beta\pi^\gamma(x),\qquad
\delta\pi^\alpha(x)=- g^{\alpha\beta\gamma}\theta^\beta\sigma^\gamma(x).
\eea
The structure constants are defined as
\bea
\[\tau^\alpha,\tau^\beta\]&=&f^{\alpha\beta\gamma} \tau^\gamma,\\
\{\tau^\alpha,\tau^\beta\}&=&g^{\alpha\beta\gamma} \tau^\gamma.
\eea

%
%
\chapter{Green functions, propagators, and proper vertices}\label{gfprpv}
In this appendix we define the Green function, connected Green functions both
in coordinate and momentum space, and we define the propagators
and proper vertices of the GNJL model.
The fermion fields are labeled with a Dirac index (first subscript $a,b,c,d$) and
a flavor index (second subscript $f,i,j,l,k$).
The scalar and pseudoscalar fields are labeled with flavor indices 
$\alpha$ and $\beta$.
Spinor indies run from $1$ to $4$, fermion flavor indices from $1$ to $N$
and scalar resp. pseudoscalar indices from $0$ to $N^2-1$.

In what follows ${\bf J}$ represents the set of sources
${\bf J}=(J,\eta,\bar\eta, J_\sigma, J_\pi)$.
\section{Green functions of the GNJL model}
The Green function are defined as time order vacuum-to-vacuum expectation values
of the fields. 
First we define the appropriate one-point Green functions:
\bea
iD^{(1)}_{^{\rm S}_\alpha}&=&\langle0|\sigma^\alpha |0\rangle
=\frac{1}{i}\frac{\delta Z[{\bf J}]}{
\delta J_\sigma^\alpha(x)}\biggr|_{{\bf J}=0},\\
iD^{(1)}_{^{\rm P}_\alpha}&=&\langle0|\pi^\alpha |0\rangle=
\frac{1}{i}\frac{\delta Z[{\bf J}]}{
\delta J_\pi^\alpha(x)}\biggr|_{{\bf J}=0}.
\eea
The one-point function are independent of coordinates $x$ due to invariance 
under space-time translation.
The two point Green functions are
\bea
iD^{(2)}_{\mu\nu}(z,z')&\equiv&\langle0|T\(A_{\mu}(z)A_{\nu}(z')\)|0\rangle\nonu\\
&=&\(\frac{1}{i}\)^2\frac{\delta^2 Z[{\bf J}]}{\delta J^{\nu}(z')\delta J^{\mu}(z)}
\biggr|_{{\bf J}=0},\\
iD^{(2)}_{^{ab}_{ij}}(x,y)&\equiv&
\langle0|T\(\psi_{ai}(x)\bar\psi_{bj}(y)\)|0\rangle\nonu\\
&=& \(\frac{1}{i}\)^2 \frac{\delta^2 Z[{\bf J}]}{\delta
\eta_{bj}(y) \delta\bar\eta_{ai}(x)}
\biggr|_{{\bf J}=0},\\
iD^{(2)}_{^{\rm SS}_{\alpha\beta}}(x,y)&\equiv&
\langle0|T\(\sigma^\alpha(x)\sigma^\beta(y)\) |0\rangle\nonu\\
&=& \(\frac{1}{i}\)^2 \frac{\delta^2 Z[{\bf J}]}{
\delta  J_\sigma^\beta(y) \delta J_\sigma^\alpha(x)}
\biggr|_{{\bf J}=0},\\
iD^{(2)}_{^{\rm PP}_{\alpha\beta}}(x,y)&\equiv&
\langle0|T\(\pi^\alpha(x)\pi^\beta(y)\) |0\rangle\nonu\\
&=& \(\frac{1}{i}\)^2 \frac{\delta^2 Z[{\bf J}]}{\delta  
J_\pi^\beta(y) \delta J_\pi^\alpha(x)}
\biggr|_{{\bf J}=0},
\eea
the three-point Green functions are
\bea
iD^{(3)}_{^{ab}_{ij}\mu}(x,y,z)&\equiv&
\langle0|T\(\psi_{ai}(x)\bar\psi_{bj}(y) A_{\mu}(z)\) |0\rangle\nonu\\
&=&\(\frac{1}{i}\)^3 \frac{\delta^3 Z[{\bf J}]}{\delta J^{\mu}(z) \delta\eta_{bj}(y) \delta\bar\eta_{ai}(x)}
\biggr|_{{\bf J}=0},\\
iD^{(3)}_{^{ab{\rm S}}_{ij\alpha}}(x,y,z)&\equiv&
\langle0|T\(\psi_{ai}(x)\bar\psi_{bj}(y)\sigma^{\alpha}(z)\) |0\rangle\nonu\\
&=& \(\frac{1}{i}\)^3 \frac{\delta^3 Z[{\bf J}]}{
\delta  J_\sigma^{\alpha}(z) \delta\eta_{bj}(y) \delta\bar\eta_{ai}(x)}
\biggr|_{{\bf J}=0},\\
iD^{(3)}_{^{ab{\rm P}}_{ij\alpha}}(x,y,z)&\equiv&
\langle0|T\(\psi_{ai}(x)\bar\psi_{bj}(y)\pi^{\alpha}(z)\) |0\rangle\nonu\\
&=& \(\frac{1}{i}\)^3 \frac{\delta^3 Z[{\bf J}]}{
\delta  J_\pi^{\alpha}(z) \delta\eta_{bj}(y) \delta\bar\eta_{ai}(x)}
\biggr|_{{\bf J}=0},
\eea
and the four-point Green function is
\bea
&&
iD^{(4)}_{^{ab,cd}_{i_1j_1,i_2j_2}}(x_1,y_1,x_2,y_2)\equiv
\langle0|T\(\psi_{a i_1}(x_1)\psi_{c i_2}(x_2)\bar\psi_{b j_1}(y_1)
\bar\psi_{d j_2}(y_2) \) |0\rangle\nonu\\
&&\qquad=\(\frac{1}{i}\)^4 \frac{\delta^4 Z[{\bf J}]}{\delta
\eta_{d j_2}(y_2)\delta\eta_{b j_1}(y_1)\delta\bar\eta_{c i_2}(x_2)\delta\bar\eta_{a i_1}(x_1)}\biggr|_{{\bf J}=0}.
\nonu\\
\eea

\section{Connected Green functions}
By making use of Wick's theorem, we derive the connected Green functions 
from the Green function of the previous section
by subtracting all disconnected parts, for instance:
\bea
\langle0|T\(\sigma^\alpha(x)\sigma^\beta(y)\) |0\rangle&=&
\langle0|\sigma^\alpha(x)|0\rangle \langle0|\sigma^\beta(y)|0\rangle\nonu\\
&&+
\langle0|T\(\sigma^\alpha(x)\sigma^\beta(y)\) |0\rangle_{\rm connected}.
\eea
Thus we define
\bea
iD_{\mu\nu}(z-z')&=&iD^{(2)}_{\mu\nu}(z,z'),\\
\delta_{ij} iS^{(i)}_{ab}(x-y)&=&iD^{(2)}_{^{ab}_{ij}}(x,y),\\
i\DelS^{(\alpha)}(x-y)
&=&iD^{(2)}_{^{\rm SS}_{\alpha\alpha}}(x,y)-\langle 0| \sigma^\alpha|0\rangle^2,\\
i\DelP^{(\alpha)}(x-y)&=&
iD^{(2)}_{^{\rm PP}_{\alpha\alpha}}(x,y)-\langle 0| \pi^\alpha|0\rangle^2,\\
\delta_{ij}iC^{(3)(i)}_{ab\mu}(x,y,z)&=&iD^{(3)}_{^{ab}_{ij}\mu}(x,y,z),\\
iC^{(3)}_{^{ab{\rm S}}_{ij\alpha}}(x,y,z)
&=&iD^{(3)}_{^{ab{\rm S}}_{ij\alpha}}(x,y,z)
-\delta_{ij}iS^{(i)}_{ab}(x-y)\langle 0| \sigma^\alpha|0\rangle,\\
iC^{(3)}_{^{ab{\rm P}}_{ij\alpha}}(x,y,z)
&=&iD^{(3)}_{^{ab{\rm P}}_{ij\alpha}}(x,y,z)
-\delta_{ij}iS^{(i)}_{ab}(x-y)\langle 0| \pi^\alpha|0\rangle,
\eea
and the connected four-point function
\bea
iC^{(4)}_{^{ab,cd}_{i_1j_1,i_2j_2}}(x_1,y_1,x_2,y_2)&=&
iD^{(4)}_{^{ab,cd}_{i_1j_1,i_2j_2}}(x_1,y_1,x_2,y_2)\nonu\\
&-&\delta_{i_2j_1} \delta_{i_1j_2} iS^{(i_2)}_{cb}(x_2-y_1)iS^{(i_1)}_{ad}(x_1-y_2)
\nonu\\
&+&\delta_{i_1j_1} \delta_{i_2j_2} iS^{(i_1)}_{ab}(x_1-y_1)
iS^{(i_2)}_{cd}(x_2-y_2).
\eea
\section{Fourier transforms}
For the two-point functions, we have the following Fourier transforms:
\bea
iD_{\mu\nu}(z-z')&=&\int\frac{{\rm d}^4k}{(2\pi)^4}{\,\rm e}^{-ik(z-z')}iD_{\mu\nu}(k),\\
iS^{(f)}_{ab}(x-y)&=&\int\frac{{\rm d}^4k}{(2\pi)^4}{\,\rm e}^{-ik(x-y)}
iS^{(f)}_{ab}(k),\label{deffpmom}\\
i\DelS^{(\alpha)}(x-y)&=&
\int\frac{{\rm d}^4k}{(2\pi)^4}{\,\rm e}^{-ik(x-y)}i\DelS^{(\alpha)}(k),\\
i\DelP^{(\alpha)}
(x-y)&=&\int\frac{{\rm d}^4k}{(2\pi)^4}{\,\rm e}^{-ik(x-y)}
i\DelP^{(\alpha)}(k).
\eea
The Fourier transform of the vertices are
\bea 
iC^{(3)(f)}_{ab\mu}(x,y,z)
&=&\int\frac{{\rm d}^4k{\rm d}^4p}{(2\pi)^8}
{\,\rm e}^{-ik(x-z)+ip(y-z)}iD_{\nu\mu}(k-p)\nonu\\
&\times&
\[iS^{(f)}(k)(-ie_0)\Gamma^{(f)\nu}(k,p)iS^{(f)}(p)\]_{ab}
,\\
iC^{(3)}_{^{ab{\rm S}}_{ij\alpha}}(x,y,z)
&=&\int\frac{{\rm d}^4k{\rm d}^4p}{(2\pi)^8}
{\,\rm e}^{-ik(x-z)+ip(y-z)}i\DelS^{(\alpha)}(k-p)\nonu\\
&\times&
\[iS^{(i)}(k)(-i)
{\Gamma_{{\rm S}ij}^{\alpha}}(k,p)iS^{(j)}(p)\]_{ab},\\
iC^{(3)}_{^{ab{\rm S}}_{ij\alpha}}(x,y,z)
&=&\int\frac{{\rm d}^4k{\rm d}^4p}{(2\pi)^8}
{\,\rm e}^{-ik(x-z)+ip(y-z)}i\DelP^{(\alpha)}(k-p)\nonu\\
&\times&
\[iS^{(i)}(k)(-i)\Gamma_{{\rm P}ij}^{\alpha}(k,p)iS^{(j)}(p)\]_{ab},
\eea
where the $\Gamma$'s represent the amputated vertices or proper vertices 
in momentum space.
Finally, the connected four-point function has the Fourier transform
\bea
&&
\!\!\!\!\!\!\!
iC^{(4)}_{^{ab,cd}_{i_1j_1,i_2j_2}}(x_1,y_1,x_2,y_2)=
\int\frac{{\rm d}^4k_1{\rm d}^4p_1{\rm d}^4k_2}{(2\pi)^{12}}
{\,\rm e}^{-ik_1(x_1-y_2)+ip_1(y_1-y_2)-ik_2(x_2-y_2)}\nonu\\
&&
\!\!\!\!\!\!\!
\times\,
iS^{(i_1)}_{aa'}(k_1)iS^{(i_2)}_{cc'}(k_2)(-ie_0^2)
K^{(0)}_{^{a'b',c'd'}_{i_1j_1,i_2j_2}}(k_1,p_1,k_2)
iS^{(j1)}_{b'b}(p_1)iS^{(j2)}_{d'd}(k_1-p_1+k_2).\nonu\\
\eea
\section{The Bethe--Salpeter scattering kernel}
The 2-fermion 1-photon-scalar-pseudoscalar 
irreducible Bethe--Salpeter kernel is defined via the
Bethe--Salpeter equation
\bea
&&(-ie_0^2)K^{(1)}_{^{ab,cd}_{i_1j_1,i_2j_2}}(k_1,p_1,k_2)
=(-ie_0^2)K^{(2)}_{^{ab,cd}_{i_1j_1,i_2j_2}}(k_1,p_1,k_2)\nonu\\
&&\qquad+
\sum_{l=1}^N\sum_{m=1}^N\int\frac{{\rm d}^4 r}{(2\pi)^4}\,
(-ie_0^2)K^{(1)}_{^{ab,c'd'}_{i_1j_1,lm}}(k_1,p_1,r+p_1)
iS^{(m)}_{d'a'}(r+k_1)\nonu\\
&&\qquad\qquad\times
(-ie_0^2)K^{(2)}_{^{a'b',cd}_{ml,i_2j_2}}(r+k_1,r+p_1,k_2)
iS^{(l)}_{b'c'}(r+p_1), \label{K2def}
\eea
where $K^{(1)}$ is defined as the 1-particle irreducible part of $K^{(0)}$, and is defined explicitly as follows:
\bea
&&(-ie_0^2)K^{(0)}_{^{ab,cd}_{i_1j_1,i_2j_2}}(k_1,p_1,k_2)
=(-ie_0^2)K^{(1)}_{^{ab,cd}_{i_1j_1,i_2j_2}}(k_1,p_1,k_2)\nonu\\
&&-\delta_{i_1j_1}(-ie_0)\Gamma^{(i_1)\mu}_{ab}(k_1,p_1)
iD_{\mu\nu}(k_1-p_1)\delta_{i_2j_2}
(-ie_0)\Gamma^{(i_2)\nu}_{cd}(k_2,k_1-p_1+k_2)\nonu\\
&&-\sum_{\alpha=0}^{N^2-1}(-i)\Gamma^{(\alpha)}_{{\rm S}^{ab}_{i_1j_1}}(k_1,p_1)
i\DelS^{(\alpha)}(k_1-p_1)
(-i)\Gamma^{(\alpha)}_{{\rm S}^{cd}_{i_2j_2}}(k_2,k_1-p_1+k_2)\nonu\\
&&-\sum_{\alpha=0}^{N^2-1}(-i)\Gamma^{(\alpha)}_{{\rm P}^{ab}_{i_1j_1}}(k_1,p_1)
i\DelP^{(\alpha)}(k_1-p_1)
(-i)\Gamma^{(\alpha)}_{{\rm P}^{cd}_{i_2j_2}}(k_2,k_1-p_1+k_2).
\label{K1def}
\eea
Note that the minus signs in the three one-particle exchange graphs 
on the righthand side come from the fact that such types of diagrams 
are related to the expectation value 
$\langle\psi_a\bar\psi_b\psi_c\bar\psi_d\rangle
=-\langle\psi_a\psi_c\bar\psi_b\bar\psi_d\rangle$.

\begin{figure}[ht!]
\epsfxsize=9cm
\epsffile[-10 440 390 540]{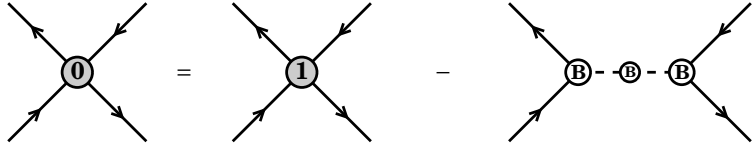}
\caption{Definition of the 1-boson irreducible scattering kernel, $K^{(1)}$.}
\label{fig_K1def}
\end{figure}

\begin{figure}[ht!]
\epsfxsize=9cm
\epsffile[-10 440 390 540]{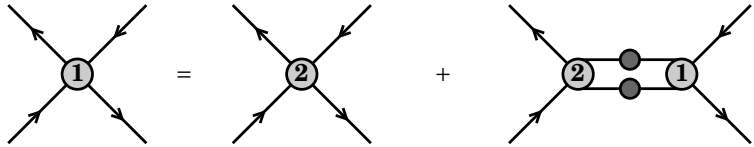}
\caption{Definition of the Bethe--Salpeter kernel, $K^{(2)}$.}
\label{fig_K2def}
\end{figure}

Eqs.~(\ref{K1def}) and (\ref{K2def}) are depicted in 
Figs.~\ref{fig_K1def} and \ref{fig_K2def}.
%
%

%
\chapter{Feynman rules}\label{feynmanrules}
\begin{itemize}
\item{The bare fermion propagator:
\bea
\barefermionprop \, =i\delta_{ij}\frac{(\pslash+m_{(i)})_{ab}}{p^2-m_{(i)}^2},
\eea
where $M_{ij}=m_{(i)} \delta_{ij}$, the diagonal mass matrix.
}
\item{
The full fermion propagator:
\bea
\fermionprop \, =
i\delta_{ij} S^{(i)}_{ab}(p).
\eea
}
\item{
The bare photon propagator:
\bea
\barephotonprop \!\!=\frac{i}{q^2}\[-g_{\mu\nu}+\frac{q_\mu q_\nu}{q^2}\],
\eea
in the Landau gauge.
}
\item{
The full photon propagator:
\bea
\photonprop \!\!=iD_{\mu\nu}(q).
\eea}
\item{
The bare scalar and pseudoscalar propagators are identical:
\bea
\barescalarprop \!\!\!\! = -iG_0.
\eea
}
\item{
The full scalar propagator:
\bea
\scalarprop =
i\Delta^{(\alpha)}_{\rm S}(q).
\eea}
\item{
The full pseudoscalar propagator:
\bea
\pseudoprop =
i\Delta^{(\alpha)}_{\rm P}(q).
\eea}
\item{
The bare photon-fermion vertex:
\bea
&&\nonu\\
&&\barevertex=(-ie_0)\delta_{ij}\gamma^\mu_{ab}.
\\
&&\nonu
\eea
}
\item{
The full photon-fermion vertex:
\bea
&&\nonu\\
&&\vertex=(-ie_0)\delta_{ij}\Gamma^{(i)\mu}_{ab}(k,p).\\
&&\nonu
\eea
}
\item{Either the bare photon-fermion, bare scalar or bare pseudoscalar vertex: 
\bea
&&\nonu\\
\barescalarvertex&=&
(-ie_0)\delta_{ij}\gamma^\mu_{ab},\quad 
(-i)\tau^\alpha_{ij}({\bf 1})_{ab},\nonu\\
&\mbox{or}& (-i)\tau^\alpha_{ij}(i\g5)_{ab}.\\
&&\nonu
\eea
}
\item{
The full scalar vertex:
\bea
&&\nonu\\
&&\scalarvertex=(-i)\Gamma^{\alpha}_{{\rm S}^{ab}_{ij}}(k,p).\\
&&\nonu
\eea
}
\item{
The full pseudoscalar vertex:
\bea
&&\nonu\\
&&\pseudovertex=(-i)\Gamma^{\alpha}_{{\rm P}^{ab}_{ij}}(k,p).\\
&&\nonu
\eea
}
\item{
The Bethe--Salpeter kernel:
\bea
&&\nonu\\
\bethesalpeter=
(-ie_0^2)K^{(2)}_{^{ab,cd}_{ij,mn}}(k,p,q).\\
&&\nonu
\eea
}
\item{A Minkowskian four-momentum integral for each closed loop:
\bea
\int_{\Lambda}\frac{{\rm d}^4k}{(2\pi)^4}
\eea 
}
\item{An extra minus sign for each closed fermion-loop:
\bea
(-1)
\eea
}
\item{For each internal fermion line a sum over all flavors: 
\bea
\sum_{i=1}^N
\eea
}
\item{For each internal `boson' line a sum
over all bosonic propagators:
\bea
\bosonprop&=&iD_{\mu\nu}(q)+
\sum_{\alpha=0}^{N^2-1}\[ i\Delta^{(\alpha)}_{\rm S}(q)
                         +i\Delta^{(\alpha)}_{\rm P}(q)\].
\eea
}
\item{The `boson' vertex, internal or external, represents either
one of the vertices:
\bea
&&\nonu\\
\bosonvertex&=&
(-ie_0)\delta_{ij}\Gamma^{(i)\mu}_{ab}(k,p),
\quad
(-i)\Gamma^{\alpha}_{{\rm S}^{ab}_{ij}}(k,p),\nonu\\
&\mbox{or}&(-i)\Gamma^{\alpha}_{{\rm P}^{ab}_{ij}}(k,p).
\\
&&\nonu
\eea
}
\item{The bare axial-vector vertex:
\bea
&&\nonu\\
&&\bareaxialvectorvertex=(-i)(\gamma^\mu\g5)_{ab}.\\
&&\nonu
\eea
}
\item{
The full axial-vector vertex ($N=1$):
\bea
&&\nonu\\
&&\axialvectorvertex=(-i)\Gamma^\mu_{5ab}(k,p).\\
&&\nonu
\eea
}
\end{itemize}

%
\chapter{Analysis of the Chebyshev expansion}\label{Chebyshev}
In this appendix we discuss the validity of the zeroth-order 
Chebyshev expansion for the Yukawa vertex function $F_1$ introduced in 
section~\ref{sec_scal_vertex}.
The problem of angular dependence in the SDEs for the Yukawa vertex functions
$F_1$ and $F_2$ is replaced by an infinite set of Chebyshev harmonics.
Subsequently this set is truncated to the lowest order harmonic of $F_1$, 
which is the only harmonic having nonhomogeneous 
ultraviolet boundary conditions because of the presence of the 
(angular independent) inhomogeneous term $1$.

As mentioned previously, the method of using expansions 
in terms of Chebyshev polynomials $U_n(x)$ (of the second kind) 
was used before \cite{fogumisi83,jamu93,mawa96}. 
These polynomials are orthogonal with respect to the angular integration 
$\int {\rm d} \Omega$. 
In the analysis of BSEs in Ref.~\cite{fogumisi83} a {\it CP} 
invariant Chebyshev expansion
was used, which has the nice property of keeping only even terms in the 
expansion. 
However, we use a slightly different expansion (not explicitly {\it CP} invariant)
which has the disadvantage of also including odd terms in the Chebyshev 
expansion, but the advantages that the integral equation for the zeroth 
order harmonic is more ``friendly'' and the zeroth order harmonic 
coincides with both the large fermion momentum limit ($p^2\gg q^2$) as well 
the large boson-momentum limit ($q^2\gg p^2$) of the Yukawa vertex, see 
Fig.~\ref{two_channel_fig}.

Thus the vertex functions satisfying the SDEs (\ref{sde_vertfies})
are expanded in the angle between fermion momentum $p$ 
and scalar boson $q$, {\em i.e.}, $p\cdot q$, in the following way:
\bea
F_1(p+q,p)&=&\sum_{n=0}^\infty f_n(p^2,q^2) U_n(\cos\alpha),\\
F_2(p+q,p)&=&\sum_{n=0}^\infty g_n(p^2,q^2) U_n(\cos\alpha),\\
\frac{1}{(r-p)^2}&=&\sum_{n=0}^\infty N_n(r^2,p^2) U_n(\cos\beta),
\label{denom}\\
A_1(r,q)&=&\sum_{n=0}^\infty a_n(r^2,q^2) U_n(\cos\gamma),\label{A1cheb}\\
A_2(r,q)&=&\sum_{n=0}^\infty b_n(r^2,q^2) U_n(\cos\gamma),\label{A2cheb}
\eea
where 
\bea
\cos \alpha=\frac{p\cdot q}{p q},\quad
\cos \beta=\frac{p\cdot r}{p r},\quad
\cos \gamma=\frac{q\cdot r}{q r}.
\eea
The vertex functions and kernels $A_1$ and $A_2$
were defined in Eq.~(\ref{vertfiesdef}), respectively, Eq.~(\ref{A1A2def}).
The coefficients $N_n$, $a_n$, and $b_n$ are 
\bea
N_n(r^2,p^2)=\frac{\theta(r^2-p^2)}{r^2}\left(\frac{p}{r}\right)^n
+\frac{\theta(p^2-r^2)}{p^2}\left(\frac{r}{p}\right)^n,\label{Nn}
\eea
and
\bea
a_0(r^2,q^2)&=&\frac{1}{2}\left[\left(2-\frac{q^2}{r^2}\right)\theta(r^2-q^2)
+\frac{r^2}{q^2} \theta(q^2-r^2)\right],\label{A0}\\
a_n(r^2,q^2)&=&(-1)^n\frac{(r^2-q^2)}{2}
\left[
\frac{\theta(r^2-q^2)}{r^2}\left( \frac{q}{r}\right)^n
+\frac{\theta(q^2-r^2)}{q^2}\left(\frac{r}{q}\right)^n\right],
\label{an}\nonu\\
&&n\geq 1, 
\eea
and 
\bea
b_0(r^2,q^2)&=&\frac{1}{2}
\left[\frac{q^2(q^2-3 r^2)}{r^2}\theta(r^2-q^2)
+\frac{r^2(r^2-3q^2)}{q^2} \theta(q^2-r^2)\right],\label{B0}\\
b_1(r^2,q^2)&=&-\frac{1}{2}
\left[
\frac{q^3(q^2-2r^2)}{r^3} \theta(r^2-q^2)
+\frac{r^3(r^2-2q^2)}{q^3} \theta(q^2-r^2)\right],\\
b_n(r^2,q^2)&=&(-1)^n\frac{(r^2-q^2)^2}{2}\left[
\frac{\theta(r^2-q^2)}{r^2}\left( \frac{q}{r}\right)^n 
+\frac{\theta(q^2-r^2)}{q^2}\left(\frac{r}{q}\right)^n\right],\nonu\\
&&n\geq 2.
\eea

The equations for the scalar vertex functions (\ref{sde_vertfies}) 
and scalar vacuum polarization (\ref{sde_vacpol2}) are expressed
in terms of an infinite set of equations between the harmonics.
Hence
\bea
\Pi_{\rm S}(q^2)=\frac{1}{4\pi^2}\int\limits_0^{\Lambda^2}
{\rm d} k^2\,\sum_{n=0}^\infty 
\left[ a_n(k^2,q^2) f_n(k^2,q^2)+b_n(k^2,q^2) g_n(k^2,q^2)\right],
\label{vacpolcheb}
\eea
and with Eqs.~(\ref{k11}) and (\ref{k12}) using 
Eqs.~(\ref{denom}) and (\ref{A2cheb}), we get for the harmonics of $F_1$
\bea
f_l(p^2,q^2)&=&\delta_{0,l}+\frac{\lambda_0}{(l+1)}
\int\limits_0^{\Lambda^2} {\rm d}r^2\,N_l(r^2,p^2)
\sum_{m=0}^\infty\sum_{n=0}^\infty C_{lmn} \nonu\\
&\times&\left[ 
a_m(r^2,q^2) f_n(r^2,q^2)+b_m(r^2,q^2) g_n(r^2,q^2)\right], 
\label{coefFl1}
\eea
where
\bea
C_{lmn}\equiv\frac{2}{\pi}\int\limits_0^\pi{\rm d} \gamma\,
\sin^2\gamma\,U_l(\cos\gamma)U_m(\cos\gamma)U_n(\cos\gamma).
\eea
In the derivation of Eq.~(\ref{coefFl1}) use has been made of 
the fact that
\bea
\frac{1}{\pi\sin\gamma}\int\limits_{-1}^1{\rm d}\cos\alpha\,
\int\limits_{\cos(\alpha+\gamma)}^{\cos(\alpha-\gamma)}
{\rm d}\cos\beta\, 
U_m(\cos\alpha)U_n(\cos\beta)=
\delta_{m,n}\frac{U_n(\cos\gamma)}{n+1}.
\eea
The symmetric index $C_{lmn}$ can be calculated using product 
properties of the Chebyshev polynomials, giving
\bea
C_{(2l) mn}&=&\sum_{k=0}^l\left(\delta_{2k,|m-n|}-\delta_{2k,n+m+2}\right),\nonu\\
C_{(2l+1) mn}&=&
\sum_{k=0}^l\left(\delta_{2k+1,|m-n|}-\delta_{2k+1,n+m+2}\right),
\qquad l=0,1,2,\dots.\label{Clmn}
\eea
Then the first two equations for the coefficients $f_n(p^2,q^2)$ 
of vertex function $F_1$ 
read
\bea
f_0(s,t)&=&1+\lambda_0
\int\limits_0^{\Lambda^2} {\rm d}u\,N_0(u,s)
\sum_{m=0}^\infty \biggr[ a_m(u,t) f_m(u,t)\nonu\\
&+&b_m(u,t) g_m(u,t)\biggr],
\label{f0eq}\\
f_1(s,t)&=&\frac{\lambda_0}{2}
\int\limits_0^{\Lambda^2} {\rm d}u\,N_1(u,s)\sum_{m=0}^\infty
\biggr[ a_{m+1}(u,t) f_m(u,t)+a_m(u,t) f_{m+1}(u,t)\nonu\\
&+& b_{m+1}(u,t) g_m(u,t)+b_m(u,t) g_{m+1}(u,t)
\biggr],
\eea
where we have introduced the variables
\bea
s=p^2,\qquad t=q^2,\qquad u=r^2.
\eea
In principle there is an equivalent set of equations for the 
coefficients $g_n$ of $F_2$.
However we did not succeed in finding explicit expression for these, 
due to our inability to compute explicitly the angular integrals given
by kernels $K_{21}$ and $K_{22}$ of Eqs.~(\ref{k21}) and(\ref{k22}).
The problem is to compute the integrals
\bea
\int\frac{{\rm d}\Omega_p}{2\pi^2}
\int\frac{{\rm d}\Omega_r}{2\pi^2}  U_n(\cos\alpha) K_{21}(p,q,r),\quad
\int\frac{{\rm d}\Omega_p}{2\pi^2}
\int\frac{{\rm d}\Omega_q}{2\pi^2} U_n(\cos\alpha) K_{21}(p,q,r).
\eea

The main approximation used in chapter~\ref{chap4} is
Eq.~(\ref{canonic}), {\em i.e.} 
the replacement of the Yukawa vertex by the zeroth-order
harmonic of $F_1$.
In what follows we estimate the error made by such an approximation.
We define the error $E(q^2)$ in computation of the scalar vacuum 
polarization Eq.~(\ref{vacpolcheb}) as follows:
\bea
\Pi_{\rm S}(q^2)=\frac{1}{4\pi^2}\int\limits_0^{\Lambda^2}
{\rm d} k^2\, a_0(k^2,q^2) f_0(k^2,q^2)+E(q^2),
\eea
where
\bea
E(q^2)\equiv
\frac{1}{4\pi^2}\int\limits_0^{\Lambda^2}
{\rm d} k^2\,\left[
\sum_{n=1}^\infty a_n(k^2,q^2) f_n(k^2,q^2)
+ \sum_{n=0}^\infty b_n(k^2,q^2) g_n(k^2,q^2)\right].\label{error}
\eea
For an estimation of $E(q^2)$ we need to know more about the harmonics
$f_n$, $n\geq 1$, and $g_n$, $n\geq 0$.
The solution to these harmonics is assumed 
to be governed by the harmonic $f_0$ only, 
since the integral over the harmonic $f_0$ acts as the largest inhomogeneous 
term in the integral equations for the higher order harmonics. 
So the equations for the higher order harmonics are approximated by
\bea
\!\!\!\!g_n(p^2,q^2)
&\approx&
\lambda_0 \int\limits_0^{\Lambda^2}{\rm d}r^2\,
\int\frac{{\rm d}\Omega_p}{2\pi^2}
\int\frac{{\rm d}\Omega_r}{2\pi^2} \,
U_n(\cos\alpha)K_{21}(p,q,r) f_0(r^2,q^2),\label{gneq}\\
\!\!\!\!f_n(p^2,q^2)&\approx&\frac{\lambda_0}{(n+1)}
\int\limits_0^{\Lambda^2} {\rm d}r^2\,
N_n(r^2,p^2) a_n(r^2,q^2) f_0(r^2,q^2),\qquad n\geq 1.\label{fneq}
\eea
Unfortunately there is no explicit expression for Eq.~(\ref{gneq}) for the reason 
described above. However it is possible to approximate the angular average by 
considering either one of the three momenta in $K_{21}(p,q,r)$
to be much smaller than the other two. Then the dependence on one of the
three angles between the momenta is lost, and the integration can be performed
explicitly. 
The result for the lowest harmonic of $F_2$, {\em i.e.}, $g_0$ given by 
Eq.~(\ref{gneq}), is
\bea
(s<t)\qquad
g_0(s,t)
&\approx&
\frac{\lambda_0}{12}
\int_0^s {\rm d}u\,\frac{u}{s^2 t}
 F_{\rm IR}(u,t)
+\frac{\lambda_0}{12}
\int_s^t {\rm d}u\,\frac{1}{s t}
 F_{\rm IR}(u,t)\nonu\\
&+&\frac{\lambda_0}{12}\int_t^{\Lambda^2} {\rm d}u\,\frac{1}{u^2}
 F_{\rm UV}(u,t),\\
(s>t)\qquad g_0(s,t)&\approx&
\frac{\lambda_0}{12}
\int_0^t {\rm d}u\,\frac{u}{s^2 t}
 F_{\rm IR}(u,t)+
\frac{\lambda_0}{12}
\int_t^s {\rm d}u\,\frac{1}{s^2}
 F_{\rm UV}(u,t)\nonu\\
&+&\frac{\lambda_0}{12}
\int_s^{\Lambda^2} {\rm d}u\,\frac{1}{u^2}
 F_{\rm UV}(u,t),
\eea
and for Eq.~(\ref{fneq}) with Eqs~(\ref{Nn}) and (\ref{an})
\bea
(s<t)\qquad
f_1(s,t)
&\approx&
-\frac{\lambda_0}{4}
\int_0^s {\rm d}u\,
\frac{u-t}{st}\sqrt{\frac{u^2}{st}} 
F_{\rm IR}(u,t)\nonu\\
&-&\frac{\lambda_0}{4}\int_s^t {\rm d}u\,
\frac{u-t}{ut}\sqrt{\frac{s}{t}} 
 F_{\rm IR}(u,t)\nonu\\
&-&\frac{\lambda_0}{4}
\int_t^{\Lambda^2} {\rm d}u\,
\frac{u-t}{u^2}\sqrt{\frac{st}{u^2}} 
F_{\rm UV}(u,t),\\
(s>t)\qquad f_1(s,t)&\approx&
-\frac{\lambda_0}{4}
\int_0^t {\rm d}u\,\frac{u-t}{st}\sqrt{\frac{u^2}{st}} 
 F_{\rm IR}(u,t)\nonu\\
&-&\frac{\lambda_0}{4}
\int_t^s {\rm d}u\,
\frac{u-t}{us}
\sqrt{\frac{t}{s}} 
F_{\rm UV}(u,t)\nonu\\
&-&\frac{\lambda_0}{4}
\int_s^{\Lambda^2} {\rm d}u\,
\frac{u-t}{u^2}\sqrt{\frac{st}{u^2}} 
 F_{\rm UV}(u,t),
\eea
where $s=p^2$, $t=q^2$, and we have used Eq.~(\ref{channelapprox}).
These equations can be analyzed in detail once the solutions
for the channel functions, Eqs.~(\ref{fir2}) and (\ref{fuv2}), are known.
But for obtaining the asymptotic behavior of the harmonics $g_0$ and $f_1$
it is sufficient to use the asymptotics of the channels, {\em i.e.}, take
$F_{\rm IR}(p^2,q^2)\rightarrow 
F_{\rm IR}(0,q^2)\propto (q^2/\Lambda^2)^{\omega/2-1/2}$,
$F_{\rm UV}(p^2,q^2)\rightarrow 
F_{\rm UV}(p^2,0)\propto (p^2/\Lambda^2)^{\omega/2-1/2}$.
This gives for the harmonics 
\bea
g_0(p^2,q^2)\propto\lambda_0 \frac{1}{p^2} F_{\rm UV}(p^2,0),&&
f_1(p^2,q^2)\propto\lambda_0 \frac{q}{p} F_{\rm UV}(p^2,0),
\quad p^2\gg q^2,\label{fguv_est}\\
g_0(p^2,q^2)\propto \lambda_0\frac{1}{p^2} F_{\rm IR}(0,q^2),
&&
f_1(p^2,q^2)\propto\lambda_0 \frac{p}{q} F_{\rm IR}(0,q^2),
\quad q^2\gg p^2.\label{fgir_est}
\eea
The above equations give the leading behavior 
(in either $q/p \ll 1$ or $p/q\ll 1$) of the harmonics 
$g_0$, $f_1$ in terms of $f_0$ up to some $\lambda_0$-dependent 
factor, which is ${\cal O}(1)$ (thus nonsingular in $\lambda_0$).
Furthermore, from Eq.~(\ref{fneq}) we get the relation
\bea
\frac{f_{n+1}(p^2,q^2)}{f_n(p^2,q^2)}
\propto-\frac{q}{p},\qquad p^2\gg q^2,\qquad
\frac{f_{n+1}(p^2,q^2)}{f_n(p^2,q^2)}
\propto
-\frac{p}{q},\qquad q^2\gg p^2, \label{fnfrac}
\eea
and we assume a similar relation to hold between the harmonics 
$g_{n+1}$ and $g_n$.
Thus the series  
\bea
S(p^2,q^2)
\equiv\sum_{m=0}^\infty \left[ a_m(p^2,q^2) f_m(p^2,q^2)
+b_m(p^2,q^2) g_m(p^2,q^2)\right],\label{S}
\eea
which occurs both in the equation for $f_0$, Eq.~(\ref{f0eq}), 
and for $\Pi_{\rm S}$, Eq.~(\ref{vacpolcheb}), will be rapidly converging for 
either $p^2\gg q^2$ or $p^2\ll q^2$, since 
\bea
\frac{a_{n+1}(p^2,q^2)f_{n+1}(p^2,q^2)}{a_n(p^2,q^2)f_n(p^2,q^2)}
\approx  \frac{\min(p^2,q^2)}{\max(p^2,q^2)}\ll 1,
\eea
and again a similar equation for the part containing the harmonics $g_n$.
At $p^2=q^2$ only three terms of the series $S$, Eq.~(\ref{S}), 
contribute, since $a_n(p^2,p^2)=0$ for $n\geq 1$ 
and $b_n(p^2,p^2)=0$ for $n\geq 2$.
Hence, a straightforward approximation for the series $S$ is
\bea
S(p^2,q^2)&\approx&a_0(p^2,q^2) 
f_0(p^2,q^2)\nonu\\
&+&{\cal O}\left(a_1(p^2,q^2) f_1(p^2,q^2)\right)
+{\cal O}\left(b_0(p^2,q^2) g_0(p^2,q^2)\right),
\eea
supporting Eq.~(\ref{canonic}).
With the expressions 
obtained for $f_1$ and $g_0$, Eqs.~(\ref{fguv_est}) and (\ref{fgir_est}), 
the leading term of the error $E$ defined 
in Eq.~(\ref{error}) can be estimated.
The leading term of the error $E(q^2)$ is given by
\bea
E(q^2)&\approx&\frac{1}{4\pi^2}\int\limits_0^{\Lambda^2}{\rm d}k^2\,
\left[a_1(k^2,q^2)f_1(k^2,q^2)+b_0(k^2,q^2)g_0(k^2,q^2)\right]\nonu\\
&\sim& \lambda_0\int\limits_{q^2}^{\Lambda^2}{\rm d}k^2\,
\frac{q^2}{k^2} F_{\rm UV}(k^2,q^2=0)
+\mbox{next-to-leading}
\nonu\\
&\sim&
\Lambda^2\left[
(1+{\cal O}(\lambda_0))\left(\frac{q^2}{\Lambda^2}\right)^{\omega/2+1/2}
-(1+{\cal O}(\lambda_0))\frac{q^2}{\Lambda^2} \right],\label{error_expr}
\eea
where we have kept only leading terms
and for $F_{\rm UV}(p^2,0)=F_1(p,p)$ given by Eq.~(\ref{GSqzero}).
Recall that $\omega=\sqrt{1-4\lambda_0}$.
The estimation of Eq.~(\ref{error_expr}) can be checked more explicitly 
by using the solutions obtained in section~\ref{sec_scal_vertex}
for the channel functions $F_{\rm IR}$, $F_{\rm UV}$, 
Eqs.~(\ref{fir2}) and (\ref{fuv2}).

Eq.~(\ref{error_expr}) shows that when $\lambda_0=0$, $\omega=1$, 
the error $E$ vanishes, and when
$\omega <1$ clearly the terms in the error can be neglected with respect to
the first two terms on the right-hand side of
$\Pi_{\rm S}(q^2)$, see Eq.~(\ref{vacasym2}).
Thus this analysis supports the assumption Eq.~(\ref{canonic})
made in section~\ref{sec_scal_vertex}
and the error $E$ contributes only to next-to-next-to-leading
order in $q^2/\Lambda^2$.
And therefore we may conclude that our 
approximation gives correct leading and next-to-leading 
behavior of $\Pi_{\rm S}(q^2)$.

%
%
\chapter{Two-loop vacuum polarization}\label{appvacpol}
In this appendix we compute 
two-loop vacuum polarization corrections including $\sigma$ and $\pi$ exchanges.
\section{Two-loop vacuum polarization}
In this section, we compute the two-loop vacuum polarization in QED.
We derive the two-loop contribution by making use of the 
one-loop computation of the photon-fermion vertex \cite{bach80,kirepe95}.

The SDE for vacuum polarization tensor reads
\bea
\Pi^{\mu\nu}(q^2)=
ie_0^2\int_\Lambda\frac{{\rm d}^4k}{(2\pi)^4}\,
\Tr\[\gamma^\mu S(k+q) \Gamma^\nu(k+q,k) S(k)\].
\eea
Assuming that the WTIs are respected, 
the vacuum polarization tensor is transverse: 
\bea
\Pi^{\mu\nu}(q^2)=\[-q^2 g^{\mu\nu}+q^\mu q^\nu\] \Pi(q^2)
\eea
If so, we find that 
\bea
\Pi(q^2)=
-\frac{ie_0^2}{3q^2}\int_\Lambda\frac{{\rm d}^4k}{(2\pi)^4}\,
\Tr\[\gamma_\mu S(k+q) \Gamma^\mu(k+q,k) S(k)\].
\eea
Let us write and denote the one-loop vertex and self-energy corrections
with a subscript $(1)$ as follows
\bea
\Gamma^\mu(k,p)=\gamma^\mu+\Gamma^{\mu}_{(1)}(k,p),\qquad 
S(p)=\frac{\pslash}{p^2}\[1+{\cal Z}_{(1)}(p^2)\]. \label{1loopcors}
\eea

The vacuum polarization up to two-loop corrections can be expressed as 
\bea
\Pi(q^2)=\Pi_{(1)}(q^2)+\Pi_{(2a)}(q^2)+\Pi_{(2b)}(q^2),
\eea
where
\bea
\Pi_{(1)}(q^2)&=&
-\frac{ie_0^2}{3q^2}\int_\Lambda\frac{{\rm d}^4k}{(2\pi)^4}\,
\frac{\Tr\[\gamma_\mu (\kslash+\qslash) \gamma^\mu \kslash\]}{(k+q)^2 k^2},\\
\Pi_{(2a)}(q^2)&=&
-\frac{ie_0^2}{3q^2}\int_\Lambda\frac{{\rm d}^4k}{(2\pi)^4}\,
\frac{\Tr\[\gamma_\mu (\kslash+\qslash)\gamma^\mu \kslash 
 \]}{(k+q)^2 k^2}\nonu\\&&\times
 \[{\cal Z}_{(1)}((k+q)^2)+{\cal Z}_{(1)}(k^2)\],\label{pi2a}\\
\Pi_{(2b)}(q^2)&=&
-\frac{ie_0^2}{3q^2}\int_\Lambda\frac{{\rm d}^4k}{(2\pi)^4}\,
\frac{\Tr\[\gamma_\mu (\kslash+\qslash) 
\Gamma_{(1)}^\mu(k+q,k) \kslash\]}{(k+q)^2 k^2}.
\eea
The one-loop vacuum polarization $\Pi_{(1)}$ 
can be computed straightforwardly
\bea
\Pi_{(1)}(q^2)=\frac{\alpha_0}{3\pi}\[\log\(\frac{\Lambda^2}{q^2}\)
-2\frac{\Lambda^2}{q^2}+{\cal O}(1)\],
\eea
where $q^2$ is an Euclidean momentum.
The quadratically divergent contribution $\Lambda^2/q^2$ is
a notorious artifact of computing vacuum polarization corrections 
in the presence of a hard cutoff
({\em i.e.} an explicit cutoff in the momentum integrations instead
of Pauli--Villars regularization, see for a recent discussion Ref.~\cite{fu94}).
The quadratically divergent term is proportional to the 
$g_{\mu\nu}$ tensor in the vacuum polarization tensor, and it
ruins the transversality of $\Pi_{\mu\nu}$.
This cutoff problem can be circumvented by making 
use of a projector $g_{\mu\nu}-4q_\mu q_\nu/q^2$, which by contraction with 
the vacuum polarization tensor eliminates 
the term in $\Pi_{\mu\nu}$ proportional to the $g_{\mu\nu}$ tensor.
With such a projector we obtain
\bea
\Pi_{(1)}(q^2)=\frac{\alpha_0}{3\pi}\[\log\(\frac{\Lambda^2}{q^2}\)
+{\cal O}(1)\],\label{pi1}
\eea
which is the well-known one-loop vacuum polarization.

We write
\bea
\Gamma_{(1)}^\mu(k,p)=\Gamma_{L(1)}^\mu(k,p)
+\Gamma_{R(1)}^\mu(k,p)+\Gamma_{I(1)}^\mu(k,p),\label{v1loopdecomp}
\eea
where $\Gamma_{L(1)}^\mu$ is the one-loop longitudinal
part of the vertex, and where $\Gamma_{R(1)}$ and $\Gamma_{I(1)}^\mu$
are one-loop transverse parts:
\bea
\Gamma_{R(1)}^\mu(k,p)&=&T^\mu_8(k,p)\tau_8(k^2,p^2,q^2),\\ 
\Gamma_{I(1)}^\mu(k,p)&=&T^\mu_2(k,p)\tau_2(k^2,p^2,q^2)
                          +T^\mu_3(k,p)\tau_3(k^2,p^2,q^2)
\nonu\\&+&T^\mu_6(k,p)\tau_6(k^2,p^2,q^2),\qquad q^2=(k-p)^2. \label{GammaI}
\eea
Furthermore, we write
\bea
\Pi_{(2b)}(q^2)=\Pi_{(2L)}(q^2)+\Pi_{(2R)}(q^2)+\Pi_{(2I)}(q^2),
\eea
where
\bea
\Pi_{(2j)}(q^2)=
-\frac{ie_0^2}{3q^2}\int_\Lambda\frac{{\rm d}^4k}{(2\pi)^4}\,
\frac{\Tr\[\gamma_\mu (\kslash+\qslash) 
\Gamma_{j(1)}^\mu(k+q,k) \kslash\]}{(k+q)^2 k^2},\label{pi2j}
\eea
with $j=L,\,R,\,I$.

As was shown by Ball and Chiu \cite{bach80}, the 
(one-loop) longitudinal vertex can be written as
\bea
\Gamma_{L(1)}^\mu(k,p)&=&
\frac{\gamma^\mu}{2}\[-{\cal Z}_{(1)}(k^2)-{\cal Z}_{(1)}(p^2)\]\nonu\\
&+&\frac{(k+p)^\mu(\kslash+\pslash)}{2(k^2-p^2)}
\[-{\cal Z}_{(1)}(k^2)+{\cal Z}_{(1)}(p^2)\]. \label{ballchiuexpr}
\eea
By construction, 
the Ball--Chiu expression for the longitudinal vertex satisfies the WTI
\bea
q_\mu\Gamma_{L(1)}^\mu(k,p)=-\kslash {\cal Z}_{(1)}(k^2)
+\pslash {\cal Z}_{(1)}(p^2).
\eea

For QED, the one-loop vertex and self-energy corrections are
\bea
\Gamma^\mu_{(1)}(k,p)&=&-ie_0^2\int_\Lambda\frac{{\rm d}^4w}{(2\pi)^4}\,
\frac{\gamma^\lambda(\kslash-\wslash)\gamma^\mu(\pslash-\wslash)\gamma_\lambda}{
(k-w)^2(p-w)^2 w^2},
\eea
and
\bea
\pslash {\cal Z}_{(1)}(p^2)&=&-i e_0^2\int_\Lambda\frac{{\rm d}^4 w}{(2\pi)^4}\,
\frac{\gamma^\lambda(\pslash-\wslash)\gamma_\lambda}{(p-w)^2 w^2},
\label{selfenergyphotmom}
\eea
in the Feynman gauge.

How can one compute the contributions given in Eq.~(\ref{pi2j})?
Since the one-loop transverse vertex functions themselves are finite,
{\em i.e.} these function are independent of the cut-off $\Lambda$, 
the leading logarithmic contributions to the vacuum polarization
result from integrations over momenta $k^2 \gg q^2$ in 
$\Pi_{(2R)}$ and $\Pi_{(2I)}$. 
These leading logarithmic contribution can be found by first
deriving the $k^2\gg q^2$ asymptotic behavior of the transverse
structure functions, after which the integration over angles can be performed.
In the Feynman gauge $\xi=1$,
the asymptotic behavior $k^2\gg q^2$ of the $\tau$'s is
\bea
\tau_2&\approx& \frac{\alpha_0}{24\pi}\frac{1}{k^4},\\
\tau_3&\approx& \frac{\alpha_0}{6\pi}\frac{1}{k^2}\,\log\(\frac{q^2}{k^2}\)
-\frac{29}{72}\frac{\alpha_0}{\pi}\frac{1}{k^2},\\
\tau_6&\approx& \frac{(2k\cdot q+q^2)}{2} \frac{\alpha_0}{24\pi}\frac{1}{k^4},\\
\tau_8&\approx& -\frac{\alpha_0}{2\pi}\frac{1}{k^2},
\eea
where $k^2$ is a Minskowskian momentum.
Using such asymptotic expressions, the integration over angles in 
$\Pi_{(2R)}$ and $\Pi_{(2I)}$ can be performed straightforwardly,
and the integrations over momenta $k^2 \geq q^2$
leads to logarithmic corrections.
Let us give an example, and compute, in this way, $\Pi_{(2R)}$.
After evaluating the trace in Eq.~(\ref{pi2j}) (with $j=R$), and
Wick rotating to Euclidean momenta, 
the expression for $\Pi_{(2R)}$ reads
\bea
\Pi_{(2R)}(q^2)&=&\frac{2\alpha_0}{3\pi}
\int\limits_0^{\Lambda^2}{\rm d}k^2\int\frac{{\rm d}\Omega_k}{2\pi^2}
\[\frac{k^2 q^2-(k\cdot q)^2}{q^2(k+q)^2}\]\tau_8(-(k+q)^2,-k^2,-q^2)\nonu\\
&\approx&
\frac{2\alpha_0}{3\pi}\int\limits_{q^2}^{\Lambda^2}
{\rm d}k^2\,   \frac{3}{4}\frac{\alpha_0}{2\pi}\frac{1}{k^2}
\approx
\frac{\alpha_0^2}{\pi^2}\[ 
\frac{1}{4}\log\(\frac{\Lambda^2}{q^2}\)+{\cal O}(1)\].\label{pi2r}
\eea

The result is
that the logarithmic corrections of $\tau_2$ and $\tau_6$ cancel each other,
and that the contributions of $\tau_3$ contain a $\log^2$ term:
\bea
\Pi_{(2I)}(q^2)\approx \frac{\alpha_0^2}{\pi^2}\[
\frac{1}{24} \log^2\(\frac{\Lambda^2}{q^2}\)
+\frac{29}{144}\log\(\frac{\Lambda^2}{q^2}\)
+{\cal O}(1)\].
\label{pi2i}
\eea

An analogous computation can be performed for the self-energy and longitudinal 
vertex corrections.

Due to the Ball--Chiu expression (\ref{ballchiuexpr}) for $\Gamma^\mu_{L(1)}$,
the contributions $\Pi_{(2a)}$ and $\Pi_{(2L)}$ depend on the 
one-loop computation of the self-energy ${\cal Z}_{(1)}$.
Since we have chosen $w$ to be the photon momentum 
in Eq.~(\ref{selfenergyphotmom}),
we find that
\bea
{\cal Z}_{(1)}(p^2)=
-\frac{\alpha_0}{4\pi}\[\log \(\frac{\Lambda^2}{-p^2}\)+\frac{3}{2}\].
\eea
After expanding ${\cal Z}_{(1)}((k+q)^2)$ for $k^2\gg q^2$,  
\bea
{\cal Z}_{(1)}((k+q)^2)&\approx& {\cal Z}_{(1)}(k^2)
+(2k\cdot q+q^2) {\cal Z}_{(1)}^{\prime}(k^2)\nonu\\
&+&\frac{1}{2}(2k\cdot q+q^2)^2{\cal Z}_{(1)}^{\prime\prime}(k^2)
+\frac{1}{6}(2k\cdot q+q^2)^3{\cal Z}_{(1)}^{\prime\prime\prime}(k^2),
\eea
and using that ${\cal Z}_{(1)}((k+q)^2) \approx{\cal Z}_{(1)}(q^2)$ 
for $q^2\gg k^2$, the angular integration can be performed, and the
logarithmic corrections can be computed.
The result is 
\bea
\Pi_{(2a)}(q^2)+\Pi_{(2L)}(q^2)\approx \frac{\alpha_0^2}{\pi^2}\[
-\frac{1}{24} \log^2\(\frac{\Lambda^2}{q^2}\)
-\frac{29}{144}\log\(\frac{\Lambda^2}{q^2}\)+{\cal O}(1)\].\label{pi2a2l}
\eea
Thus, comparing this expression with Eq.~(\ref{pi2i}), we see that  
the ``overlapping divergencies'' ({\em i.e.} the $\log^2$) 
cancel
\bea
\Pi_{(2a)}(q^2)+\Pi_{(2L)}(q^2)+\Pi_{(2I)}(q^2)\approx 
(\alpha_0^2/\pi^2) {\cal O}(1).
\eea
Such a cancellation occurs in a similar manner in any
covariant gauge $\xi$.
Thus, the two-loop contribution contribution to $\Pi$ is described
solely by the part of the transverse vertex containing the 
$T^\mu_8$ tensor, {\em i.e.} $\Pi_{(2R)}$,
and, after adding all the pieces, we find that
\bea
\Pi(q^2)&\approx& \frac{\alpha_0}{2\pi}
\(\frac{2}{3}+\frac{\alpha_0}{2\pi}\)\log\(\frac{\Lambda^2}{q^2}\)
+(\alpha_0/\pi){\cal O}(1).
\eea
As was shown in \cite{kirepe95}, the particular transverse structure 
function $\tau_8$ does not depend on the gauge parameter $\xi$.
This observation agrees with the JBW computation of vacuum polarization effects.
\section{Scalar and pseudoscalar contributions}
The two-loop contributions of scalars and pseudoscalars to the vacuum 
polarization can be computed rather easily by making use of the results
of the previous section.
Besides the photon exchange, we now take into account also
scalar and pseudoscalar exchanges in the one-loop vertex, and self-energy
defined in Eq.~(\ref{1loopcors}), {\em i.e.},
\bea
\Gamma_{(1)}^\mu(k,p)&=&\Lambda^\mu_{(V)}(k,p)+\Lambda^\mu_{(S)}(k,p)
+\Lambda^\mu_{(P)}(k,p),\\
\pslash {\cal Z}_{(1)}(p^2)&=&-\Sigma_{(V)}(p)-\Sigma_{(S)}(p)-\Sigma_{(P)}(p).
\eea
With one-loop vertex corrections
\bea
(-ie_0) \Lambda^\mu_{(V)}(k,p)&=&
\int_\Lambda\frac{{\rm d}^4 w}{(2\pi)^4}\,
(-ie_0)\gamma^\lambda iS(k-w)\nonu\\&&\times (-ie_0)\gamma^\mu iS(p-w)
(-ie_0)\gamma^\sigma iD_{\lambda\sigma}(w),\\
(-ie_0) \Lambda^\mu_{(S)}(k,p)&=&
\int_\Lambda\frac{{\rm d}^4 w}{(2\pi)^4}\,
(-ig_Y){\bf 1} iS(k-w)\nonu\\&&\times (-ie_0)\gamma^\mu iS(p-w)
(-ig_Y){\bf 1} i\DelS(w),\\
(-ie_0) \Lambda^\mu_{(P)}(k,p)&=&
\int_\Lambda\frac{{\rm d}^4 w}{(2\pi)^4}\,
(-ig_Y)i\g5 iS(k-w)\nonu\\&&\times (-ie_0)\gamma^\mu iS(p-w)
(-ig_Y)i\g5 i\DelP(w),\label{pseudooneloop}
\eea
and
the self-energies
\bea
i\Sigma_{(V)}(p)&=&\int_\Lambda\frac{{\rm d}^4 k}{(2\pi)^4}\,
(-ie_0)\gamma^\mu iS(k) (-ie_0)\gamma^\nu iD_{\mu\nu}(k-p),\\
i\Sigma_{(S)}(p)&=&\int_\Lambda\frac{{\rm d}^4 k}{(2\pi)^4}\,
(-ig_Y){\bf 1} iS(k) (-ig_Y){\bf 1} i\DelS(k-p),\\
i\Sigma_{(P)}(p)&=&\int_\Lambda\frac{{\rm d}^4 k}{(2\pi)^4}\,
(-ig_Y)i\g5 iS(k) (-ig_Y) i\g5 i\DelP(k-p),
\eea
Taking free massless scalar and pseudoscalar propagator,
and  the photon propagator in the Feynman gauge,
\bea
D_{\mu\nu}(q)=-\frac{g_{\mu\nu}}{q^2},\qquad 
\DelS(q)=\DelP(q)=\frac{1}{q^2}.
\eea
the one-loop vertices can be expressed as
\bea
\Lambda^\mu_{(V)}(k,p)&=&-ie_0^2\int_\Lambda\frac{{\rm d}^4w}{(2\pi)^4}\,
\frac{\gamma^\lambda(\kslash-\wslash)\gamma^\mu(\pslash-\wslash)\gamma_\lambda}{
(k-w)^2(p-w)^2 w^2},\\
\Lambda^\mu_{(S)}(k,p)&=&ig_Y^2\int_\Lambda\frac{{\rm d}^4w}{(2\pi)^4}\,
\frac{(\kslash-\wslash)\gamma^\mu(\pslash-\wslash)}{
(k-w)^2(p-w)^2 w^2},\\
\Lambda^\mu_{(P)}(k,p)&=&\Lambda^\mu_{(S)}(k,p),
\eea
where the last identity is obtained form Eq.~(\ref{pseudooneloop}) 
by using $\g5 \gamma^\mu=-\gamma^\mu \g5$.
Using Eq.~(\ref{selfenergyphotmom}), the self-energy contributions read 
\bea
\Sigma_{(V)}(p)+\Sigma_{(S)}(p)+\Sigma_{(P)}(p)&=&
2(e_0^2+g_Y^2)\[-i\int_\Lambda\frac{{\rm d}^4 w}{(2\pi)^4}\,
\frac{(\pslash-\wslash)}{(p-w)^2 w^2}\]\nonu\\
&=&
2(e_0^2+g_Y^2)\frac{\pslash}{32\pi^2}\[\log\(\frac{\Lambda^2}{-p^2}\)
+\frac{3}{2}\],
\eea
The sum of one-loop vertex corrections can be rewritten as
\bea
\Lambda^\mu_{(V)}(k,p)+2\Lambda^\mu_{(S)}(k,p)
=2\[e_0^2-g_Y^2\]R^\mu(k,p)+2\[e_0^2+g_Y^2\] S^\mu(k,p), \label{vertRS}
\eea
where
\bea
R^\mu(k,p)&=&-i\int_\Lambda\frac{{\rm d}^4w}{(2\pi)^4}\,
\frac{\gamma^\mu(\pslash-\wslash)(\kslash-\wslash)/2
-(\kslash-\wslash)(\pslash-\wslash)\gamma^\mu/2 }{
(k-w)^2(p-w)^2 w^2},\\
S^\mu(k,p)&=&-i\int_\Lambda\frac{{\rm d}^4w}{(2\pi)^4}\,
\biggr[
\frac{(k-w)\cdot(p-w)\gamma^\mu}{(k-w)^2(p-w)^2 w^2}
-\frac{(k-w)^\mu(\pslash-\wslash)}{(k-w)^2(p-w)^2 w^2}\nonu\\
&&\qquad
-\frac{(p-w)^\mu(\kslash-\wslash)}{(k-w)^2(p-w)^2 w^2}\biggr].
\eea
The vertex part $R^\mu$ is defined so that it is proportional to the transverse 
vertex with the $T^\mu_8$ tensor:
\bea
2 e_0^2 R^\mu(k,p)=\Gamma_{R(1)}^\mu(k,p)=
\tau_8(k^2,p^2,q^2) T^\mu_8(k,p),\quad q^2=(k-p)^2.
\eea
Thus we see that the part of Eq.~(\ref{vertRS}), that is
proportional to $\Gamma_{R(1)}^\mu$, and 
which turned out to be responsible for the two-loop contribution to the
vacuum polarization, has a difference in sign between terms corresponding to 
photon exchanges, and terms corresponding to (pseudo)scalar exchanges. 

Clearly the vertex part $S^\mu$ must be
\bea
2 e_0^2 S^\mu(k,p)=\Gamma_{L(1)}^\mu(k,p)+\Gamma_{I(1)}^\mu(k,p),
\eea
with $\Gamma_{L(1)}^\mu$ and $\Gamma_{I(1)}^\mu$ given by 
Eqs.~(\ref{ballchiuexpr}) and (\ref{GammaI}).
Since we already computed the contributions
of $\Sigma_{(V)}$, $\Gamma_{L(1)}^\mu$, $\Gamma_{R(1)}^\mu$, 
and $\Gamma_{I(1)}^\mu$ to the vacuum polarization in Eqs.~(\ref{pi2a2l}),
(\ref{pi2i}), and (\ref{pi2r}),
we deduce that 
\bea
\Pi(q^2)&\approx& \frac{\alpha_0}{2\pi}
\(\frac{2}{3}+\frac{\alpha_0}{2\pi}-\frac{\lambda_Y}{2\pi}\)
\log\(\frac{\Lambda^2}{q^2}\)
+(\alpha_0/\pi){\cal O}(1),\label{2loopwscalar}
\eea
with
\bea
\lambda_Y=\frac{g_Y^2}{4\pi}.
\eea
So from the point of view of two-loop correction to the vacuum polarization,
the negative sign of the scalar and pseudoscalar interactions
in Eq.~(\ref{2loopwscalar}) suggests that these type of interactions tend
to reduce the screening, {\em i.e.} represents the attractive nature of
forces between virtual fermion pairs in the vacuum polarization.  
\section{Perturbative four-fermionic contributions}
In this section, 
we compute the two-loop contributions of four-fermion 
interactions to the vacuum polarization. 
The four-fermion interactions are treated in a perturbation expansion
in the dimensionless four-fermion coupling $g_0$ defined in Eq.~(\ref{g0def}). 
Then in the lowest order approximation, 
the $\sigma$ and $\pi$ exchanges corresponding to the four-fermion
interactions are described by the bare or free 
propagators
\bea
\DelS(q)=\DelP(q)=-G_0.
\eea
With such $\sigma$ and $\pi$ propagators,
the one-loop vertex and self-energy corrections
are given by
\bea
\Gamma^\mu_{(1)}(k,p)&=&-iG_0\int_\Lambda \frac{{\rm d}^4w}{(2\pi)^4}\,
\frac{(\kslash -\wslash)\gamma^\mu (\pslash -\wslash)}{(k-w)^2(p-w)^2},\\
\pslash {\cal Z}_{(1)}(p^2)
&=&
-i G_0\int_\Lambda\frac{{\rm d}^4 w}{(2\pi)^4}\,
\frac{(\pslash-\wslash)}{(p-w)^2}.
\eea
The one-loop vertex can be decomposed into 
\bea
\Gamma^\mu_{(1)}(k,p)=\frac{G_0}{16\pi^2}\[
\kslash \gamma^\mu \pslash I(k,p)
-(\gamma_\lambda\gamma^\mu\pslash +\kslash \gamma^\mu \gamma_\lambda)
I_\lambda(k,p)+I_{\lambda\sigma}(k,p)\],
\eea
where
\bea
I(k,p)&=&-i\int_\Lambda \frac{{\rm d}^4w}{\pi^2}\,
\frac{1}{(k-w)^2(p-w)^2},\\
I_\lambda(k,p)&=&-i\int_\Lambda \frac{{\rm d}^4w}{\pi^2}\,
\frac{w_\lambda}{(k-w)^2(p-w)^2},\\
I_{\lambda\sigma}(k,p)&=&
-i\int_\Lambda \frac{{\rm d}^4w}{\pi^2}\,
\frac{w_\lambda w_\sigma}{(k-w)^2(p-w)^2}.
\eea
Let us write
\bea
&&I=I_0,\qquad I_\lambda=(k_\lambda+p_\lambda) J_0,\\
&&I_{\lambda\sigma}=g_{\lambda\sigma}K_0
+(k_\lambda k_\sigma+p_\lambda p_\sigma)K_1
+(p_\lambda k_\sigma+k_\lambda p_\sigma)K_2,
\eea
and with the help of Ref.~\cite{akbe65},
we find 
\bea
&&I_0=\log\(\frac{\Lambda^2}{-q^2}\)+1,\qquad
J_0=\frac{1}{2}\[\log\(\frac{\Lambda^2}{-q^2}\)+\frac{1}{2}\],\\
&&K_0=-\frac{\Lambda^2}{4}-\frac{q^2}{2}\log\(\frac{\Lambda^2}{-q^2}\)
+\frac{5 q^2}{72}-\frac{(k^2+p^2)}{12},\\
&&K_1=\frac{1}{3}\log\(\frac{\Lambda^2}{-q^2}\)+\frac{1}{9},\qquad
K_2=\frac{1}{6}\log\(\frac{\Lambda^2}{-q^2}\)-\frac{1}{36},
\eea

Again it is useful to decompose the one-loop vertex into transverse and 
longitudinal parts, just as in Eq.~(\ref{v1loopdecomp}). The result is 
\bea
\Gamma_{R(1)}^\mu(k,p)&=& 
\frac{G_0}{16\pi^2}\[2J_0-I_0\]T^\mu_8(k,p)=c_8 T^\mu_8(k,p),\\
\Gamma_{I(1)}^\mu(k,p)&=& \frac{G_0}{16\pi^2}\[\frac{I_0}{2}-K_1+K_2\]
T^\mu_3(k,p)\nonu\\
&=&\frac{G_0}{16\pi^2}\[
\frac{1}{3}\log\(\frac{\Lambda^2}{-q^2}\)+c_3 \]T^\mu_3(k,p),
\eea
where $c_3$ and $c_8$ are constants whose exact values are irrelevant 
for the computation.
Since the four-fermion interactions are gauge-invariant, we assume
that the Ball--Chiu Ansatz is also valid in this case.
We find for the longitudinal vertex part
\bea
\Gamma_{L(1)}^\mu(k,p)&=&\frac{G_0}{16\pi^2}
\[(k^2+p^2)(2J_0-\frac{I_0}{2}-K_1-K_2)+q^2(K_1-J_0)-2K_0\]\gamma^\mu\nonu\\
&+&\frac{G_0}{16\pi^2}\[\frac{I_0}{2}-2J_0+K_1+K_2\](k+p)^\mu(\kslash+\pslash)
\nonu\\
&\approx& \frac{G_0}{16\pi^2}\biggr\{\[\frac{\Lambda^2}{2}
+{\cal O}\(q^2\log\(\frac{\Lambda^2}{-q^2} \)\)\]\gamma^\mu
+\frac{1}{12}(k+p)^\mu(\kslash+\pslash)\biggr\}.\nonu\\
\eea
The self-energy contribution is
\bea
 {\cal Z}_{(1)}(p^2)
&\approx& 
\frac{G_0}{16\pi^2}\[-\frac{\Lambda^2}{2} 
+{\cal O}\(p^2\log\(\frac{\Lambda^2}{-p^2}\)\)\]\approx -\frac{g_0}{8}, 
\label{fvself}
\eea
where $G_0=4\pi^2 g_0/\Lambda^2$.
With the one-loop vertex and self-energy corrections given above,
we can proceed in the same manner as in the previous section,
and obtain from Eqs.~(\ref{pi2a}), (\ref{pi1}), (\ref{pi2j}), 
and (\ref{fvself}) that:
\bea
\Pi_{(2a)}(q^2)+\Pi_{(2L)}(q^2)\approx
-\frac{g_0}{8} \Pi_{(1)}(q^2)=
\frac{\alpha_0 g_0}{\pi}\[-\frac{1}{24}\log\(\frac{\Lambda^2}{q^2}\)
+{\cal O}(1)\].
\eea
The contributions of the one-loop transverse vertex parts 
to the vacuum polarization can be performed exactly, since
the angular integral can be done, the result is
\bea
\Pi_{(2I)}(q^2)=
\frac{\alpha_0 g_0}{\pi}\[\frac{1}{24}\log\(\frac{\Lambda^2}{q^2}\)
+{\cal O}(1)\],\qquad
\Pi_{(2R)}(q^2)=(\alpha_0 g_0/\pi)  {\cal O}(1).
\eea
Hence
\bea
\Pi(q^2)=\frac{\alpha_0}{3\pi}\log\(\frac{\Lambda^2}{q^2}\)+
(\alpha_0 g_0/\pi)  {\cal O}(1),\label{ffpertvacpol}
\eea
and the four-fermion interaction treated in a perturbative way 
do not give contributions to the $\beta$ function.
Thus again in this case too, the contributions
of the longitudinal vertex cancel the contributions of the transverse vertex 
part which are proportional to the $T^\mu_3$ tensor.
In this case, however, the transverse part connected with $T^\mu_8$ does not give
a logarithmic contribution to screening.

The four-fermion interactions treated in a perturbative way are irrelevant,
and, 
according to the RG of Wilson, these interactions do not contribute to the 
low-energy effective dynamics such as charge screening.

%
%
\chapter{A derivation of the JWB equation}\label{dervJWB}
In this appendix we review the derivation of equation (\ref{PHIK2}) 
for the $f$ function given by Eq.~(\ref{fdef})
(and thus the $\beta$ function) of QED introduced by Johnson, 
Willey, and Baker \cite{jowiba67}.
Since their result is formulated in terms of the BS fermion-fermion scattering 
kernel $K^{(2)}$
their method is also applicable to the GNJL model.

In order to derive the JWB result, the following is assumed:
\begin{itemize}
\item{The fermion wave function equals one, 
${\cal Z}=1/A=1$, in the Landau gauge.
In principle, this assumption is redundant
since the JWB result is valid in any gauge, as long as the WTI is satisfied.}
\item{The quenched approximation is assumed to be consistent. 
Internal photon propagators are quenched. Only a single fermion loop, 
thus a single power of $\log \Lambda$ contributes to the 
vacuum polarization. Such an approximation is assumed to be valid 
close to an ultraviolet fixed point.}
\item{Translational invariance of 
naively logarithmically divergent and finite momentum space integrals is assumed.
This is an important point, since such translational invariance is 
lost in naively linear divergent integrals, such a the integral in the SDE
for the fermion wave function ${\cal Z}$.}
\item{Also it is assumed that we are in the scaling region of the theory, 
where the only relevant dimensionless variable is $q^2/\Lambda^2$. 
Thus the momenta are assumed to be larger than the particle masses or bound states masses 
in the model.}
\end{itemize}
Hence,
\bea
{\cal Z}(k^2)=1/A(k^2)=1,\qquad S(k)=\frac{\kslash}{k^2},\qquad
\Gamma^{\mu}(k,k)&=&\gamma^\mu.\label{assum1}
\eea
The vacuum polarization tensor is
\bea
\Pi^{\mu\nu}(q)=\frac{iN\alpha_0}{4\pi^3}
\int
{\rm d}^4 k\,\Tr\[ S(k+q)\Gamma^\mu(k+q,k) S(k)\gamma^\nu\].\label{vacpoltens}
\eea
Since vacuum polarization tensor is transverse, 
$\Pi_{\mu\nu}(q)=(-g_{\mu\nu}q^2+q_\mu q_\nu)\Pi(q^2)$,
and the only relevant momentum variable is $q^2/\Lambda^2$, 
the equation for the vacuum polarization can be written as
\bea
\Pi(\mu^2)=-\lim_{q^2\rightarrow \mu^2}
\frac{1}{6}\frac{q^\mu q^\nu}{q^2}
\frac{\partial^2}{\partial q_\alpha q^\alpha}\Pi_{\mu\nu}(q)
+{\cal O}(1)+{\cal O}\( (\mu/\Lambda)^\sigma \),
\eea
where $\mu$ is some infrared reference scale, $\mu^2\ll \Lambda^2$,
{\em e.g.} $\mu\sim m_\sigma$, and $\sigma$ is some positive power.
After inserting Eq.~(\ref{vacpoltens}), and setting $\mu^2=0$ in the integrand, 
and using $\mu^2$ as the infrared cutoff in the momentum integral, 
we obtain 
\bea
\Pi(\mu^2)&=&-\lim_{q^2\rightarrow \mu^2}
\frac{1}{6}\frac{q_\mu q_\nu}{q^2}
\frac{iN\alpha_0}{4\pi^3}
\int_\mu
{\rm d}^4 k\,\Tr\big[S_\alpha^\alpha(k)\Gamma^\mu(k) S(k)\gamma^\nu\nonu\\
&&\qquad+2S_\alpha(k)\Gamma^{\mu,\alpha}(k) S(k)\gamma^\nu
+S(k)\Gamma^{\mu,\alpha}_\alpha(k) S(k)\gamma^\nu\big]\nonu\\
&+&{\cal O}(\mu^2/\Lambda^2),
\label{intder1}
\eea
where the derivatives are defined as follows:
\bea
S_\alpha(k)&\equiv&\frac{\partial}{\partial k^\alpha}S(k)=
-\frac{\kslash\gamma_\alpha\kslash}{k^4},\quad
S_\alpha^\alpha(k)\equiv\frac{\partial^2}{\partial k_\alpha\partial k^\alpha}S(k)=
-\frac{4\kslash}{k^4},\\
\Gamma_{\mu,\alpha}(k)&\equiv&
\frac{\partial}{\partial q^\alpha}\Gamma_\mu(k+q,k)\biggr|_{q=0},\quad
\Gamma_{\mu,\alpha}^\alpha(k)\equiv
\frac{\partial^2}{\partial q_\alpha \partial q^\alpha}
\Gamma_\mu(k+q,k)\biggr|_{q=0},
\eea
and $\Gamma^\mu(k)\equiv \Gamma^\mu(k,k)$.
Since the integral Eq.~(\ref{intder1}) can only be proportional to 
$g^{\mu\nu}$, it reduces to
\bea
\Pi(\mu^2)&=&-\lim_{q^2\rightarrow \mu^2}
\frac{iN\alpha_0}{96\pi^3}
\int_\mu
{\rm d}^4 k\,\Tr\big[ S_\alpha^\alpha(k)\Gamma^\mu(k) S(k)\gamma_\mu\nonu\\
&&\qquad+2S_\alpha(k)\Gamma^{\mu,\alpha}(k) S(k)\gamma_\mu
+S(k)\Gamma^{\mu,\alpha}_\alpha(k) S(k)\gamma_\mu
\big]\nonu\\
&+&{\cal O}(\mu^2/\Lambda^2). \label{vpbjwderv}
\eea
The SDE for the vertex $\Gamma^\mu$ reads, in terms of  $K^{(2)}$
\bea
\Gamma^{\mu}_{ab}(k+q,k)&=&
\gamma^\mu_{ab}+\sum_{{\rm flavors}}
(ie_0^2)
\int_{\Lambda}\frac{{\rm d}^4 p}{(2\pi)^4}\,
\[S(p+q)\Gamma^{\mu}(p+q,p)S(p)\]_{dc}\nonu\\
&\times&
K^{(2)}_{cd,ab}(p,p+q,k+q).\label{vdsek2}
\eea

The first derivative of the vertex is anti-symmetric in $\alpha$ 
and $\mu$, because of the assumption Eq.~(\ref{assum1}).
Furthermore, $CP$-invariance implies that the only nonzero contribution
to the first derivative of $\Gamma^\mu$ must be proportional
to the tensor $(\gamma_\mu\kslash \gamma_\alpha-\gamma_\alpha\kslash\gamma_\mu)$.
Thus we write
\bea
\Gamma_{[\mu,\alpha]}(k)&\equiv&
\Gamma_{\mu,\alpha}(k)-\Gamma_{\alpha,\mu}(k)
=\frac{(\gamma_\mu\kslash \gamma_\alpha-\gamma_\alpha\kslash\gamma_\mu)}{k^2} 
\Gamma^\prime,\\
\Gamma_{(\mu,\alpha)}(k)&\equiv&
\Gamma_{\mu,\alpha}(k)+\Gamma_{\alpha,\mu}(k)=S^{-1}_{\mu\alpha}(k).
\eea
The dimensionless scalar function $\Gamma^\prime$ 
is related to the transverse structure function
$\tau_8(k^2,k^2,0)$, see ref.~\cite{kirepe95},
\bea
\Gamma^\prime=-k^2 \tau_8(k^2,k^2,0).
\eea
Since $S^{-1}_{\nu\mu}(k)=0$ due to the Ward-Takahashi identity 
for the vertex, and Eq.~(\ref{assum1}).
We find that
\bea
\Gamma_{\mu,\alpha}(k)=\frac{1}{2}\Gamma_{[\mu,\alpha]}(k)
+\frac{1}{2}\Gamma_{(\mu,\alpha)}(k)=
\frac{1}{2}\Gamma_{[\mu,\alpha]}(k)
=\frac{(\gamma_\mu\kslash \gamma_\alpha-\gamma_\alpha\kslash\gamma_\mu)}{2k^2} 
\Gamma^\prime.
\eea
Differentiating now the SDE (\ref{vdsek2}) 
with respect to $q$, and setting $q=0$, we obtain  
\bea
\Gamma_{\mu,\alpha}(k)
&=&
\sum_{{\rm flavors}}
(ie_0^2)
\int_{\Lambda}\frac{{\rm d}^4 p}{(2\pi)^4}
\biggr[S(p)\Gamma_{\mu,\alpha}S(p)K^{(2)}(p,k)\nonu\\
&+&S_\alpha(p)\Gamma_{\mu}(p)S(p)K^{(2)}(p,k)
+S(p)\Gamma_{\mu}(p)S(p)K^{(2)}_\alpha(p,k)\biggr].\label{firderv2}
\eea
This gives
\bea
\Gamma_{\mu,\alpha}(k)
&=&\[\phi_1(\alpha_0)+\phi_2(\alpha_0)
+\Gamma^\prime \phi_1(\alpha_0)\]]
\frac{(\gamma_\mu\kslash \gamma_\alpha-\gamma_\alpha\kslash\gamma_\mu)}{2k^2}.
\eea
Thus
\bea
\Gamma^\prime=\phi_1+\phi_2+\Gamma^\prime \phi_1.\label{gammaprime}
\eea
The following identities have been used
\bea
\Tr\[(\gamma_\mu\kslash \gamma_\alpha-\gamma_\alpha\kslash\gamma_\mu)
(\gamma^\mu\kslash \gamma^\alpha-\gamma^\alpha\kslash\gamma^\mu)\]&=&-96 k^2,\\
S_\alpha(p)\Gamma_\mu(p)S(p)=-S(p)\Gamma_\mu(p)S_\alpha(p)&=&\frac{(
\gamma_\mu\pslash \gamma_\alpha-\gamma_\alpha\pslash\gamma_\mu)}{2p^4},\\
S(p)\Gamma_{\mu,\alpha}(p)S(p)&=&\Gamma^\prime
\frac{(\gamma_\mu\pslash \gamma_\alpha-\gamma_\alpha\pslash\gamma_\mu)}{2p^4}.
\eea
Furthermore, the functions $\phi_j$ are defined as follows:
\bea
\!\!\!\!\!\phi_1(\alpha_0)&\equiv& \sum_{{\rm flavors}}-\frac{ie_0^2}{48}
\int\frac{{\rm d}^4p}{(2\pi)^4}\,\nonu\\
\!\!\!\!\!&\times&
\Tr\[
\frac{(\gamma^\mu\pslash \gamma^\alpha-\gamma^\alpha\pslash\gamma^\mu)}{2p^4}
K^{(2)}(p,k)(\gamma_\mu\kslash \gamma_\alpha-\gamma_\alpha\kslash\gamma_\mu)
\],\label{phi1i}\\
\!\!\!\!\!\phi_2(\alpha_0)&\equiv& \sum_{{\rm flavors}}-\frac{ie_0^2}{48}
\int\frac{{\rm d}^4p}{(2\pi)^4}\,
\Tr\[
\frac{\pslash\gamma^\mu \pslash}{p^4}
K^{(2)\alpha}(p,k)(\gamma_\mu\kslash \gamma_\alpha-\gamma_\alpha\kslash\gamma_\mu)\],\label{phi2i}\\
\!\!\!\!\!\phi_3(\alpha_0)&\equiv& 
\sum_{{\rm flavors}}
\frac{ie_0^2}{48}
\int\frac{{\rm d}^4p}{(2\pi)^4}\,
\Tr\[
\frac{\pslash\gamma^\mu \pslash}{p^4}
K_\alpha^{(2)\alpha}(p,k) \kslash\gamma_\mu\kslash\],\label{phi3i}
\eea
where the trace over spinor indices is defined as
\bea
\Tr\[L(p)K(p,k)R(k)\]\equiv L_{dc}K_{cd,ab}(p,k)R_{ba}(k),
\eea
with $L$ and $R$ some projectors.
The derivatives of the BS kernel $K^{(2)}$ are
\bea
K^{(2)}_{cd,ab}(p,k)&\equiv& K^{(2)}_{cd,ab}(p,p+q,k+q)\biggr|_{q=0},\\
{K^{(2)}_\alpha}_{cd,ab}(p,k)
&\equiv& \frac{\partial}{\partial q^\alpha} 
K^{(2)}_{cd,ab}(p,p+q,k+q)\biggr|_{q=0},\label{k1derdef}
\\
{K_\alpha^{(2)\alpha}}_{cd,ab}(p,k)
&\equiv& 
\frac{\partial^2}{\partial q^\alpha\partial q_\alpha}
K^{(2)}_{cd,ab}(p,p+q,k+q)\biggr|_{q=0}.
\eea
Assuming that $K^{(2)}$ is translationally invariant,
we can derive the following properties:
\bea
\int{\rm d}^4p\,\Tr\[L(p)K(p,k)R(k)\]&=&\int{\rm d}^4p\,\Tr\[R(k)K(k,p)L(p)\],
\label{inverseK}\\
\int{\rm d}^4p\,\Tr\[L(p)K_\alpha (p,k)R(k)\]&=&-
\int{\rm d}^4p\,\Tr\[R(k)K_\alpha(k,p)L(p)\],
\label{inverseKa}\\
\int{\rm d}^4p\,\Tr\[L(p)K_\alpha^\alpha(p,k)R(k)\]&=&\int{\rm d}^4p\,
\Tr\[R(k)K_\alpha^\alpha(k,p)L(p)\].\label{inverseKaa}
\eea

By making use of the full SDE for the vertex, the second derivate 
$\Gamma^{\mu,\alpha}_\alpha(k)$ can be eliminated in Eq.~(\ref{intder1}). 
The second derivative of the vertex is
\bea
\Gamma^{\mu,\alpha}_\alpha(k)
&=&\sum_{{\rm flavors}}(ie_0^2)
\int_{\Lambda}\frac{{\rm d}^4 p}{(2\pi)^4}
\biggr[S(p)\Gamma^{\mu,\alpha}_\alpha(p)S(p)K^{(2)}(p,k)\nonu\\
&+&2S_\alpha(p)\Gamma^{\mu,\alpha}(p)S(p)K^{(2)}(p,k)
+S_\alpha^\alpha(p)\Gamma^{\mu}(p)S(p)K^{(2)}(p,k)\nonu\\
&+&2S(p)\Gamma^{\mu,\alpha}(p)S(p)K^{(2)}_\alpha(p,k)
+2S^\alpha(p)\Gamma^{\mu}(p)S(p)K^{(2)}_\alpha(p,k)\nonu\\
&+&S(p)\Gamma^{\mu}(p)S(p)K^{(2)\alpha}_\alpha(p,k)\biggr].\label{secderv2}
\eea
Since $\Gamma^\mu(k)=\gamma^\mu$, and using Eq.~(\ref{secderv2}) 
for $\Gamma^{\mu,\alpha}_\alpha(k)$, the second derivative of 
$\Gamma^{\mu,\alpha}_\alpha$ in Eq.~(\ref{vpbjwderv}) can be eliminated.
The result is
\bea
\Pi(\mu^2)&=&\frac{N\alpha_0}{2\pi}\big[
\sum_{n=1}^{5} I_n(\alpha_0,\mu^2/\Lambda^2)+{\cal O}(\mu^2/\Lambda^2)\big],
\eea
where 
\bea
I_1&\equiv& -\frac{i}{48}\int_\mu\frac{{\rm d}^4k}{\pi^2}
\Tr\[S_\alpha^\alpha(k)\gamma^\mu S(k)\gamma_\mu\],\\
I_2&\equiv& -\frac{i}{24}\int_\mu\frac{{\rm d}^4k}{\pi^2}
\Tr\[S_\alpha(k)\Gamma^{\mu,\alpha}(k) S(k)\gamma_\mu\],\\
I_3&\equiv& \frac{e_0^2}{24}\int_\mu\frac{{\rm d}^4k}{\pi^2}
\int_\mu\frac{{\rm d}^4p}{(2\pi)^4}\,\nonu\\
&\times&\Tr\[
S(p)\Gamma_{\mu,\alpha}(p)S(p)
K^{(2)\alpha}(p,k) S(k)\gamma^\mu S(k) \],\\
I_4&\equiv& \frac{e_0^2}{24}\int_\mu\frac{{\rm d}^4k}{\pi^2}
\int_\mu\frac{{\rm d}^4p}{(2\pi)^4}\,\Tr\[S_\alpha(p)\gamma_\mu S(p)
K^{(2)\alpha}(p,k)S(k)\gamma^\mu S(k) \],\\
I_5&\equiv& \frac{e_0^2}{48}\int_\mu\frac{{\rm d}^4k}{\pi^2}
\int_\mu\frac{{\rm d}^4p}{(2\pi)^4}\,\Tr\[
S(p)\gamma_\mu S(p)
K^{(2)\alpha}_\alpha(p,k)
S(k)\gamma^\mu S(k)\].
\eea
Using Eqs.~(\ref{inverseK})--(\ref{inverseKaa}),
we find 
\bea
I_1&=&\frac{2}{3}\int\limits_{\mu^2}^{\Lambda^2}\frac{{\rm d}k^2}{k^2}
+{\cal O}(1),\quad 
I_2=\Gamma^\prime\int\limits_{\mu^2}^{\Lambda^2}\frac{{\rm d}k^2}{k^2}
+{\cal O}(1),\nonu\\
I_3&=&\Gamma^\prime \phi_2
\int\limits_{\mu^2}^{\Lambda^2}\frac{{\rm d}k^2}{k^2}+{\cal O}(1),
\quad 
I_4=\phi_2\int\limits_{\mu^2}^{\Lambda^2}\frac{{\rm d}k^2}{k^2}
+{\cal O}(1),\\
I_5&=&\phi_3\int\limits_{\mu^2}^{\Lambda^2}\frac{{\rm d}k^2}{k^2}
+{\cal O}(1).\nonu
\eea
Thus 
\bea
\Pi(\mu^2)&=&\frac{N\alpha_0}{2\pi}\[\frac{2}{3}+\Phi(\alpha_0)\]
\log\frac{\Lambda^2}{\mu^2}+\mbox{finite},
\eea
where, inserting Eq.~(\ref{gammaprime}),
\bea
\Phi&=&
\frac{\phi_1+\phi_2(2+\phi_2)}{1-\phi_1}+\phi_3.
\label{jwbres}
\eea
In fact, Eq.~(\ref{jwbres}) is the main result of \cite{jowiba67}.
The entire derivation did not yet specify the BS kernel $K^{(2)}$, and
the JWB equation is applicable to the GNJL model as well.
The crucial approximation is that internal gauge bosons 
are taken to be quenched, {\em i.e.},
$D_{\mu\nu}(p)=(-g_{\mu\nu}+p_\mu p_\nu/p^2)/p^2$, which is considered to be 
reasonable near a fixed point.

\newif\ifdraft
\newif\ifaps
\def\optbibfield#1{\ifdraft{#1}\else\ifaps\else{#1}\fi\fi}
\bibliographystyle{myunsrt}
\bibliography{bib/refs,bib/gusmir,bib/qft_gen,bib/books,bib/chap4,bib/curpen,bib/adlbar,bib/lattice,bib/konyam,bib/apphol,bib/atkmar}

\appendix
\hyphenation{quan-tum-mecha-ni-sche}
\hyphenation{ele-men-taire}
\hyphenation{be-we-gings-ver-ge-lij-king-en}
\hyphenation{ver-ge-lij-king-en}
\hyphenation{mo-del-len}
\hyphenation{be-na-dering-en}
\hyphenation{be-na-dering}
\hyphenation{niet-sto-rings-ach-tige}
\hyphenation{niet-sto-rings-ach-tig}
\hyphenation{be-schrijft}
\hyphenation{dy-nami-sche}
\hyphenation{deel-tjes}
\hyphenation{re-pre-senteert}
\hyphenation{toe-stand-en}
\hyphenation{af-han-ke-lijk}
\nonumchapter{Samenvatting}
In dit proefschrift wordt het {\em Gauged Nambu--Jona-Lasinio 
{\rm (GNJL)} model} behandeld.
Het GNJL model is een quantumvelden-theoretisch model voor
dynamische chirale symmetriebreking.
De  begrippen {\em chirale symmetrie} en {\em chirale symmetriebreking}
worden gebruikt in de hoge energie fysica.
Chirale symmetriebreking 
verschaft een mechanisme om massa's
van elementaire deeltjes (zoals bijv. het elektron, quarks, neutrino's (?))
te genereren.

In de jaren dertig en veertig bleek dat voor het modelleren
van het gedrag van een elementair deeltje, zoals het elektron,
zowel een quantummechanische beschrijving als een relativistische 
beschrijving nodig was.
Dit heeft aanleiding gegeven tot de quantumvelden-theorie.
In de jaren zeventig zijn de quantum-velden-theoretische 
modellen van de electromagnetische, de zwakke 
en de sterke wisselwerking verenigd in het Standaard Model. 
Het Standaard Model beschrijft de interacties tussen de elementaire deeltjes
(elementair voor zover we dat experimenteel kunnen testen) 
in termen van zogenaamde ijktheorie\"en.
Deze ijktheorie\"en hebben een bijzondere eigenschap
die {\em renormaliseerbaarheid} wordt genoemd en die hen toepasbaar maakt
over een heel groot energiegebied.
 
E\'en van de vraagstukken in de natuurkunde is de oorsprong 
van de massa's van de elementaire deeltjes en hun verscheidenheid. 
Deze massa's volgen niet uit het Standaard Model. 
Hoewel hierover veel idee\"en zijn, bestaat er nog geen succesvol
model waarmee de massa's berekend kunnen worden.
Het enige dat we nu kunnen is   
de massa's van de elementaire deeltjes meten met behulp van deeltjesversnellers
en de gemeten waarden gebruiken als uitgangspunt in specifieke 
modellen. 
Om massa's aan elementaire deeltjes toe te kennen gebruiken we in het 
Standaard Model het zogenaamde Higgs mechanisme.
Dit Higgs mechanisme voorspelt tevens het bestaan van een 
Higgs boson, een deeltje dat nog niet is waargenomen.

De wiskundige formulering van quantumvelden-theorie\"en 
is erg ingewikkeld, en niet compleet.
De interessantste en fysisch meest relevante modellen
leiden tot een oneindig groot systeem 
van gekoppelde bewegingsvergelijkingen.
Deze worden ook wel Schwinger--Dyson vergelijkingen genoemd. 
Dergelijke vergelijkingen zijn praktisch onoplosbaar en dus zijn we gedwongen
benaderingen te ontwikkelen waarmee we iets concreets kunnen uitrekenen. 

Het Standaard Model is erg succesvol.
Dat hebben we te danken aan het feit dat de koppelingsconstanten 
(die de sterkte van de fundamentele natuurkrachten weergeven)
in de meeste gevallen vrij klein zijn. Daardoor kunnen we gebruik maken van 
een expansietechniek die storingstheorie heet.

Een direct gevolg van renormalisatie is dat koppelingsconstanten 
energie-afhan-kelijk zijn.
Zo hangt de sterkte van een bepaalde interactie tussen deeltjes
af van de afstand tussen die deeltjes.
Meestal is het zo dat een interactie vrij zwak is voor lange afstanden 
en sterker wordt naarmate de afstand tussen de deeltjes kleiner wordt
(het omgekeerde is ook mogelijk).
Een aantal fysische verschijnselen is het gevolg 
van een sterke wisselwerking tussen deeltjes, en kan dus
niet, i.v.m. grote koppelingsconstanten, 
met storingstheorie beschreven worden.
Voorbeelden hiervan zijn massageneratie en de formatie van gebonden toestanden
zoals hadronen.
Deze hadronen zijn gebonden toestanden van twee of drie quarks.
De {\em sterke wisselwerking} tussen quarks wordt beschreven door
QCD (quantumchromodynamica).
Hoewel we denken dat QCD de juiste theorie is voor de dynamica van quarks
is het erg moeilijk om uit QCD de hadronen te herleiden.

Zulke niet-storingsachtige fenomenen gaan vaak samen met het optreden van 
een faseovergang in het model als functie van de koppelingsconstante.
Als de koppelingsconstante boven een bepaalde critische waarde komt, 
ontstaan gebonden toestanden van fermionen 
en wordt een massa gegenereerd ({\em dit is dynamische chirale symmetriebreking}).
Dergelijke chirale faseovergangen hebben veel overeenkomsten met faseovergangen 
in modellen voor bijvoorbeeld ferromagnetisme in de statistische mechanica.

Wat het GNJL model interessant maakt is het feit
dat een combinatie van voldoend sterke en attractieve 
{\em vier-fermion}
interacties
en een zogenaamde {\em Abelse} ijkinteractie\footnote{Dit in tegenstelling tot 
QCD dat gebaseerd is op een niet-Abelse ijktheorie.}
zo'n chirale faseovergang veroorzaakt. De faseovergang hangt samen met
de formatie van gebonden toestanden en massageneratie.
Het blijkt dat de gebonden toestanden in het GNJL model veel compacter zijn en
sterker gebonden dan in QCD.
Dit suggereert dat de gebonden toestanden relevant zijn
voor de beschrijving van de dynamica van de fermionen over een heel groot 
energiegebied. 
Technisch gesproken betekent dit dat de vier-fermion interactie 
renormaliseerbaar is; de dracht van de vier-fermion interacties
is veel langer dan op grond van storingstheorie berekend kan worden.

Het GNJL model zegt daarom veel over de mogelijke rol die 
vier-fermion interacties kunnen spelen voor het 
modelleren van massa's van elementaire deeltjes.

Hieronder volgt een korte samenvatting per hoofdstuk.

\paragraph{Hoofdstuk 1.}
Hierin introduceren we het pad-integraal formalisme
en de daaraan gerelateerde Schwinger--Dyson vergelijking.
De renormalisatiegroep-methode wordt uitvoerig behandeld
en de connectie tussen faseovergangen in statische mechanische modellen en de
renormalisatie van quantumvelden theorie\"en wordt gemaakt.
Aan het eind introduceren we
het Goldstone mechanisme en de Lagrangeaan van het GNJL model.

\paragraph{Hoofdstuk 2.}
Dit hoofdstuk is een noodzakelijk kwaad en nogal technisch van aard.
De Schwinger--Dyson vergelijkingen voor een aantal specifieke
Green functies (correlaties functies) worden afgeleid.
Ook worden de zogeheten Ward--Takahashi identiteiten behandeld.
Dit zijn vergelijkingen die de symmetrie\"en van het model
representeren en uitermate belangrijk zijn in het bepalen 
van een geloofwaardige benaderingsmethode.
De Ward--Takahashi identiteiten 
zijn het centrale uitgangspunt voor het formuleren
van niet-storingsachtige benaderingen.

\paragraph{Hoofdstuk 3.}
In hoofdstuk 3 worden een aantal specifieke 
niet-storingsachtige benaderingen ge{\"\i}ntroduceerd en onder de loep genomen: 
de ``ladder'' benadering, de ``quenched'' benadering en 
de ``mean-field'' (Hartree--Fock) benadering voor
de Schwinger--Dyson vergelijking (de {\em gap-vergelijking})
voor de massa van het fermion in het GNJL model.
De basiseigenschappen van de chirale faseovergang die de gap-vergelijking
beschrijft worden behandeld.
De resultaten van deze specifieke benaderingen worden vergeleken
met andere niet-storingsachtige technieken zoals numerieke roostersimulaties
en niet-storingsachtige renormalisatiegroep technieken.
Hoewel dit hoofdstuk geen nieuwe resultaten beschrijft,
geeft het een overzicht van de literatuur en vormt het een raamwerk 
voor het begrijpen van hoofdstuk 4 en 5.

\paragraph{Hoofdstuk 4.}
Dit hoofdstuk is gewijd aan het berekenen van de Yukawa vertex
en de scalaire instabiele gebonden toestanden (resonanties)
in de quenched ladder benadering.

We ontwikkelen een methode waarmee we analytische uitdrukkingen kunnen vinden
voor de Yukawa-vertex en de scalaire propagator die deze resonantie 
representeert.
De scalaire propagator is te beschouwen als een soort Higgs deeltje.
De Yukawa vertex beschrijft de interactie tussen
fermionen en gebonden toestanden en is een functie van de impulsen
van de ingaande fermionen en de uitgaande gebonden toestand.

De resultaten worden uitvoerig vergeleken met resultaten van andere auteurs.
Het nieuwe aan onze berekeningen is dat het impuls gedrag
van de Yukawa-vertex wordt uitgebreid en dat het fasediagram
uitvoeriger wordt behandeld dan in eerdere studies.
Een van de conclusies die uit hoofdstuk~4 getrokken kan worden 
is het feit dat de Hartree--Fock benadering voor vier-fermion interacties
in het GNJL model in het algemeen inconsistent is.

\paragraph{Hoofdstuk 5.}

In dit hoofdstuk wordt de {\em continuum limiet} behandeld.
Dit komt er kortweg op neer dat we onderzoeken of het GNJL model
een niet-triviale renormaliseerbare theorie is.
De vier-fermion interacties worden nu benaderd in een $1/N$ 
expansie\footnote{Een andere niet-storingsachtige techniek waar $N$ een getal is dat het aantal 
verschillende typen fermionen aangeeft.} 
in plaats van de Hartree--Fock benadering.

Gekeken wordt naar de vacuumpolarisatie.
Dat is een functie die de effectieve koppelingscontante 
oftewel de effectieve lading van de fermionen beschrijft.
Uiteindelijk bepaalt het gedrag van de vacuumpolarisatie
of het GNJL model een {\em niet-triviaal} renormaliseerbaar model is.
De vacuumpolarisatie is een functie van zowel de Yukawa-vertex 
als de scalaire propagator waarvoor we in hoofdstuk 4 expliciete
uitdrukkingen hebben afgeleid.

\paragraph{Conclusie.}
Het doel van dit proefschrift 
was om het sterke koppelingsgedrag en de chirale faseovergang
in het GNJL model beter te begrijpen en de conclusie van dit 
proefschrift kan als volgt worden samengevat.
We hebben aangetoond binnen het kader van een aantal 
niet-storingsachtige benaderingen (nl. de ladder benadering en de $1/N$ expansie)
dat het GNJL model een niet-triviale renormaliseerbare theorie 
is, in de buurt van de chirale faseovergang, 
mits
het aantal typen fermionen (aangeduid met $N$) groter 
is dan een critische waarde ($N_c$).

Deze critische waarde is nogal groot ($N_c\approx 50$), en is 
veel groter dan het aantal typen fermionen dat tot nu toe bekend is.
Daarom is er wat dat betreft nog geen directe toepassing voor het model.
Toch is het resultaat belangrijk omdat de enige tot nu bekende niet-triviale
renormaliseerbare theorie\"en gebaseerd zijn op {\em niet-Abelse} ijktheorie\"en
terwijl het door ons beschouwde GNJL model is gebaseerd 
op een {\em Abelse} ijktheorie.


\nonumchapter{Acknowledgements}

This thesis would never have seen the light of day if it wasn't for
the help of Valery Gusynin who has been 
my tower of strength.
For this I cannot thank him enough.
The countless e-mail exchanges containing ideas, computations, and discussions
have formed the foundation of this work. 
With great pleasure I recall his visit to our institute and my visit to Kiev.
The visit to Kiev was very important and it has given me the 
necessary courage and perspective.
I thank him, his family, and his colleagues at the Bogolyubov institute
for their hospitality.

Marinus Winnink has given me hope in anxious days. 
I thank him for his intensive and humorous guidance and
for acting as supervisor.
The, on beforehand, not so obvious combination
of high energy physics and mathematical physics turned out to 
be very interesting and fruitful for both of us.

Also I would like thank Aernout van Enter for the interesting discussions
on the theory of phase transitions and for his useful suggestions concerning
the literature.

I thank Tony Dorlas, Volodya Miransky, and Mike Pennington for
their willingness to form the reading committee.

Special thanks to Ayse Kizilers\"u and Mike Pennington
for the successful e-mail collaboration in 1994--1995.

Anthony Hams I thank for the numerous discussions we had and I wish
him good luck and success with his PhD research. 

Thanks to Anne and Titus for helping with the ceremonial fuss.

Thanks also to my colleagues, secretaries, and ``voluntary'' system managers
who, all in their own way, have contributed to the making of this thesis.

Last but not least I thank Karin, my family, and friends for their
unconditional patience, support, and encouragements.

\end{document}